\documentclass[final,fleqn,sort&compress,3p,times]{elsarticle} 
\usepackage{geometry}
\usepackage{mathrsfs,amsmath,amssymb}
\usepackage{natbib}
\usepackage{txfonts}

\usepackage{bm}
\usepackage{cases}
\usepackage{longtable}
\usepackage{multirow}
\usepackage[bookmarks=true,colorlinks,%
            citecolor=blue,linkcolor=blue, %
            breaklinks=true]{hyperref}

\usepackage{ulem}

\newcommand{\delete}{\bgroup\markoverwith{\textcolor{red}{\rule[0.5ex]{2pt}{1pt}}}\ULon}

\journal{Progress in Particle and Nuclear Physics}

\begin{document}

\begin{frontmatter}



\title{Towards an \textit{ab initio} covariant density functional theory for nuclear structure}


\author[PKU,Milan,INFN]{Shihang Shen}
\author[RIKEN,UT]{Haozhao Liang}
\author[LZU,LZ]{Wen Hui Long}
\author[PKU,YITP,SA]{Jie Meng\corref{cor1}}
\cortext[cor1]{Corresponding author}
\ead{mengj@pku.edu.cn}
\author[PKU,TUM]{Peter Ring}

\address[PKU]{State Key Laboratory of Nuclear Physics and Technology,
              School of Physics,Peking University, Beijing 100871, China}
\address[Milan]{Dipartimento di Fisica, Universit\`{a} degli Studi di Milano, Italy}
\address[INFN]{INFN, Sezione di Milano, via Celoria 16, I-20133 Milano, Italy}
\address[RIKEN]{RIKEN Nishina Center, Wako 351-0198, Japan}
\address[UT]{Department of Physics, Graduate School of Science, The University of Tokyo, Tokyo 113-0033, Japan}
\address[LZU]{School of Nuclear Science and Technology, Lanzhou University, Lanzhou 730000, China}
\address[LZ]{Key Laboratory of Special Function Materials and Structure Design, Ministry of Education,
    Lanzhou 730000, China}
\address[YITP]{Yukawa Institute for Theoretical Physics, Kyoto University, Kyoto 606-8502, Japan}
\address[SA]{Department of Physics, University of Stellenbosch, Stellenbosch, South Africa}
\address[TUM]{Physik-Department der Technischen Universit\"{a}t M\"{u}nchen, D-85748 Garching, Germany}

\begin{abstract}

Nuclear structure models built from phenomenological mean fields, the effective nucleon-nucleon interactions (or Lagrangians), and the realistic bare nucleon-nucleon interactions are reviewed. The success of covariant density functional theory (CDFT) to describe nuclear properties and its influence on Brueckner theory within the relativistic framework are focused upon. The challenges and ambiguities of predictions for unstable nuclei without data or for high-density nuclear matter, arising from relativistic density functionals, are discussed. The basic ideas in building an \textit{ab initio} relativistic density functional for nuclear structure from \textit{ab initio} calculations with realistic nucleon-nucleon interactions for both nuclear matter and finite nuclei are presented. The current status of fully self-consistent relativistic Brueckner-Hartree-Fock (RBHF) calculations for finite nuclei or neutron drops (ideal systems composed of a finite number of neutrons and confined within an external field) is reviewed. The guidance and perspectives towards an \textit{ab initio} covariant density functional theory for nuclear structure derived from the RBHF results are provided.
\end{abstract}


\end{frontmatter}


\tableofcontents

\section{Introduction}\label{Sect:1}



\subsection{Brief introduction on nuclear theory}\label{Sect:1.1}


The discoveries of radioactivity by Becquerel \cite{Becquerel1896_CR122-501} and the Curies \cite{Curie1898_CR127-175, Curie1898_CR127-1215} and the existence of a compact nucleus at the center of an atom by Rutherford \textit{et al.} \cite{Rutherford1911_PM21-669} opened the door of nuclear physics.
During the hundred years of development in nuclear physics, there emerged several significant milestones, including the discovery of the neutron by Chadwick \cite{Chadwick1932_Nature129-312} which verified
the composition of the nucleus as protons and neutrons, the meson-exchange theory for the strong interaction between nucleons by Yukawa \cite{Yukawa1935_PPMSJ17-48}, the independent-particle shell model of the nucleus by Goeppert-Mayer \cite{Mayer1949_PR75-1969}, Haxel, Jensen, and Suess \cite{Haxel1949_PR75-1766}, and the collective Hamiltonian for nuclear rotation and vibration by Rainwater \cite{Rainwater1950_PR79-432}, Bohr and Mottelson \cite{Bohr1952_MFMDVS26-14, Bohr1953_MFMDVS27-16}, etc.

With the understanding of the composition of a nucleus as protons and neutrons \cite{Chadwick1932_Nature129-312} and the meson-exchange theory for the strong interaction between the nucleons \cite{Yukawa1935_PPMSJ17-48},
nuclear physicists hoped to describe the nucleus, a quantum many-body system, from the underlying nucleon-nucleon interaction.
Euler, a student of Heisenberg, assumed the nuclear force as a two-body (2N) interaction with a Gaussian shape and calculated the infinite nuclear system, i.e., homogeneous nuclear matter, using second-order perturbation theory \cite{Euler1937_ZP105-553}.
However, the strong repulsive core of the realistic nuclear force \cite{Jastrow1951_PR81-165} prevents the application of  perturbation theory.

On the other hand, the nuclear structure model with a phenomenological mean field achieved great success.
Goeppert-Mayer \cite{Mayer1949_PR75-1969}, Haxel, Jensen, and Suess \cite{Haxel1949_PR75-1766} introduced a strong spin-orbit potential and proposed the nuclear independent-particle shell model that successfully explained the conventional magic numbers in nuclei.
Rainwater \cite{Rainwater1950_PR79-432}, Bohr and Mottelson \cite{Bohr1952_MFMDVS26-14, Bohr1953_MFMDVS27-16} explored the nuclear deformation and proposed the nuclear collective model that provided successful descriptions for nuclear collective excitations.
Although the independent-particle shell model could describe the single-particle motion in a nucleus with a phenomenological mean-field potential, it could not provide even a qualitative description for nuclear bulk properties.
On the contrary, a unified phenomenological description of nuclear vibrations and rotations could be achieved by the collective Hamiltonian, whereas it was helpless in understanding the motion of a single nucleon.

Since the 1970s, nuclear density functionals that start from effective nucleon-nucleon interactions (or Lagrangians) emerged, such as the Skyrme-Hartree-Fock theory \cite{Vautherin1972_PRC5-626}, the relativistic mean-field (RMF) theory \cite{Walecka1974_APNY83-491}, and the Gogny-Hartree-Fock theory \cite{Decharge1980_PRC21-1568}.
The parameters in these density functionals are determined by fitting to the properties of nuclear matter and the ground or even excited states of selected nuclei, and thus they are called the phenomenological density functionals or effective interactions.
These effective interactions can describe the ground and excited states for almost all nuclei in the whole nuclear chart very well and became important tools to study heavy nuclei and exotic nuclei far away from the stability region \cite{Bender2003_RMP75-121,Vretenar2005_PR409-101,Meng2016}.

The efforts to describe nuclear structure from realistic bare nucleon-nucleon interactions, later on known as nuclear \textit{ab initio} calculations, have been started in the 1950s. Brueckner, Levinson, and Mahmoud \cite{Brueckner1954_PR95-217} defined the reaction matrix in the nuclear medium, the so-called $G$-matrix, which took into account the two-body short-range correlations induced by the strong repulsive core of the nuclear force, and reproduced in the Hartree-Fock approximation qualitatively the saturation properties of nuclear matter. Different from that, Jastrow \cite{Jastrow1955_PR98-1479} introduced the correlation function to take into account correlation effects induced by the strong repulsive core, and to determine the nuclear ground state using a variational method.
By starting from a Slater determinant, Coester \cite{Coester1958_NP7-421} proposed the coupled-cluster method.
Inspired by Brueckner's $G$-matrix, Kuo and Brown \cite{Kuo1966_NP85-40} developed the configuration-interaction shell model. After the 1980s, with improvements of the nuclear force and with increasing computational resources, \textit{ab initio} calculations starting from realistic nucleon-nucleon interactions have been largely promoted with more and more advanced many-body methods, such as the quantum Monte-Carlo method \cite{Carlson1988_PRC38-1879}, the self-consistent Green's function method \cite{Ramos1989_NPA503-1}, the no-core shell model \cite{Zheng1993_PRC48-1083}, the Monte-Carlo shell model \cite{Honma1996_PRL77-3315}, the nuclear lattice effective field theory \cite{Mueller2000_PRC61-044320}, or the in-medium similarity renormalization group \cite{Tsukiyama2011_PRL106-222502}.
For recent reviews, see Refs. \cite{Carlson2015_RMP87-1067,Hagen2014_RPP77-096302,Barrett2013_PPNP69-131,Dickhoff2004_PPNP52-377,Lee2009_PPNP63-117,Hergert2016_PR621-165}.

Recently, relativistic Brueckner-Hartree-Fock (BHF) theory, a fully self-consistent relativistic version of \textit{ab inito} calculations has been successfully applied to finite nuclei \cite{Shen2016_CPL33-102103, Shen2017_PRC96-014316}, and this will be reviewed in more detail in the following text.
On the one hand, \textit{ab initio} calculations are of fundamental significance by themselves. On the other hand, they can also be used to guide the development of nuclear density functionals, which was one of the original ideas of nuclear density functional theory \cite{Negele1970_PRC1-1260}. See, in particular, Sections \ref{Sect:1.3}-\ref{Sect:1.5} for the motivations of RBHF and CDFT as well as the connections in between.


\subsection{The realistic nucleon-nucleon interaction}\label{Sect:1.2}


The nuclear force models can be dated back to 1935 when Yukawa proposed the meson-exchange model \cite{Yukawa1935_PPMSJ17-48}.
The one-pion-exchange model achieved great successes in explaining the nucleon-nucleon scattering and deuteron properties. It provided a good description of the long-range part of the nuclear force.
However, investigations of multi-pion-exchange processes in the 1950s encountered severe difficulties, and there existed considerable ambiguities in the theoretical models \cite{Taketani1952_PTP7-45,Brueckner1953_PR90-699, Phillips1959_RPP22-314}.
On the other hand, experiments on the nuclear force had made much progress at that period.
By analyzing the proton-proton scattering data at $340$~MeV, Jastrow pointed out that there is a very strong repulsive core in the short-range part of the nuclear force and introduced the hard-core model \cite{Jastrow1951_PR81-165}.

In parallel, many efforts were made to construct phenomenological nuclear forces according to symmetries.
In 1941, Eisenbud and Wigner \cite{Eisenbud1941_PNAS27-281} considered possible interaction forms which linearly depended on the relative momentum of the two nucleons, including
the central, spin-spin, tensor, and spin-orbit terms.
If a quadratic dependence of the momentum is allowed, there will be a quadratic spin-orbit term as well \cite{Okubo1958_APNY4-166}.
With such interaction forms, Gammel and Thaler obtained in 1957 a quantitative nuclear force model by fitting to the nucleon-nucleon scattering data for the first time \cite{Gammel1957_PR107-291}. In the Gammel-Thaler potential, a hard-core was used for the short-range part ($r \leq 0.4$~fm).
During the 1960s, phenomenological nuclear forces were further developed.
The Yale group \cite{Lassila1962_PR126-881} and Hamada and Johnston \cite{Hamada1962_NP34-382} independently developed their nuclear force models with a quadratic spin-orbit term.
Reid not only improved further the hard-core potential but also proposed a soft-core potential \cite{Reid1968_APNY50-411}.
Comparing with the hard-core potential, the short-range part of the nuclear force in the soft-core potential is no longer infinite repulsive but of finite values.
One of the Reid soft-core potentials became one of the most widely used nuclear force
later \cite{Reid1968_APNY50-411}.

With more mesons discovered in the 1960s \cite{Erwin1961_PRL6-628, Maglicc1961_PRL7-178}, the nuclear force based on the meson-exchange picture was been revived again.
The nuclear force based on the one-boson-exchange (OBE) model was proposed and achieved great success \cite{Green1967_RMP39-594, Erkelenz1974_PR13-191}. It is still one of the most economical models in describing the nucleon-nucleon interaction quantitatively.
It provides a simple parametrization and can describe the nucleon-nucleon scattering data well using only a few parameters.

However, the scalar isoscalar $\sigma$ boson, which provides the middle-range attraction, has a very large width ($400\sim700$ MeV as summarized by the Particle Data Group \cite{Tanabashi2018_PRD98-030001}).
It is considered to be related to the two-pion exchange, and many efforts were made to derive the middle-range attraction from the correlated two-pion exchange, including the one-boson-exchange model of the Stony Brook group~\cite{Jackson1975_NPA249-397}, the Paris potential~\cite{Lacombe1980_PRC21-861} derived from dispersion relations, the Partovi-Lomon model~\cite{Partovi1970_PRD2-1999} and the Bonn potential \cite{Machleidt1987_PR149-1} from field theory.  

With the establishment of quantum chromodynamics (QCD) as the basic theory of strong interaction, the meson-exchange theory was no longer the basic starting point for the nuclear force.
However, because of its non-perturbative nature at low energies, it is difficult to derive the nuclear force directly from QCD.
The quark model which started from QCD began to be developed \cite{Detar1980_NPA335-203, Myhrer1988_RMP60-629}. It tried to describe the hadron structure and the hadron-hadron interaction in a unified way from the quark level, and many features of the nuclear force were explained successfully.
Nevertheless, the quark model is still a phenomenological model with a set of parameters and therefore it cannot be viewed as a basic theory either.

One breakthrough appeared in 1990 when Weinberg applied the chiral effective field theory \cite{Weinberg1979_PA96-327} to the low-energy region of QCD \cite{Weinberg1990_PLB251-288, Weinberg1991_NPB363-3}.
The idea is that the effective degrees of freedom of the nuclear system in the low-energy region should be nucleons and pions instead of quarks and gluons. The pions are, as Goldstone bosons, connected with the violation of chiral symmetry in QCD.
Then the nuclear force can be described by a Lagrangian of nucleons and pions satisfying the chiral symmetry of QCD.
The chiral Lagrangian provides a perturbative expansion framework in deriving the nucleon-nucleon interaction, where the small quantity in the expansion is $Q/\Lambda_\chi$ with the low momentum $Q$ and the scale of chiral symmetry breaking $\Lambda_\chi \sim 1$~GeV.
  This low-energy perturbation theory is also called chiral perturbation theory, with the potential being iterated at least to the leading order. 
The perturbative order is characterized by the power of $Q$, which defines a power counting scheme \cite{Weinberg1991_NPB363-3}.
Ord\'o\~nez and van Kolck \cite{Ordonez1992_PLB291-459} constructed the next-to-next-to-the-leading order (NNLO) chiral force in time-ordered perturbation theory reproducing most features of the nuclear force in a quantitative way.
Using the unitary transformation method, Epelbaum, Gl\"ockle, and Meissner (Bochum-J\"ulich group) solved the energy-dependent problem in time-ordered perturbation theory and constructed the NNLO chiral force \cite{Epelbaum1998_NPA637-107, Epelbaum2000_NPA671-295}.
Afterward, the three-body force from NNLO was introduced \cite{Epelbaum2002_PRC66-064001}.
Entem and Machleidt (Idaho group) \cite{Entem2003_PRC68-041001} and the Bochum-J\"ulich group \cite{Epelbaum2005_NPA747-362} constructed the chiral force up to N$^3$LO, and then up to N$^4$LO \cite{Epelbaum2015_PRL115-122301, Entem2015_PRC91-014002}.
By investigating the major terms in N$^5$LO, it was found that the expansion of the chiral force achieved good convergence at N$^4$LO \cite{Entem2015_PRC92-064001}.
Recently, a relativistic chiral force was developed, and here the description for the nucleon-nucleon scattering data at the leading order is comparable with that of the non-relativistic NLO chiral force \cite{Ren2018_CPC42-014103, Li2018_CPC42-014105}.
This indicates that the expansion of the chiral force may achieve faster convergence in the relativistic framework.
For recent reviews on the nuclear chiral force, see Refs.~\cite{Epelbaum2009_RMP81-1773, Machleidt2011_PR503-1}.

Since the 1990s, several versions of  high-precision nuclear forces have been constructed.
With the effort of the Nijmegen group, the analysis of nucleon-nucleon scattering phase shifts has been improved notably \cite{Stoks1993_PRC48-792}. They made a multi-energy partial-wave analysis for all nucleon-nucleon scattering data below $T_{\rm lab} = 350$~MeV, published between 1955 and 1992.
The final database consists of $1787$ proton-proton and $2514$ neutron-proton scattering data after careful examination.
The fitting result is $\chi^2/N_{\rm df} = 1.08$, with the total number of degrees of freedom $N_{\rm df} = 3945$.
This result is remarkably better than the preceding phase-shift fitting accuracy \cite{Stoks1993_PRC48-792} and forms the ground for a new generation of high-precision nuclear forces.
Other well known high-precision nuclear forces include Nijmegen \cite{Stoks1994_PRC49-2950}, Reid93 \cite{Stoks1994_PRC49-2950}, and Argonne V18 \cite{Wiringa1995_PRC51-38} potentials based on phenomenological models, or CD-Bonn \cite{Machleidt2001_PRC63-024001} based on the one-boson-exchange model.
 For the forces based on chiral perturbation theory, for example, the $\chi^2$/datum for the description of the Nijmegen phase shifts has achieved 0.3 (np) and 0.6 (pp) for data with an energy less than 200 MeV calculated by N$^4$LO (Bochum group) \cite{Epelbaum2015_PRL115-122301}, and it has achieved 1.15 for data with an energy less than 290 MeV by N$^4$LO (Idaho group) \cite{Entem2017_PRC96-024004}. 
These high-precision nuclear forces can reproduce experimental nucleon-nucleon scattering data accurately and provide the foundation for nuclear many-body calculations.

The lattice QCD approach solving the QCD on a lattice provides a promising way to derive the nuclear force directly from QCD. There are two major approaches to study hadron-hadron interactions with lattice QCD.
One is based on L\"uscher's finite volume formula \cite{Luscher1991_NPB354-531}, which extracts the ground-state energy using the temporal correlation functions on a finite lattice space, see Ref.~\cite{Beane2011_PPNP66-1} for a review.
Another one is the HAL QCD method \cite{Ishii2007_PRL99-022001}, which defines and extracts the interaction between hadrons through the Nambu-Bethe-Salpeter wave function, see Ref.~\cite{Aoki2012_PTEP2012-01A105} for a review.
Recent lattice simulations, extensively developed by the HAL QCD collaboration, have obtained successfully the baryon-baryon interactions with a realistic pion mass \cite{Doi2015_LATTICE20-086}.
This provides a new possibility to describe the nuclear system from QCD.


\subsection{Description of nucleus as a relativistic system}\label{Sect:1.3}




The relativistic description for nuclear systems goes in parallel with the non-relativistic one. In 1955, Johnson and Teller found that a good description of nuclear matter saturation properties can be obtained if a velocity-dependent nuclear force is introduced \cite{Johnson1955_PR98-783}.
Duerr reformulated this theory in a covariant form, avoided the collapse of the nucleus occurring in the non-relativistic theory for high kinetic energies, obtained similar saturation properties, and predicted naturally a large spin-orbit splitting \cite{Duerr1956_PR103-469}. Rozsnyai extended Duerr's Hamiltonian to study finite nuclei \cite{Rozsnyai1961_PR124-860}. By readjusting the coupling strengths and the range of the force, he obtained a reasonable binding energy and density distribution. After the discovery of scalar and vector mesons, the one-boson-exchange potential was gradually established \cite{Green1967_RMP39-594}. Miller and Green developed the relativistic Hartree-Fock (RHF) theory based on a model of scalar and vector mesons exchange, which produces strong attractive and repulsive interactions, respectively \cite{Miller1972_PRC5-241}.

In 1974, Walecka proposed a renormalizable relativistic mean-field $\sigma$-$\omega$ model that gave good saturation properties of nuclear matter \cite{Walecka1974_APNY83-491}. In the calculation of the densities, only the particles in the Fermi sea are considered. Within the no-sea approximation~\cite{Serot1986_ANP16-1}, the contribution from the Dirac sea is neglected, because they would lead to couplings to high energy nucleon-antinucleon pairs. From the point of view of effective field theory~\cite{Weinberg1990_PLB251-288}, it can be understood that the high-energy intermediate states, including nucleon-antinucleon pairs, can be integrated out using an appropriate renormalization scheme~\cite{Chin1977_APNY108-301}. In practice, all successful relativistic calculations based on this model have used the no-sea approximation. 

The success of the simple Walecka model promoted the development of relativistic many-body theory. Boguta and Bodmer \cite{Boguta1977_NPA292-413} solved the problem of too large incompressibility in Walecka's model by introducing nonlinear self-couplings of the $\sigma$ meson.
Serot extended Walecka's $\sigma$-$\omega$ model by including the isovector $\pi$ and $\rho$ mesons \cite{Serot1979_PLB86-146}. In nuclei with open valence shells the Munich group presented a successful relativistic description of axial \cite{Gambhir1990_APNY198-132} and triaxial \cite{Koepf1988_PLB212-397} deformations, pairing correlations have been introduced in the framework of relativistic Hartree-Bogolubov (RHB) theory \cite{Kucharek1991_ZPA339-23} and rotating nuclei are well described by relativistic cranking theory
\cite{Koepf1989_NPA493-61,Afanasjev1996_NPA608-107}.
For the development of relativistic nuclear many-body theory in this period, one can see the review papers \cite{Serot1986_ANP16-1, Malfliet1988_PPNP21-207, Reinhard1989_RPP52-439,Savushkin2004}. In these RMF models, the coupling strengths were determined by fitting to the saturation properties of nuclear matter and to the selected data of finite nuclei, similar to the zero-range Skyrme force \cite{Vautherin1972_PRC5-626} or the finite-range Gogny force  \cite{Decharge1980_PRC21-1568} in the non-relativistic framework.

The phenomenological relativistic mean-field theory \cite{Walecka1974_APNY83-491,Ring1996_PPNP37-193}, also called covariant density functional theory (CDFT) in recent years, is able to reproduce very successfully basic properties for most of the nuclei in the nuclear chart. Along the branch with the nonlinear self-couplings of the meson fields, also referred as  NLRMF, the parameter set NL1 \cite{Reinhard1986_ZPA323-13} gives a good description of stable nuclei, the sets NL-SH \cite{Sharma1993_PLB312-377}, NL3 \cite{Lalazissis1997_PRC55-540}, and NL3* \cite{Lalazissis2009_PLB671-36} extended this for nuclei far from stablity, TM1 \cite{Sugahara1994_NPA579-557} further considered the nonlinear self-couplings of $\omega$ mesons, and PK1 \cite{Long2004_PRC69-034319} takes into account a microscopic center-of-mass (c.m.) correction and neutron-proton mass differences. The idea of nonlinear meson couplings goes back to the 1970s, a time when the relativistic models were considered as a first step to a fully fledged quantum field theory  \cite{Serot1986_ANP16-1} for the description of nuclei, called Quantum Hadrodynamics (QHD) \cite{Serot1992_RPP55-1855}. For a review see \cite{Serot1997_IJMPE6-515}. Such a theory had to be renormalizable \cite{Chin1977_APNY108-301} and therefore Boguta and Bodmer \cite{Boguta1977_NPA292-413} introduced a $\phi^4$-coupling scheme in the meson sector. When it turned out that there were serious problems with renormalization in the second order \cite{Furnstahl1989_PRC40-321}, the RMF theory was interpreted as a relativistic density functional theory based on the \textit{ab initio} model of Brockmann and Toki \cite{Brockmann1992_PRL68-3408} with density-dependent coupling strengths adjusted to relativistic Brueckner-Hartree-Fock calculations. This model was too simple to compete with the very successful phenomenological nonlinear meson-coupling models, but a phenomenological density dependence (referred as DDRMF) with several parameters, adjusted in a similar way to experimental data as the nonlinear models, turned out to be very successful. Moreover, DDRMF is closer to the spirit of density functional theory. Therefore, it become more and more popular with extensive applications and developments of effective Lagrangians, such as TW99 \cite{Typel1999_NPA656-331}, PKDD \cite{Long2004_PRC69-034319},  DD-ME1 \cite{Niksic2002_PRC66-024306}, DD-ME2 \cite{Lalazissis2005_PRC71-024312}, etc.

In the RMF Lagrangians mentioned above, the nuclear force is propagated by massive mesons, thus keeping the feature of finite range. This needs some numerical efforts, in particular for excited states in the density dependent relativistic Random Phase Approximation and for methods going beyond mean field approach. Therefore, similar to the zero-range Skyrme force in the non-relativistic case, the relativistic point-coupling model, a new branch of the RMF approach, was developed by replacing the meson exchange with the local many-body contact interactions. Compared to NLRMF and DDRMF, the relativistic point-coupling model avoids solving the complicated meson fields and therefore has the advantages of a simple form, fast computing, and easy to be extended to methods beyond mean field approach. Commonly used point-coupling RMF sets include the  nonlinear PC-LA \cite{Nikolaus1992_PRC46-1757}, PC-F1 \cite{Burvenich2002_PRC65-044308}, PC-PK1 \cite{Zhao2010_PRC82-054319}, and the density-dependent DD-PC1 \cite{Niksic2008_PRC78-034318}.

With these effective interactions, the relativistic many-body theory has achieved great success in describing properties of nuclear ground-states and excited-states. The related reviews include, for example, the description of nuclear bulk properties, exotic nuclei, pairing correlations, high-spin states, and time-dependent problems \cite{Ring1996_PPNP37-193, Vretenar2005_PR409-101}, relativistic continuum Hartree-Bogoliubov theory for exotic nuclei \cite{Meng2006_PPNP57-470}, relativistic quasiparticle random phase approximation for collective modes in exotic nuclei \cite{Paar2007_RPP70-691}, relativistic density functional theory beyond mean field \cite{Ring2009_PAN72-1285,Niksic2011_PPNP66-519}, the study of nuclear magnetic and antimagnetic rotation with the tilted axis cranking covariant density functional theory \cite{Meng2013_FPC8-55}, the relativistic description of pseudospin and spin symmetries \cite{Liang2015_PR570-1}, the investigation of halo phenomena in medium-heavy and heavy nuclei \cite{Meng2015_JPG42-093101}. The current status in the relativistic many-body theory for nuclear structure has been summarized in Ref. \cite{Meng2016}.

Although the relativistic many-body theory achieved great success in describing many properties of nuclear structure, it is based on phenomenological interactions, which are determined by fitting to ground-state properties of stable nuclei. Substantial ambiguities exist in the prediction for unstable nuclei without data and for nuclear matter at high densities. On the other hand, some effects which do not show up clearly in stable nuclei, such as the tensor-force effect, are difficult to study by phenomenological interactions. Meanwhile, the tensor force is very important in describing the properties of neutron-rich exotic nuclei, in particular, the evolution of the shell-structure \cite{Otsuka2005_PRL95-232502}. Particularly in systems with non-spin-saturated configurations it has a strong influence on the single-particle structure.

One has to distinguish between the tensor part of the bare nucleon-nucleon interaction, which is in principle determined by the scattering data, and the tensor term in the effective interaction in the nuclear medium. In microscopic nuclear matter calculations, the bare tensor force plays a crucial role. The second-order diagram containing a two-pion exchange coupled to $T=0$ is the origin of the major part of the middle-range attraction, which is, in phenomenological models, usually described by the exchange of the isoscalar $\sigma$ meson. Nuclear matter is spin-saturated. There is no spin-orbit splitting, and thus the so-called ``first-order'' tensor in the effective force, which acts between particles in open spin configurations, can be neglected. This is very different in finite heavy nuclei with open spin configurations, where the two spin-orbit partners are on different sides of the Fermi surface. Here the tensor force can lead to considerable shifts in the single-particle energies, and these in turn can lead to new magic numbers, to changes of the deformation, and to quantum phase transitions. In most of the successful relativistic and non-relativistic density functionals, effective tensor terms are not taken into account. The phenomenological fit of such functionals to bulk properties includes such effects in a somewhat averaged way, and as a consequence these functionals are relatively successful in the region of stable nuclei. Systematic investigations of relativistic density functionals~\cite{Agbemava2014_PRC89-054320,Agbemava2019_PRC99-014318} have shown that the various types of functionals produce rather similar results in the region of stable nuclei, where their parameters have been adjusted. There are, however, essential differences far from stability, and they are often caused by differences in the single-particle energies in those regions~\cite{Afanasjev2015_PRC91-014324}. There is hope that a proper treatment of the tensor force~\cite{Sagawa2014_PPNP76-76} can help to increase the predictive power of such future functionals.

In such cases, \textit{ab initio} calculations become very important.
By solving the nuclear system on the basis of realistic bare nucleon-nucleon forces, one can establish the link between the microscopic nucleon-nucleon interaction and nuclear structure properties.
This will provide valuable information in describing exotic nuclei, nuclear matter under extreme conditions, and the effects of the tensor-force. In this way one will be able improve the nuclear density functional \cite{Bogner2013_CPC184-2235}.



\subsection{Relativistic Brueckner-Hartree-Fock theory}\label{Sect:1.4}


Brueckner theory \cite{Brueckner1954_PR95-217} has been introduced in the 1950s to overcome the difficulties caused by the extremely strong short-range repulsion of the bare nuclear force \cite{Jastrow1951_PR81-165} in many-body calculations. For the description of nucleon-nucleon scattering, the scattering amplitude calculated in first-order, i.e., in Born approximation, is totally wrong. The correct scattering amplitude can be obtained \cite{Day1967_RMP39-719} only by the exact solution of the two-particle Lippmann-Schwinger equation \cite{Lippmann1950_PR79-469}.
In multiple scattering of hadrons on the nucleus, instead of
using the realistic nuclear force, Watson \textit{et al.} \cite{Watson1953_PR89-575} constructed an optical potential via the scattering matrix ($T$ matrix) derived from the solution of the Lippmann-Schwinger equation and found rather good descriptions of the experimental results.
By replacing the bare nuclear force with an effective interaction in the medium, later known as the $G$-matrix, Brueckner  \textit{et al.} \cite{Brueckner1954_PR95-217} extended this idea to describe nuclear structure in a mean-field
approximation. By summing all the ladder diagrams and taking into account the Pauli principle in nuclear medium, the obtained $G$-matrix incorporates the two-body short-range correlation and describes the saturation properties of nuclear matter within the Hartree-Fock approximation qualitatively.

Bethe elaborated the Brueckner theory by constructing a model of the nucleus, in which each nucleon moves in a self-consistent potential, which is obtained from the reaction matrix for two nucleons derived from scattering theory in the nuclear medium \cite{Bethe1956_PR103-1353}.
This provides the foundation for the extension of Brueckner theory from nuclear matter to realistic finite nuclei.
A formal derivation of Brueckner theory was provided by Goldstone \cite{Goldstone1957_PRSA239-267}. These are breakthroughs in microscopic nuclear many-body theory, and many developments have been made along this direction  in the 1960s \cite{Day1967_RMP39-719,Rajaraman1967_RMP39-745,Baranger1969_Varenna40-511}.

In Brueckner theory, the $G$-matrix is obtained by solving the scattering equation in the nuclear medium.
In contrast to two-body scattering in free space, the Pauli principle must be taken into account, because all the states below the Fermi surface are occupied in the nuclear medium and therefore they cannot be used in the intermediate scattering processes. This effect of the Pauli principle was already pointed out by Brueckner very early  \cite{Brueckner1954_PR95-217}, but, at the beginning, it was neglected as an approximation.
Later, Brueckner and Wada \cite{Brueckner1956_PR103-1008} took this effect into account in perturbation theory.
A full treatment of the Pauli principle was first given by Bethe and Goldstone \cite{Bethe1957_PRSA238-551}, and therefore, the scattering equation in Brueckner theory is also called the Bethe-Goldstone (BG) equation.

Because of the Pauli operator and the energy denominator, the Bethe-Goldstone equation represents a nonlinear integral equation, which was very difficult to solve even for nuclear matter.
By replacing the spectrum of the intermediate states with a "reference spectrum", Bethe, Brandow, and Petschek reduced the Brueckner integral equation to a differential one which can be solved easily \cite{Bethe1963_PR129-225}. Afterwards they took the Pauli operator into account order by order in a perturbative method. Separating the two-nucleon interaction into short-range and long-range parts, Moszkowski and Scott \cite{Moszkowski1960_APNY11-65} proposed a method, which is very useful in solving the diagonal matrix elements of the $G$-matrix but it turns out to be relatively difficult for the non-diagonal parts. These methods are mainly applicable to the static nuclear force.
For a velocity dependence, one usually needs to work in momentum space.
Brown, Jackson, and Kuo \cite{Brown1969_NPA133-481} as well as Haftel and Tabakin \cite{Haftel1970_NPA158-1} independently developed the matrix inversion method to solve the Bethe-Goldstone equation in momentum space. It accurately reproduces the results obtained in coordinate space.

Compared with the bare nuclear force, the $G$-matrix has included some two-body short-range correlations and can be viewed as an effective nuclear interaction suitable to be used in a perturbation expansion.
For a hard-core potential, the matrix elements of the bare interaction are divergent, but the elements of the $G$-matrix are finite \cite{Ring1980}.
This indicates that one can use the $G$-matrix for a perturbative expansion.
 The result is called the Brueckner-Bethe-Goldstone expansion, or the hole-line expansion as the expansion order is indicated by the number of independent hole lines in each diagram \cite{Rajaraman1967_RMP39-745}. The convergence of this expansion is roughly determined by the particle density times the wound integral of the two-body wave function \cite{Day1967_RMP39-719}. The lowest order gives the Brueckner-Hartree-Fock framework. 
In the BHF or general Hartree-Fock framework, a good choice of the single-particle mean-field potential can cancel contributions of certain higher order diagrams. This is good for a rapid convergence of the expansion.
For a review of the three-hole-line expansion beyond BHF theory, see Ref.~\cite{Rajaraman1967_RMP39-745}, and for applications for finite nuclei, see Ref.~\cite{Baranger1969_Varenna40-511}.

Early theoretical works showed that BHF theory was a good approximation for the nuclear matter system near or below the saturation density.
However, from a quantitative point of view, the saturation properties given by BHF were not good enough compared with the empirical data.
Coester \textit{et al.} \cite{Coester1970_PRC1-769} found that all the BHF results with various nuclear forces were located on or near the now-called Coester line, which deviated from the empirical saturation region systematically.
Neither can BHF theory give a good description of binding energies and charge radii of finite nuclei at the same time.
On the other hand, in the late 1970s, a totally different approach, the variational method based on a correlation operator acting on a Slater determinant was developed to deal with the short-range repulsion of the nuclear force, and the results obtained with this method were similar as those of the Brueckner method \cite{Day1978_RMP50-495}.
This leaded to the discussion on the missing of a three-body (3N) force, which has been proposed at earlier times  \cite{Primakoff1939_PR55-1218,Drell1953_PR91-1527,Fujita1957_PTP17-360, Brown1969_NPA137-1}.
By introducing a three-body force, good descriptions of nuclear matter saturation properties as well as the ground state and a few excited states properties of light nuclei were obtained \cite{Pieper2001_ARNPS51-53}.
Including a three-body force, BHF has been extended to study neutron star and nuclear matter properties \cite{Lejeune1986_NPA453-189,Baldo1997_AA328-274,Zuo2002_NPA706-418}.



Inspired by the success of the relativistic phenomenological many-body theory in the 1970s, different research groups tried to extend Brueckner theory to the relativistic framework.
In the pioneering work of the Brooklyn group, the single-particle wave function was chosen as the Dirac spinor in free space, and the relativistic effect was taken into account in first-order perturbation theory \cite{Anastasio1980_PRL45-2096}.
The results obtained in this way showed a significant improvement compared to non-relativistic investigations. Both binding energy and saturation density were much closer to the empirical values.
Horowitz and Serot \cite{Horowitz1984_PLB137-287,Horowitz1987_NPA464-613}, Brockmann and Machleidt \cite{Brockmann1984_PLB149-283, Brockmann1990_PRC42-1965}, and ter Haar and Malfliet \cite{terHaar1986_PRL56-1237, terHaar1987_PR149-207} independently solved the nuclear matter system using relativistic Brueckner-Hartree-Fock (RBHF) theory self-consistently, see also Refs.~\cite{Anastasio1983_PR100-327,Celenza1986_WSLN2,Malfliet1988_PPNP21-207,terHaar1987_PR149-207}.
With the one-boson-exchange potentials Bonn A, B, and C \cite{Machleidt1989_ANP19-189}, the RBHF results improved the non-relativistic Coester line greatly \cite{Brockmann1990_PRC42-1965}.
RBHF theory was also extended to study asymmetric nuclear matter \cite{terHaar1987_PRL59-1652}, the optical potential \cite{Nuppenau1990_NPA511-525}, neutron stars \cite{Huber1996_NPA596-684}, neutron stars with hyperon degree of freedom \cite{Katayama2015_PLB747-43}, etc. Latest reviews can be found in
Refs.~\cite{VanGiai2010_JPG37-064043,Sammarruca2010_IJMPE19-1259,VanDalen2010_IJMPE19-2077,Muther2017_IJMPE26-1730001,Lenske2018_PPNP98-119}.

 The fact, that relativistic effects play such an important role, is somehow beyond expectations, as the Fermi momentum of a nuclear system near saturation density is small compared with the rest mass of the nucleon. It raises questions of where this relativistic effect comes from and why it is essential in improving the theoretical description.
In the relativistic framework, the exchange of scalar mesons induces a strong attraction between nucleons. For the particle states with positive energy, it is to a large extent compensated by the repulsion caused by the strong zeroth component of the vector field induced by the omega exchange. For the negative-energy states, the effects of both fields add up. This has two essential consequences: (i) A very strong velocity-dependent part of the field in the form of a strong spin-orbit term. It cannot be neglected even for the relatively small velocities in the Fermi sea, because the factor in front of the velocity is very large. (ii) An additional repulsion between the nucleons increasing with the density. The scalar field  is proportional to the Dirac effective mass of the nucleon. This effective mass decreases as the density increases. Therefore, the average attraction felt by the nucleons decreases too. On the other hand, the vector mesons producing repulsion have no such feature. The net effect is that the repulsion felt by nucleons increases with increasing density. 

This density-dependent effect leads to the saturation of nuclear matter in the relativistic framework \cite{Brown1986_AIPC142-2, Brown1987_CNPP17-39, Brown1987_PS36-209, Brockmann1990_PRC42-1965}. In non-relativistic density functionals such as Skyrme \cite{Vautherin1972_PRC5-626} or Gogny \cite{Decharge1980_PRC21-1568}, this effect does not exist and one needs a strongly repulsive phenomenological density-dependent term $t_3$ to find proper saturation.
By expanding the Dirac spinor for plane waves with an effective mass $M^*/M$ in nuclear matter in terms of the free Dirac spinor with effective mass $M^*/M=1$, Brown \textit{et al.} \cite{Brown1987_CNPP17-39} pointed out that, in a relativistic description, one effectively takes into account nucleon-antinucleon excitations (or the so-called $Z$-diagram).
 Zuo \textit{et al.} \cite{Zuo2002_NPA706-418} showed in a non-relativistic BHF investigation of the possible contributions to three-body forces in nuclear matter that this $Z$-diagram contributes a large part to the total three-body forces, at least for densities not much higher than the saturation density. It has also been shown in nuclear matter that the results of RBHF with 2N interactions is very similar to BHF with chiral 2N + 3N interactions~\cite{Sammarruca2012_PRC86-054317}. Furthermore, a good description of both nuclear matter and finite nuclei has been achieved in a non-relativistic framework with a 3N interaction that simulates the relativistic effects ~\cite{Muther2017_IJMPE26-1730001}. 
This explains clearly the reason why, without three-body forces, the relativistic BHF calculations lead to a much better description of the saturation properties of nuclear matter than the non-relativistic ones.

Although RBHF theory with only two-body forces can already give rather good saturation properties of nuclear matter, there are still several open problems \cite{VanGiai2010_JPG37-064043}.
Because of the cluster effects, the computation at low density becomes unstable \cite{Typel2010_PRC81-015803}.
Since the solution of the Bethe-Goldstone equation in most of the RBHF calculations of nuclear matter do not include the antinucleon degrees of freedom, the scalar and vector channels of the nucleon self-energy cannot be decomposed uniquely ~\cite{Fuchs2004_LNP641-111}. Such a decomposition is necessary for the solution of the RHF equations in the next step of the iteration. Different decompositions produce different results.
The method proposed by Brockmann and Machleidt \cite{Brockmann1990_PRC42-1965} gives reasonable results for symmetric nuclear matter, but not for asymmetric nuclear matter \cite{Ulrych1997_PRC56-1788, Shen1997_PRC55-1211}.
Another possibility is the mapping of the $G$-matrix onto 5 (or 6 for asymmetric nuclear matter) complete Lorentz invariants \cite{Horowitz1987_NPA464-613}, but such a mapping is not unique either because of the missing  antinucleon degrees of freedom.
Nowadays, a commonly used method introduced by M\"uther and his collaborators \cite{Ulrych1997_PRC56-1788, Schiller2001_EPJA11-15} is to decompose the $G$-matrix in the form $G = V + \Delta G$ where $V$ is the bare nuclear force. This is similar to the optimal $T$-matrix representation developed by the T\"ubingen group \cite{VanDalen2004_NPA744-227,VanDalen2005_PRC72-065803,VanDalen2007_EPJA31-29}.
Boersma \textit{et al.} \cite{Boersma1994_PRC49-233} parameterized the $G$-matrix in the form of a Yukawa function with form factors by fitting the masses and the coupling constants.
De Jong and Lenske \cite{deJong1998_PRC58-890} and Huber \textit{et al.} \cite{Huber1995_PRC51-1790} tried to solve this problem by projecting in a full Dirac space where both the positive-energy and negative-energy solutions are included. However, because the bare nucleon-nucleon interaction was obtained by considering the positive-energy states only, such a full-space projection is not consistent \cite{deJong1998_PRC58-890}.
More discussions on RBHF theory in nuclear matter can be found in the review \cite{VanGiai2010_JPG37-064043}.

With the success of RBHF theory in nuclear matter, it is natural to extend RBHF for finite nuclei.
However, due to considerable numerical difficulties, such a project was extremely challenging and
a full solution was for a long time not possible.
M\"uther \textit{et al.} \cite{Muther1988_PLB202-483, Muther1990_PRC42-1981} started with a non-relativistic BHF code for finite nuclei \cite{Muther1992-CompNuclPhys2-43} and took relativistic effects into account using the {\it effective density approximation}.
More investigations were mainly based on the local density approximation (LDA) \cite{Negele1970_PRC1-1260}.
The idea of LDA is to find an effective interaction which can reproduce the RBHF results for nuclear matter at a fixed  density. This lead to an effective density-dependent interaction which is used to study finite nuclei.
However, because of the uncertainties of RBHF theory in nuclear matter mentioned above, the mapping from nuclear matter to finite nuclei is far from unique \cite{Brockmann1992_PRL68-3408, Fritz1993_PRL71-46, Fritz1994_PRC49-633,
Boersma1994_PRC49-233, Shi1995_PRC52-144, Fuchs1995_PRC52-3043, Ulrych1997_PRC56-1788,Shen1997_PRC55-1211,Hofmann2001_PRC64-034314,Ma2002_PRC66-024321,VanDalen2011_PRC84-024320}.



\subsection{Connections between RBHF and CDFT}\label{Sect:1.5}

Based on the meson exchange picture of the nuclear force, the RMF approach, consistent with the spirit of density functional theory \cite{Hohenberg1964_PR136-B864, Kohn1965_PR140-A1133}, has achieved many successes in describing properties of nuclear ground and excited states. In the RMF approach, i.e. in the relativistic Hartree approach, some important ingredients are missing, such as the $\pi$- and $\rho$-tensor couplings which enter via the exchange (Fock) terms. Almost in parallel to the RMF approach, the relativistic Hartree-Fock (RHF) theory was initialized by Miller and Green \cite{Miller1972_PRC5-241} and applied systematically to nuclear matter \cite{Bouyssy1985_PRL55-1731} and finite nuclei \cite{Bouyssy1987_PRC36-380}. Similar to the RMF approach, the problem of the too large incompressibility was solved by nonlinear self-couplings of the $\sigma$ meson \cite{Bernardos1993_PRC48-2665} or, in the zero-range limit, by three- and four-body contact terms \cite{Marcos2004_JPG30-703}. However, due to the numerical complexity of the Fock terms, it remained a long-standing problem for RHF theory to provide a quantitative description of nuclear properties comparable with RMF.

The situation changed when computer power was increasing and when an explicit density-dependence of the meson-nucleon coupling strengths was introduced into the RHF scheme, leading to the so-called density-dependent relativistic Hartree-Fock (DDRHF) theory \cite{Long2006_PLB640-150, Long2007_PRC76-034314, Long2010_PRC81-024308}.
In this case it became possible, at least in the spherical case, to carry out extensive calculations,
such that reasonable new RHF Lagrangians have been obtained by the Peking-Orsay group, such as PKO1 \cite{Long2006_PLB640-150}, PKO2 \cite{Long2008_EPL82-12001}, and PKO3 \cite{Long2008_EPL82-12001}, and by the Peking-Aizu group, such as PKA1 \cite{Long2007_PRC76-034314}. Within these models, similar accuracy as in standard RMF has been achieved for a quantitative description of nuclear properties. Pairing correlations can been included in relativistic Hartree-Fock-Bogoliubov (RHFB) theory \cite{Kucharek1991_ZPA339-23}. The corresponding equation can now be solved also for finite spherical nuclei \cite{Long2010_PRC81-024308}.

The explicit treatment of the Fock terms brings significant improvements.
The missing degrees of freedom associated are taken into account naturally with the $\pi$ and $\rho$-tensor couplings via the Fock terms, which improve the self-consistent description of nuclear shell evolution \cite{Long2008_EPL82-12001,Long2009_PLB680-428,Wang2013_PRC87-047301}.
The inclusion of $\rho$-tensor couplings in PKA1, which corresponds to a strong attractive potential, leads to a better preserved pseudo-spin symmetry \cite{Long2006_PLB639-242, Long2007_PRC76-034314, Long2010_PRC81-031302}, and eliminates the spurious shell closures $N,Z=58$ and $92$ that commonly exist in the conventional covariant density functional calculations \cite{Geng2006_CPL23-1139,Long2007_PRC76-034314,Long2009_PLB680-428}.
With the presence of Fock terms, the  nuclear tensor force, an important ingredient of the nuclear force, can be naturally taken into account \cite{Long2008_EPL82-12001,Jiang2015_PRC91-025802,Jiang2015_PRC91-034326,Wang2018_PRC98-034313,Zong2018_CPC42-024101} in a  phenomenological way. Substantial effects due to the Fock terms were also revealed in extending the RHF scheme to describe exotic nuclei \cite{Long2010_PRC81-024308, Long2010_PRC81-031302,Lu2013_PRC87-034311}, superheavy magic shells \cite{Li2014_PLB732-169}, the nuclear symmetry energy, and the equation of state \cite{Sun2008_PRC78-065805,Long2012_PRC85-025806,Li2018_EPJA54-133}.

Besides the nuclear ground-state properties, the Fock terms also lead to essential improvements in the investigations of spin- and isospin excitations in nuclei. Combined with the random phase approximation (RPA), the RHF+RPA approach provides a fully self-consistent description for nuclear Gamow-Teller and spin-dipole excitations. Here the Fock terms of the isoscalar $\sigma$ and $\omega$ meson fields were found to play an essential role \cite{Liang2008_PRL101-122502, Liang2012_PRC85-064302}. Based on RHFB theory, a fully self-consistent quasiparticle random phase approximation (QRPA) was developed \cite{Niu2013_PLB723-172, Niu2017_PRC95-044301} for the study the $\beta$-decay half-lives of spherical neutron-rich nuclei. The available data are well reproduced by including an isospin-dependent proton-neutron pairing interaction in the isoscalar channel of the RHFB+QRPA approach \cite{Niu2013_PLB723-172}. On the other side, with all these successes of RHF theory including Fock terms, one has to emphasize that the solution of the RHF equations with finite meson masses in deformed systems is still connected with enormous numerical difficulties even on the most modern computers. There exist codes for axially deformed nuclei working in an oscillator  basis \cite{Ebran2011_PRC83-064323}, but the number of oscillator shells is limited and therefore such calculations have only been carried out for relatively light nuclei. The RHF applications for triaxial or rotating nuclei and for investigations beyond mean field are not yet feasible at the moment.

As discussed in Sections~\ref{Sect:1.3} and \ref{Sect:1.4}, the success of the early CDFT investigations in the 1970s that take into account Lorentz invariance for the nuclear system stimulated relativistic Brueckner-Hartree-Fock theory for nuclear matter which has achieved great success.
However, for a long period, RBHF studies for finite nuclei have stayed on the level of the local density approximation. This situation has changed recently.
In Ref.~\cite{Shen2016_CPL33-102103}, the RBHF equations were solved directly for finite nuclei in the Dirac Woods-Saxon (DWS) basis using the Bonn A interaction \cite{Machleidt1989_ANP19-189}.
The DWS basis is obtained by solving the Dirac equation with fixed scalar and vector potentials of Woods-Saxon type \cite{Zhou2003_PRC68-034323}. It keeps the full relativistic structure of the Dirac spinors.
Therefoe the solution of the BG equation in this basis does not need approximations like angle averaging \cite{Schiller1999_PRC59-2934,Suzuki2000_NPA665-92}.
Taking $^{16}$O as an example, convergent results are achieved for an energy cutoff of around $1.1$ GeV.
The resulting binding energies, charge radii, and spin-orbit splittings are considerably improved in comparison with non-relativistic BHF results.
In Ref. \cite{Shen2017_PRC96-014316}, the fully self-consistent RBHF theory for finite nuclei was realized using a self-consistent RHF basis. In particular, the Pauli operator in the BG equation is defined in this basis.
A detailed discussion of the RBHF formalism and its application for finite nuclei is given in Ref. \cite{Shen2017_PRC96-014316}.
Spin symmetry in the Dirac sea (or equivalently pseudospin symmetry in the Fermi sea) was discussed in Ref. \cite{Shen2018_PLB781-227}.

It is noticeable that the fully self-consistent RBHF theory, without any adjustable parameters, has achieved such good results even at the Hartree-Fock level, using only the two-body interaction fitted to the nucleon-nucleon scattering and deuteron properties.
Of course, the accuracy of RBHF calculations (as any other \textit{ab initio} methods) cannot be compared with that of phenomenological CDFTs. On the other hand RBHF calculations can provide essential information to guide the construction of DFTs. For example, the difficulty of nuclear DFTs to fix some terms of the functional by experimental data, such as the tensor term \cite{Sagawa2014_PPNP76-76}, can be solved by nuclear \textit{ab initio} calculations.
As a simple ideal system, neutron drops have been studied using RBHF theory and a specific pattern due to the tensor force was found in the evolution of spin-orbit splittings as a function of the neutron number \cite{Shen2018_PLB778-344,Shen2018_PRC97-054312}. This forms an important guide for determining the tensor force and microscopic derivations of relativistic and non-relativistic DFTs.
As an example, recently a new Skyrme functional with a tensor force, SAMi-T, has been developed along this line \cite{Shen2019_PRC99-034322}.

\subsection{Structure of the Review}\label{Sect:1.6}

In this Review, we will focus on the progress of the relativistic Brueckner-Hartree-Fock theory for finite nuclei, and its guidance  towards an \textit{ab initio} covariant density functional theory.
The paper is organized as follows.
The basic concepts connecting relativistic Brueckner-Hartree-Fock theory and covariant density functional theory will be given in Section~\ref{Sect:2}. The relativistic Brueckner-Hartree-Fock calculations for finite nuclei and neutron drops will be discussed in Sections~\ref{Sect:3} and \ref{Sect:4}.
Finally, a summary and perspectives will be given in Section~\ref{Sect:5}.


\section{Basic Concepts}\label{Sect:2}


\subsection{Hohenberg-Kohn and Kohn-Sham density functional theory}

Density functional theory is one of the most important and successful theories in dealing with quantum many-body systems \cite{Hohenberg1964_PR136-B864, Kohn1965_PR140-A1133}.
It was developed first in atomic systems with the Hohenberg-Kohn (HK) theorem \cite{Hohenberg1964_PR136-B864}: for an electromechanical many-body system in an external potential $U_{\rm ext}(\mathbf{r})$, there exists a density functional $E_U[\rho]$ of the density $\rho(\mathbf{r})$,
\begin{equation}
  E_U[\rho] = F_{\rm HK}[\rho] + \int d^3r\, U_{\rm ext}(\mathbf{r}) \rho(\mathbf{r}),
\end{equation}
  where the functional $F_{\rm HK}[\rho]$ does not depend on the external potential $U_{\rm ext}(\mathbf{r})$ nor on the particle number of the system (the particle-number dependence is included implicitly in the normalization of the density). 
It only depends on the interaction between the particles.
It is, therefore, identical for all Coulombic systems, such as atoms, molecules, and solids.
In this sense, this density functional is supposed to be universal.
The variational principle determines the exact ground-state density $\rho_{\rm gs}(\mathbf{r})$ and the corresponding exact ground-state energy by minimizing $E_U[\rho]$ with respect to the density $\rho(\mathbf{r})$.

 Although this theorem is exact, it does not show explicit clues on how to construct this functional $F_{\rm HK}[\rho]$. Furthermore, it is highly non-trivial, if not impossible, to derive shell effects from the variation of a density functional $E_U[\rho]$ with respect to the local density $\rho(\mathbf{r})$.
This problem of HK theory has been solved by Kohn and Sham (KS) \cite{Kohn1965_PR140-A1133} in a very elegant scheme, by mapping the exact local density $\rho(\mathbf{r})$ in a unique way to a local auxiliary single-particle potential $v_{\rm KS}(\mathbf{r})$ defined in such a way that the exact ground-state density $\rho_{\rm gs}(\mathbf{r})$ of the interacting system is the same as the ground-state density of an auxiliary non-interacting system
\begin{equation}
  \rho_{\rm gs}(\mathbf{r}) = \sum_{i=1}^N |\phi_i(\mathbf{r})|^2,
\end{equation}
which is expressed in terms of the $N$ lowest occupied single-particle orbitals of the Kohn-Sham equations
\begin{equation}
  \left[-\frac{\hbar^2}{2M}\nabla^2 + v_{\rm KS}(\mathbf{r})\right] \phi_i(\mathbf{r}) = \varepsilon_i \phi_i(\mathbf{r}).
\end{equation}
To date, most practical applications of density functional theory use this Kohn-Sham scheme.

The exact density functional $F_{\rm HF}[\rho]$ can be then decomposed into three separate terms:
\begin{equation}
\label{eq:decomp}
  F_{\rm HK}[\rho] = T_s[\rho] + E_H[\rho] + E_{xc}[\rho].
\end{equation}
  Here $T_s$ is the kinetic energy of the auxiliary non-interacting system, which is supposed to capture the main part of the kinetic energy of the realistic interacting system as well as the shell effects;
$E_H$ is the Hartree energy, and all the rest is left in the exchange-correlation energy $E_{xc}$ which contains, by definition, the Fock term as well as all the other many-body effects. It is clear that the robustness of the Kohn-Sham scheme crucially depends on how accurately the universal exchange-correlation energy functional can be built.

The extension of the Hohenberg-Kohn theorem \cite{Hohenberg1964_PR136-B864} to relativistic systems was first formulated in Ref.~\cite{Rajagopal1973_PRB7-1912} by utilizing a quantum electrodynamics (QED)-based Hamiltonian with a four-current. Within the no-sea approximation, which is also called no-pair approximation in atomic physics, i.e. neglecting vacuum polarization, i.e. all effects due to the creation of particle-antiparticle pairs, the total energy of the system can be expressed as a functional with respect
to the four-current $j^\mu(\mathbf{r}) = (\rho(\mathbf{r}), \mathbf{j}(\mathbf{r}))$ \cite{Rajagopal1978_JPC11-L943, MacDonald1979_JPC12-2977} instead of only $\rho(\mathbf{r})$ in the non-relativistic case.
 In the early development, however, questions related to zero-point energies, radiative corrections, and UV-divergences were not discussed. The UV-divergences showed up in the derivation of gradient corrections to the kinetic energy \cite{Gross1981_PLA81-447}. Detailed discussions of the renormalization and related issues can be found in Ref. \cite{Engel1995_PRA51-1159}.

At the local density approximation level, it is found that the relativistic LDA gives bigger errors than the non-relativistic counterpart. This comes from the fact that, in the non-relativistic LDA, there is a cancellation of errors between the exchange and the correlation energy, which does not hold in the relativistic framework \cite{Engel1996_PRA53-1367}. In Ref. \cite{Shadwick1989_CPC54-95}, the optimized-potential-method has been extended to the relativistic framework on the longitudinal no-pair level. Later on, this method was applied to the study of atoms, it was found that the relativistic optimized-potential-method agrees well with the relativistic Hartree-Fock calculation and relativistic effects become drastic for high-$Z$ atoms \cite{Engel1995_PRA52-2750}. Though it achieved an accurate description, the optimized-potential-method is limited by its computational cost, because, besides the kinetic energy, the exchange energy is also treated in orbital representations. Later, the relativistic generalized gradient approximation has been developed, and a good description could be achieved with rather small computational cost \cite{Engel1996_PRA53-1367}. For more recent developments and investigations of relativistic density functional theory in atomic systems, see Ref. \cite{Engel2011_DFT8-351}.

\subsection{Challenges in nuclear DFT}

DFT in nuclear physics was introduced in the early 1970s in a slightly different way.
First of all, there is an essential difference between the DFTs of Coulombic and nuclear systems..
The Hohenberg-Kohn theorem was derived under the assumption that there is an external potential $U_{\rm ext}$, which, in atoms, is the Coulomb potential provided by the nucleus.
However, the nucleus itself is a self-bound system, and such an external potential does not exist .
It is still not clear whether the Hohenberg-Kohn theorem holds in nuclei as an exact statement, or only as a reasonable approximation \cite{Giraud2008_PRC77-014311}.
Recent discussions on the Hohenberg-Kohn theorem for a self-bound wave-packet state can be found in the reviews~\cite{Nakatsukasa2012_PTEP01A207,Nakatsukasa2016_RMP88-045004}.
On the other hand, the shell structure in nuclei pointed out by Goeppert-Mayer and Jensen \cite{Mayer1949_PR75-1969,Haxel1949_PR75-1766} indicates that there should be an effective mean field, in which eventually the Hohenberg-Kohn theorem and the Kohn-Sham scheme could be applied at least in a very good approximation.

The second challenge comes from the fact that the microscopic origin of such an effective mean field is not apparent at all. Because of the existence of a strong repulsive core in the nucleon-nucleon interaction, or in general because of the non-perturbative nature of this interaction, the attempt of deriving the mean field using Hartree-Fock theory or using the decomposition (\ref{eq:decomp}) into Hartree and exchange-correlation terms fails in nuclear physics, whereas  it is very successful in Coulombic systems.

It was only the work of Brueckner \textit{et al.} \cite{Brueckner1954_PR95-217} that gave a better understanding, why the concept of a mean field works so well in nuclei.
They treated the scattering process between two nucleons in the nuclear medium similar to the case in the vacuum summing up the ladder diagrams in the Born series.
Furthermore, they took into account the Pauli principle, that two nucleons in the nuclear medium can only be scattered to the unoccupied states above the Fermi levels.
The obtained scattering matrix is the so-called $G$-matrix, which is treated as an effective interaction between two nucleons in the nuclear medium and gives a reasonable description for nuclear systems even at the Hartree-Fock level.
This is the BHF theory mentioned in the introduction.

BHF calculations in nuclear matter show, that the $G$-matrix is strongly density dependent and that Hartree-Fock calculations with this effective interactions failed \cite{Negele1970_PRC1-1260}.
Based on this observation, Vautherin and Brink \cite{Vautherin1972_PRC5-626} adopted a simple phenomenological interaction introduced by Skyrme \cite{Skyrme1958_NP9-615} containing a zero-range three-body term.
At the mean-field level, this leads to an effective density-dependent two-body interaction.
By an appropriate fit, they were able to reproduce not only the experimental binding energies but also the radii at the same time.
This illustrated the importance of density dependence in effective nuclear interactions.
Therefore, from the beginning, even in many cases today, nuclear DFT does not start with a density functional, but rather with a density-dependent interaction $V[\rho]$.
The underlying relation is clear as the corresponding functional can be derived from the Hartree-Fock approximation,
\begin{equation}
  E[\rho] = \langle \Phi|T + V[\rho]|\Phi \rangle,
\end{equation}
where $|\Phi\rangle$ is a single Slater determinant and $T$ is the kinetic operator.

Besides the above difference, the way to derive the functional is also quite different in atomic and nuclear systems.
Because the interaction in atoms, the Coulomb force, is well known and has a simple form, one can derive the density functional microscopically without any additional parameters \cite{Perdew2004_LNP620-269}.
On the contrary it is very difficult for nuclear systems because of the complexity of nuclear force \cite{Epelbaum2009_RMP81-1773, Machleidt2011_PR503-1}.
Essential degrees of freedom in nuclear interaction include the spin, isospin, pairing, and relativistic degrees of freedom.
Furthermore, the components of the nuclear interaction include not only the central force but also the spin-orbit force, tensor terms, and so on.

\subsection{Current status in nuclear density functional}

Because of the above challenges, all current successful nuclear density functionals are based on phenomenological parameters, which are determined by fitting to selected experimental ground-state (and excited-state) properties.
The form of the functional, i.e. the structure of the various terms, is chosen based on the consideration of symmetries and simplicity, for example, the translational and rotational symmetry in the non-relativistic case, and the Poincare symmetry (including the Lorentz invariance) in the relativistic case.

The importance of some terms, such as the central force and the spin-orbit force, is easy to be seen from the experimental observables.
This might not be the case for other terms, as  for example, for the tensor force.
It is known that the tensor force is an essential component in the bare nucleon-nucleon interaction from the scattering data and deuteron properties \cite{Machleidt1989_ANP19-189}.
However, the role of the tensor force in nuclear density functional  is much less clear \cite{Sagawa2014_PPNP76-76}.
In non-relativistic functionals, the zero-range tensor terms can be easily included \cite{Stancu1977_PLB68-108}, but their strengths are hard to be adjusted as they have little influences on the binding energies or charge radii.
Even though the tensor terms have strong influences on the single-particle energies, according to the Kohn-Sham concept, in the framework of DFT these single-particle energies are purely artificial quantities and should not be used in a fit. Indeed, the single-particle energies in nuclei are usually fragmented, and thus beyond-mean-field effects are crucial, such as particle-vibrational coupling \cite{Colo1994_PRC50-1496, Litvinova2006_PRC73-044328}.
Therefore, it is difficult to take into account such effects in the fitting both conceptually and technically.

\subsection{Towards \textit{ab initio} nuclear density functional }

In such a case mentioned above, the derivation of the density functional  from \textit{ab initio} calculations is important although such a derivation for nuclear systems is much more difficult than its counterpart for atomic systems and it is still at the infancy stage.

Following the concepts in Coulombic systems, firstly the nuclear matter problem is solved, where the volume term can be studied.
Then one could try to derive exact results for gradient terms, surface terms, and so on.
Along with this direction, Fayans proposed to start with a density functional for nuclear matter from an \textit{ab initio} calculation and to add on top it surface, Coulomb and spin-orbit terms, reducing in this way the total number of phenomenological parameters \cite{Fayans1998_JETPL68-169}.
This idea has been realized in Ref.~\cite{Baldo2008_PLB663-390} together with further developments \cite{Vinas2009_IJMPE18-935,Robledo2010_PRC81-034315,Baldo2010_JPG37-064015,Baldo2013_PRC87-064305,Baldo2017_PRC95-014318}, where the functional is based on the results of microscopic nuclear and neutron matter calculations with the Brueckner $G$-matrix at various densities.
With only four to five adjustable parameters, the authors have shown that it is enough to reproduce nuclear binding energies and radii with the same quality as obtained with other commonly used effective interactions.
Progress has also been made for the relativistic density functional.
In Ref.~\cite{RocaMaza2011_PRC84-054309}, the scalar-isovector term, which in most cases has been neglected for simplicity, is included in the functional DD-ME$\delta$ by fitting to the isovector effective mass difference derived from the RBHF theory in nuclear matter. Together with the constraints from relativistic and non-relativistic BHF results in nuclear matter, only four phenomenological parameters in DD-ME$\delta$ need to be adjusted to experimental data in finite nuclei. Of course, at this moment, it does not necessarily mean that this functional is superior to other relativistic density functionals.  For example, the DD-ME$\delta$ functional does not predict octupole deformation in actinide nuclei which are known to be octupole deformed \cite{Agbemava2016_PRC93-044304}, and it gives significantly lower inner fission barriers in super-heavy nuclei \cite{Agbemava2015_PRC92-054310}.

These promising results demonstrate the feasibility in deriving a microscopic \textit{ab initio} density functional for nuclei.
However, the concept to derive functionals only from the properties of infinite nuclear matter cannot teach us much in cases, where one wants to describe effects which do not show up in nuclear matter calculations, e.g., the influence of tensor terms in spin non-saturated systems. For that, we need \textit{ab initio} calculations for finite nuclei.
In particular, for covariant density functional theory, we need the \textit{ab initio} calculations for finite nuclei in a relativistic framework.
Therefore, in the following Section~\ref{Sect:3}, we will introduce in detail the progress of RBHF theory for both nuclear matter and finite nuclei, and in Section~\ref{Sect:4} we will discuss explicitly how the RBHF results are able to guide the developments of covariant DFT.

\section{Relativistic Brueckner-Hartree-Fock Theory and Applications}\label{Sect:3}

\subsection{BHF formalism}\label{Sect:3.1}


Reviews for non-relativistic Brueckner theory can be found in Refs. \cite{Day1967_RMP39-719,Rajaraman1967_RMP39-745,Brandow1967_RMP39-771,Bethe1971_ARNS21-93,Sprung1972_ANP5-225,Kohler1975_PR18-217}.
%
For a nuclear system composed of $A$ nucleons, the Hamiltonian including a one-body kinetic-energy term $T$ and a two-body interaction term $V$ reads as,
\begin{equation}
  H = T + V = \sum_{k'k} \langle k'|T|k \rangle a_{k'}^\dagger a_k^{~} + \frac{1}{2}\sum_{k'l'kl} \langle k'l'|V|kl \rangle a_{k'}^\dagger a_{l'}^\dagger a_l^{~} a_k^{~},
\end{equation}
where $a_k^\dagger$ and $a^{}_k$ form a complete set of creation and annihilation operators, referring to the state $|k\rangle$ with the single-particle wave function $\psi_k(\mathbf{r})$, and the one-body and two-body matrix elements are
\begin{align}
  \langle k'|T|k \rangle &= \int d^3r \psi_{k'}^\dagger(\mathbf{r}) \left( -\frac{\nabla^2}{2M}\right) \psi_k(\mathbf{r}), \\
  \langle k'l'|V|kl \rangle &= \int d^3r_1 d^3r_2 \psi_{k'}^\dagger(\mathbf{r}_1) \psi_{l'}^\dagger(\mathbf{r}_2) V(\mathbf{r}_1,\mathbf{r}_2) \psi_k(\mathbf{r}_1)\psi_l(\mathbf{r}_2).
\end{align}


Within BHF theory, the ground state of the nuclear system is given by the Hartree-Fock approximation,
\begin{equation}\label{Eq:3.1.Phi0}
  |\Phi_0 \rangle = \prod_k^A a_k^\dagger |0\rangle,
\end{equation}
where $|0\rangle$ is the vacuum and $a_k|0\rangle \equiv 0$. The total energy of the system can be derived as
\begin{equation}\label{Eq:3.1.EHF}
  E^{\rm HF} = \langle \Phi_0|H|\Phi_0 \rangle = \sum_{k}^A \langle k|T|k \rangle + \frac{1}{2}\sum_{kl}^A \langle kl|\bar{V}|kl \rangle,
\end{equation}
with the antisymmetrized two-body matrix element $\langle kl|\bar{V}|kl \rangle = \langle kl|V|kl \rangle - \langle kl|V|lk \rangle$.
The single-particle states $|k\rangle$ satisfy the HF equation
\begin{equation}\label{Eq:3.1.HFeq}
  (T+U^{\rm HF}) |k\rangle = e_k |k\rangle,
\end{equation}
where $e_k$ is the eigenenergy of the state $|k\rangle$, and $U^{\rm HF}$ is the HF single-particle potential
\begin{equation}\label{Eq:3.1.UHF}
  \langle k'|U^{\rm HF}|k \rangle = \sum_l^A \langle k'l|\bar{V}|kl \rangle.
\end{equation}
However, as the bare nuclear interaction has a very strong repulsive core \cite{Jastrow1951_PR81-165}, it cannot be used directly in the HF approximation. In modern \textit{ab initio} calculations, the renormalization group method has been widely used to decouple the low-energy physics, which is of the interest, from the high-energy degrees of freedom, which is induced by the repulsive core. This greatly simplifies the numerical problem and benefits much the modern \textit{ab initio} calculations \cite{Bogner2010_PPNP65-94}.


In BHF theory, the bare interaction $V$ is replaced by an effective interaction, the $G$-matrix in the nuclear medium. It satisfies the Bethe-Goldstone equation,
\begin{equation}\label{Eq:3.1.BG}
  \langle k'l'|\bar{G}(W)|kl \rangle = \langle k'l'|\bar{V}|kl \rangle + \frac{1}{2}
  \sum_{mn} \langle k'l'|\bar{V}|mn \rangle \frac{Q(m,n)}{W-e_m-e_n} \langle mn|\bar{G}(W)|kl \rangle,
\end{equation}
where $W$ is the starting energy, and $e_m,e_n$ are the single-particle energies of the two particles in the intermediate states.
The Pauli operator $Q(m,n)$ forbids the two particles being scattered to the already occupied states below the Fermi surface, i.e., $Q(m,n) = 1$ for $e_m > e_F$ and $e_n > e_F$, otherwise $Q(m,n) = 0$.
Thus, in the BHF framework, the ground-state energy can be expressed as,
\begin{equation}\label{Eq:3.1.EBHF}
  E^{\rm BHF} = \sum_{k}^A \langle k|T|k \rangle + \frac{1}{2}\sum_{kl}^A \langle kl|\bar{G}(W=e_k+e_l)|kl \rangle.
\end{equation}
The basis and correctness of the replacement of the bare interaction with an effective interaction of this form in the nuclear medium is rooted in many-body perturbation theory, or more specifically, in the Goldstone expansion \cite{Goldstone1957_PRSA239-267}.
Using Goldstone diagrams to represent the contributions of the full perturbation series, such a replacement is equivalent to consider the series of ladder diagrams in all orders of of $V$.
Numerically, the matrix elements of $G$ are much smaller than those of $V$. Therefore $G$-matrix is more suitable to be applied in perturbation theory. Already at the first order the expression (\ref{Eq:3.1.EBHF}) gives most of the contributions \cite{Rajaraman1967_RMP39-745}.

Following the replacement of $V$ by $G$, the definition of the HF single-particle potential in Eq.~(\ref{Eq:3.1.UHF}) has to be modified. Unlike the expression of the total energy in Eq.~(\ref{Eq:3.1.EBHF}) where the starting energy is well defined, the single-particle potential in Eq.~(\ref{Eq:3.1.UHF}) is an auxiliary quantity from the beginning and its definition depends on specific choices.
Even if the form of the HF single-particle potential is used, i.e.,
\begin{equation}\label{Eq:UBHF}
  \langle k|U^{\rm BHF}|k \rangle = \sum_{l}^A \langle kl|\bar{G}(W)|kl \rangle,
\end{equation}
it still remains a problem how to choose the starting energy $W$.
Fortunately, it was proven \cite{Bethe1963_PR129-225} that for occupied states $k$ a large amount of higher-order diagrams cancel with each other, if the starting energy is chosen on shell: $W = e_k + e_l$. This is good for a faster convergence and improves the BHF approximation.

However, there is no exact guideline for the choice of the starting energy when $k$ is an unoccupied state.
At this point, there exists an uncertainty in the BHF scheme, and different recipes have been proposed, e.g., the \textit{gap choice} ($U = 0$) \cite{Brandow1966_PR152-863}, the \textit{continuous choice} (the same form as the one when $k$ is occupied) \cite{Jeukenne1976_PR25-83}, or choices in between the above two.
In principle, the auxiliary single-particle potential is introduced to make the convergence of the perturbation expansion faster.
If the expansion is carried out to high enough order, the potential itself does not affect the final result.
In Ref.~\cite{Song1998_PRL81-1584} it is shown that in the next order beyond the BHF, i.e., the three hole-line expansion, the equation of state for nuclear matter near or below the saturation density calculated with the gap choice and the continuous choice agree with each other. Therefore, it seems reasonable to use a choice in between the gap choice and continuous choice.

Summarizing these discussions, the single-particle potential in the BHF framework can be chosen as
\begin{equation}\label{Eq:3.1.UBHF}
  \langle k'|U^{\rm BHF}|k\rangle=
  \begin{cases}
  \frac{1}{2}\sum_{l=1}^A \langle k'l|\bar{G}(e_{k'}+e_l)+ \bar{G}(e_k+e_l)|kl\rangle,
   & e_{k'}, e_k \leq e_F \\
  \sum_{l=1}^A \langle k'l|\bar{G}(e_k+e_l)|kl\rangle,
  & e_{k'} \leq e_F,~e_k > e_F  \\
  \sum_{l=1}^A \langle k'l|\bar{G}(e'+e_l)|kl\rangle,
  & e_{k'}, e_{k} > e_F,
  \end{cases}
\end{equation}
where $e'$ is uncertain and its choice for finite nuclei has been discussed both in the non-relativistic BHF \cite{Baranger1969_Varenna40-511,Davies1969_PR177-1519} and relativistic BHF \cite{Shen2017_PRC96-014316,Shen2018_PLB781-227} theories.
In nuclear matter, only the diagonal matrix elements of $U$ exist, that is $k' = k$ in the above expression.
Whereas the off-diagonal matrix elements appear in finite nuclei, and the related discussion can be seen in Ref.~\cite{Baranger1969_Varenna40-511}.

\subsection{RBHF formalism and its application for nuclear matter}\label{Sect:3.2}



\subsubsection{RBHF formalism}

In the relativistic Brueckner-Hartree-Fock framework, the nucleon-nucleon (NN) interaction is of covariant form.
Using the one-boson-exchange Bonn interaction as an example, the nuclear interaction is mediated by six bosons \cite{Machleidt1989_ANP19-189},
\begin{align}
  \mathscr{L}_{\rm NNs} &= g_s \bar{\psi}\psi\phi^{(s)}, \quad
  \mathscr{L}_{\rm NNpv} = -\frac{f_{ps}}{m_{ps}} \bar{\psi}\gamma^5\gamma^\mu
  \psi \partial_\mu \phi^{(ps)},  \notag \\
  \mathscr{L}_{\rm NNv} &= -g_v \bar{\psi}\gamma^\mu\psi\phi_\mu^{(v)} - \frac{
  f_v}{4M}\bar{\psi}\sigma^{\mu\nu}\psi \left( \partial_\mu\phi_\nu^{( v)}
  - \partial_\nu \phi_\mu^{(v)} \right), \quad\quad \mu,\nu = 0,1,2,3,
\label{Eq:3.2.Lagrangian}
\end{align}
including the scalar (s) $(\sigma,\delta)$, the pseudovector (pv) $(\eta,\pi)$, and the vector (v) $(\omega,\rho)$ meson-nucleon couplings,
and each pair contains isoscalar or isovector character.
For the isovector mesons, an additional isospin operator $\vec{\tau}$ is present.
The nucleon field is denoted by $\psi$, and operator $\sigma^{\mu\nu} = \frac{i}{2}[\gamma^\mu,\gamma^\nu]$.
Without specification, bold letters will be used for three-vectors (space only) such as $\mathbf{x},\mathbf{q},\bm{\gamma}$, and usual letters will be used for four-vectors (time and space) such as $x,q,\gamma$.
By default, the Einstein summation convention is adopted. For the isovector couplings, the densities or currents in Eqs. (\ref{Eq:3.2.Lagrangian}) are replaced by the isovector ones, namely $\bar\psi\vec\tau\Gamma\psi$ with $\Gamma = 1, \gamma^\mu, \gamma^5\gamma^\mu$ and $\sigma^{\mu\nu}$.

In free space, the above Lagrangians lead to the following OBE two-body matrix elements (or OBE amplitudes) in the plane-wave helicity representation \cite{Machleidt1989_ANP19-189}
\begin{subequations}\label{Eq:3.2.VOBE}\begin{align}
  \langle \mathbf{q}'\lambda_1'\lambda_2'|V_{s}|\mathbf{q}\lambda_1\lambda_2 \rangle
  &= -g_s^2 \bar{u}(\mathbf{q}',\lambda_1') u(\mathbf{q},\lambda_1)
  \frac{1}{(\mathbf{q}'-\mathbf{q})^2 + m_{s}^2 }
  \bar{u}(-\mathbf{q}',\lambda_2') u(-\mathbf{q},\lambda_2), \label{Eq:3.2.VOBEs} \\
  \langle \mathbf{q}'\lambda_1'\lambda_2'|V_{pv}|\mathbf{q}\lambda_1\lambda_2 \rangle
  &= \frac{f_{ps}^2}{m_{ps}^2} \bar{u}(\mathbf{q}',\lambda_1') \gamma^5\gamma^\mu
  i(q'-q)_\mu u(\mathbf{q},\lambda_1)
  \frac{1}{(\mathbf{q}'-\mathbf{q})^2 + m_{ps}^2 }
  \bar{u}(-\mathbf{q}',\lambda_2') \gamma^5\gamma^\mu i(q'-q)_\mu u(-\mathbf{q},\lambda_2), \label{Eq:3.2.VOBEpv} \\
  \langle \mathbf{q}'\lambda_1'\lambda_2'|V_{v}|\mathbf{q}\lambda_1\lambda_2 \rangle
  &= \left[ g_v^2 \bar{u}(\mathbf{q}',\lambda_1') \gamma^\mu u(\mathbf{q},\lambda_1)
  +\frac{f_v}{2M} \bar{u}(\mathbf{q}',\lambda_1') \sigma^{\mu\nu} i(q'-q)_\nu u(\mathbf{q},\lambda_1) \right]
  \frac{1}{(\mathbf{q}'-\mathbf{q})^2 + m_{v}^2 } \notag \\
  &~~~\times \left[ g_v^2 \bar{u}(-\mathbf{q}',\lambda_2') \gamma_\mu u(-\mathbf{q},\lambda_2)
  +\frac{f_v}{2M} \bar{u}(-\mathbf{q}',\lambda_2') \sigma_{\mu\nu} i(q'-q)^\nu u(-\mathbf{q},\lambda_2) \right] ,\label{Eq:3.2.VOBEv}
\end{align}\end{subequations}
where the reference frame is chosen at zero center-of-mass momentum.
The momenta for two initial (final) states are $\mathbf{p}_1 = \mathbf{q}$ and $\mathbf{p}_2 = -\mathbf{q}$ ($\mathbf{p}_1' = \mathbf{q}'$ and $\mathbf{p}_2' = -\mathbf{q}'$), respectively.
The Dirac spinor with positive energy is labelled as $u$, which in the helicity representation is
\begin{equation}\label{Eq:3.2.uHelicity}
  u(\mathbf{q},\lambda) = \sqrt{\frac{E_{\mathbf{q}}+M}{2M}}
  \left(\begin{array}{c}
  1 \\ \frac{2\lambda |\mathbf{q}|}{E_{\mathbf{q}}+M}
  \end{array}\right)
  |\lambda\rangle,
\end{equation}
with $E_{\mathbf{q}}^2 = M^2 + \mathbf{q}^2$.
Notice that with this form, the Dirac spinor is normalized covariantly as
  $\bar{u}(\mathbf{q},\lambda)u(\mathbf{q},\lambda) = 1$.
This is convenient for solving the scattering equation as the two-body matrix elements in Eqs.~(\ref{Eq:3.2.VOBE}) are Lorentz scalars.
For calculating physical observables in many-body theory, we should have creation and annihilation operators obeying Fermi commutation relations and the normalization condition should be $u^\dagger u = 1$, and the spinor becomes 
\begin{equation}\label{Eq:3.2.udagger}
  u(\mathbf{q},\lambda) = \sqrt{\frac{E_{\mathbf{q}}+M}{2E}}
  \left(\begin{array}{c}
  1 \\ \frac{2\lambda |\mathbf{q}|}{E_{\mathbf{q}}+M}
  \end{array}\right)
  |\lambda\rangle.
\end{equation}


In the Bonn interaction, a form factor is applied to each meson-nucleon vertex \cite{Machleidt1989_ANP19-189}:
\begin{equation}\label{Eq:3.2.FormFac}
  \frac{\Lambda_\phi^2 - m_\phi^2}{\Lambda_\phi^2 + (\mathbf{q}'-\mathbf{q})^2},
\end{equation}
where $m_\phi$ is the mass for meson $\phi$.
The cut-off parameters $\Lambda_\phi$, together with the coupling strengths $g_\phi$, are determined by fitting to the NN scattering data and deuteron properties.
Three examples of nuclear interactions, Bonn A,  Bonn B, and Bonn C, have been obtained with the Thompson equation \cite{Machleidt1989_ANP19-189}, in which the average nucleon mass $M = 938.926$ MeV is used. For the pseudoscalar mesons, the coupling strength of the pseudovector channel ($f_{ps}$) is related to the pseudoscalar one ($g_{ps}$) by
\begin{equation}
  f_{ps} = g_{ps} \frac{m_{ps}}{2M}.
\end{equation}

In the relativistic framework, the two-nucleon scattering equation is the Bethe-Salpeter equation \cite{Salpeter1951_PR84-1232}, which can be written in the operator form as
\begin{equation}\label{Eq:3.2.BSeq}
  \mathscr{M} = \mathscr{V} + \mathscr{V}\mathscr{G}\mathscr{M},
\end{equation}
where $\mathscr{M}$ is the relativistic two-nucleon scattering amplitude and $\mathscr{G}$ is the relativistic free two-nucleon propagator.
In this equation, $\mathscr{V}$ is the infinite sum of all irreducible diagrams and is usually replaced by the one-boson exchange contribution within the ``ladder'' approximation, $\mathscr{V}\approx V_{\rm OBE}$ (the subscript ``OBE'' will be omitted without causing ambiguity).

The Bethe-Salpeter equation (\ref{Eq:3.2.BSeq}) is a four-dimensional integral equation, and it is very difficult to solve.
In practice three-dimensional reductions are used, and such reductions still remain covariant and the relativistic elastic unitarity is satisfied.
Since unitarity does not uniquely determine the Green's function, the three-dimensional reduction is not unique and in principle an infinite number of choices exist \cite{Yaes1971_PRD3-3086}.
Typically a propagator $g$ is introduced,
\begin{subequations}
\begin{align}
  \mathscr{M} =& \mathscr{W} + \mathscr{W} g \mathscr{M}, \label{Eq:3.2.BS3a}\\
  \mathscr{W} =& V + V (\mathscr{G}-g) \mathscr{W},
\end{align}
\end{subequations}
so that the scattering equation (\ref{Eq:3.2.BS3a}) involves only a three-dimensional integration.
With the approximation $\mathscr{W}\approx V$, Eq.~(\ref{Eq:3.2.BS3a}) becomes
\begin{equation}
  \mathscr{M} = V + Vg\mathscr{M},
\end{equation}
which can be easily solved in momentum space.

Take the Thompson choice as an example, one finds the Thompson equation \cite{Thompson1970_PRD1-110} (the spin indices have been omitted) as,
\begin{equation}\label{Eq:3.2.Thompson}
  \mathscr{M}(\mathbf{q}',\mathbf{q}|\mathbf{P}) = V(\mathbf{q}',\mathbf{q})
  + \int \frac{d^3k}{(2\pi)^3} V(\mathbf{q}',\mathbf{k})
  \frac{M^2}{E_{\mathbf{k}} E_{\mathbf{P}/2+\mathbf{k}} }
  \frac{1}{2E_{\mathbf{q}} - 2E_{\mathbf{k}} + i\epsilon } \mathscr{M}(\mathbf{k},\mathbf{q}|\mathbf{P}),
\end{equation}
where $\mathbf{P}$ is the total momentum of two particles, and $\mathbf{q}', \mathbf{k}$, and $\mathbf{q}$ are the relative momenta of the initial, intermediate, and final states respectively.
For the OBE interaction, the expressions of the two-body interaction matrix element $V(\mathbf{q}',\mathbf{q})$ are given in Eqs.~(\ref{Eq:3.2.VOBE}).
By solving the Thompson equation (\ref{Eq:3.2.Thompson}), the scattering amplitudes and phase shifts can be calculated and compared with experimental data.
Three examples for the bare nuclear interaction, Bonn A,  Bonn B, and Bonn C, have been provided by Ref. \cite{Machleidt1989_ANP19-189}.

The Hamiltonian density is obtained by a Legendre transformation, and a three-dimensional integration leads to the total Hamiltonian $H$ in the stationary case,
\begin{equation}
  \mathscr{H} = \sum_i \frac{\partial \mathscr{L}}{\partial(\partial^0\phi_i)}
  \partial^0\phi_i - \mathscr{L}, \quad H = \int d^3r \mathscr{H}(\mathbf{r}),
\end{equation}
where $\phi_i$ represents the nucleon field $\psi$, the meson fields $\phi$, and the photon field $A$.

Eliminating the meson fields, the many-body Hamiltonian for nuclear systems is derived as \cite{Brockmann1978_PRC18-1510},
\begin{equation}
H= \int \mathrm{d}^{3}r \hat{\bar{\psi}}\left( -i\bm{\gamma }\cdot \bm{\nabla}+M\right) \hat{\psi}
 +\frac{1}{2}\sum_{\phi }\int \mathrm{d}^{3}r_{1}\mathrm{d}^{3}r_{2}
 \hat{\bar{\psi}}(\mathbf{r}_{1})\Gamma _{\phi }^{(1)}\hat{\psi}(\mathbf{r}_{1})
 D_{\phi }(\mathbf{r}_{1},\mathbf{r}_{2})
 \hat{\bar{\psi}}(\mathbf{r}_{2})\Gamma _{\phi }^{(2)}\hat{\psi} (\mathbf{r}_{2}),
\end{equation}
where the field operators $\hat{\psi}$ (and $\hat{\bar{\psi}} = \hat{\psi}^\dagger \gamma^0$) have been hatted in order to distinguish them from the single-particle wave functions, and $\Gamma _{\phi }^{(1)}$ and $\Gamma _{\phi }^{(2)}$ are the interaction vertices for particles at the coordinates $\mathbf{r}_{1}$ and $\mathbf{r}%
_{2}$, respectively,
\begin{subequations}\label{Eq:3.2.gamma12}
\begin{align}
\Gamma _{s}=& g_{s}, \\
\Gamma _{pv}=& \frac{f_{ps}}{m_{ps}}\gamma ^{5}\gamma ^{i}\partial _{i}, \\
\Gamma _{v}^{\mu }=& g_{v}\gamma ^{\mu }+\frac{f_{v}}{2M}\sigma ^{i\mu
}\partial _{i}.
\end{align}%
\end{subequations}
In Minkowski space, the meson propagators $D_{\phi }(x_1,x_2)$ are the retarded solutions of the
Klein-Gordon equations,
\begin{equation}
D_{\phi }(x_1,x_2)=\pm \int \frac{d^{4}q}{(2\pi )^{4}}\frac{1}{m_{\phi}^{2}-q^{2}}e^{-iq\cdot (x_1-x_2)},
\end{equation}%
where $q$ is the four-momentum transfer between the two particles, and the sign $-$
holds for the scalar and pseudoscalar mesons and $+$ for the
vector ones. The dependence on the zero-component of momentum transfer $q_0$ (energy) reflects the retardation of the interaction.
For the Bonn interaction in Ref.~\cite{Machleidt1989_ANP19-189}, this effect was deemed to
be small and was ignored from the beginning. In this way, the meson propagators are just Yukawa
functions:
\begin{equation}
D_{\phi }(\mathbf{r}_1,\mathbf{r}_2)=
\pm \int \frac{d^{3}q}{(2\pi )^{3}}\frac{1}{m_{\phi }^{2}+\mathbf{q}^{2}}e^{i\mathbf{q}\cdot(\mathbf{r}_1-\mathbf{r}_2)}
=\pm \frac{1}{4\pi}\frac{e^{-m_{\phi }|\mathbf{r}_1-\mathbf{r}_2|}}{|\mathbf{r}_1-\mathbf{r}_2|}.
\end{equation}
Notice that, with the form factor in Eq.~(\ref{Eq:3.2.FormFac}), the meson
propagators are no longer simple Yukawa functions.

The nucleon-field operators $\hat{\psi} (\mathbf{r})$ and $\hat{\psi} ^{\dag }(\mathbf{r})$ can be expanded on a static relativistic basis $|k\rangle $,
\begin{equation}
\hat{\psi} ^{\dag }(\mathbf{r})=\sum_{k}\psi _{k}^{\dag }(\mathbf{r})a_{k}^{\dag },\text{
\ \ \ \ }\hat{\psi} (\mathbf{r})=\sum_{k}\psi _{k}(\mathbf{r})a_{k}^{{}},
\end{equation}
where $a_{k}^{\dagger }$ and $a^{}_{k}$ form a complete set of creation and
annihilation operators of nucleons, referred to the state $|k\rangle $ with positive or negative energies, and $\psi _{k}(\mathbf{r})$ is the
corresponding Dirac spinor. {The quantum number $k$ characterizing the state $|k\rangle$ contains also the
isospin $t=n,\,p$ for neutrons and protons}. Then the Hamiltonian for the nuclear
system in second quantized form reads,
\begin{equation}\label{Eq:3.2.H}
  H=\sum_{k^{\prime }k}\langle k^{\prime }|T|k\rangle a_{k^{\prime }}^{\dagger
  }a_{k^{{}}}^{{}}+\frac{1}{2}\sum_{k^{\prime }l^{\prime }kl}\langle
  k^{\prime }l^{\prime }|V|kl\rangle a_{k^{\prime }}^{\dagger }a_{l^{\prime }}^{\dagger
  }a_{l}^{{}}a_{k}^{{}},
\end{equation}
where the matrix elements are given by
\begin{subequations}\label{Eq:3.2.meH}\begin{align}
  \langle k^{\prime }|T|k\rangle & =\int d^{3}r\,\bar{\psi}_{k^{\prime }}(\mathbf{r})\left( -i%
  \bm{\gamma}\cdot \nabla +M\right) \psi _{k}(\mathbf{r}), \label{Eq:3.2.meT} \\
  \langle k^{\prime }l^{\prime }|V_{\phi }|kl\rangle & =\int
  d^{3}r_{1}d^{3}r_{2}\,\bar{\psi}_{k^{\prime }}(\mathbf{r}_{1})\Gamma _{\phi }^{(1)}\psi_{k}(\mathbf{r}_{1})
  D_{\phi }(\mathbf{r}_{1},\mathbf{r}_{2})\bar{\psi}_{l^{\prime }}(\mathbf{r}_{2})\Gamma_{\phi }^{(2)}\psi _{l}(\mathbf{r}_{2}) \label{Eq:3.2.meV}.
\end{align}\end{subequations}
The two-body interaction $V$ contains the contributions from various meson (photon) coupling channels $\phi$.
The indices $k,l$ run over an complete basis of Dirac spinors with positive and negative energies, for instance, the plane wave states $u(\mathbf{k},s)$ and $v(\mathbf{k},s)$ in momentum space~\cite{Itzykson1980} or the eigensolutions of a Dirac equation with Woods-Saxon potentials \cite{Koepf1991_ZPA339-81,Zhou2003_PRC68-034323}.
The two-body interaction matrix elements for the OBE model in the plane-wave helicity basis have been given in Eqs.~(\ref{Eq:3.2.VOBE}), and the retardation is ignored in the Thompson choice so that $(q'-q)_0 = 0$ in these equations.

\subsubsection{Nuclear matter}\label{Sect:3.2.2}

In the relativistic scheme, there are solutions with both positive and negative single-particle energies.
The \textit{no-sea approximation} \cite{Ring1996_PPNP37-193} is usually adopted, i.e. the ground state of nuclear system is taken as a simple Slater determinant composed of the single-particle states in the Fermi sea.
The nucleon Dirac equation for nuclear matter reads
\begin{equation}\label{Eq:3.2.RHFeq}
  (\bm{\gamma}\cdot \mathbf{p} + M + U) u(\mathbf{p},s) = \gamma^0 E_{\mathbf{p}} u(\mathbf{p},s).
\end{equation}
The single-particle energy in momentum space is represented by a capital letter ``$E_{\mathbf{p}}$'', which is equivalent to the lower case ``$e_k$'' in Eq.~(\ref{Eq:3.1.HFeq}).
Another convention is the single-particle potential $U$, which is usually referred as \textit{self-energy} labelled by $\Sigma$.
Although Eq.~(\ref{Eq:3.2.RHFeq}) does not distinguish solutions with positive or negative energy, usually the spinor $u(\mathbf{p},s)$ is used to represent a positive-energy solution and the spinor $v(\mathbf{p},s)$ a solution with negative energy.

In free space where $U = 0$, the plane-wave solutions for Eq.~(\ref{Eq:3.2.RHFeq}) in helicity representation are given in Eq.~(\ref{Eq:3.2.uHelicity}).
For nuclear matter with good parity, time-reversal invariance, and hermiticity, the single-particle potential has in general the form \cite{Serot1986_ANP16-1}
\begin{equation}\label{Eq:3.2.UDef}
  U(\mathbf{p}) = U_s(\mathbf{p}) + \gamma^\mu U_\mu(\mathbf{p}) = U_s(\mathbf{p}) + \gamma^0 U_0(\mathbf{p}) - \bm{\gamma}\cdot \hat{\mathbf{p}}U_v(\mathbf{p}).
\end{equation}
Introducing the starred quantities,
\begin{equation}\label{Eq:3.2.MEpstar}
  M^*(\mathbf{p}) = M + U_s(\mathbf{p}), \quad
  E^*(\mathbf{p}) = E - U_0(\mathbf{p}), \quad
  \mathbf{p}^* = \mathbf{p} - \hat{\mathbf{p}}U_v(\mathbf{p}),
\end{equation}
where ${E^*}^2 = {M^*}^2 + {\mathbf{p}^*}^2$ and $\hat{\mathbf{p}}$ is the unit vector of $\mathbf{p}$, the positive-energy solution of the Dirac equation (\ref{Eq:3.2.RHFeq}) is
\begin{equation}\label{Eq:3.2.uRHF}
  u(\mathbf{p},s) = \sqrt{\frac{E_{\mathbf{p}}^*+M_{\mathbf{p}}^*}{2E_{\mathbf{p}}^*}}
  \left(\begin{array}{c}
  1 \\ \frac{\bm{\sigma}\cdot \mathbf{p}^*}{E_{\mathbf{p}}^*+M_{\mathbf{p}}^*}
  \end{array}\right)
  \chi_s.
\end{equation}

Unlike the covariantly normalized spinor in Eq.~(\ref{Eq:3.2.uHelicity}), the spinor in Eq.~(\ref{Eq:3.2.uRHF}) is normalized in the standard way $u^\dagger u = 1$.
Different from the free space, the spinors in nuclear matter (\ref{Eq:3.2.uRHF}) depend on the single-particle potentials (self-energies). They are often called the ``dressed'' spinors.
Changing the covariantly normalized spinor to the standardly normalized spinor will give an extra factor $\sqrt{E_{\mathbf{p}}^*/M_{\mathbf{p}}^*}$.

The RHF single-particle potential in nuclear matter can be obtained from Eq. (\ref{Eq:3.1.UHF}) as
\begin{equation}
  U_{s_1t_1}(\mathbf{p}_1) = \sum_{t_2s_2} \int^{k_{F{t_2}}} \frac{d^3p_2}{(2\pi)^3} \langle \mathbf{p}_1s_1t_1,\mathbf{p}_2s_2t_2|\bar{V}|\mathbf{p}_1s_1t_1,\mathbf{p}_2s_2t_2 \rangle,
\end{equation}
where the integration is over the Fermi sphere with radius ${k_{F}}_{t_2}$, the Fermi momentum of particle $2$.

With the relativistic two-body matrix element (\ref{Eq:3.2.VOBE}), the in-medium spinor (\ref{Eq:3.2.uRHF}) and the components (\ref{Eq:3.2.UDef}),  the single-particle potentials for the isoscalar mesons
are \cite{Bouyssy1987_PRC36-380},
\begin{subequations}\label{Eq:3.2.usuvRHF}\begin{align}
  U_{s,t_1}(p) =& - \left( \frac{g_s}{m_s} \right)^2 \rho_s + \frac{1}{(4\pi)^2} \frac{1}{p}
  \sum_{t_2} \delta_{t_1t_2} \int_0^{k_{F2}} qdq \frac{M^*(q)}{E^*(q)}
  \Bigg\{ g_s^2\Theta_s(p,q) - f_{ps}^2 \Theta_{ps}(p,q)
  - 4g_v^2\Theta_v(p,q) \notag \\
  &-3\left( \frac{f_{v}}{2M} \right)^2 m_{v}^2 \Theta_v(p,q)
  + 6 \frac{q^*}{M^*(q)} \frac{f_{v}g_{v}}{2M} \left[ p\Theta_v(p,q)-2q\Phi_v(p,q) \right]\Bigg\}, \\
  U_{0,t_1}(p) =& \left( \frac{g_v}{m_v} \right)^2 \rho_b + \frac{1}{(4\pi)^2} \frac{1}{p}
  \sum_{t_2} \delta_{t_1t_2} \int_0^{k_{F2}} qdq
  \left[ g_s^2\Theta_s(p,q) - f_{ps}^2 \Theta_{ps}(p,q)
  + 2g_v^2\Theta_v(p,q)-\left( \frac{f_{v}}{2M} \right)^2 m_{v}^2 \Theta_v(p,q)\right], \\
  U_{v,t_1}(p) =& \frac{1}{(4\pi)^2} \frac{1}{p}
  \sum_{t_2} \delta_{t_1t_2} \int_0^{k_{F2}} qdq \frac{q^*}{E^*(q)}
  \left\{ g_s^2\Phi_s(p,q) -2\left( \frac{f_{ps}}{m_{ps}} \right)^2
  \left[ (p^2+q^2) \Phi_{ps}(p,q) -pq \Theta_{ps}(p,q) \right] \right. \notag \\
  &\left. + 4g_v^2\Phi_v(p,q) -4\left( \frac{f_{v}}{2M} \right)^2
  \left[ \left( p^2+q^2- \frac{m_v^2}{2}\right) \Phi_{v}(p,q) -pq \Theta_{v}(p,q) \right]
  - 6 \frac{M^*(q)}{q^*} \frac{f_{v}g_{v}}{2M} \left[ p\Theta_v(p,q)-2q\Phi_v(p,q) \right]\right\}.
\end{align}\end{subequations}
In these expressions, the convention for the isospin is $t = 0$ for proton and $t = 1$ for neutron.
The scalar density $\rho_s$ and baryon density $\rho_b$ are defined as
\begin{equation}\label{Eq:3.2.rhosb}
  \rho_s = \sum_i \psi_i^\dagger \gamma^0 \psi_i = \sum_{t} \frac{1}{\pi^2} \int_0^{k_{Ft}} q^2 dq \frac{M^*(q)}{E^*(q)},\quad\quad
  \rho_b = \sum_i \psi_i^\dagger \psi_i = \sum_{t} \frac{k_{Ft}^3}{3\pi^2}.
\end{equation}
For meson $i$, the definitions of $\Theta_i(p,q)$ and $\Phi_i(p,q)$ are
\begin{equation}
  \Theta_i(p,q) = \ln \left[ \frac{m_i^2+(p+q)^2}{m_i^2+(p-q)^2} \right],\quad\quad
  \Phi_i(p,q) = \frac{p^2+q^2+m_i^2}{4pq} \Theta_i(p,q) - 1.
\end{equation}

For the isovector mesons, the single-particle potentials are almost identical to Eqs.~(\ref{Eq:3.2.usuvRHF}) with the replacements:
  $\rho_s \to \rho_{s,t_1} - \rho_{s,1-t_1},\, \rho_b \to \rho_{b,t_1} - \rho_{b,1-t_1},\, \delta_{t_1t_2} \to 2-\delta_{t_1t_2}$.
For the Bonn potentials \cite{Machleidt1989_ANP19-189}, there is no isoscalar-tensor channel, i.e., $f_\omega = 0$.

In the relativistic framework, the single-particle wave functions in nuclear matter must be solved iteratively.
As shown in Eqs.~(\ref{Eq:3.2.uRHF}) and (\ref{Eq:3.2.MEpstar}), the single-particle wave functions depend on the potentials $U_s(p), U_0(p),$ and $U_v(p)$, which are calculated from Eqs.~(\ref{Eq:3.2.usuvRHF}).
After solving the RHF equation iteratively, the total energy (\ref{Eq:3.1.EHF}) is
\begin{equation}\label{Eq:3.2.ERHF}
  E^{\rm RHF} = \sum_{t,s} \int^{k_{Ft}} \frac{d^3p}{(2\pi)^3}  \bar{u}(\mathbf{p},s,t) (\bm{\gamma}\cdot\mathbf{p}+M) u(\mathbf{p},s,t)
  + \frac{1}{2}\sum_{t_1t_2,s_1s_2} \int^{k_{F{t_1}}} \frac{d^3p_1}{(2\pi)^3} \int^{k_{F{t_2}}} \frac{d^3p_2}{(2\pi)^3} \langle \mathbf{p}_1s_1t_1,\mathbf{p}_2s_2t_2|\bar{V}|\mathbf{p}_1s_1t_1,\mathbf{p}_2s_2t_2 \rangle.
\end{equation}


In the RBHF calculation, the scattering equation in the nuclear medium has to be solved, and then the bare interaction can be replaced by the effective one, i.e., scattering matrix in the nuclear medium.
Following the work by Brockmann and Machleidt \cite{Brockmann1990_PRC42-1965}, replacing the plane wave in free space (\ref{Eq:3.2.uHelicity}) by the plane wave in nuclear medium (\ref{Eq:3.2.uRHF}) and adding the Pauli operator, the relativistic scattering Thompson equation in the nuclear medium can be obtained as
\begin{equation}\label{Eq:3.2.ThompsonNM}
  G(\mathbf{q}',\mathbf{q}|\mathbf{P},W) = V(\mathbf{q}',\mathbf{q})
  + \int \frac{d^3k}{(2\pi)^3} V(\mathbf{q}',\mathbf{k})
  \frac{Q(\mathbf{k},\mathbf{P})}{W - 2E_{\mathbf{P}/2+\mathbf{k}}^* + i\epsilon } G(\mathbf{k},\mathbf{q}|\mathbf{P},W).
\end{equation}
Several points should be commented for obtaining Eq.~(\ref{Eq:3.2.ThompsonNM}) from the Thompson equation in the free space~(\ref{Eq:3.2.Thompson}):
\begin{enumerate}
  \item The scattering matrix $\mathscr{M}$ has been relabelled as $G$ to be consistent with the BHF convention.
  \item The two-body matrix elements $V(\mathbf{q}',\mathbf{q})$ have the same forms as free space (\ref{Eq:3.2.VOBE}) but with the dressed spinors (\ref{Eq:3.2.uRHF}).
  \item The starting energy is denoted as $W$, for on-shell definition $W = 2E_{\mathbf{P}/2+\mathbf{q}}^*$.
  \item The Pauli operator $Q(\mathbf{k},\mathbf{P})$ ensures that the two intermediate particles are not scattered into the occupied states.
  \item Same as in Ref.~\cite{Brockmann1990_PRC42-1965}, the considered approximations include replacing $E_{\mathbf{k}}$ by $E_{\mathbf{P}/2+\mathbf{k}}$, $E_{\mathbf{q}}$ by $E_{\mathbf{P}/2+\mathbf{q}}$, and the angle-averaging approximation, i.e., the energy of the intermediate particle $E_{\mathbf{P}/2+\mathbf{k}} = E_{\mathbf{P}/2-\mathbf{k}}$ and the energy of the initial particle $E_{\mathbf{P}/2+\mathbf{q}} = E_{\mathbf{P}/2-\mathbf{q}}$. Then the energy denominator is the difference between two initial single-particle energies and two intermediate ones.
  \item Because of the choices of three-dimensional reduction for the Bethe-Salpeter equation, and the approximations mentioned above, the relativistic Bethe-Goldstone (RBG) equation is not unique.
\end{enumerate}

Equation~(\ref{Eq:3.2.ThompsonNM}) is formally identical to the non-relativistic BG equation in nuclear matter , and
usually solved in the helicity basis. By decomposing the two-body matrix elements in different partial-wave channels $V^{JT}(q',q)$, with $J$ the total angular momentum and $T$ the total isospin, the scattering equation is solved in different $J,T$ channels \cite{Erkelenz1974_PR13-191}.

In Ref.~\cite{Brockmann1990_PRC42-1965}, the RBG equation is solved in the rest frame.
Alternatively, one can solve the $G$ matrix in the two-nucleon center-of-mass frame, as usually done in the calculations of two-nucleon scattering, and then transform to the rest frame to calculate the single-particle potential (\ref{Eq:3.1.UBHF}) and ground-state energy (\ref{Eq:3.1.EBHF}).
This has been adopted, for example, by the Brooklyn group \cite{Anastasio1980_PRL45-2096,Anastasio1983_PR100-327}, Horowitz and Serot \cite{Horowitz1984_PLB137-287,Horowitz1987_NPA464-613}, and the Groningen group \cite{terHaar1986_PRL56-1237,terHaar1987_PR149-207}.

In the relativistic framework, the single-particle wave functions are not known from the beginning and must be solved iteratively.
The difficulty is that the single-particle wave functions (\ref{Eq:3.2.uRHF}), which are needed to calculate the two-body matrix elements $V(\mathbf{q}',\mathbf{q})$, and the single-particle energies appearing in the denominator of RBG equation (\ref{Eq:3.2.ThompsonNM}), depend explicitly on different components of the single-particle potential $U_s,U_0,U_v$ (\ref{Eq:3.2.MEpstar}).
On the other hand, the effective interaction $G$-matrix has mixed all the channels in solving the RBG equation (\ref{Eq:3.2.ThompsonNM}) and the different components of the single-particle potential cannot be easily calculated using the form of Eqs.~(\ref{Eq:3.2.usuvRHF}).
In RBHF theory, several methods have been proposed to deal with this problem, including the solution in the full basis
\cite{Anastasio1980_PRL45-2096,Anastasio1981_PRC23-2273,Anastasio1983_PR100-327,Huber1993_PLB317-485,Huber1995_PRC51-1790,
deJong1998_PRC58-890,Poschenrieder1988_PRC38-471,Poschenrieder1988_PLB200-231,Sehn1997_PRC56-216,Tjon1985_PRC32-1667},
projection \cite{Horowitz1984_PLB137-287,Horowitz1987_NPA464-613,terHaar1986_PRL56-1237,terHaar1987_PR149-207,
Boersma1994_PRC49-233,Fuchs1998_PRC58-2022,Gross-Boelting1999_NPA648-105,deJong1998_PRC57-3099,Schiller2001_EPJA11-15,VanDalen2004_NPA744-227},
and a momentum-dependence analysis \cite{Brockmann1984_PLB149-283,Brockmann1990_PRC42-1965,Lee1997_PLB412-235}.

 In Table~\ref{tab1}, the saturation properties of symmetric nuclear matter calculated by RBHF using the interactions Bonn A, B, and C \cite{Brockmann1990_PRC42-1965} are shown for different methods. For the results of the projection method, because of the uncertainty mentioned above, different schemes of projections have been used. Details are given in the corresponding references. 
Neglecting the very different results from Ref.~\cite{Katayama2015_PLB747-43}, the energies per nucleon by different methods are receptively  $-16.49 \sim -15.59$ for Bonn A, $-15.73 \sim -13.60$ for Bonn B, and $-14.38 \sim -12.26$ for Bonn C, and the saturation densities $0.174 \sim 0.189$ for Bonn A, $0.159 \sim 0.174$ for Bonn B, and $0.138 \sim 0.170$ for Bonn C. It is clear that the uncertainties with the different methods exist and an accurate calculation of the single-particle potential is still an open problem for RBHF theory in nuclear matter.
\begin{table}
  \caption{Saturation properties of symmetric nuclear matter calculated by RBHF using the Bonn A, B, and C interactions \cite{Machleidt1989_ANP19-189} by different methods:
  Saturation density $\rho_0$ in fm$^{-3}$ and energy per nucleon $E/A$ in MeV.}
  \label{tab1}
  \centering
  \begin{tabular}{ccccccll}
  \hline\hline
  \multicolumn{2}{c}{Bonn A} & \multicolumn{2}{c}{Bonn B} & \multicolumn{2}{c}{Bonn C} & \\
  $\rho_0$ & $E/A$ & $\rho_0$ & $E/A$ & $\rho_0$ & $E/A$ & Description & Ref. \\
  \hline
  $0.174$ & $-16.49$ & $0.172$ & $-15.73$ & $0.170$ & $-14.38$ & Full basis, momentum-averaged s.p. potential & \cite{Huber1995_PRC51-1790} \\
  $0.174$ & $-15.72$ & $0.170$ & $-14.81$ & $0.162$ & $-13.73$ & Full basis, momentum-dependent s.p. potential & \cite{Huber1995_PRC51-1790} \\
  $0.149$ & $-10.5\,\,\,$ & $0.130$ & $-7.3$ & $0.112$ & $-5.2$ & Full basis &
  \cite{Katayama2015_PLB747-43} \\
  $0.181$ & $-16.15$ & $0.163$ & $-14.59$ & $0.145$ & $-13.69$ & Projection, $ps$ for $T_{\rm Sub}$ & \cite{Gross-Boelting1999_NPA648-105} \\
  $0.181$ & $-15.72$ & $0.159$ & $-13.99$ & $0.138$ & $-13.00$ & Projection, complete $pv$ for $T_{\rm Sub}$ & \cite{Gross-Boelting1999_NPA648-105} \\
  $0.189$ & $-15.81$ & $0.166$ & $-13.70$ & $0.148$ & $-12.31$ & Projection, conventional $pv$ & \cite{Gross-Boelting1999_NPA648-105} \\
  $0.185$ & $-15.59$ & $0.174$ & $-13.60$ & $0.155$ & $-12.26$ & Momentum-dependence analysis & \cite{Brockmann1990_PRC42-1965} \\
  \hline\hline
  \end{tabular}
\end{table}

Besides the uncertainty in calculating different components of the single-particle potential, other uncertainties may exist because of the adopted approximations, such as the angle-average approximation of the Pauli operator \cite{Schiller1999_PRC59-2934,Schiller1999_PRC60-059901,Suzuki2000_NPA665-92,Sammarruca2000_PRC62-014614} or the average approximation of total momentum. Recently, an exact treatment for the total momentum in RBHF has been reported \cite{Tong2018_PRC98-054302}.

 Table \ref{tab2} shows properties of symmetric and asymmetric nuclear matter at saturation density obtained by RBHF theory using the Bonn A, B, and C interactions for the exact and for averaged center-of-mass momentum. In Ref. \cite{Tong2018_PRC98-054302}, the calculation of the self-energy is based on a momentum-dependence analysis similar to that of Ref. \cite{Brockmann1990_PRC42-1965} used in Table \ref{tab1}, but there are uncertainties, such as which momenta are chosen to calculate the self-energy. The values given by the different calculations may also slightly depend on numerical details.

The incompressibility $K_\infty$ and skewness parameter $Q_0$ are defined by the expansion of the binding energy per nucleon $E/A$ in symmetric nuclear matter around the saturation density $\rho_0$ \cite{RocaMaza2018_PPNP101-96},
\begin{equation}
  E(\rho) = E(\rho_0) + \frac{K_\infty}{2} \left( \frac{\rho-\rho_0}{3\rho_0} \right)^2 + \frac{Q_0}{6} \left( \frac{\rho-\rho_0}{3\rho_0} \right)^3 + \dots
\end{equation}
The slope of the incompressibility is $M_0 = 3\rho \frac{\partial K(\rho)}{\partial \rho} |_{\rho = \rho_0}$.
The slope parameter $L$ and curvature parameter $K_{\rm sym}$  are defined by the expansion of the symmetry energy $E_{\rm sym}$ around the saturation density \cite{RocaMaza2018_PPNP101-96},
\begin{equation}
  E_{\rm sym}(\rho) = E_{\rm sym}(\rho_0) + L \left( \frac{\rho-\rho_0}{3\rho_0} \right)^2 + \frac{K_{\rm sym}}{2} \left( \frac{\rho-\rho_0}{3\rho_0} \right)^3 + \dots
\end{equation}
The incompressibility for finite nuclei is parameterized as,
\begin{equation}
  K_A \approx K_\infty + K_{\rm surf}A^{-1} + K_{\tau} \alpha^2 + K_{\rm Coul} \frac{Z^2}{A^{4/3}},
\end{equation}
with $\alpha = (\rho_n-\rho_p)/\rho, K_{\tau} = K_{\rm sym} - 6L - \frac{Q_0}{K_\infty}L,$ and $K_{\rm Coul} = \frac{3}{5}\frac{e^2}{r_0} \left( -8-\frac{Q_0}{K_\infty} \right)$.
When $Q_0 \approx 0$, one defines parameter $K_\tau \approx K_{\rm asy} = K_{\rm sym} - 6L$.

The results of RBHF in Table \ref{tab2} are compared with those obtained by non-relativistic BHF theory with and without three-body force (TBF)~\cite{Brockmann1990_PRC42-1965,Li2006_PRC74-047304,Vidana2009_PRC80-045806}.
Results of two sets of TBF, namely TBFa \cite{Baldo1999_PRC59-682} and TBFb \cite{Vidana2009_PRC80-045806}, are presented. The empirical values are also given for comparison \cite{Margueron2018_PRC97-025805}.
It can be seen that the average approximation is acceptable for properties like $\rho_0, E/A, K_\infty, E_{\rm sym}$, and $L$, but not necessarily for the higher-order parameters such as $Q_0, M_0, K_{\rm sym}, K_{\rm asy}$, and $K_{\rm Coul}$.

\begin{table}
\caption{Bulk parameters (see text for definitions) of symmetric and asymmetric nuclear matter at saturation density
$\rho_0$ obtained by RBHF theory using the Bonn A, B, and C interactions, with the exact and averaged
center-of-mass (c.m.) momentum. The quantities $\Delta$ are defined as the differences between the exact
and the averaged treatments of the c.m. momentum. Results are compared with those obtained by non-relativistic BHF theory with and without three-body forces (TBF)~\cite{Brockmann1990_PRC42-1965,Li2006_PRC74-047304,Vidana2009_PRC80-045806} and with empirical values~\cite{Margueron2018_PRC97-025805}. Table taken from Ref.~\cite{Tong2018_PRC98-054302}.
}
\scalebox{0.88}{
\begin{tabular}{l|ccrrrrrrrrrrr}
\hline\hline
\multicolumn{3}{l}{\multirow{2}{*}{Model} \multirow{2}{*}{Potential}} & \multicolumn{1}{c}{$\rho_0$} & \multicolumn{1}{c}{$E/A$} & \multicolumn{1}{c}{$K_{\infty}$} & \multicolumn{1}{c}{$Q_0$} & \multicolumn{1}{c}{$M_0$} & \multicolumn{1}{c}{$E_{\mathrm{sym}}$} & \multicolumn{1}{c}{$L$} & \multicolumn{1}{c}{$K_{\mathrm{sym}}$} & \multicolumn{1}{c}{$K_{\mathrm{asy}}$} & \multicolumn{1}{c}{$K_{\tau}$} & \multicolumn{1}{c}{$K_{\mathrm{Coul}}$} \\
\multicolumn{2}{c}{} & & \multicolumn{1}{c}{(fm$^{-3}$)} & \multicolumn{1}{c}{(MeV)} & \multicolumn{1}{c}{(MeV)} & \multicolumn{1}{c}{(MeV)} & \multicolumn{1}{c}{(MeV)} & \multicolumn{1}{c}{(MeV)} & \multicolumn{1}{c}{(MeV)} & \multicolumn{1}{c}{(MeV)} & \multicolumn{1}{c}{(MeV)} & \multicolumn{1}{c}{(MeV)} & \multicolumn{1}{c}{(MeV)} \\
\hline
&  & exact&    $0.180$ & $-15.38$ & $286$ & $731$ & $4163$ & $33.7$ & $75.8$ & $-57.0$ & $-512$ & $-705$ & $-8.30$ \\
& A & average& $0.182$ & $-15.04$ & $289$ & $650$ & $4118$ & $32.6$ & $74.7$ & $-53.1$ & $-501$ & $-669$ & $-8.09$ \\
&  &$\Delta$& $-0.002$ &  $-0.34$ &  $-3$ &  $81$ &   $45$ &  $1.1$ &  $1.1$ &  $-3.9$ &  $-11$ &  $-36$ & $-0.21$ \\
\cline{2-14}
&  & exact&    $0.164$ & $-13.44$ & $222$ & $547$ & $3211$ & $29.9$ & $63.0$ & $-56.3$ & $-434$ & $-590$ & $-7.98$ \\
RBHF
& B & average& $0.165$ & $-13.08$ & $220$ & $791$ & $3431$ & $28.7$ & $65.3$ & $-47.5$ & $-439$ & $-674$ & $-8.86$ \\
&  &$\Delta$& $-0.001$ & $-0.36$  &   $2$ &$-244$ & $-220$ &  $1.2$ & $-2.3$ &  $-8.8$ &    $5$ &   $84$ &  $0.88$ \\
\cline{2-14}
&  & exact&    $0.149$ & $-12.12$ & $176$ & $260$ & $2372$ & $26.8$ & $51.7$ & $-55.6$ & $-366$ & $-442$ & $-7.00$ \\
& C & average& $0.150$ & $-11.75$ & $168$ & $638$ & $2654$ & $25.6$ & $58.8$ & $-41.1$ & $-394$ & $-618$ & $-8.74$ \\
&  &$\Delta$& $-0.001$ &  $-0.37$ &   $8$ &$-378$ & $-282$ &  $1.2$ & $-7.1$ & $-14.5$ &   $28$ &  $176$ &  $1.74$ \\
\hline
& A  & &       $0.428$ & $-23.55$ & $204$ &       &        & $32.1$ \\
\multirow{4}{*}{BHF}
& B  & &       $0.309$ & $-18.30$ & $160$ &       &        & $31.8$ \\
& C  & &       $0.247$ & $-15.75$ & $143$ &       &        & $28.5$ \\
&AV18& no TBF& $0.240$ & $-17.30$ & $214$ &$-225$ & $2343$ & $35.8$ & $63.1$ & $-27.8$ & $-406$ & $-340$ & $-6.01$ \\
&AV18& TBFa&   $0.187$ & $-15.23$ & $196$ &$-281$ & $2071$ & $34.3$ & $66.5$ & $-31.3$ & $-430$ & $-335$ & $-5.23$ \\
&AV18& TBFb&   $0.176$ & $-14.62$ & $186$ &$-225$ & $2007$ & $33.6$ & $66.9$ & $-23.4$ & $-425$ & $-344$ & $-5.30$ \\
\hline
\multicolumn{2}{c}{\multirow{2}{*}{Empirical}}
& &            $0.155$ & $-15.8~\,$ & $230$ & 300 & & $32~\,\,\,$ & $60~\,\,\,$ & $-100~~~$ &  & $-400$  \\
\multicolumn{2}{c}{} & & $\pm0.005$ & $\pm0.3~\,$ & $\pm20$ & $\pm 400$ & & $\pm 2~\,\,\,$ & $\pm15~\,\,\,$ & $\pm100~~~$ &  & $\pm100$  \\
\hline\hline
\end{tabular}
}
\label{tab2}
\end{table}

\begin{figure}
  \centering
  \includegraphics[width=7cm]{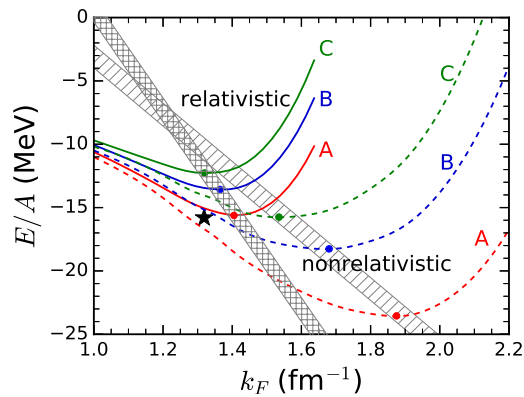}
  \caption{(Color online) Equation of state of symmetric nuclear matter calculated with the Bonn A, B, and C interactions by RBHF theory (relativistic) and by BHF theory (non-relativistic). The black star denotes the empirical saturation value \cite{Margueron2018_PRC97-025805}.}
  \label{fig1}
\end{figure}

In Fig.~\ref{fig1}, the equations of state for symmetric nuclear matter calculated with Bonn A, B, and C interactions by RBHF theory are shown in comparison with the results by non-relativistic BHF theory \cite{Brockmann1990_PRC42-1965}.
  The empirical saturation properties $E/A = -15.8\pm 0.3$ MeV and $k_F = 1.319\pm0.014$ fm$^{-1}$ are also plotted as the black star \cite{Margueron2018_PRC97-025805}. The shadowed bands, which connect the three saturation points obtained by the interactions Bonn A, B, and C, are a schematic representation of the Coester lines. It was originally found in Ref.~\cite{Coester1970_PRC1-769} that the saturation points obtained in BHF calculations with interactions of different tensor strength locate on a line that systematically deviates from the empirical region. As the three Bonn interactions are different in their tensor strengths, their nuclear matter saturation points are at different locations on the Coester line, as seen in Fig.~\ref{fig1}.
For the results of both RBHF and BHF, Bonn A with the weakest tensor force gives the largest binding energy and saturation density, and Bonn C with the strongest tensor force gives the smallest.

Looking at the non-relativistic results, the Coester line connecting the saturation points given by different interactions misses the empirical region systematically. Compared to the non-relativistic results, the relativistic framework leads to considerably improved results and the saturation points given by different interactions form a new ``Coester'' line, which is much closer to the empirical region.
The relativistic effect produces a strong density-dependent repulsion, which can be explained by the following simple estimation in Ref.~\cite{Brown1987_CNPP17-39}.

Considering the simplest case, the effective interaction $G$-matrix can be viewed as the effective interaction in the Walecka model with the $\sigma$ and $\omega$ mesons in the mean-field (Hartree) approximation.
The single-particle potentials in Eqs.~(\ref{Eq:3.2.usuvRHF}) become
\begin{equation}
  U_s = - \left( \frac{g_s}{m_s} \right)^2 \rho_s,\quad
  U_0 = \left( \frac{g_v}{m_v} \right)^2 \rho_b.
\end{equation}
Typical values of the scalar and vector potentials are $U_s = -400 (\rho/\rho_0)$ MeV and $U_0 = 300 (\rho/\rho_0)$ MeV around the saturation density.
From free space to nuclear matter, the positive-energy wave functions change as
\begin{equation}\label{Eq:3.2.u0uRHF}
  u_0(\mathbf{p},s) = \sqrt{\frac{p_0+M}{2p_0}}
  \left(\begin{array}{c}
  1 \\ \frac{\bm{\sigma}\cdot \mathbf{p}}{p_0+M}
  \end{array}\right)
  \chi_s \quad\to \quad
  u(\mathbf{p},s) = \sqrt{\frac{E^*+M^*}{2E^*}}
  \left(\begin{array}{c}
  1 \\ \frac{\bm{\sigma}\cdot \mathbf{p}}{E^*+M^*}
  \end{array}\right)
  \chi_s,
\end{equation}
with $M$ being replaced by the effective mass $M^* = M + U_s$.
Here $p_0 = \sqrt{\mathbf{p}^2+M^2}$.

In free space, the relativistic effect reveals itself as a kinetimatic effect: the larger the velocity (or equivalently the momentum $\mathbf{p}$) the larger the relativistic effect (i.e. the small component becomes larger).
For $\mathbf{p} = 0$ the spinor becomes trivial unity and the wave function is the same as the non-relativistic plane wave. As $\mathbf{p}$ increases, the small component becomes larger and the relativistic wave function becomes more different from the non-relativistic plane wave.

In nuclear matter, situation gets more complicated as there is an additional origin of the relativistic effect, the dynamic effect: the larger the scalar potential $U_s$, the larger the relativistic effect.
In other words, the small component can become non-negligible even in the small velocity limit, as $M^*$ can be significantly smaller than the bare mass $M$.
  As discussed in Section 1.4 , this has the consequence that in nuclear physics where the average velocity of a nucleon is generally small, effects of relativity can still be important.

To evaluate the size of the relativistic effect caused by replacing $M$ with $M^*$, one can expand the positive single-particle energy as
\begin{equation}\label{Eq:3.2.Ep}
  E_p = U_0 + E_p^* = U_0 + \sqrt{\mathbf{p}^2+(M+U_s)^2} = U_0 + p_0 + \frac{MU_s}{p_0} + \frac{\mathbf{p}^2}{2p_0^3} U_s^2 + \dots.
\end{equation}
In the small velocity limit $|\mathbf{p}| << M$, the leading relativistic correction to the energy is
\begin{equation}\label{Eq:3.2.deltaEp}
  \delta E_p = \left( \frac{U_s}{M} \right)^2 \frac{\mathbf{p}^2}{2M}.
\end{equation}
Given the average kinetic energy per nucleon is $\langle \mathbf{p}^2/(2M) \rangle \approx 23 (\rho/\rho_0)^{2/3}$ MeV, the average correction can be evaluated as \cite{Brown1987_CNPP17-39}
\begin{equation}
  \overline{\delta E} = \left( \frac{U_s}{M} \right)^2 \bigg\langle \frac{\mathbf{p}^2}{2M} \bigg\rangle \approx 4.2 \left( \frac{\rho}{\rho_0} \right)^{8/3}~\text{MeV}.
\end{equation}
Taking the results of the Bonn B interaction in Fig.~\ref{fig1} as an example, the relativistic correction to the energy per nucleon is well fitted by this ansatz with a different factor $\overline{\delta E} \approx 2 (\rho/\rho_0)^{8/3}$ MeV \cite{Brockmann1990_PRC42-1965}.

The Dirac spinor in nuclear matter $u(\mathbf{p},s)$ can be decomposed in the complete basis of free Dirac spinor (both positive-energy $u_0(\mathbf{p},s)$ and negative-energy solutions $v_0(\mathbf{p},s)$) \cite{Anastasio1980_PRL45-2096},
\begin{equation}
  u(\mathbf{p},s) = a(p) u_0(\mathbf{p},s) + b(p)\sum_{s'} \langle s|\bm{\sigma}\cdot\hat{\mathbf{p}}|s' \rangle v_0(-\mathbf{p},-s'),
\end{equation}
with the expansion coefficients $a(p)$ and $b(p)$ normalized to $a^2(p)+b^2(p) = 1$.
In first-order perturbation theory, the correction to the positive-energy wave function is \cite{Brown1987_CNPP17-39}
\begin{equation}
  \delta u = u(\mathbf{p},s)-u_0(\mathbf{p},s) = \frac{\Lambda_-}{2\sqrt{\mathbf{p}^2+M^2}} (\beta U_s + U_0) u_0(\mathbf{p},s),
\end{equation}
where $\Lambda_-(p) = (p_0-H_D)/(2p_0)$ is the Casimir projection operator, $H_D = \bm{\alpha}\cdot\mathbf{p} + \beta M$ is the single-particle Dirac Hamiltonian.
In the small velocity limit and ignoring the correction to the large component, one has the first-order correction to $u(\mathbf{p},s)$,
\begin{equation}
  \delta u = -
  \left(\begin{array}{c}
  0 \\ \frac{U_s\bm{\sigma}\cdot \mathbf{p}}{2M^2}
  \end{array}\right) \chi_s.
\end{equation}

\begin{figure}
  \centering
  \includegraphics[width=4cm]{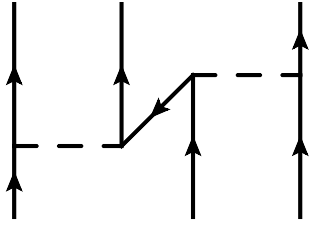}
  \caption{Three-body interaction of relativistic orgin.}
  \label{fig2}
\end{figure}

The mixing of negative-energy states in nuclear matter can be illustrated by the so-called Z-diagram in Fig.~\ref{fig2}.
Representing the matrix element of each dashed line as $-(U_s/M)\bm{\sigma}\cdot\mathbf{p}$ and the energy denominator $(2M)^{-1}$, the correction to the energy can be calculated to have the same value as in Eq.~(\ref{Eq:3.2.deltaEp}).
This correction is included naturally in the dressed relativistic spinor $u$ (\ref{Eq:3.2.uRHF}) when one evaluates the full energy (\ref{Eq:3.2.Ep}).
If one wants to calculate this in a non-relativistic basis, where the spinor $u_0$ is used, an additional term (with matrix element $-(U_s/M)\bm{\sigma}\cdot\mathbf{p}$) has to be added to the non-relativistic energy $E_0$.
This matrix element is caused by the difference between the relativistic dressed spinor and the non-relativistic spinor $\delta u$: $u = u_0 + \delta u = a u_0 + b v_0$, with the momentum-dependent expansion coefficients $a$ and $b$ of the positive- and negative-energy free spinors.
Such a mixing of the free spinors with positive and negative energies in the dressed spinor $u$ is interpreted in the non-relativistic framework as an admixture of nucleon-antinucleon excitations.
In other words, in leading order the relativistic effect caused by the strong scalar potential can be expressed in the non-relativistic framework as a three-body interaction caused by an intermediate nucleon-antinucleon excitation.

In Ref.~\cite{Grange1989_PRC40-1040}, a three-body interaction was constructed from the two-body Paris interaction, and among different contributions, the effect of the nucleon-antinucleon excitation term was confirmed with the above discussed properties~\cite{Lacombe1980_PRC21-861}.
The same three-body interaction has been used together with the AV18 two-body interaction \cite{Wiringa1995_PRC51-38} in a non-relativistic BHF calculation and the results have been compared with RBHF using the Bonn B interaction \cite{Zuo2002_NPA706-418}.
After including the $2\sigma$-exchange three-body interaction with  the intermediate nucleon-antinucleon excitation in the BHF framework, the results are very close to those of full RBHF theory.
Of course, one should also notice that in the given three-body interaction \cite{Grange1989_PRC40-1040}, there are other important contributions such as the $\sigma\omega$ nucleon-antinucleon excitation, Roper resonances, etc.~\cite{Zuo2002_NPA706-418}. The other contributions cancel each other more or less and the net effect is slight attraction.

\begin{figure}
  \centering
  \includegraphics[width=8cm]{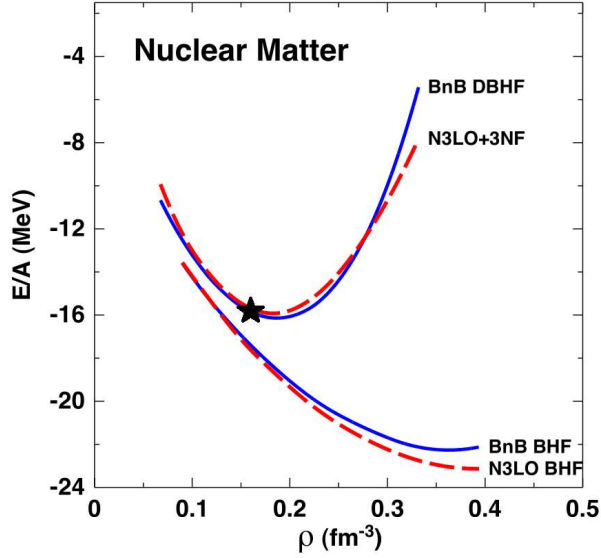}
  \caption{(Color online) Equation of state of symmetric nuclear matter calculated by BHF (RBHF) with the Bonn B interaction (blue lines), in comparison with the BHF results using N$^{3}$LO interaction without and with the three-body interaction (dashed red lines).
  The black star denotes the empirical saturation value \cite{Margueron2018_PRC97-025805}.
  Figure redrawn from Ref.~\cite{Sammarruca2012_PRC86-054317}.}
  \label{fig3}
\end{figure}

 In the modern nuclear forces derived from chiral effective field theory, three-body interactions, as well as higher many-body interactions, appear on the same footing as the two-body interaction in the chiral expansions. It is therefore interesting to compare relativistic effects and the chiral three-body interaction. In Ref.~\cite{Sammarruca2012_PRC86-054317}, RBHF and non-relativistic BHF calculations with Bonn B have been compared with non-relativistic BHF calculations with N$^{3}$LO without and with the three-body interaction \cite{Entem2003_PRC68-041001}.
The results are shown in Fig.~\ref{fig3}. By comparing the results of BHF with Bonn B and with N$^{3}$LO, it can be seen that these two interactions give very similar results in the same framework. By comparing the changes from BHF to RBHF and the changes from N$^{3}$LO to N$^{3}$LO plus three-body interaction, it is evident that the effect of a relativistic treatment is very similar to the effect of the three-body interaction in the non-relativistic framework.

\subsection{RBHF formalism for finite nuclei}\label{Sect:3.3}


Using the Thompson choice, the relativistic Bethe-Goldstone (RBG) equation for finite nuclei has the form,
\begin{equation}\label{Eq:3.3.BGeq}
  \langle a'b'|\bar{G}(W)|ab\rangle =
  \langle a'b'|\bar{V}|ab\rangle +\frac{1}{2}\sum_{cd}
  \langle a'b'|\bar{V}|cd\rangle \frac{Q(c,d)}{W-e_{c}-e_{d}}
  \langle cd|\bar{G}(W)|ab\rangle.
\end{equation}
The quantum numbers $a,b,c,\dots$ represent the single-particle states being the solutions of RBHF equations.
The indices $k,l,m,\dots$ will be used for an arbitrary complete relativistic single-particle basis.
The Pauli operator $Q$ has the same definition as in BHF theory, i.e., $Q(c,d) = 1$ for $e_c > e_F$ and $e_d > e_F$, otherwise $Q(c,d) = 0$.

Adopting the choice in between the gap choice and the continuous choice, the single-particle potential of RBHF theory is calculated similar as in Eq.~(\ref{Eq:3.1.UBHF}),
\begin{equation}\label{Eq:57}
  \langle a|U^{\rm RBHF}|b\rangle=
  \begin{cases}
  \frac{1}{2}\sum_{c=1}^A \langle ac|\bar{G}(e_{a}+e_c)+ \bar{G}(e_b+e_c)|bc\rangle,
   & e_{a}, e_b \leq e_F \\
  \sum_{c=1}^A \langle ac|\bar{G}(e_a+e_c)|bc\rangle,
  & e_{a} \leq e_F,~e_b > e_F  \\
  \sum_{c=1}^A \langle ac|\bar{G}(e'+e_c)|bc\rangle,
  & e_{a}, e_{b} > e_F,
  \end{cases}
\end{equation}
where some uncertainty exists in $e'$ and its choice has been discussed in Refs.~\cite{Shen2017_PRC96-014316,Shen2018_PLB781-227}.
This is exactly the same as Eq.~(\ref{Eq:3.1.UBHF}) with only different labels for quantum states, except that now the single-particle energies $e_a$ and $e_b$ may correspond also to states with negative energies, to states in the Dirac sea in the relativistic framework.

With this single-particle potential, one can solve the RHF equation,
\begin{equation}\label{Eq:3.3.RHFeq}
  (T+U^{\rm RBHF})|a\rangle = e_a|a\rangle.
\end{equation}
which is also formally the same as the non-relativistic HF equation (\ref{Eq:3.1.HFeq}).
The matrix elements of the kinetic energy are given in Eq.~(\ref{Eq:3.2.meT}), and the RHF equation can be solved by the standard diagonalization.
The RBHF equations for finite nuclei (\ref{Eq:3.3.BGeq},\ref{Eq:57},\ref{Eq:3.3.RHFeq}) will be solved iteratively until convergence is reached. Different from the RBHF equations for nuclear matter where different components of single-particle potential must be decomposed before the solution can be obtained, the decomposition of single-particle potential is not necessary for finite nuclei. Only the matrix elements of single-particle potential (\ref{Eq:57}) are needed, which can be calculated directly from the $G$-matrix.

This difference between the calculations in finite nuclei presented here and
those in nuclear matter discussed in Section~\ref{Sect:3.2.2} is caused by the fact that the solution of the RBG equations~(\ref{Eq:3.3.BGeq}) in finite nuclei involves not only the matrix elements between pairs of states $|ab\rangle$ with positive energies ($e_a>0, e_b > 0$), but also matrix elements between all possible pairs ($e_a>0, e_b < 0$), ($e_a<0, e_b < 0$) etc., which are needed for the solution of the corresponding RHF equation~(\ref{Eq:3.3.RHFeq}). In the usual nuclear matter calculations discussed in Section~\ref{Sect:3.2.2}, the RBG equations~(\ref{Eq:3.2.ThompsonNM}) are only solved for pairs with positive energies. The matrix elements of the $G$-matrix for all other pairs, which are needed for the solution of the RHF equations~(\ref{Eq:3.2.RHFeq}), are directly calculated from the potential $U$, i.e., one needs the decomposition of $U$ into the various relativistic channels, scalar, vector, etc., and this decomposition is not unique. Of course, a full solution of the RBG equations in nuclear matter for all  pairs with arbitrary energies would solve this problem. However, this is complicated and usually not done. The fact that one solves in finite nuclei the full set of RBG equations increases the numerical efforts considerably.

The total energy in the RBHF framework is (\ref{Eq:3.1.EBHF})
\begin{equation}\label{Eq:3.3.ERBHF}
  E = \sum_{a}^A \langle a|T|a \rangle + \frac{1}{2}\sum_{ab}^A \langle ab|\bar{G}(W=e_a+       e_b)|ab \rangle.
\end{equation}

To solve the RBHF equations, one must start with a complete initial basis $\{|k\rangle\}$, as the solution $\{|a\rangle\}$ is not known from the beginning.
The RBHF single-particle states are expressed as linear combinations of the basis states
\begin{equation}
  |a\rangle = \sum_k D_{ka} |k\rangle,
\end{equation}
with $D_{ka}$ the expansion coefficients.

  As the requirement of the completeness of the basis, $|k\rangle$ runs not only over the positive-energy states but also over the negative-energy states in the Dirac sea \cite{Zhou2003_PRC68-034323}. One should not mix this with the no-sea approximation adopted in the RBHF or RHF calculations, which refers to neglecting the Dirac sea in calculating densities and currents. For the RBHF calculation, the input Bonn NN interaction was constructed without considering antinucleon degrees of freedom \cite{Machleidt1989_ANP19-189}, and this will be kept in the many-body calculation.

For the case with spherical symmetry, the eigenfunctions of the Dirac equation can be written as
\begin{equation}\label{Eq:3.3.wf}
  |a\rangle = \frac{1}{r} \left(
  \begin{array}{c}
  F_{n_a\kappa_a}(r) \Omega_{j_am_a}^{l_a}(\theta,\varphi) \\
  iG_{n_a\kappa_a}(r) \Omega_{j_am_a}^{\tilde{l}_a}(\theta,\varphi)
  \end{array}
  \right),
\end{equation}
where the radial, orbital angular momentum, total angular momentum, and magnetic quantum numbers are denoted by $n,l,j,$ and $m$, respectively, and the spinor spherical harmonics are
\begin{equation}
  \Omega_{jm}^{l}(\theta,\varphi) = \sum_{m_lm_s} C_{lm_l\frac{1}{2}m_s}^{jm} Y_{lm_l}(\theta,\varphi) \chi_{m_s}.
\end{equation}
In practice, the relativistic quantum number $\kappa$ is often used, i.e., $\kappa = \pm \left( j+1/2 \right)$, for $j = l\mp 1/2$.
The orbital angular moment for the lower component in Eq.~(\ref{Eq:3.3.wf}), $\tilde{l} = 2j-l$, is different from the one for the upper component.
The wave functions $F(r)$ and $G(r)$ satisfy the radial equation
\begin{equation}\label{Eq:3.3.RDeq}
  \left( \begin{array}{cc}
  M+\Sigma(r) & -\frac{d}{dr}+\frac{\kappa}{r} \\
  \frac{d}{dr}+\frac{\kappa}{r} & -M+\Delta(r)
  \end{array} \right)
  \left( \begin{array}{c}
  F_a(r) \\ G_a(r) \\
  \end{array} \right)
  =e_a \left( \begin{array}{c}
  F_a(r) \\ G_a(r) \\
  \end{array} \right),
\end{equation}
where $\Sigma = V+S$ and $\Delta = V-S$ are the sum and the difference of vector and scalar potentials.
The initial basis is chosen as the Dirac Woods-Saxon (DWS) basis which is obtained by solving the radial Dirac equation (\ref{Eq:3.3.RDeq}) with the Woods-Saxon potentials $\Sigma(r)$ and $\Delta(r)$ \cite{Zhou2003_PRC68-034323}.

Being consistent with the no-sea approximation, the index $c$ in Eq.~(\ref{Eq:57}) is restricted to the $A$ occupied states in the Fermi sea. The
RHF equation (\ref{Eq:3.3.RHFeq}) in the basis $\{|k\rangle\}$ reads
\begin{equation}  \label{Eq:rhf2}
\sum_{l} \left( T_{kl} + U_{kl} \right) D_{la} = e_{a} D_{ka}.
\end{equation}

As both the RBG equation~(\ref{Eq:3.3.BGeq}) and
the single-particle potential (\ref{Eq:57}) are defined in the
RBHF single-particle basis $\{|a\rangle\}$, it requires a
nested iteration procedure.

\begin{enumerate}
\item Start with a set of trial single-particle states, e.g., a discrete DWS basis \cite{Zhou2003_PRC68-034323} with the single-particle wave functions $|k\rangle$ and the corresponding energies $e_k$.

\item Calculate the matrix elements of the kinetic energy $T_{k'k}$ and the antisymmetrized two-body matrix elements of the bare interaction $\bar{V}_{k'l'kl}$ in this basis.

\item Solve the the RBG equation (\ref{Eq:3.3.BGeq}) in this basis by matrix inversion. This yields a set of $G$-matrix elements $\bar{G}_{k'l'kl}$.

\item For the fixed $G$-matrix, solve the non-linear RHF equation (\ref{Eq:3.3.RHFeq}) by a nested inner iteration:
\begin{enumerate}
     \item Use these $G$-matrix elements to calculate the single-particle potentials $U_{k'k}$ in Eq.~(\ref{Eq:57}). With the basis transformation coefficient $D_{kk'}$, it becomes,
         \begin{equation}\label{eq}
           U_{k'k} = \sum_{c=1}^A \sum_{l'l} \bar{G}_{k'l'kl}(W) D_{l'c}^* D_{lc}^{}.
         \end{equation}
         In the first iteration, $D_{kk'}$ is a unity matrix.
\item Solve the RHF equation (\ref{Eq:3.3.RHFeq}) with these matrix elements $T_{k'k}$ and $U_{k'k}$:
      \begin{equation}
         \sum_{k} \left( T_{k'k} + U_{k'k} \right) D_{ka} = e_{l} D_{k'a},
         \label{Eq:rhf3}
      \end{equation}
      and a new set of single-particle states $\{|a\rangle\}$ with single-particle energies
      $e_a$ is obtained.
\item Calculate a new relativistic single-particle density by occupation of the A orbits in
      the local Fermi sea ( $0 < e_a  \leq e_F $ ) for the next step of the RHF-iteration and a new potential
      $U$ with the same $G$-matrix, and continue this inner RHF-iteration till convergence.
      \end{enumerate}
\item Compare the converged RHF solution $\{|a\rangle\}$ with the initial basis $\{|k\rangle\}$ in which the $G$-matrix is calculated. If $\{|k\rangle\} = \{|a\rangle\}$ then the whole RBHF iteration is converged. Otherwise the RHF basis $\{|a\rangle\}$ will be used in the next iteration in Step 2. Alternatively, basis transformation can be used instead of calculating the matrix elements as in Step 2,
    \begin{align}
      T^{}_{a'a} &= \sum_{k'k} D_{k'a'}^* D^{}_{ka} T^{}_{k'k}, \\
      \bar{V}^{}_{a'b'ab} &= \sum_{k'l'kl}D_{k'a'}^* D_{l'b'}^* D^{}_{ka}D^{}_{lb} \bar{V}^{}_{k'l'kl} .
      \label{Eq:transV}
    \end{align}
\end{enumerate}

\subsection{Local density approximation on RBHF and its promotion to CDFT}\label{Sect:3.4}


From the early 1990s, one of the links between the RBHF theory and the covariant density functional theory involves the local density approximation (LDA), which was first introduced in the non-relativistic framework by Brueckner \textit{et al.} \cite{Brueckner1958_PR110-431}.
The $G$-matrix in a finite nucleus is assumed depend on the local density $\rho(\mathbf{r})$ and it is the same as the $G$-matrix in nuclear matter with the same density. Then the $G$-matrix is parameterized as a density-dependent (DD) effective interaction in the HF framework.
In order to account for the discrepancy between the RHF results and experimental data, on top of the effective interaction derived from the $G$-matrix, higher-order correction are then parameterized \cite{Nemeth1970_PLB32-561,Negele1970_PRC1-1260,Sprung1971_NPA168-273,Campi1972_NPA194-401}.

With the success of RBHF theory in describing nuclear matter, it is natural to describe finite nuclei in the same framework with the LDA.
In general, an effective Lagrangian is constructed in a similar way as the bare Bonn interaction in Eqs.~(\ref{Eq:3.2.Lagrangian}), but now with DD coupling strengths,
\begin{subequations}\label{Eq:3.4.Lagrangian}\begin{align}
  \mathscr{L}_{\rm NNs}^{\rm (eff)} &= g_s(\rho) \bar{\psi}\psi\varphi^{(s)}, \\
  \mathscr{L}_{\rm NNpv}^{\rm (eff)} &= -\frac{f_{ps}(\rho)}{m_{ps}} \bar{\psi}\gamma^5\gamma^\mu
  \psi \partial_\mu \varphi^{(ps)},  \\
  \mathscr{L}_{\rm NNv}^{\rm (eff)} &= -g_v(\rho) \bar{\psi}\gamma^\mu\psi\varphi_\mu^{(v)} - \frac{
  f_v(\rho)}{4M}\bar{\psi}\sigma^{\mu\nu}\psi \left( \partial_\mu\varphi_\nu^{( v)}
  - \partial_\nu \varphi_\mu^{(v)} \right), \quad\quad \mu,\nu = 0,1,2,3.
\end{align}\end{subequations}
By fitting to the results of RBHF for nuclear matter, the effective interaction, i.e., the DD coupling strengths $g(\rho)$, can be determined.
In practice, one can define the effective interaction in the RMF or in the RHF approximation. Accordingly the obtained effective interactions will be used in DDRMF or DDRHF calculations.
There are many choices to fit the RBHF results, e.g., fits to the single-particle potential or fits to the $G$-matrix elements.

Although self-consistent RBHF theory for finite nuclei has been improved recently \cite{Shen2016_CPL33-102103, Shen2017_PRC96-014316}, self-consistent RBHF calculations for all nuclei in the nuclear chart are still far above
the numerical capabilities of the most modern computers.
Definitely it is expected that RBHF results for both nuclear matter and finite nuclei will provide more and more insight for the nuclear density functional.

The LDA in the RBHF theory was initiated by the Brooklyn group \cite{Ai1987_PRC35-2299,Ai1989_PRC39-236}. The relativistic $G$-matrix in nuclear matter was calculated using the interactions HEA \cite{Holinde1972_NPA198-598} and HM2 \cite{Holinde1976_NPA256-479} by taking into account relativistic effects in first-order perturbation theory.
The single-particle potential calculated by RBHF is then fitted to determine the effective interaction in the RHF framework with the mesons, $\sigma,\omega$, and $\delta$ \cite{Ai1987_PRC35-2299}. The density-dependence of the effective interactions is parameterized by analytical functions, and nuclear matter saturation curve can be well reproduced \cite{Ai1987_PRC35-2299,Ai1989_PRC39-236}.

Marcos \textit{et al.} \cite{Marcos1989_PRC39-1134} improved the input by self-consistent RBHF calculation for asymmetric nuclear matter \cite{terHaar1987_PR149-207}. The total energy and different components of the single-particle potential calculated by RBHF are fitted in the RHF framework by an effective interaction with four mesons, $\sigma,\omega,\pi$, and $\rho$, including form factors.
Further work using self-consistent RBHF results for nuclear matter~\cite{Brockmann1984_PLB149-283} as an input can be found in Ref.~\cite{Brockmann1988_ZPA331-367}.

In the above works, the single-particle potential and optionally the total energy has been used in the fit to determine the effective interaction. The $G$-matrix elements can also be used to determine the effective interaction, and in principle they should contain more information \cite{Elsenhans1990_NPA515-715}.

Adopting a parametrization of the $G$-matrix calculated by RBHF using the Bonn C interaction in Ref.~\cite{Brockmann1990_PRC42-1965} as the form $G = V + \Delta G$, the first part $V$ was determined by fitting to the matrix elements of the original bare interaction, and the second part $\Delta G$ containing four mesons, $\sigma,\omega,\delta$, and $\rho$, together with the first part were then fitted to the $G$-matrix \cite{Elsenhans1990_NPA515-715}. The correlation term, $\Delta G$, mainly reduces sizeably the repulsive $\omega$-exchange \cite{Elsenhans1990_NPA515-715}. By discarding the form factors in the parametrization, the resulting effective interaction ~\cite{Elsenhans1990_NPA515-715} became simpler, but it was still difficult in DDRHF calculations for the study of finite nuclei. Therefore, a simple and accurate, in the sense of reproducing the RBHF results, effective interaction is important in describing properties of finite nuclei.

With the input of RBHF  \cite{terHaar1987_PR149-207} and the non-relativistic variational method \cite{Friedman1981_NPA361-502},
Gmuca \cite{Gmuca1991_JPG17-1115,Gmuca1992_ZPA342-387,Gmuca1992_NPA547-447} found that effective interactions with a $\sigma$-$\omega$ model with nonlinear $\sigma$-couplings, and a $\sigma$-$\omega$ model with a DD $m_\sigma$ \cite{Jackson1983_NPA407-495} give similar results for nuclear matter, but quite different results for finite nuclei.
Extending the nonlinear $\sigma$ model to include a  nonlinear $\omega$ term, the effective RMF interaction is found to reproduce better the EoS derived from RBHF \cite{Gmuca1992_ZPA342-387}.
In Ref.~\cite{Gmuca1992_NPA547-447}, the finite nuclei $^{16}$O and $^{40}$Ca were studied using the effective interaction determined by RBHF calculations using the Bonn A, B, and C interactions \cite{Brockmann1990_PRC42-1965} and the Groningen interaction \cite{deJong1991_PRC44-998}.
Nonlinear effective interactions including $\sigma,\omega,\rho$, and $\pi$ mesons were further developed by Savushkin \textit{et al.} \cite{Savushkin1997_PRC55-167}.

Self-consistent RBHF calculations for nuclear matter by Brockmann and Machleidt using the newly developed Bonn A, B, and C interactions presented more detailed and more systematic results \cite{Brockmann1990_PRC42-1965}.
Jong and Malfliet improved the treatment of the singularity in the scattering matrix \cite{deJong1991_PRC44-998}.
These results became very popular and were used directly to construct effective interactions.

A DD effective interaction with $\sigma$-$\omega$ mesons was developed in Refs.~\cite{Brockmann1992_PRL68-3408,Brockmann1990_PRC42-1965} by fitting to the single-particle potential of nuclear matter derived from RBHF calculations with the Bonn A interaction. An extension to the DDRHF framework can be found in Ref. \cite{Fritz1993_PRL71-46,Fritz1994_PRC49-633}. Furthermore, the isovector $\rho$ meson without tensor was taken into account with a fixed coupling constant \cite{Ma1994_PRC50-3170}.
The Groningen group \cite{deJong1991_PRC44-998} developed a DDRHF effective interaction parameterized in the invariant Lorentz space by fitting to the $G$-matrix derived from RBHF calculations \cite{Boersma1994_PRC49-233,Boersma1994_PRC50-1253}.
In Ref.~\cite{Boersma1994_PRC49-233} an explicit density dependence was introduced in the fitting procedure, therefore such a parametrization can be applied to finite nuclei easily \cite{Boersma1994_PRC49-1495}.

Up to this point, DDRMF and DDRHF do not include the so-called rearrangement term.
This term appears if the single-particle potential is defined by the variation of the total energy with respect to the single-particle state,
\begin{equation}
  U \psi = \frac{\delta \langle V \rangle}{\delta \psi^\dagger}.
\end{equation}
For a usual non-density-dependent two-body interaction, this will give the results of Eq.~(\ref{Eq:3.1.UHF}).
However, for a interaction with DD coupling strengths, the variation will give an additional term,
\begin{equation}
  \frac{\delta \langle V \rangle}{\delta \psi^\dagger} =
  \frac{\partial \langle V \rangle}{\partial \psi^\dagger} +
  \frac{\partial \langle V \rangle}{\partial \rho} \frac{\delta \rho}{\delta \psi^\dagger},
\end{equation}
where the last term is the so-called rearrangement potential.
In the HF total energy (\ref{Eq:3.1.EHF}), the rearrangement term does not appear explicitly, but
it has significant influence on the single-particle energies and wave functions and thus will change the total energy indirectly.

In (R)BHF theory, as discussed in Section~\ref{Sect:3.1}, the definition of single-particle potential is not unique. It is chosen usually for a better convergence of the hole-line expansion.
With the choice of Eq.~(\ref{Eq:3.1.UBHF}), which is similar to the HF potential (\ref{Eq:3.1.UHF}), there is no rearrangement term.

The concept of a ``rearrangement term'' in (R)BHF theory is similar but more complicated.
It was first used by Brueckner to refer to the difference between the binding of a $A$-nucleon system and a ($A-1$)-nucleon system plus the 1 nucleon removal energy, due to the change of the Pauli operator in the BG equation when removing a particle \cite{Brueckner1955_PR97-1353}.
Considering the single-particle energy derived in perturbation theory, the leading correction from Eq.~(\ref{Eq:3.1.UBHF}) corresponds to another definition of the single-particle energy as \cite{Brueckner1960_PR117-207,Brueckner1960_PR118-1438}
\begin{equation}
  e_i = \frac{\partial E_{\rm BHF}}{\partial n_i},
\end{equation}
where the occupation number $n_i$ is the expectation value of the number operator for the state $i$.
This is the energy required to remove a particle from the system, leaving a hole in the state $i$. Therefore it satisfies the separation energy theorem of Hugenholtz and van Hove \cite{Hugenholtz1958_Physica24-363}.
Another type is sometimes referred as the ``orbital rearrangement'', in contrast to the ``Brueckner rearrangement'', caused by the change of the self-consistent potential (or the wave function) \cite{Kohler1965_PR137-B1145,Kohler1965_PR138-B831,Kohler1966_NP88-529,Faessler1969_ZP223-192}.
See also a discussion in Ref.~\cite{Muther1973_NPA215-213} for these two rearrangement terms in BHF theory and the relation to Brandow's definition of single-particle energy \cite{Brandow1963_PL4-8,Brandow1967_RMP39-771}, and the renormalized BHF theory \cite{Becker1970_PRL24-400,Becker1970_PLB32-263}.

Back to the LDA for RBHF in nuclear matter, the missing rearrangement term in the calculation of finite nuclei using the DD effective interactions was investigated in detail in Refs.~\cite{Lenske1995_PLB345-355,Fuchs1995_PRC52-3043}.
The DDRMF effective interaction used  \cite{Haddad1993_PRC48-2740} was based on the single-particle potential of RBHF in nuclear matter \cite{Brockmann1990_PRC42-1965}, and two density dependence schemes were used. For the vector density dependence (VDD), the baryon density $\rho_b=\rho_0$ (the $0$-component of vector) density) is used and for both the $\sigma$-coupling and $\omega$-coupling. For the scalar density dependence (SDD),  the scalar density $\rho_s$ is used for $\sigma$-coupling while vector density for $\omega$-coupling.
Both schemes including the rearrangement term improved the description of binding energy and charge radius simultaneously. Later, the effect of the rearrangement term was studied \cite{Ineichen1996_PRC53-2158} using the DDRMF effective interaction based on RBHF calculations for nuclear matter in a full basis \cite{Huber1993_PLB317-485,Huber1995_PRC51-1790}, and similar results were found.

Most of the studies discussed so far are based on RBHF calculation for symmetric nuclear matter, and the isovector mesons were not included or only included with fixed phenomenological coupling constants.
With RBHF results for asymmetric nuclear matter \cite{Engvik1994_PRL73-2650,Engvik1996_AJ469-794}, the DDRMF effective interaction with $\sigma,\omega,\rho$, and $\pi$ mesons were determined and applied to finite nuclei in Ref.~\cite{Shen1997_PRC55-1211}.
However, the resulting coupling strength $g_\rho(\rho)$ decreased too much with the density and even became negative \cite{Shen1997_PRC55-1211}.
Using a parametrization of the $G$-matrix of Ref. \cite{Boersma1994_PRC49-233,Boersma1994_PRC50-1253}, it was demostrated that the single-particle potential determined by analyzing the momentum dependence in Ref.~\cite{Brockmann1990_PRC42-1965} might predict a ``wrong sign'' for the isovector dependence of $U_s$ and $U_0$ \cite{Ulrych1997_PRC56-1788}.
The RBHF calculation for asymmetric nuclear matter of the Giessen group adopted the projection method \cite{deJong1998_PRC57-3099}, and therefore it did not have the sign problem in the LDA study \cite{Hofmann2001_PRC64-034314}.
Another slightly different approach was adopted by the T\"ubingen group which used the so-called substracted $T$-matrix method to study asymmetric nuclear matter with RBHF \cite{VanDalen2004_NPA744-227,VanDalen2007_EPJA31-29}, and the results were later used for an LDA study \cite{Gogelein2008_PRC77-025802}.
In Ref.~\cite{VanDalen2011_PRC84-024320}, by choosing all the mesons as density-dependent and determining the coupling strengths by fitting to the RBHF results of asymmetric nuclear matter, the DDRHF model with $\sigma,\omega,\rho,\pi$, and $\delta$ mesons yielded the best reproduction of the RBHF results in nuclear matter and provided a good description for finite nuclei.

As shown in the above discussion, RBHF calculations have made substantial progress and deepened our understanding for relativistic \textit{ab initio} calculations of the nuclear system over the past decades.
However, it is still a challenge to realize a fully self-consistent RBHF for nuclear matter in the full Dirac space with high precision. An exact treatment and the estimation of the approximations for the angle-average of total momentum \cite{Tong2018_PRC98-054302} and Pauli operator \cite{Schiller1999_PRC59-2934,Suzuki2000_NPA665-92,Schiller1999_PRC60-059901} in the RBHF framework are also quite important.

On the other hand, the \textit{ab initio} RBHF calculations provides valuable information and stimulate the development of nuclear (covariant) density functional theory. In the following, some of the examples on the guidance of the CDFT through RBHF will be discussed. Possible future opportunities to improve CDFT through RBHF calculation in finite nuclei will be discussed in Section~\ref{Sect:4}.

The importance of the nonlinear coupling for the $\omega$ meson has been pointed out in the parametrization of effective interactions in nuclear matter \cite{Gmuca1992_ZPA342-387}.
By comparing the single-particle potentials derived from the RBHF calculations \cite{Brockmann1990_PRC42-1965}, Sugahara and Toki showed the necessity to introduce a nonlinear coupling for the $\omega$ meson \cite{Sugahara1994_NPA579-557}.

Suggested by the microscopic \textit{ab initio} RBHF calculations for nuclear matter \cite{terHaar1987_PR149-207,Brockmann1990_PRC42-1965,Sehn1997_PRC56-216,deJong1998_PRC57-3099} and the related LDA studies \cite{Marcos1989_PRC39-1134,Brockmann1992_PRL68-3408,Fritz1993_PRL71-46,Haddad1993_PRC48-2740,Fritz1994_PRC49-633,Boersma1994_PRC49-233,Boersma1994_PRC50-1253}, the DD effective interaction, later often referred as TW99, was proposed in Ref.~\cite{Typel1999_NPA656-331}.
After that, based on these general considerations, phenomenological CDFTs with density dependent coupling constants have been developed. By a careful adjustment of the parameters new DD effective interactions were developed
which reproduced with high precision the available data not only for ground state properties, but also for collective rotational and vibrational excitations. They also have been successfully applied for density functional applications beyond mean field. For reviews see Refs. \cite{Vretenar2005_PR409-101,Meng2006_PPNP57-470,Niksic2011_PPNP66-519,Meng2016}

\begin{figure}
  \centering
  \includegraphics[angle=270,width=6cm]{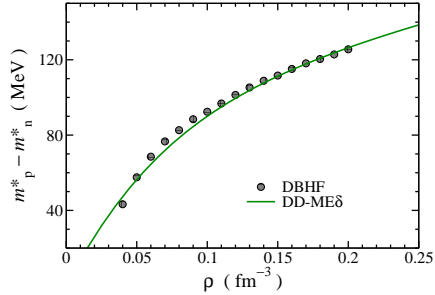}
  \caption{(Color online) Proton-neutron effective mass difference as a function of the nucleon density in pure neutron matter. The dots represent the results of RBHF calculations \cite{VanDalen2005_PRL95-022302,VanDalen2007_EPJA31-29} and the line is the fit of DD-ME$\delta$. Figure modified from Ref.~\cite{RocaMaza2011_PRC84-054309}.}
  \label{fig4}
\end{figure}

Among these DDRMF functionals, DD-ME$\delta$ is another example how the \textit{ab initio} RBHF studies can guide the development of nuclear density functionals \cite{RocaMaza2011_PRC84-054309}.
Before that, most of relativistic functionals do not include the scalar-isovector $\delta$ meson, because a general fit to the binding energies and charge radii of finite nuclei is not sensitive to this degree of freedom. One needs {\it ab initio} information. Guided by the proton-neutron effective mass splitting in neutron matter calculated by RBHF with the Bonn A interaction \cite{VanDalen2005_PRL95-022302,VanDalen2007_EPJA31-29}, the strength of the $\delta$ coupling can be determined, as shown in Fig.~\ref{fig4}.
  Although the inclusion of the $\delta$ meson does not improve the accuracy of the properties of finite nuclei such as masses and radii (for example, an earlier parameter set DD-ME2 which does not include the $\delta$ meson outperforms DD-ME$\delta$ in a global description of the properties of finite nuclei~\cite{Agbemava2014_PRC89-054320}), the proton-neutron effective mass splitting is correctly incorporated by DD-ME$\delta$, and its EoS at higher densities is in a better agreement with experimental data derived from heavy-ion reactions than that of DD-ME2~\cite{RocaMaza2011_PRC84-054309}.

The first DDRHF effective interaction PKO1, which has comparable precision for nuclear ground-state properties with the DDRMF ones, was developed in Ref.~\cite{Long2006_PLB639-242}.
By including the Fock terms and the contribution of $\pi$ meson, it improves the descriptions of the nucleon effective mass and its isospin and energy dependence, while it describes well for bulk properties like binding energies and radii.

However, although some of the studies of LDA from RBHF in nuclear matter included $\pi$ or the $\rho$ tensor \cite{Ma1994_PRC50-3170, Shi1995_PRC52-144,VanDalen2011_PRC84-024320}, their values could not be constrained from the properties of nuclear matter and were fixed to phenomenological values.
In such a case, the RBHF study for finite nuclei is indispensable and we will show in Section~\ref{Sect:4} how nuclear density functional can be further improved through the information of finite nuclei.

\subsection{Self-consistent RBHF calculations for finite nuclei}\label{Sect:3.5}

In this Section, we will introduce the details of the self-consistent RBHF calculations for finite nuclei, including the convergence check, the treatment of center-of-mass motion, the self-consistent basis, and the choice of the single-particle potential.

The numerical settings for the RBHF calculation for finite nuclei has been given in Ref.~\cite{Shen2017_PRC96-014316} in detail.
For the nucleus $^{16}$O, the realistic bare $NN$ interaction Bonn A which is adjusted to the $NN$ scattering data~\cite{Machleidt1989_ANP19-189} is used as the only input.
The DWS basis, the solution of the Dirac equation with a Woods-Saxon potential, is then obtained by solving the spherical Dirac equation~(\ref{Eq:3.3.RDeq}) in a box with the box size $R=7$~fm and mesh size $dr=0.05$~fm \cite{Zhou2003_PRC68-034323}.
The procedure to solve the full RBHF equations has been introduced in Section~\ref{Sect:3.3}.
As one works always in the RHF basis during the iteration, the Pauli operator in Eq.~(\ref{Eq:3.3.BGeq}) and all its relativistic structure is fully taken into account.
In particular, there is no angle averaging involved, as such an averaging is involved in most Brueckner calculations for nuclear matter \cite{Schiller1999_PRC59-2934,Suzuki2000_NPA665-92}.
The RBG equation~(\ref{Eq:3.3.BGeq}) is solved for four different values of the starting energy $W$, equally distributed between the lowest and the highest single-particle energies in the Fermi sea.
The $G$-matrix with specific starting energy $W$ is obtained by a four point Lagrange polynomial interpolation \cite{Davies1969_PR177-1519}.

\subsubsection{Convergence check}\label{Sect:3.5.1}

It is well known that the bare $NN$ interaction contains a strong repulsive core and a strong tensor part which can connect the nucleons below the Fermi surface to the states with high momentum in the continuum.
In most of the non-relativistic \textit{ab initio} calculations, such a bare interaction is usually renormalized to a soft one while keeping the low-momentum properties unchanged.
The renormalized interaction is then applied to the nuclear-structure calculations, see, e.g., the $V_{{\rm low}~k}$ \cite{Bogner2003_PR386-1} or the similarity renormalization group method \cite{Bogner2007_PRC75-061001}.
The RBHF study for finite nuclei developed in Refs.~\cite{Shen2016_CPL33-102103,Shen2017_PRC96-014316} uses directly the bare Bonn interactions \cite{Machleidt1989_ANP19-189}, and therefore it is crucial to investigate the corresponding convergence of the RBHF calculations with respect to the basis space.

\begin{figure}
  \centering
  \includegraphics[width=8cm]{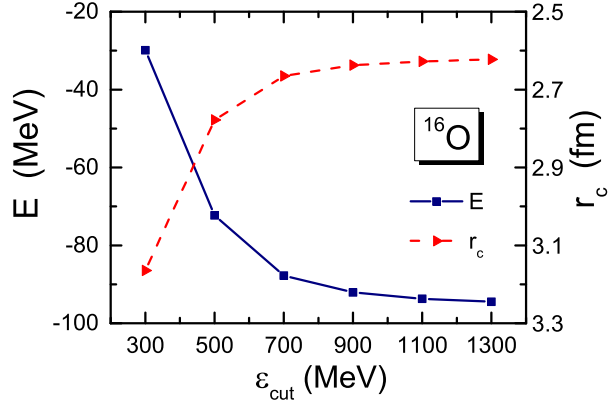}
  \caption{(Color online) Total energy $E$ and charge radius $r_c$ of $^{16}$O as a function of the energy cut-off $\varepsilon_{\rm cut}$ calculated in RBHF theory. The center-of-mass correction has not been included. Figure taken from Ref.~\cite{Shen2017_PRC96-014316}.}
  \label{fig5}
\end{figure}

The total energy $E$ and the charge radius $r_c$ of $^{16}$O calculated by RBHF theory are shown in Fig.~\ref{fig5} as a function of the energy cut-off $\varepsilon_{\rm cut}$.
In usual RMF (similarly in RHF) calculations with a DWS basis expansion for nuclei like $^{16}$O, a cut-off $\varepsilon_{\rm cut} = 300$ MeV would be more than enough to achieve convergence with very high precision \cite{Zhou2003_PRC68-034323}.
However, as seen in Fig.~\ref{fig5}, this is not the case for RBHF calculations.
To achieve a precision $\Delta E < 1$ MeV, one needs a cut-off of at least $\varepsilon_{\rm cut} = 1100$ MeV.
This is due to the short-range correlation induced by the strong repulsive core of the bare interaction, which involves contributions of the states with high momentum in the continuum.

Other cut-offs, such as the box size, angular momentum, etc., have been carefully checked in Ref.~\cite{Shen2017_PRC96-014316} and will not be repeated here.
For $^{16}$O, the final basis space used in the calculation is as following: the single-particle angular momentum cut-off $l_{\rm cut} = 20 \hbar$, the energy cut-off in the Fermi sea $\varepsilon_{\rm cut} = 1100$ MeV and that in the Dirac sea $\varepsilon_{\rm cut} = -1700$ (with the condition that at least 2 negative-energy states are included in each block), the box size $R = 7$ fm, the total angular momentum cut-off $J_{\rm cut} = 6 \hbar$ for $jj$-coupled two-body matrix elements, and the pair cut-off for the single-particle angular momentum and energy is applied when $J > 3~\hbar$ as $l_1+l_2 < 20~\hbar$, $\varepsilon_{1}+\varepsilon_{2} < 1100$ MeV.
With these conditions, the total size of the two-body matrix elements stored in the computer is about 256 GB, and the computation time is about several days depends on specific machines.
With the development of computational resources and possible optimization of parallelizing the RBHF code, the study of heavy nuclei such as $^{208}$Pb by the relativistic \textit{ab initio} calculations is promising in the foreseen future.

\subsubsection{Center-of-mass motion}\label{Sect:3.5.2}

For light finite nuclei, the treatment of the spurious center-of-mass (c.m.) motion is very important.
Since the Hamiltonian is invariant against translations, the exact many-body eigenstates of the system should be eigenfunctions of the total momentum $ \mathbf{P} = \sum_i^A \mathbf{p}_i$.
It has been shown in Refs.~\cite{Beck1970_ZP231-26,Ring1980} that, for large values of the particle number $A$, the projected energy is obtained in a good approximation by removing the c.m. energy
\begin{equation}\label{Eq:3.3.Ecm}
   E_{cm} =  \frac{{\langle\bf{P}\rangle}^2}{2AM}
\end{equation}
from the total energy $\langle H\rangle=\langle T+V\rangle$.

In most of the (R)HF calculations, the variation is carried out without projection, i.e., the (R)HF equations are solved for the total Hamiltonian $H$, and the spurious c.m. energy in Eq.~(\ref{Eq:3.3.Ecm}) is removed after the variation Ref.~\cite{Long2004_PRC69-034319}.
This is a projection after variation (PAV).
A strict treatment would be to exclude this term also in the (R)HF equation, i.e., to carry out a projection before variation (PBV) \cite{Zeh1965_ZP188-361,Chabanat1998_NPA635-231}.
In none of these cases, i.e., PAV and PBV, the c.m. term is included in the solution of the BG equation (\ref{Eq:3.3.BGeq}).
These two approaches have been discussed in the non-relativistic BHF calculation long time ago, which were found to give similar results \cite{Becker1974_PRC9-1221}.
However, in Ref.~\cite{Becker1974_PRC9-1221}, the BHF equations were not solved fully self-consistently, since the BG equation was only solved at the beginning and the $G$-matrix was not recalculated during the iteration.

In Table \ref{tab3}, the total energy, charge radius, and c.m. correction (\ref{Eq:3.3.Ecm}) for $^{16}$O calculated by RBHF for PBV and PAV are shown.
It can be seen that although the values of the c.m. correction energy are similar for these two approaches, the corresponding total energies are quite different. PBV gives $9$ MeV more binding than PAV.
This difference is much larger than that reported in Ref.~\cite{Becker1974_PRC9-1221}.
From Eq.~(\ref{Eq:3.3.Ecm}), one can conclude that the wave functions given by PBV and PAV are similar to each other, since the c.m. correction energies are similar.
The difference in the total energy then can only be due to the difference in $G$-matrix for the two cases.

\begin{table}
\caption{Total energy, charge radius, and c.m. correction for $^{16}$O calculated by RBHF for PBV and PAV.}
\label{tab3}
\centering
\begin{tabular}{lccc}
\hline\hline
& $E$ (MeV) & $r_c$ (fm) & $E_{cm}$ (MeV) \\
\hline
PBV & $-110.1$ & $2.566$ & $-11.83$ \\
PAV & $-101.4$ & $2.577$ & $-11.12$ \\
\hline\hline
\end{tabular}
\end{table}

\begin{figure}[!htbp]
  \centering
  \includegraphics[width=6cm]{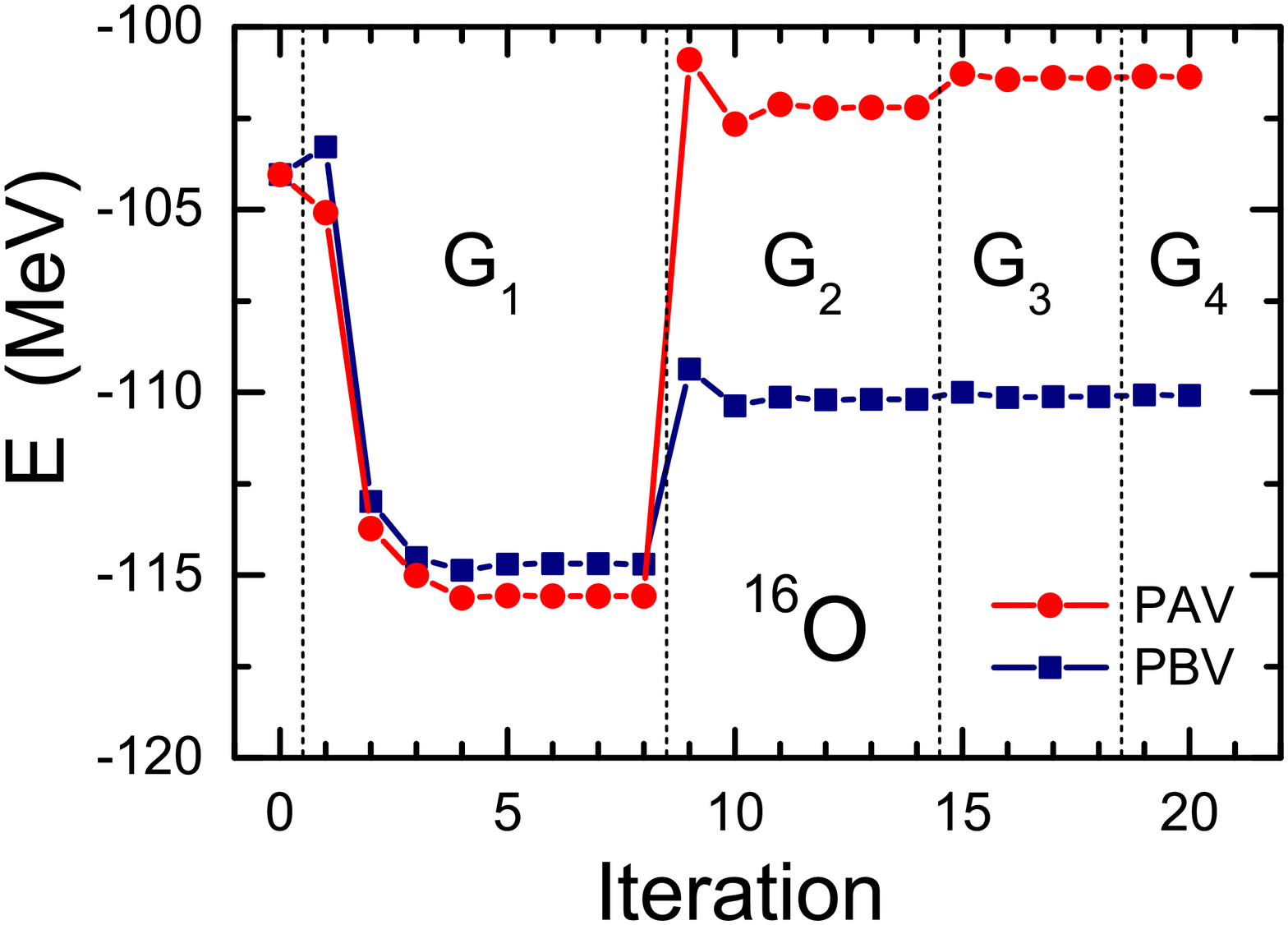}
  \hspace{2cm}
  \includegraphics[width=6cm]{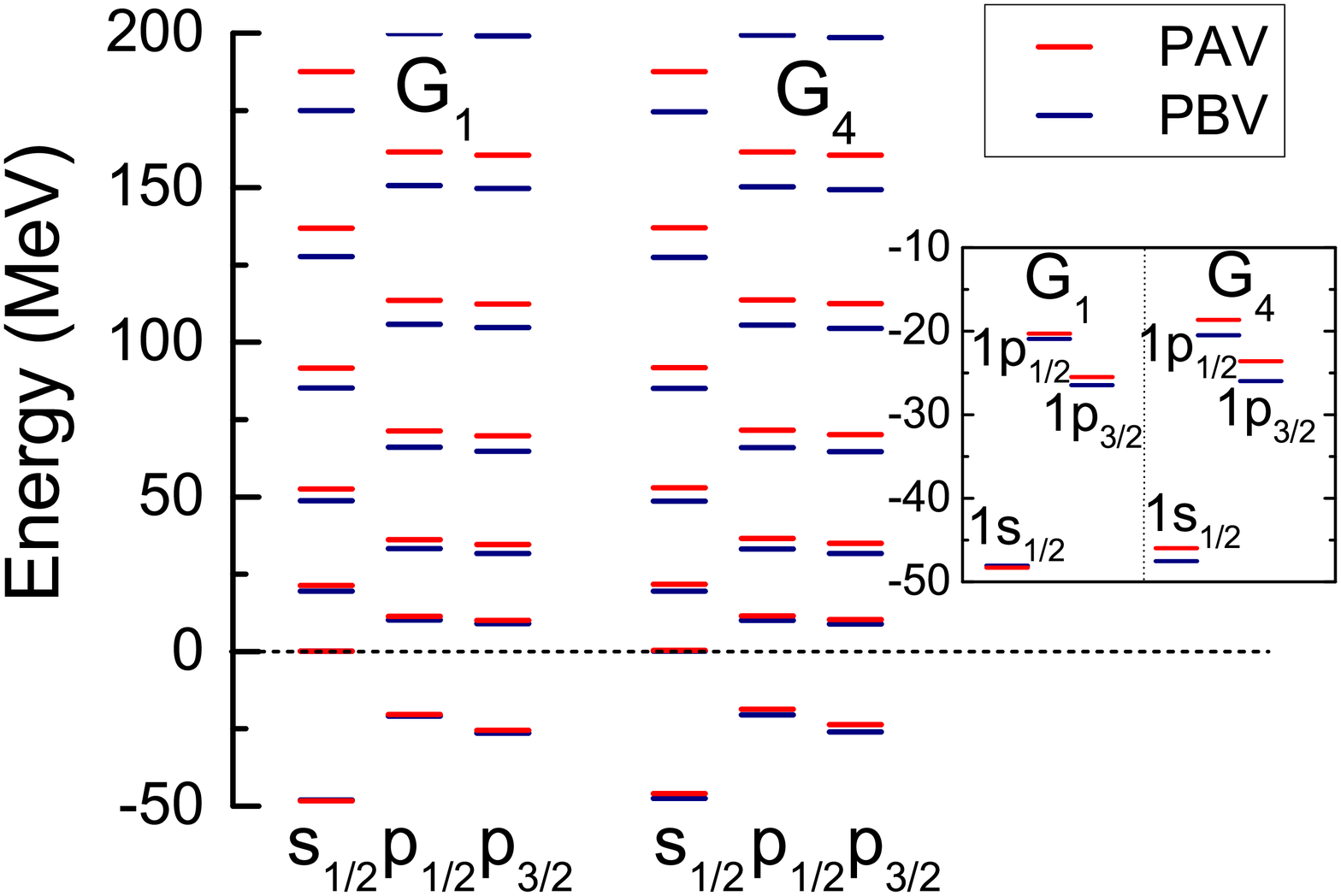}
  \caption{(Color online) (Left) Total energies at each RBHF iteration step for PAV and PBV. (Right) Single-particle spectra in the $s$ and $p$ blocks at each RBHF iteration step for PAV and PBV. Figures taken from Ref.~\cite{Shen2017_PRC96-014316}.}
  \label{fig6}
\end{figure}

In order to understand this more clearly, at each step of the iteration  the total energy is shown in the left panel of Fig.~\ref{fig6}.
It can be seen that there is no significant difference between PBV and PAV during the first RBHF iteration step, where the $G$-matrix $G_1$ is calculated from the initial DWS basis.
One may say that with the same interaction, PAV can be viewed as a good approximation of PBV.
While in the next RBHF iterations, the $G$-matrices $G_2,\,G_3,$ and $G_4$ are calculated each time in the converged RHF basis of the previous step, and we observe that the energy of PBV becomes smaller than that of PAV.
The reason for such a sudden change can be understood from the right panel of Fig.~\ref{fig6}, where the single-particle energies of the $s$ and $p$ blocks are given after the first ($G_1$) and the last ($G_4$) RBHF iterations.
It can be seen that in the RBHF calculation the PBV approach leads in general to lower single-particle energies than the PAV, especially for the high-lying states.
In the conventional RHF calculations, PBV and PAV approaches give similar results because only the occupied states are involved and they are similar for PAV and PBV.
Whereas for the RBHF calculations, the difference in the single-particle spectra, in particular in the high-lying states, will lead to different $G$-matrices in next iterations, $G_2, G_3$, and $G_4$.
Lower unoccupied states generally coincide with more attractive $G$-matrix elements of occupied states.
As a result, the $G$-matrix in PBV is more attractive and the total binding energy becomes larger.

\subsubsection{Self-consistent basis}\label{Sect:3.5.3}

In the above discussion for the c.m. correction, two approaches, PAV and PBV, make a big difference if the RBHF equations are solved in the self-consistent RHF basis.
This indicates the importance of using the self-consistent basis in the (R)BHF calculations for finite nuclei.
Another way to see the importance of the self-consistent RHF basis is to examine the dependence of the final results on the chosen initial basis.

At the beginning, one does not know the final RBHF solution and therefore a trial basis is needed.
To take into account the relativistic structure of the Pauli operator in Eq.~(\ref{Eq:3.3.BGeq}),
a relativistic basis is preferred.
In Refs.~\cite{Shen2016_CPL33-102103,Shen2017_PRC96-014316}, the relativistic DWS basis~\cite{Zhou2003_PRC68-034323} has been chosen as the initial trial basis. In comparison with the harmonic oscillator basis~\cite{Gambhir1990_APNY198-132}, it has advantages like a proper asymptotic behavior of nuclear density distribution, which is crucial for describing, e.g., halo nuclei.
More important here is that the nucleon single-particle potential is close to the DWS shape, which serves as a good approximation for the final converged RBHF single-particle states.

To show the importance of the self-consistent RHF basis, two types of calculations have been performed, i.e., solving the RBHF equations with a fixed initial basis \cite{Shen2016_CPL33-102103} or with the self-consistent RHF basis  \cite{Shen2017_PRC96-014316}.
In both calculations, the DWS basis is obtained by solving the radial Dirac equation ~(\ref{Eq:3.3.RDeq}) with the Woods-Saxon potentials.

\begin{figure}
  \centering
  \includegraphics[width=8cm]{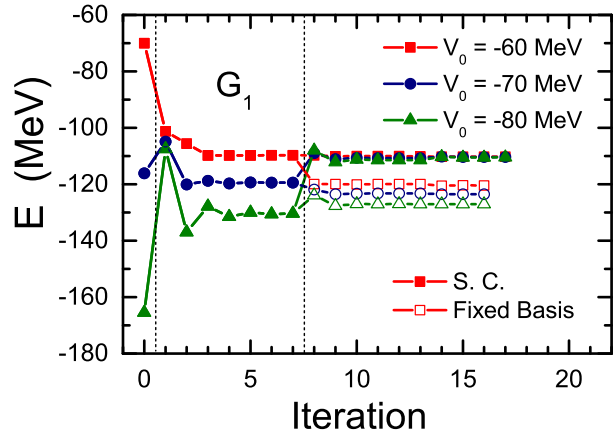}
  \caption{(Color online) Total energies at each iteration step calculated by RBHF in the self-consistent RHF basis (solid symbols) and in the fixed DWS basis (open symbols). Figure taken from Ref.~\cite{Shen2017_PRC96-014316}.}
  \label{fig7}
\end{figure}

The total energies of both calculations at each iteration step are plotted in Fig.~\ref{fig7}.
The filled symbols stand for the self-consistent RBHF calculations, and the open ones for the fixed DWS basis calculations, and different types of symbols represent different DWS potential depths $V_0$.
Similar to Fig.~\ref{fig6}, the first RBHF iteration is represented by $G_1$.
In Fig.~\ref{fig6} or Fig.~\ref{fig7}, different calculations may have different number of iteration steps, but to have better comparison it has been adjusted slightly in the figures to make the number of iterations the same without losing too much precision.
For the first RBHF iteration $G_1$, the self-consistent RBHF calculations are identical to the ones with fixed DWS basis for various $V_0$ values, because in both cases the same basis is used.
After the first iteration $G_1$, there appears a distinct difference between the self-consistent calculations and those with the fixed basis.
Moreover, the self-consistent RBHF calculations get converged with respect to the potential depths $V_0$, while those with a fixed basis do not.
As expected, self-consistency is very important to get unambiguous results.

\subsubsection{Choice of the single-particle potential}\label{Sect:3.5.4}

One uncertainty in (R)BHF theory is the choice of the single-particle potential for the particle states in Eq. (\ref{Eq:57}).
Different choices have been discussed in BHF for finite nuclei and more details can be found in Ref. \cite{Davies1969_PR177-1519}.
Those were non-relativistic investigations and therefore one only had matrix elements $\langle a|U|b\rangle$ for the single-particle states $|a\rangle$ and $|b\rangle$ in the Fermi sea and above the Fermi level.
In the relativistic framework, it becomes more complicated as the single-particle states $|a\rangle$ and $|b\rangle$ can have negative energies (in the Dirac sea) and the definition of single-particle potential for these states is another open question.

In Eq.~(\ref{Eq:57}), the single-particle states $|a\rangle$ and $|b\rangle$ in the Dirac sea were treated as occupied.
This choice seems to be reasonable since in the Bethe-Goldstone equation (\ref{Eq:3.3.BGeq}) the intermediate states $|c\rangle$ and $|d\rangle$ are only allowed to be states above the Fermi surface $e_c,\,e_d > e_F$.
From this point of view, the single-particle states in the Dirac sea are ``occupied'' hole states.

In Ref. \cite{Shen2017_PRC96-014316}, these states were treated as unoccupied, i.e. in a similar way as the states above the Fermi level, and the formula is slightly different from that of Eq.~(\ref{Eq:57}) as
\begin{equation}\label{Eq:74}
\langle a|U|b\rangle=
\begin{cases}
\frac{1}{2}\sum_{i=1}^A \langle ai|\bar{G}(e_a+e_i)+ \bar{G}(e_b+e_i)|bi\rangle,
 & 0 < (e_a, e_b) \leq e_F, \\
\sum_{i=1}^A \langle ai|\bar{G}(e_a+e_i)|bi\rangle,
& 0 < e_a \leq e_F,~e_b > e_F ~\text{or}~e_b < 0,  \\
\sum_{i=1}^A \langle ai|\bar{G}(e'+e_i)|bi\rangle,
& e_a, e_b > e_F ~\text{or}~ < 0,
\end{cases}
\end{equation}
where the index $i$ runs over the occupied states in the Fermi sea (\textit{no-sea} approximation).

As discussed in Ref.~\cite{Shen2018_PLB781-227}, there is no ``right'' or ``wrong'' choice for the single-particle potential in (R)BHF theory, since (R)BHF theory can be viewed as the two-hole-line expansion in the more general hole-line expansion (or the Brueckner-Bethe-Goldstone expansion) \cite{Day1967_RMP39-719} and as the expansion goes to higher order the results become independent of the choice of $U$ \cite{Song1998_PRL81-1584}.
On the other hand, there do exist ``better'' choices of $U$ as they will affect the convergence rate of the hole-line expansion.
It has been shown that the definition for hole states ($0 < e \leq e_F$) in Eq.~(\ref{Eq:3.1.UBHF}) cancels a certain large amount of higher-order diagrams and this accelerates the convergence of the hole-line expansion and improves the BHF approximation~\cite{Bethe1963_PR129-225,Baranger1969_Varenna40-511}.
However, there is no similar proof for the particle states nor for the states in the Dirac sea.

The effect of different choices are shown in Table \ref{tab4} \cite{Shen2017_PRC96-014316,Shen2018_PLB781-227}  including,
\begin{enumerate}
  \item Formula I, Eq.~(\ref{Eq:74}), and for $e_a, e_b > e_F ~\text{or}~ < 0$ the gap choice $\langle a|U|b \rangle = 0$ is chosen.
  \item Formula I, Eq.~(\ref{Eq:74}), and $e'$ fixed as the energy of the lowest occupied states, $e_{\nu 1s1/2}$.
  \item Formula I, Eq.~(\ref{Eq:74}), and $e'$ fixed as the energy of the highest occupied states, $e_{\pi 1p1/2}$.
  \item Formula II, Eq.~(\ref{Eq:57}), and $e'$ fixed as the energy of the highest occupied states, $e_{\pi 1p1/2}$.
\end{enumerate}

\begin{table}
\caption{Binding energy per nucleon ($B/A$ in MeV), charge radius ($r_c$ in fm), and proton $1p$ spin-orbit splitting ($\Delta E_{\pi1p}^{ls}$ in MeV) of $^{16}$O in the RBHF calculations with different choices for $U_{ab}$.
Formula I for Eq.~(\ref{Eq:74}), and formula II for Eq.~(\ref{Eq:57}).
The experimental data are from \cite{Wang2017_CPC41-030003,Angeli2013_ADNDT99-69,Coraggio2003_PRC68-034320}.}
\label{tab4}
\centering
\begin{tabular}{lccccc}
  \hline\hline
     & I, Gap & I, $e'=e_{\nu 1s1/2}$ & I, $e'=e_{\pi 1p1/2}$ & II, $e'=e_{\pi 1p1/2}$ & Exp. \\
  \hline
  $B/A$                    & $5.41$ & $6.88$ & $7.10$ & $7.51$ & $7.98$ \\
  $r_c$                     & $2.64$ & $2.57$ & $2.56$ & $2.53$ & $2.70$ \\
  $\Delta E_{\pi1p}^{ls}$  & $5.4\,\,\,$ & $5.4\,\,\,$ & $5.4\,\,\,$ & $5.3\,\,\,$ & $6.3\,\,\,$ \\
  \hline\hline
\end{tabular}
\end{table}

For Formula I, it can be seen that the gap choice gives the least bound system, while fixing $e'$ as an energy among the occupied states gives $1.4$ to $1.7$ MeV per nucleon more binding.
This is consistent with the non-relativistic BHF calculations.
In the framework of the Bethe-Brueckner-Goldstone expansion (or hole-line expansion, see for instance Ref.~\cite{Day1967_RMP39-719}), the gap choice and the continuous choice have been discussed in nuclear matter in Ref.~\cite{Song1998_PRL81-1584}. The continuous choice gives more binding than the gap choice at the BHF level. By performing the BBG expansion up to the three hole-line level,
the above two choices give similar results. They lie between the results of these two choices at the BHF level~\cite{Song1998_PRL81-1584}.
From this point of view, the choice of $e'$ as an energy among the occupied states is a reasonable choice.

For Formula II, fixing $e'=e_{\pi 1p1/2}$ gives $0.4$ MeV per nucleon more binding than Formula I and is in better agreement with the data.
On the other hand, the rms charge radius is by $0.03$ fm smaller, and the SO splitting is smaller by $0.1$ MeV.
The two formulas will have another major effect on the spin symmetries in the Dirac sea, and this will be shown in Section~\ref{Sect:3.7}.

\subsection{Ground-state properties for finite nuclei}\label{Sect:3.6}


  In this Section, the ground-state properties for $^4$He, $^{16}$O, and $^{40}$Ca are studied by RBHF. Here we will show RBHF results for $^{16}$O and $^{40}$Ca using formula II (\ref{Eq:57}) for the choice of the single-particle potential, instead of using formula I (\ref{Eq:74}) as it was done in Ref.~\cite{Shen2017_PRC96-014316}. The formula II is preferred as the treatment of the single-particle states in the Dirac sea is similar to that of occupied states, and the resulting spin symmetry properties are more consistent with the findings in the studies of CDFT (see the discussion of next Section). The results of $^4$He in Ref. \cite{Shen2017_PRC96-014316} will not be updated as RBHF theory is more focusing on medium-heavy and heavy nuclei, and for $^4$He there are exact solutions by solving the Faddeev-Yakubovsky equation, as for instance in Ref. \cite{Binder2016_PRC93-044002}. Without specification, the center-of-mass correction is treated by the PBV method.

\begin{figure}
  \centering
  \includegraphics[width=8cm]{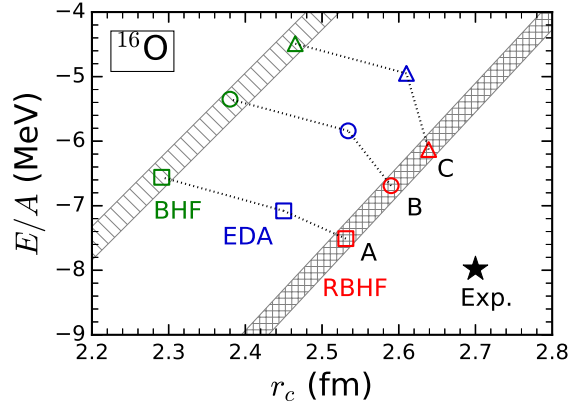}
  \caption{(Color online) Energy per nucleon $E/A$ for $^{16}$O as a function of the charge radius $r_c$ calculated by RBHF using the interactions Bonn A, B, and C, in comparison with BHF and the relativistic effective density approximation (EDA).}
  \label{fig8}
\end{figure}

  In Fig.~\ref{fig8} we show results of the RBHF calculations with the interactions Bonn A, B, and C~\cite{Machleidt1989_ANP19-189}, the energy per nucleon and the charge radius of $^{16}$O, and compare them with the renormalized BHF \cite{Muther1990_PRC42-1981} and the relativistic effective density approximation (EDA)  \cite{Muther1990_PRC42-1981} using the same interactions. Similar as for the results of nuclear matter, we have connected the results of Bonn A, B, and C to indicate schematically the ``Coester lines''. 
The experimental data \cite{Wang2017_CPC41-030003,Angeli2013_ADNDT99-69} is denoted by the black star.

By comparing the results of BHF and EDA with RBHF, it can be seen that the relativistic effects improve the description considerably, both for the energy and the radius.
The ``relativistic Coester line'' is much closer to the experimental data, consistent with the results in nuclear matter. By comparing the results of EDA and RBHF, it can be seen that self-consistentcy  is very important and it further improves the descriptions.
One interesting point to be noted is that the three points of RBHF (with the Bonn A, B, C interactions) are closer to each other than those for BHF or EDA.
This is similar as the results for nuclear matter, as shown in Fig.~\ref{fig1}, where the three saturation points given by RBHF with Bonn A, B, and C are closer to each other than those of BHF.
It should be mentioned that both calculations of BHF and EDA have adopted renormalized occupation probabilities, which takes into account certain rearrangement effects \cite{Becker1970_PRL24-400}.
This has not been done for RBHF, and therefore the comparison in Fig.~\ref{fig8} is not fully on the equal grounds.
If the results of the investigations of renormalized BHF hold also for RBHF, renormalized RBHF should further improve the description for finite nuclei, especially for the radii.

\begin{figure}
  \centering
  \includegraphics[width=8cm]{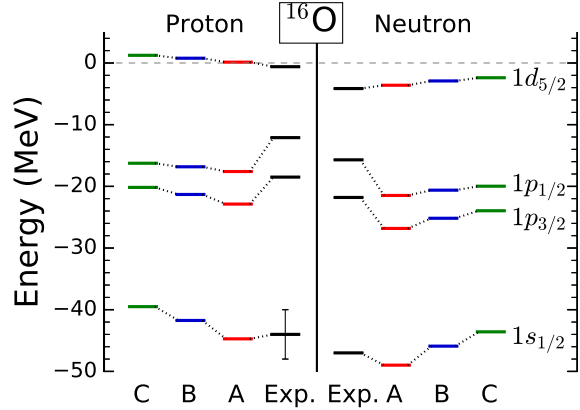}
  \caption{(Color online) Single-particle energies of $^{16}$O calculated by RBHF using the interactions Bonn A, B, and C, in comparison with experimental data \cite{Coraggio2003_PRC68-034320}.}
  \label{fig9}
\end{figure}

In Fig.~\ref{fig9}, the single-particle energies for $^{16}$O calculated by RBHF using the interactions Bonn A, B, and C \cite{Machleidt1989_ANP19-189} are shown in comparison with the experimental data \cite{Coraggio2003_PRC68-034320}.
The results of RBHF are in a reasonable agreement with the data, especially for the spin-orbit splittings.
However, the $p$-levels are slightly low.
It might be due to the lack of more complicated configurations such as particle-vibration coupling \cite{Litvinova2006_PRC73-044328} in the RBHF framework, where only the ladder diagrams have been included.

In Table~\ref{tab5} we list the total energy, charge radius $r_c$, matter radius $r_m$, and proton spin-orbit splitting for the $1p$ shell of $^{16}$O calculated by RBHF \cite{Shen2017_PRC96-014316,Shen2018_PLB781-227} with the interactions Bonn A, B, and C \cite{Machleidt1989_ANP19-189}, in comparison with the experimental data~\cite{Wang2017_CPC41-030003,Angeli2013_ADNDT99-69,Ozawa2001_NPA691-599,Coraggio2003_PRC68-034320}.
The RBHF results are different form Ref.~\cite{Shen2017_PRC96-014316} as they are updated with formula II for the single-particle potential (\ref{Eq:57}) instead of formula I (\ref{Eq:74}).
We also include the corresponding results from the density-dependent relativistic Hartree-Fock (DDRHF) with
the phenomenological interactions PKO1~\cite{Long2006_PLB639-242} and PKA1~\cite{Long2007_PRC76-034314}, the non-relativistic BHF results~\cite{Hu2017_PRC95-034321} with $V_{{\rm low}-k}$ derived from Argonne $v_{18}$~\cite{Wiringa1995_PRC51-38}, the coupled-cluster (CC) method~\cite{Hagen2009_PRC80-021306},
the in-medium similarity renormalization group (IM-SRG) method \cite{Hergert2013_PRC87-034307}, the no-core shell model (NCSM) \cite{Roth2011_PRL107-072501}, and the self-consistent Green's function (SCGF) method \cite{Cipollone2013_PRL111-062501} with N$^3$LO~\cite{Entem2003_PRC68-041001}, nuclear lattice effective field theory (NLEFT)~\cite{Lahde2014_PLB732-110} with N$^2$LO~\cite{Epelbaum2009_EPJA41-125}, and the quantum Monte-Carlo (QMC) method \cite{Lonardoni2018_PRC97-044318} with the local chiral force N$^2$LO \cite{Gezerlis2013_PRL111-032501}. All the results listed in Table~\ref{tab5} do not include bare three-body interactions.

  For the results of IM-SRG and NCSM, the uncertainties induced by the extrapolation procedure are given in parentheses. For the result of NLEFT, the combined statistical and extrapolation errors are given. For QMC, the first error is statistical, and the second is based on the EFT expansion uncertainty. For RBHF, the errors of extrapolation with respect to the box size were estimated as 0.3 MeV for the total energy and 0.02 fm for the charge radius \cite{Shen2017_PRC96-014316}. Other major uncertainties arise for the RBHF method due to cut-offs with respect to the single-particle basis space, the choice of starting energy, and three hole-line contributions. These shall be investigated in more detail in the future. 

  In the above references of non-relativistic \textit{ab initio} calculations with chiral interactions, some have reported also results with 3N interactions. However, there are still large uncertainties, and the results with 3N forces obtained by different many-body methods do not agree as well as the results with 2N interactions. For completeness, we also cite some of the values. For example, IM-SRG gives $E = -130.5(1)$ MeV, similar to $E = -130.8(1)$ MeV for SCGF; but NCSM gives $E = -143.7(2)$ to $-147.8(1)$ MeV when changing the flow parameter from $0.05$ to $0.08$; NLEFT gives $E = -138.8(5)$ MeV; QMC gives $E = -117(5)(16)$ MeV and $r_c = 2.71(5)(13)$ when the cutoff parameter is chosen as 1.0 fm, but the results are affected by this parameter to some extent.

\begin{table}
\caption{Total energy, charge radius, matter radius, and $\pi1p$ spin-orbit splitting of $^{16}$O calculated by RBHF theory \cite{Shen2017_PRC96-014316,Shen2018_PLB781-227} with the interactions Bonn A, B, and C, in comparison with experimental data. The corresponding results from DDRHF with effective interactions PKO1 \cite{Long2006_PLB639-242} and PKA1 \cite{Long2007_PRC76-034314}, non-relativistic BHF~\cite{Hu2017_PRC95-034321} with $V_{{\rm low}-k}$ derived from Argonne $v_{18}$, CC method \cite{Hagen2009_PRC80-021306}, IM-SRG \cite{Hergert2013_PRC87-034307}, NCSM \cite{Roth2011_PRL107-072501}, and SCGF method \cite{Cipollone2013_PRL111-062501} with N$^3$LO, NLEFT~\cite{Lahde2014_PLB732-110} with N$^2$LO, and QMC method \cite{Lonardoni2018_PRC97-044318} with local chiral force N$^2$LO are also included.}
\centering
\begin{tabular}{llclcc} 
\hline\hline
&\multicolumn{1}{c}{$E$ (MeV)} & \multicolumn{1}{c}{$r_c$ (fm)} & \multicolumn{1}{c}{$r_m$ (fm)} &
\multicolumn{1}{c}{$\Delta E_{\pi1p}^{ls}$ (MeV)} & Ref. \\
\hline
Exp. & $-127.6$ & $2.70$ & $2.54(2)$ & $~~6.3$ & \cite{Wang2017_CPC41-030003,Angeli2013_ADNDT99-69,Ozawa2001_NPA691-599,Coraggio2003_PRC68-034320} \\
RBHF, Bonn A & $-120.2$ & $2.53$ & $2.39$ & $\,\,\,5.3$ & \cite{Shen2017_PRC96-014316} \\
RBHF, Bonn B & $-107.1$ & $2.59$ & $2.45$ & $\,\,\,4.5$ & \cite{Shen2017_PRC96-014316} \\
RBHF, Bonn C & $\,\,\,-98.0$ & $2.64$ & $2.50$ & $\,\,\,3.9$ & \cite{Shen2017_PRC96-014316} \\
DDRHF, PKO1 & $-128.3$ & $2.68$ & $2.54$ & $\,\,\,6.4$ & \cite{Long2006_PLB639-242} \\
DDRHF, PKA1 & $-127.0$ & $2.80$ & $2.67$ & $\,\,\,6.0$ & \cite{Long2007_PRC76-034314} \\
BHF, AV18  & $-134.2$ &  & $1.95$ & $13.0$ & \cite{Hu2017_PRC95-034321} \\
CC, N$^3$LO  & $-120.9$ &  & $2.30$ & & \cite{Hagen2009_PRC80-021306} \\
IM-SRG, N$^3$LO & $-122.71(9)$ & & & & \cite{Hergert2013_PRC87-034307} \\
NCSM, N$^3$LO  & $-119.7(6)$ &  &  &  & \cite{Roth2011_PRL107-072501} \\
SCGF, N$^3$LO & $-122$ &  & & & \cite{Cipollone2013_PRL111-062501} \\
NLEFT, N$^2$LO  & $-121.4(5)$ &  &  & & \cite{Lahde2014_PLB732-110} \\
QMC, N$^2$LO & $\,\,\,-87(3)(11)$ & $2.76(5)(12)$ & & & \cite{Lonardoni2018_PRC97-044318} \\
\hline\hline
\end{tabular}
\label{tab5}
\end{table}

The phenomenological functional PKO1 includes the tensor correlations induced by the $\pi$-exchange and PKA1 includes in addition the tensor correlations by the $\rho$-exchange. As expected, the phenomenological results, which are both fitted to the experimental data \cite{Long2006_PLB639-242,Long2007_PRC76-034314}, are in much better agreements with data than all the \textit{ab initio} results.

The total energies given by the RBHF calculation with Bonn A, B, and C interactions are underbound by $7.4, 20.5,$ and $29.6$ MeV (or by $6\%, 16\%,$ and $23\%$), respectively.
The charge radius is smaller by $0.17,0.11,$ and $0.06$ fm (or by $6\%,4\%,$ and $2\%$) compared to the experimental value, respectively.
In the self-consistent non-relativistic BHF calculations~\cite{Hu2017_PRC95-034321}, the interaction $V_{{\rm low}-k}$ derived from the Argonne interaction $v_{18}$ was used.
The result shows an overbinding by $6.6$ MeV (or by 5\%) and the rms radius is too small.
In comparison, the value of
$E=-120.9$ MeV is obtained with CC~\cite{Hagen2009_PRC80-021306} using the chiral $NN$ interaction N$^{3}$LO, $E=-122.8$ MeV is obtained with IM-SRG~\cite{Hergert2013_PRC87-034307}, $E=-119.7(6)$ MeV is obtained with NCSM~\cite{Roth2011_PRL107-072501},
and $E=-122$ MeV is obtained with SCGF~\cite{Cipollone2013_PRL111-062501}.
 Using the same interaction N$^{3}$LO, the calculated results agree with each other quite well and are similar to the results obtained by RBHF with Bonn A.
The one obtained with the NLEFT~\cite{Lahde2014_PLB732-110} using the interaction N$^2$LO, $E=-121.4(5)$, is also similar to the above values.
On the other hand, with the two-body force only, QMC~\cite{Lonardoni2018_PRC97-044318} gives $E = -87$ MeV, much less binding than the others.
This might be because the N$^2$LO force used there is a local one and different from that of Ref.~\cite{Epelbaum2009_EPJA41-125}.

The spin-orbit splittings of the $1p$ proton shell obtained in RBHF theory with the Bonn A interaction $\Delta E_{\pi1p}^{ls}=5.3$ MeV is slightly smaller than the data but still in reasonable agreement ($16\%$ deviation).
Those with Bonn B and C, which have stronger tensor force, give a further decreased spin-orbit splitting.

\begin{figure}
  \centering
  \includegraphics[width=8cm]{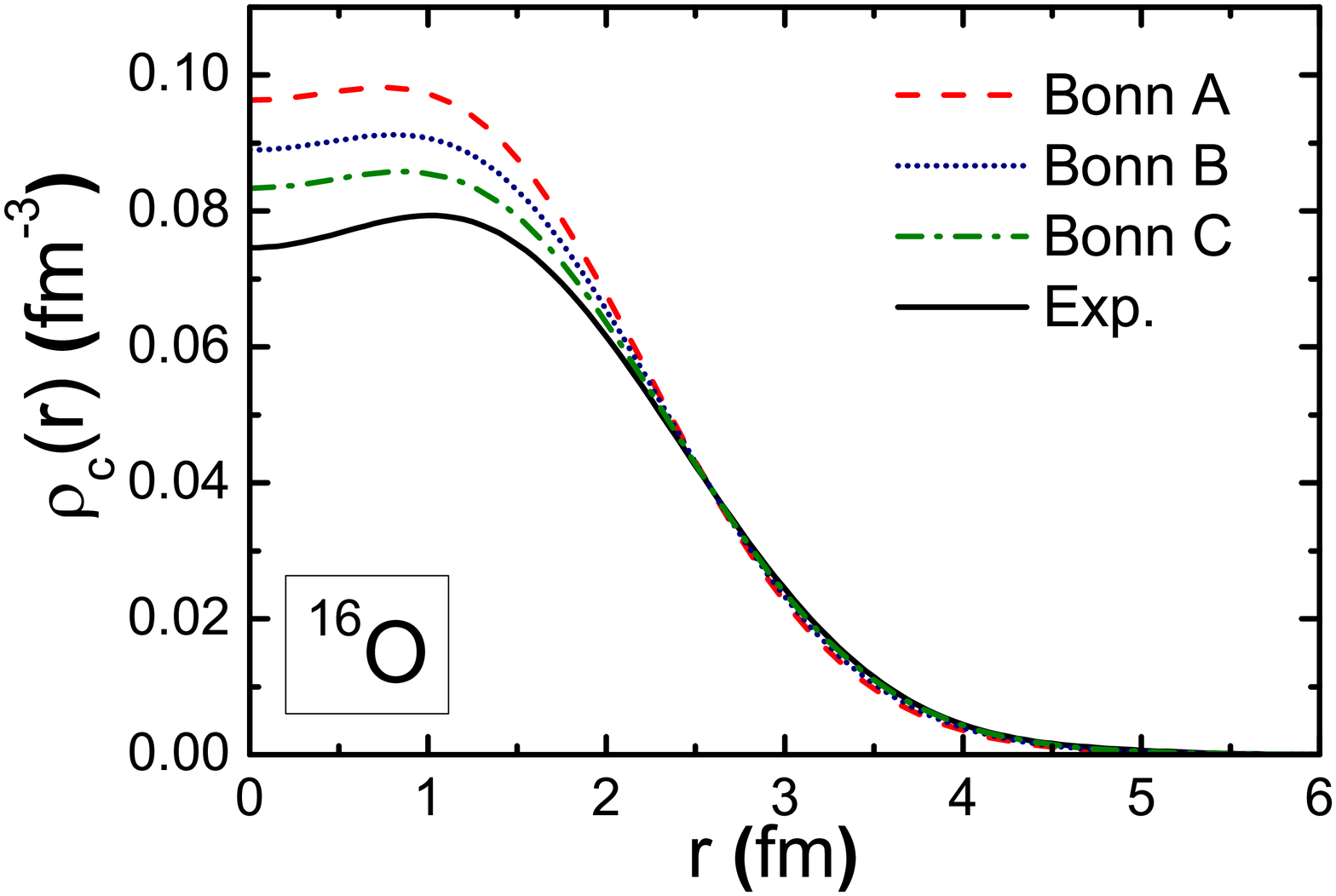}
  \caption{(Color online) Charge density distributions of $^{16}$O calculated by RBHF theory with the interactions Bonn A, B, and C, in comparison with experimental data \cite{DeVries1987_ADNDT36-495}.}
  \label{fig10}
\end{figure}

Figure \ref{fig10} shows the charge-density distributions of $^{16}$O calculated by RBHF theory with the interactions Bonn A, B, and C, in comparison with experimental data \cite{DeVries1987_ADNDT36-495}.
As already reflected in the radius (see Fig. \ref{fig8}), here the central density distributions are too large. While Bonn C (with the strongest tensor force) gives the largest radius, it also describes the density distribution best.

  Similar problems have also been found in the non-relativistic \textit{ab initio} calculations, where the radii are too small even when the 3N interaction is included \cite{Lapoux2016_PRL117-052501}. Only by using revised chiral interactions, for instance NNLO$_{\rm sat}$~\cite{Ekstrom2015_PRC91-051301}, which also includes the radii of finite nuclei such as $^{16}$O in the fitting, the situation is improved. However, this interaction cannot describe well the NN scattering data above 35 MeV, and thus the concept of \textit{ab initio} calculation using this interaction is still questionable. Further investigations are needed for a better understanding of this problem.

 Besides the radii, other physical observables, such as the single-particle energies and spin-orbit splittings, are also closely related to the properties of the density distribution. In the non-relativistic BHF calculations, the single-particle energies of $^{16}$O are shown to be more bound than the data \cite{Hu2017_PRC95-034321}. For example, the single-particle energies of proton in the $1s_{1/2}, 1p_{3/2}, 1p_{1/2}$ orbits given by Ref. \cite{Hu2017_PRC95-034321} are $-72.8, -37.2, -24.2$ MeV, respectively. On the other hand, this implies that the nucleons concentrate too much in the center and the mean-field potential is deeply overbound there. As can be expected, a large slope of the density leads to too large spin-orbit splittings: the proton $1p$ spin-orbit splitting given by BHF is 13 MeV (Table \ref{tab5}), considerably deviating from the experiment with 6.3 MeV. In RBHF, the situation has been improved much: the central density increases only slightly, and the single-particle energies are, on average, only slightly overbound as compared to the data (Fig. \ref{fig9}). Obviously, in a relativistic theory, the slope of density distribution is not the only decisive factor, because the spin-orbit splittings given by RBHF with the Bonn interactions are smaller than the data even though the density slope is larger. 

\begin{figure}
  \centering
  \includegraphics[width=8cm]{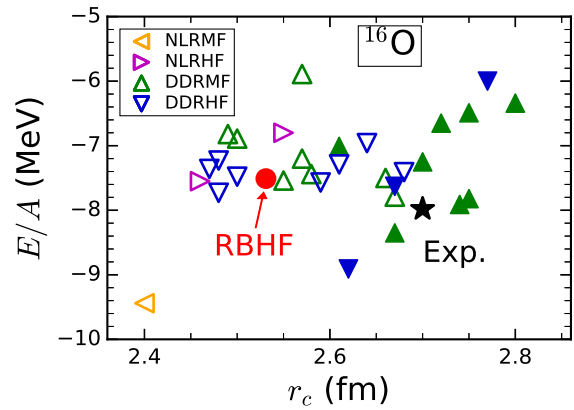}
  \caption{(Color online) Energy per nucleon $E/A$ and charge radius $r_c$ of $^{16}$O calculated by RBHF using Bonn A interaction, in comparison with experimental data \cite{Wang2017_CPC41-030003,Angeli2013_ADNDT99-69} and LDA form RBHF in nuclear matter.
  See text for the meanings of different symbols.}
  \label{fig11}
\end{figure}

As there are many investigations for LDA based on RBHF using the Bonn A interaction, it is interesting to compare the self-consistent and the LDA results.
In Fig. \ref{fig11}, the energy per nucleon and charge radius calculated by RBHF using Bonn A interaction are compared with LDA derived from RBHF in nuclear matter using the same interaction.
The LDA results include: effective interactions defined in the nonlinear relativistic mean field (NLRMF) model \cite{Gmuca1992_NPA547-447}, the nonlinear relativistic Hartree-Fock (NLRHF) model \cite{Savushkin1997_PRC55-167}, the DDRMF model \cite{Brockmann1992_PRL68-3408,Fritz1993_PRL71-46,Fritz1994_PRC49-633,Ma1994_PRC50-3170,Lenske1995_PLB345-355,
Ineichen1996_PRC53-2158,Shen1997_PRC55-1211,Gogelein2008_PRC77-025802,VanDalen2011_PRC84-024320}, and the DDRHF model \cite{Fritz1993_PRL71-46,Fritz1994_PRC49-633,Ma1994_PRC50-3170,Shi1995_PRC52-144,VanDalen2011_PRC84-024320}.
The filled (open) symbols are used for calculations with (without) the rearrangement term.

As ambiguities exist in the mappings from RBHF calculations for nuclear matter to the effective interactions for finite nuclei in LDA, the results for finite nuclei are very different.
In this case, the self-consistent RBHF calculation for finite nuclei \cite{Shen2016_CPL33-102103,Shen2017_PRC96-014316,Shen2018_PLB781-227} is indispensable to provide a benchmark.

In Fig. \ref{fig11}, the difference between LDA results with and without the rearrangement term can also be seen.
The filled symbols including the rearrangement term are generally located on the right-hand side of the corresponding open ones without the rearrangement term.
In other words, the rearrangement term improves the description of the radius substantially. It also improves the description of the binding energy, but to a smaller extent.


Both $^{4}$He and $^{40}$Ca are calculated in RBHF theory with the Bonn A interaction as well \cite{Shen2017_PRC96-014316}.
In Table \ref{tab6}, the results for $^{4}$He are shown. The c.m. correction is taken into account with both PBV and PAV as introduced in Section \ref{Sect:3.5.2}.
The corresponding results are compared with the experimental data~\cite{Wang2017_CPC41-030003,Angeli2013_ADNDT99-69,Lu2013_RMP85-1383}, as well as those from DDRHF with PKO1 \cite{Long2006_PLB639-242} and PKA1 \cite{Long2007_PRC76-034314}, the Faddeev-Yakubovsky (FY) equation~\cite{Nogga2000_PRL85-944} with CD-Bonn~\cite{Machleidt1996_PRC53-R1483}, FY~\cite{Binder2016_PRC93-044002} with N$^4$LO~\cite{Epelbaum2015_PRL115-122301}, no-core shell model (NCSM)~\cite{Navratil2007_FBS41-117} with N$^3$LO~\cite{Entem2003_PRC68-041001}, nuclear lattice effective field theory (NLEFT)~\cite{Lahde2014_PLB732-110} with N$^2$LO, and BHF~\cite{Hu2017_PRC95-034321} with $V_{{\rm low}-k}$ derived from Argonne $v_{18}$~\cite{Wiringa1995_PRC51-38}.
In all the above calculations, and also for the later results for $^{40}$Ca, only the two-body bare $NN$ interactions are used.

\begin{table}
\caption{Total energy, charge radius, and proton radius of $^{4}$He calculated by RBHF theory using the interaction Bonn A with PBV and PAV~\cite{Shen2017_PRC96-014316}, in comparison with experimental data~\cite{Wang2017_CPC41-030003,Angeli2013_ADNDT99-69,Lu2013_RMP85-1383}.
The corresponding results from DDRHF with PKO1 \cite{Long2006_PLB639-242} and PKA1 \cite{Long2007_PRC76-034314}, the FY equation~\cite{Nogga2000_PRL85-944} with CD-Bonn, FY~\cite{Binder2016_PRC93-044002} with N$^4$LO, NCSM~\cite{Navratil2007_FBS41-117} with N$^3$LO, NLEFT~\cite{Lahde2014_PLB732-110} with N$^2$LO, and BHF~\cite{Hu2017_PRC95-034321} with $V_{{\rm low}-k}$ derived from Argonne $v_{18}$ are also included.}
\label{tab6}
\centering
\begin{tabular}{llclc} 
\hline\hline
&\multicolumn{1}{c}{$E$ (MeV)} & \multicolumn{1}{c}{$r_c$ (fm)} & \multicolumn{1}{c}{$r_p$ (fm)} & Ref. \\
\hline
Exp. & $-28.30$ & $1.68$ & $1.46$ & \cite{Wang2017_CPC41-030003,Angeli2013_ADNDT99-69,Lu2013_RMP85-1383} \\
RBHF (PBV), Bonn A & $-35.05$ & $1.83$ & $1.64$ & \cite{Shen2017_PRC96-014316} \\
RBHF (PAV), Bonn A & $-26.31$ & $1.90$ & $1.73$ & \cite{Shen2017_PRC96-014316} \\
DDRHF, PKO1 & $-28.45$ & $1.90$ & $1.72$ & \cite{Long2006_PLB639-242} \\
DDRHF, PKA1 & $-28.28$ & $2.06$ & $1.90$ & \cite{Long2007_PRC76-034314} \\
FY, CD-Bonn  & $-26.26$ &  &  & \cite{Nogga2000_PRL85-944} \\
FY, N$^4$LO  & $-24.27(6)$ &  & 1.547(2) & \cite{Binder2016_PRC93-044002} \\
NCSM, N$^3$LO  & $-25.39(1)$ &  & $1.515(2)$ & \cite{Navratil2007_FBS41-117} \\
NLEFT, N$^2$LO  & $-25.60(6)$ &  &  & \cite{Lahde2014_PLB732-110} \\
BHF, AV18  & $-25.90$ &  &  & \cite{Hu2017_PRC95-034321} \\
\hline\hline
\end{tabular}
\end{table}

As expected, the center-of-mass correction plays an important role for such light nuclei.
The binding energy given by PBV is much overbinding, and the one given by PAV is slightly underbinding.
The radii of both are much larger than the data, and this is also observed in other calculations, including DDRHF with PKO1 and PKA1, FY equation, and NCSM.


In Table~\ref{tab7}, the total energy, charge radius, matter radius,
and proton spin-orbit splitting for the $1d$ orbit are listed for $^{40}$Ca calculated by RBHF with the interaction Bonn A \cite{Machleidt1989_ANP19-189}, in comparison with the experimental data~\cite{Wang2017_CPC41-030003,Angeli2013_ADNDT99-69,Coraggio2003_PRC68-034320}.
The RBHF results are different form the values given in Ref.~\cite{Shen2017_PRC96-014316} as they are updated with formula II for the single-particle potential (\ref{Eq:57}), instead of formula I (\ref{Eq:74}).
The corresponding results from DDRHF with PKO1~\cite{Long2006_PLB639-242} and PKA1~\cite{Long2007_PRC76-034314}, BHF~\cite{Hu2017_PRC95-034321}, NCSM \cite{Roth2007_PRL99-092501}, and CC \cite{Hagen2007_PRC76-044305} with $V_{{\rm low}-k}$ derived from Argonne $v_{18}$~\cite{Wiringa1995_PRC51-38}, CC \cite{Hagen2010_PRC82-034330} and IM-SRG \cite{Hergert2013_PRC87-034307} with N$^3$LO~\cite{Entem2003_PRC68-041001} are also included.

\begin{table}
\caption{Total energy, charge radius, matter radius,
and proton spin-orbit splitting for the $1d$ shell of $^{40}$Ca calculated by RBHF \cite{Shen2017_PRC96-014316,Shen2018_PLB781-227} with the interaction Bonn A, in comparison with experimental data~\cite{Wang2017_CPC41-030003,Angeli2013_ADNDT99-69,Coraggio2003_PRC68-034320}.
The corresponding results from DDRHF with PKO1~\cite{Long2006_PLB639-242} and PKA1~\cite{Long2007_PRC76-034314}, BHF~\cite{Hu2017_PRC95-034321}, NCSM \cite{Roth2007_PRL99-092501}, and CC \cite{Hagen2007_PRC76-044305} with $V_{{\rm low}-k}$ derived from Argonne $v_{18}$, CC \cite{Hagen2010_PRC82-034330} and IM-SRG \cite{Hergert2013_PRC87-034307} with N$^3$LO are also included.}
\centering
\begin{tabular}{lccccc} 
\hline\hline
&\multicolumn{1}{c}{$E$ (MeV)} & \multicolumn{1}{c}{$r_c$ (fm)} & \multicolumn{1}{c}{$r_m$ (fm)} & $\Delta E_{\pi1d}^{ls}$ (MeV) & Ref. \\
\hline
Exp. & $-342.1$ & $3.48$ &   & $6.6\pm 2.5$ & \cite{Wang2017_CPC41-030003,Angeli2013_ADNDT99-69,Coraggio2003_PRC68-034320} \\
RBHF, Bonn A & $-306.1$ & $3.22$ & $3.10$ & $\,\,\,5.8$ & \cite{Shen2017_PRC96-014316} \\
DDRHF, PKO1 & $-343.3$ & $3.44$ & $3.33$ & $\,\,\,6.6$ & \cite{Long2006_PLB639-242} \\
DDRHF, PKA1 & $-341.7$ & $3.53$ & $3.41$ & $\,\,\,7.2$ & \cite{Long2007_PRC76-034314} \\
BHF, AV18  & $-552.1$ &  & $2.20$ & $24.9$ & \cite{Hu2017_PRC95-034321} \\
NCSM, AV18  & $-461.8$ &  & $2.27$ & & \cite{Roth2007_PRL99-092501} \\
CC, AV18  & $-502.9$ &  &  & & \cite{Hagen2007_PRC76-044305} \\
CC, N$^3$LO  & $-345.2$ &  &  & & \cite{Hagen2010_PRC82-034330} \\
IM-SRG, N$^3$LO & $-358.5$ & & & & \cite{Hergert2013_PRC87-034307} \\
\hline\hline
\end{tabular}
\label{tab7}
\end{table}

Similar as for $^{16}$O, the total energy of $^{40}$Ca calculated by RBHF with the interaction Bonn A is underbound by 36 MeV (or by 11\%) and the charge radius is smaller by 0.26 fm (or by 7\%) as compared to the experimental values.
For the proton $1d$ spin-orbit splitting, RBHF with Bonn A gives a very good description of the data.
 Most of the non-relativistic results, because of the missing three-body force, give a too-large binding energy and a too-small radius, except for the CC and IM-SRG methods with N$^3$LO which reproduce the experimental binding energy well. In the calculation by IM-SRG, the result with inclusion of chiral 3N interaction has also been reported as $E = -376.1$~\cite{Hergert2013_PRC87-034307}, giving more overbinding than that with 2N interactions only. Again, as has been mentioned in the discussion of $^{16}$O, there are still large uncertainties in the 3N interactions and more investigations are called for to get a better understanding.

\begin{figure}
  \centering
  \includegraphics[width=8cm]{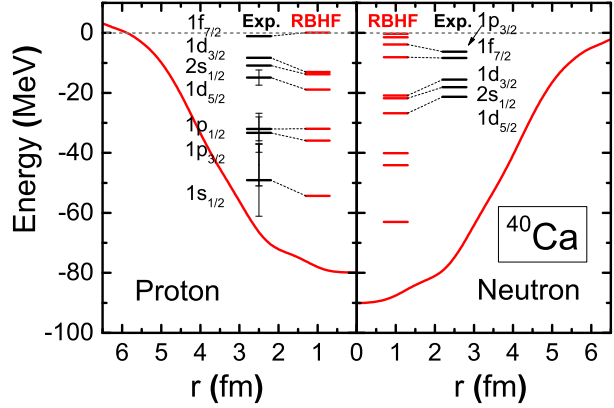}
  \caption{(Color online) Localized single-particle potential with $1s_{1/2}$ wave function and single-particle energy levels calculated by RBHF \cite{Shen2017_PRC96-014316} with Bonn A interaction, in comparison with experimental data \cite{Coraggio2003_PRC68-034320}.}
  \label{fig12}
\end{figure}

In Fig.~\ref{fig12}, we show the localized single-particle potential and the single-particle energy levels calculated by RBHF \cite{Shen2017_PRC96-014316} with Bonn A, in comparison with the experimental data \cite{Coraggio2003_PRC68-034320}.
In RBHF, the single-particle potential is indeed nonlocal, as seen in Eq.~(\ref{Eq:57}).
Therefore, the one shown in Fig.~\ref{fig12} is obtained for a specific state $|a\rangle$ by the radial Dirac equation (\ref{Eq:3.3.RDeq}),
\begin{equation}\label{Eq:3.6.sigma}
  \Sigma_a(r) = e_a - M + \frac{\frac{dG_a(r)}{dr}-\frac{\kappa}{r}G_a(r)}{F_a(r)}.
\end{equation}
In this way, the single-particle potential can be illustrated intuitively, but of course, it depends on the chosen state $|a\rangle$.
In Fig.~\ref{fig12}, the proton (neutron) $1s_{1/2}$ wave function with positive energy is used to obtain the localized single-particle potential. Similar to the case of $^{16}$O, RBHF with Bonn A gives reasonable SO splittings.
Comparing with the experimental data, the gap in the Fermi surface is relatively large, which might be due to the lack of high-order configurations like the particle-vibration coupling \cite{Litvinova2006_PRC73-044328}.

\begin{figure}
  \centering
  \includegraphics[width=8cm]{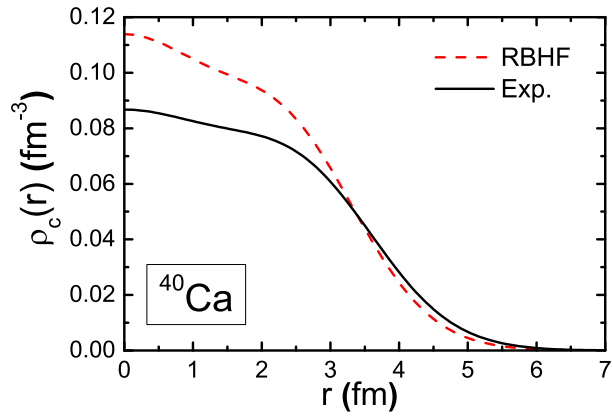}
  \caption{(Color online) Charge density distribution of $^{40}$Ca calculated by RBHF \cite{Shen2017_PRC96-014316} with Bonn A, in comparison with experimental data \cite{DeVries1987_ADNDT36-495}.}
  \label{fig13}
\end{figure}

In Fig.~\ref{fig13}, the charge-density distribution calculated by RBHF with Bonn A interaction \cite{Shen2017_PRC96-014316} is shown, in comparison with the experimental data \cite{DeVries1987_ADNDT36-495}.
Similar to $^{16}$O, the central density given by RBHF for $^{40}$Ca is too large (about $1.4$ times larger than the experimental value at $r = 0$ fm).

\begin{figure}
  \centering
  \includegraphics[width=8cm]{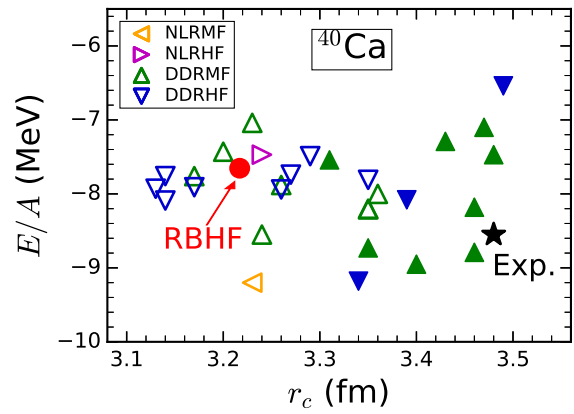}
  \caption{(Color online) Energy per nucleon $E/A$ and radius $r_c$ of $^{40}$Ca obtained by self-consistent RBHF with Bonn A, in comparison with various LDA results.}
  \label{fig14}
\end{figure}

In Fig.~\ref{fig14}, the energy and the radius of $^{40}$Ca obtained by self-consistent RBHF with Bonn A are compared with various LDA results. Similar to Fig.~\ref{fig14}, the LDA results are organized as: effective interactions defined in the NLRMF model \cite{Gmuca1992_NPA547-447}, the NLRHF model \cite{Savushkin1997_PRC55-167}, the DDRMF model \cite{Brockmann1992_PRL68-3408,Fritz1993_PRL71-46,Fritz1994_PRC49-633,Ma1994_PRC50-3170,Lenske1995_PLB345-355,
Ineichen1996_PRC53-2158,Shen1997_PRC55-1211,Gogelein2008_PRC77-025802,VanDalen2011_PRC84-024320}, and the DDRHF model \cite{Fritz1993_PRL71-46,Fritz1994_PRC49-633,Ma1994_PRC50-3170,Shi1995_PRC52-144,VanDalen2011_PRC84-024320}.
Filled (open) symbols are used for calculations with (without) the rearrangement term.
The self-consistent result (red point) sits in the middle of those LDA results without the rearrangement term.
Among them, the LDA results of Refs.~\cite{Shen1997_PRC55-1211,Savushkin1997_PRC55-167} are very close to the self-consistent one.

\subsection{Pseudospin and spin symmetries}\label{Sect:3.7}

The spin symmetry and its breaking in the nuclear system is very important for nuclear physics.
The large spin-orbit (SO) splitting, which was introduced by Goeppert-Mayer~\cite{Mayer1949_PR75-1969} and Haxel \textit{et al.}~\cite{Haxel1949_PR75-1766} in 1949, formed the ground for the nuclear shell model.
Twenty years later, the so-called pseudospin symmetry was proposed to explain the near degeneracy between two single-particle states with the quantum numbers $(n,l,j=l+1/2)$ and $(n-1,l+2,j=l+3/2)$ \cite{Arima1969_PLB30-517,Hecht1969_NPA137-129}.
Such two states are regarded as pseudospin doublets with the pseudospin quantum numbers  $(\tilde{n} = n-1, \tilde{l}=l+1,j=\tilde{l}\pm 1/2)$.
It was found that the angular momentum of the pseudospin doublets $\tilde{l}$ is nothing but the orbital angular momentum of the lower component of the Dirac spinor (\ref{Eq:3.3.wf}), and the pseudospin symmetry is exact when the sum of vector and scalar potentials $V+S$ vanishes \cite{Ginocchio1997_PRL78-436}, or a more general condition $d(V+S)/dr = 0$ which can be approximately fulfilled in exotic nuclei \cite{Meng1998_PRC58-R628,Meng1999_PRC59-154}.
Since then, pseudospin symmetry as a relativistic symmetry has been realized and much work has been done for investigating its origin and its properties using phenomenological single-particle Hamiltonians, relativistic mean field theory, or relativistic Hartree-Fock (RHF) theory \cite{Marcos2001_PLB513-30,Chen2003_CPL20-358,Lisboa2004_PRC69-024319,Long2006_PLB639-242, Long2010_PRC81-031302,Liang2011_PRC83-041301,Liang2013_PRC87-014334,Shen2013_PRC88-024311,
Guo2014_PRL112-062502,Shi2014_PRC90-034318,Zhao2014_PRC90-054326,Li2015_PRC91-024311, Li2016_PRC93-054312, Gao2017_PLB769-77,Sun2017_PRC96-044312}.

In the Dirac equation, there exist solutions not only with positive energy but also with negative energy, i.e., the states in the Dirac sea.
It was pointed out in Ref.~\cite{Zhou2003_PRL91-262501} that the spin symmetry in the Dirac sea has the same origin as the pseudospin symmetry in the Fermi sea.
In other words, the spin doublets in the Dirac sea have the quantum numbers $(n,\tilde{l},j=\tilde{l}\pm 1/2)$, and the spin symmetry breaking term is proportional to $d(V+S)/dr$, similar to the pseudospin symmetry in the Fermi sea.
See the related works in Refs.~\cite{He2006_EPJA28-265,Song2009_CPL26-122102,Liang2010_EPJA44-119,Lisboa2010_PRC81-064324, Hamzavi2014_APNY341-153,Sun2017_IJMPE26-1750025} and reviews~\cite{Ginocchio2005_PR414-165,Liang2015_PR570-1}.

In most of these investigations, the pseudospin symmetry in the Fermi sea or the spin symmetry in the Dirac sea has been studied by starting from a phenomenological single-particle Hamiltonian, or the relativistic density functional \cite{Ring1996_PPNP37-193,Vretenar2005_PR409-101,Meng2006_PPNP57-470, Niksic2011_PPNP66-519,Meng2016}.
It is therefore interesting to see to what extent the spin symmetry in the Dirac sea is found in RBHF calculations starting from the bare nucleon-nucleon interaction.
As the single-particle properties of states in the Dirac sea is concerned, the two formulas on the choice of single-particle potential in the Dirac sea, namely, formula I (\ref{Eq:74}) and formula II (\ref{Eq:57}) should be investigated \cite{Shen2018_PLB781-227}.

\begin{figure}
  \centering
  \includegraphics[width=12cm]{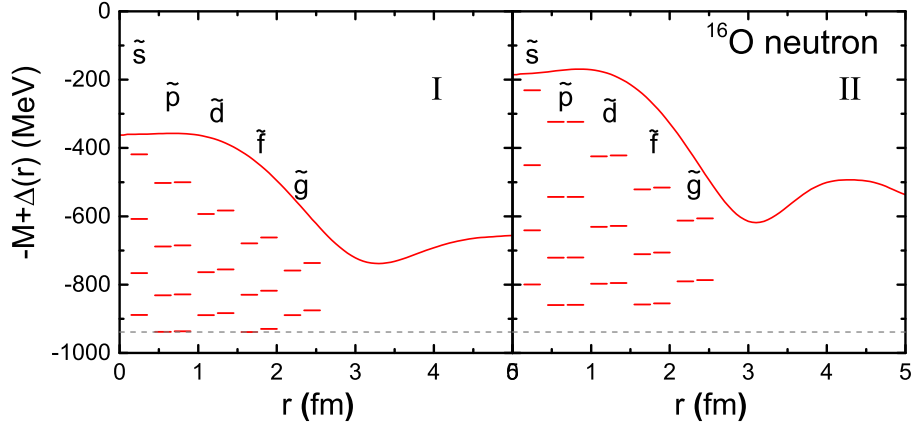}
  \caption{(Color online) Single-particle spectrum in the Dirac sea and the effective single-particle potential for the $\nu 1s1/2$ state in the Dirac sea calculated by RBHF theory using the interaction Bonn A with different choice of the single-particle potential in the Dirac sea: I for Eq.~(\ref{Eq:74}) and II for Eq.~(\ref{Eq:57}).
  Figure redrawn from Ref.~\cite{Shen2018_PLB781-227}.}
  \label{Fig:3.8.pso1}
  \label{fig15}
\end{figure}

The single-particle spectrum and the effective single-particle potential $\Delta(r) = V(r) - S(r)$ in the Dirac sea calculated by RBHF theory with different choices of the single-particle potential $U$ are shown in Fig.~\ref{fig15}.
Similar to Eq.~(\ref{Eq:3.6.sigma}), the effective single-particle potential $\Delta(r)$ is obtained by the radial Dirac equation~(\ref{Eq:3.3.RDeq}) for a specific state $|a\rangle$,
\begin{equation}
  \Delta_a(r) = e_a + M - \frac{\frac{dF_a(r)}{dr}-\frac{\kappa}{r}F_a(r)}{G_a(r)}.
\end{equation}
The localized potentials shown in Fig.~\ref{fig15} are calculated from the wave function $\nu 1s_{1/2}$ in the Dirac sea (or $\nu 1\tilde{p}_{1/2}$ if labelled with the angular momentum of the lower component in Eq.~(\ref{Eq:3.3.wf})).
The single-particle levels are grouped by the angular momentum $\tilde{l}$ of the lower component in the Dirac spinor (\ref{Eq:3.3.wf}) with negative energy, thus, $\tilde{s}, \tilde{p}, \tilde{d}, \dots$ means $\tilde{l} = 0, 1, 2, \dots$.
The potentials in both panels are not approaching 0 when $r\to \infty$ as usually found in the RMF study \cite{Zhou2003_PRL91-262501} because of the nonlocality of the RBHF single-particle potential $U$ in Eq.~(\ref{Eq:57}).

By comparing the two panels in Fig.~\ref{fig15}, it can be seen that formula II gives a deeper single-particle potential in the Dirac sea, and the spectrum is higher by $100\sim 200$ MeV.
Even though the two choices give only $0.4$ MeV per nucleon difference for the binding energy (see Table \ref{tab4}), the difference in the Dirac sea is much more significant.
This is understandable as the difference between these two choices lies in the definition of states in the Dirac sea only, and the effect on the states in the Fermi sea is indirect.
Beside the shift of single-particle levels upward as a whole with formula II in Fig.~\ref{fig15}, the SO splittings given by II are also generally smaller.

\begin{figure}
  \centering
  \includegraphics[width=12cm]{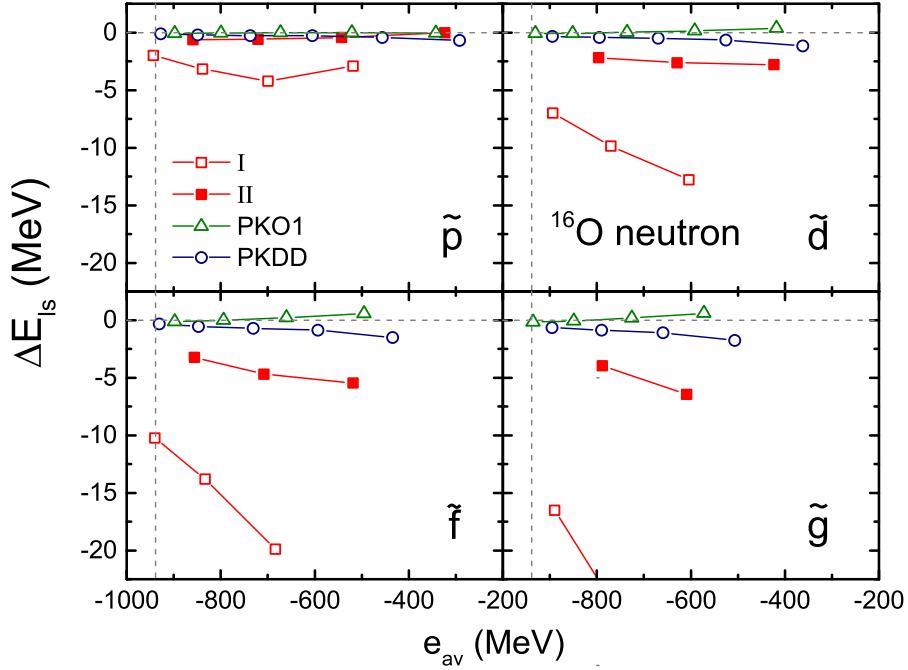}
  \caption{(Color online) { SO splittings $\Delta E_{\rm ls} = e_{j_<} - e_{j_>}$ versus the average energy of the spin doublets calculated by RBHF with formula I (open squares)~\cite{Shen2017_PRC96-014316} and formula II (filled squares)~\cite{Shen2018_PLB781-227} for the single-particle potential in the Dirac sea, in comparison with the results of the relativistic density functionals PKO1 (open triangles)~\cite{Long2006_PLB640-150} and PKDD (open circles)~\cite{Long2004_PRC69-034319}}.
  }
 \label{fig16}
\end{figure}
To see the SO splittings more clearly, in Fig.~\ref{fig16} the SO splittings $\Delta E_{\rm ls} = e_{j_<} - e_{j_>}$ versus the average energy of the spin doublets $e_{\rm av} = (e_{j_<} + e_{j_>})/2$ are shown, with $j_< = \tilde{l} - 1/2$ and $j_> = \tilde{l} + 1/2$.
The results are compared with those of the phenomenological relativistic density functionals PKO1 \cite{Long2006_PLB640-150} and PKDD \cite{Long2004_PRC69-034319}.
It can be seen that the SO splittings calculated by RBHF using formula II are much smaller than those of I, thus the spin symmetry is better conserved and is in better agreement with phenomenological relativistic density functional findings.
This may give some hints that formula II of the single-particle potential (\ref{Eq:57}), where the states in Dirac sea are treated as occupied states, is preferred.

 In short, the prediction of the spin symmetry in the single-particle spectrum in the Dirac sea is supported by the relativistic \textit{ab initio} calculations by starting from the bare nucleon-nucleon interaction. In future, it would also be interesting to continue this investigation considering particle-vibration coupling, as it has been shown to have an important effect on the pseudospin symmetry \cite{Litvinova2011_PRC84-014305}.
%
%
\section{RBHF Calculations for Neutron Drops}\label{Sect:4}


\subsection{Neutron drops}\label{Sect:4.1}

A neutron drop is an ideal system, composed of a finite number of neutrons and confined in an external field to keep the neutrons bound.
Similar to nuclear matter, it is not a real system existing in nature, but by studying it one can obtain rich knowledge of nuclear structure and various properties of droplets \cite{RocaMaza2018_PPNP101-96}.
Since in such systems there exists only the neutron-neutron interaction, the equations for neutron drops are much easier to solve and therefore they have been investigated by many \textit{ab initio} methods \cite{Pudliner1996_PRL76-2416,Bogner2011_PRC84-044306,Maris2013_PRC87-054318,Potter2014_PLB739-445,Shen2018_PLB778-344}, as well as by DFT \cite{Pudliner1996_PRL76-2416,Zhao2016_PRC94-041302}.
In this way, different \textit{ab initio} methods can be benchmarked and useful information can be extracted for the nuclear density functional.

The neutron drops were first studied by the quantum Monte-Carlo method~\cite{Pudliner1996_PRL76-2416} for $N = 7$ and $8$ using the two-nucleon interaction Argonne $v_{18}$~\cite{Wiringa1995_PRC51-38} and the three-nucleon interaction Urbana IX ~\cite{Pudliner1995_PRL74-4396}.
It was found that the commonly used Skyrme functionals overestimate the central density of these drops and the spin-orbit splitting of the $N=7$ neutron drop~\cite{Pudliner1996_PRL76-2416}.
In Ref.~\cite{Smerzi1997_PRC56-2549}, the ground-state energy was studied for a $N = 6$ neutron drop, and the neutron pairing energy was discussed by comparison with Ref.~\cite{Pudliner1996_PRL76-2416}.
Later, more systematic studies were performed for larger $N$ values with different external fields and different interactions using the quantum Monte-Carlo method~\cite{Pederiva2004_NPA742-255,Gandolfi2011_PRL106-012501,Maris2013_PRC87-054318,Tews2016_PRC93-024305}.
The studies with the modern high-precision chiral 2N interaction N$^3$LO \cite{Entem2003_PRC68-041001} and the 3N interaction N$^2$LO~\cite{Epelbaum2002_PRC66-064001} have been benchmarked with different \textit{ab initio} methods, including the no-core shell model~\cite{Barrett2013_PPNP69-131} and the coupled-cluster theory~\cite{Hagen2014_RPP77-096302}, and it was found that the results were consistent with each other~\cite{Potter2014_PLB739-445}.
However, by comparing these \textit{ab initio} calculations, one found a significant dependence on the selected interactions, especially on the 3N interactions~\cite{Maris2013_PRC87-054318,Potter2014_PLB739-445,Tews2016_PRC93-024305}.

Recently, the effect of the tensor force in neutron drops was studied by \textit{ab initio} relativistic Brueckner-Hartree-Fock (RBHF) theory with the Bonn A interaction \cite{Shen2018_PLB778-344,Shen2018_PRC97-054312}.
  An evidence of the tensor force was found in the evolution of the spin-orbit splittings with increasing neutron number. This finding can provide a guidance for determining the strengths of effective tensor force in the nuclear medium. 

Various non-relativistic and relativistic density functionals have been used to study neutron drops, and a strong linear correlation between the rms radii of neutron drops and the neutron-skin thickness of $^{208}$Pb and $^{48}$Ca has been pointed out in Ref.~\cite{Zhao2016_PRC94-041302}.
Because of the uncertainty in the isovector part of the density functionals, there is also an uncertainty in the results of neutron drops for different functionals.
In Ref. \cite{Bonnard2018_PRC98-034319}, \textit{ab initio} calculations for neutron drops were used to further constrain the density functionals by the microscopic results of effective field theories.

\subsection{General properties}\label{Sect:4.2}


Without further specification, all the results of neutron drops shown below are calculated in an external harmonic oscillator field
\begin{equation}
  U_{\rm HO} = \frac{1}{2}M \omega^2 r^2.
\end{equation}

\begin{figure}
  \centering
  \includegraphics[width=12cm]{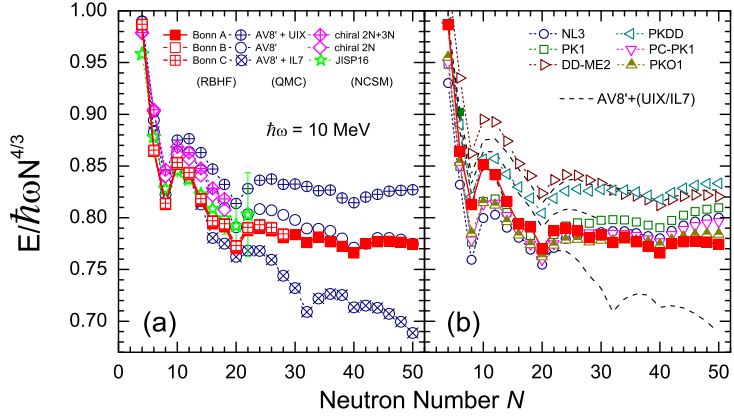}
  \caption{(Color online) Total energy in units of $\hbar\omega N^{4/3}$ for $N$-neutron drops in a HO trap ($\hbar\omega = 10$ MeV) calculated by RBHF theory using the interactions Bonn A, B, and C: (a) in comparison with QMC calculations~\cite{Gandolfi2011_PRL106-012501,Maris2013_PRC87-054318} using the interactions AV8' + UIX, AV8', and AV8' + IL7, with NCSM calculations~\cite{Potter2014_PLB739-445,Maris2013_PRC87-054318} using chiral 2N + 3N forces, chiral 2N forces, and the interaction JISP16. (b) in comparison with results based on the relativistic density functionals \cite{Zhao2016_PRC94-041302,Long2006_PLB640-150}. The dashed lines indicates the QMC results.
  Figure redrawn from Ref.~\cite{Shen2018_PRC97-054312}.}
  \label{fig17}
\end{figure}

Figure~\ref{fig17} shows the total energy divided by $\hbar\omega N^{4/3}$ for $N$-neutron drops ($N$ from $4$ to $50$ for Bonn A and from $4$ to $28$ for Bonn B and C) in a HO trap ($\hbar\omega = 10$ MeV) calculated by RBHF theory \cite{Shen2018_PLB778-344,Shen2018_PRC97-054312}.
The factor $\hbar\omega N^{4/3}$ is based on the consideration that, in the Thomas-Fermi approximation \cite{Ring1980}, the total energy for a non-interacting $N$-fermion system in a HO trap is given by
\begin{equation}
  E = \frac{3^{4/3}}{4} \hbar\omega N^{4/3} \approx 1.082 \hbar\omega N^{4/3}.
\end{equation}
In other words, the energy below the line $E/\hbar\omega N^{4/3} \approx 1.082$ corresponds to the binding due to the nuclear interactions.
This intrinsic binding energy grows linearly with the HO strength $\hbar\omega$.
The energy of the drops grows with the neutron number $N^{4/3}$, in contrast to the nuclear case where the binding energy grows with mass number $A$.
For the cases of open shells, the filling approximation is used.
The results are compared with quantum Monte-Carlo (QMC) calculations \cite{Gandolfi2011_PRL106-012501,Maris2013_PRC87-054318} based on the interactions AV8' + UIX, AV8', and AV8' + IL7, with no-core shell model (NCSM) calculations~\cite{Potter2014_PLB739-445,Maris2013_PRC87-054318} based on chiral 2N + 3N forces, chiral 2N forces, and the JISP16 force, and finally with calculations using relativistic density functionals~\cite{Zhao2016_PRC94-041302,Long2006_PLB640-150}.

It can be seen that the results of Bonn A, B, and C are very similar.
This is because the main difference among these three Bonn interactions is the strength of the $T=0$ tensor force \cite{Machleidt1989_ANP19-189}, which has no influence on the $T = 1$ neutron-neutron states.
It is also consistent with the results in pure neutron matter, where the EoS calculated by RBHF with these three interactions are very close to each other \cite{Li1992_PRC45-2782}.
Therefore, in later discussions only the results of Bonn A will be shown.

By comparing with the QMC and NCSM calculations, the results of RBHF with the interaction Bonn A are similar to the results with the JISP16 interaction, and AV8' + IL7 (for $N\leq 14$), and getting closer to AV8' (for $N\geq 20$).
This similarity is favourable as JISP16 is a phenomenological nonlocal $NN$ interaction which can reproduce the scattering data and works well for light nuclei \cite{Shirokov2007_PLB644-33,Maris2009_PRC79-014308}.
On the other hand, AV8' + IL7 gives a better description of light nuclei up to A = 12 than AV8' or AV8' + UIX, but it gives too much over-binding for pure neutron matter at higher densities \cite{Sarsa2003_PRC68-024308,Maris2013_PRC87-054318}.

In comparison with the relativistic density functional calculations in the right panel of Fig.~\ref{fig17}, four types of functionals have been chosen as the following,
\begin{enumerate}
  \item RH nonlinear meson-exchange models: NL3~\cite{Lalazissis1997_PRC55-540}, PK1~\cite{Long2004_PRC69-034319};
  \item RH density-dependent meson-exchange models: DD-ME2~\cite{Lalazissis2005_PRC71-024312}, PKDD~\cite{Long2004_PRC69-034319};
  \item RH nonlinear point-coupling models: PC-PK1~\cite{Zhao2010_PRC82-054319};
  \item RHF density-dependent meson-exchange models: PKO1~\cite{Long2006_PLB640-150} (which includes the tensor force from the $\pi$ meson).
\end{enumerate}
As there is no pairing in the RBHF calculations, pairing is also excluded in the relativistic density functional calculations.
It can be seen that the binding energies (attraction between neutrons) given by RBHF are generally larger than those given by DD-ME2 and PKDD.
For $N = 8$, RBHF is close to PKDD, but getting closer to PK1 from $N = 14$ to 26, and closer to PC-PK1, NL3, and PKO1 from $N = 28$ to 36.
From $N = 20$ on, the results of RBHF and DD-ME2 are close to a horizontal line, while the others have a small tendency of increasing.
 The microscopic results obtained by RBHF can be a guidance for future density functionals. For example, the neutron-neutron interaction might be in general too repulsive in DD-ME2, whereas in NL3 it might be too attractive when the neutron number N is small and then becomes too repulsive as N becomes large. This is consistent with the findings in the equation-of-state of neutron matter, where NL3 gives overbinding for small densities but underbinding for larger densities \cite{Niksic2002_PRC66-024306}.


Some clues on the shell structure are seen in Fig. \ref{fig17}.
For example, the energies of neutron drops at $N = 8, 20, 40$ decrease suddenly comparing with their neighbours.
This can be understood easily as the external HO potential has the magic numbers 8, 20, and 40.

\begin{figure}
  \centering
  \includegraphics[width=12cm]{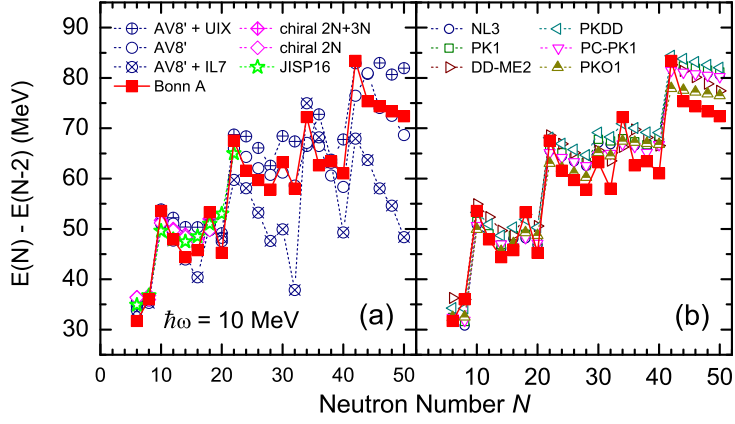}
  \caption{(Color online) Two-neutron energy difference of $N$-neutron drops in a HO trap ($\hbar\omega = 10$ MeV) calculated by RBHF theory using the Bonn A interaction. (a) In comparison with QMC calculations~\cite{Gandolfi2011_PRL106-012501,Maris2013_PRC87-054318} using the interactions AV8' + UIX, AV8', and AV8' + IL7, with NCSM calculations using the interactions chiral 2N + 3N force, chiral 2N force~\cite{Potter2014_PLB739-445}, and JISP16~\cite{Maris2013_PRC87-054318}. (b) In comparison with relativistic density functionals \cite{Zhao2016_PRC94-041302,Long2006_PLB640-150}.
  Figure redrawn from Ref.~\cite{Shen2018_PRC97-054312}.}
  \label{fig18}
\end{figure}

In order to see the shell structure more clearly, the two-neutron separation energies for the above calculations are presented in Fig.~\ref{fig18}.
The HO magic numbers 8, 20, and 40 are clearly shown in all calculations.
Beside the above magic numbers, the results of RBHF indicate sub-shell closures at $N = 16$ and $N = 32$, similar as the results of AV8' + IL7.
However, the $N = 32$ gap given by AV8' + IL7 might be too strong.
The sub-shell closure at $N = 32$ is not significant for AV8', and does not exist for AV8' + UIX.
For the $N = 28$ sub-shell closure, the results of Bonn A and AV8' + UIX show a small hint, while AV8' and AV8' + IL7 do not show it.
On the other hand, all the relativistic density functionals only show the HO magic numbers 8, 20, and 40, but no clear sub-shell closures for $N = 16, 28$ or $32$.

\begin{figure}
  \centering
  \includegraphics[width=6cm]{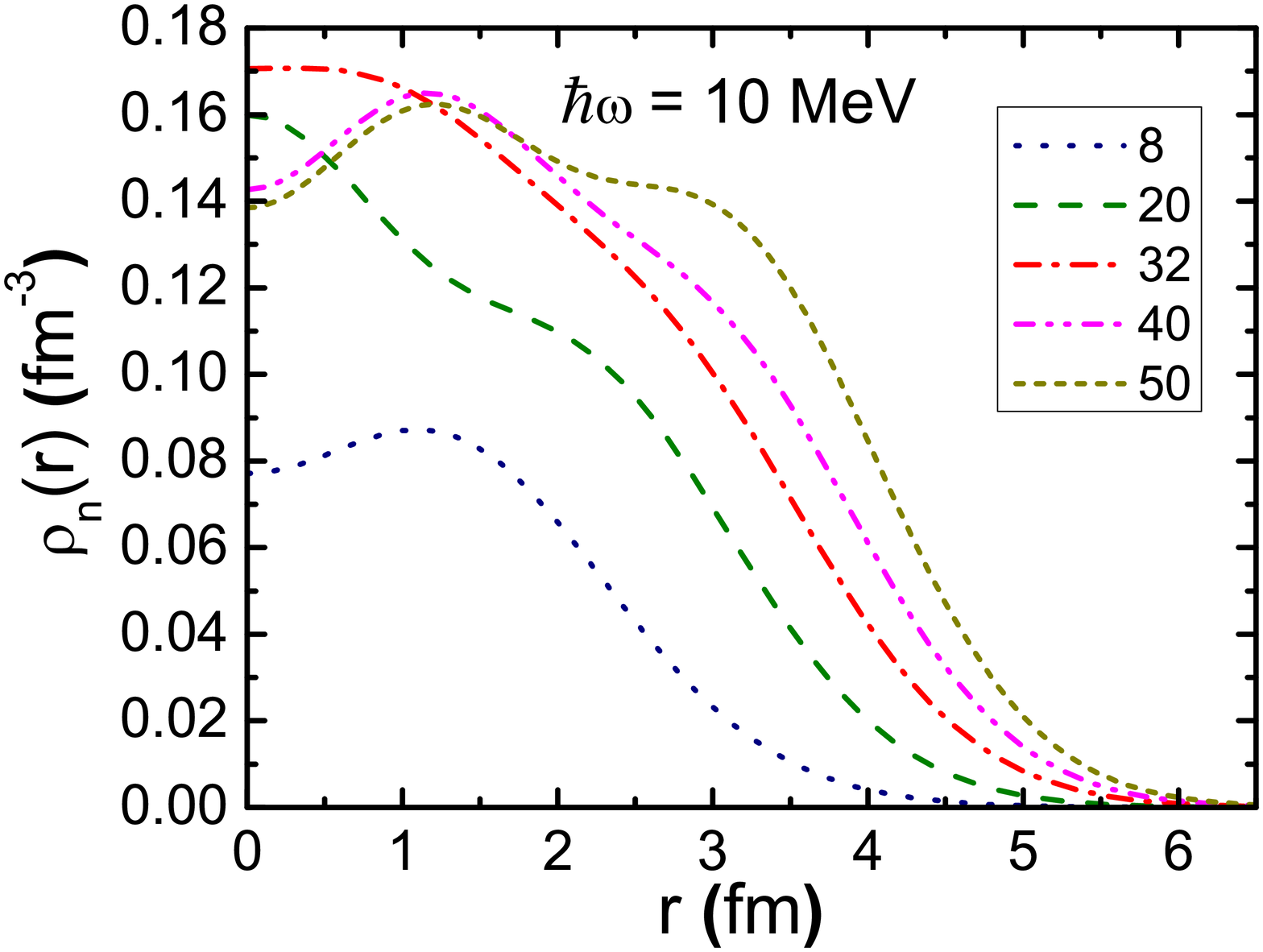}
  \hspace{2cm}
  \includegraphics[width=6.5cm]{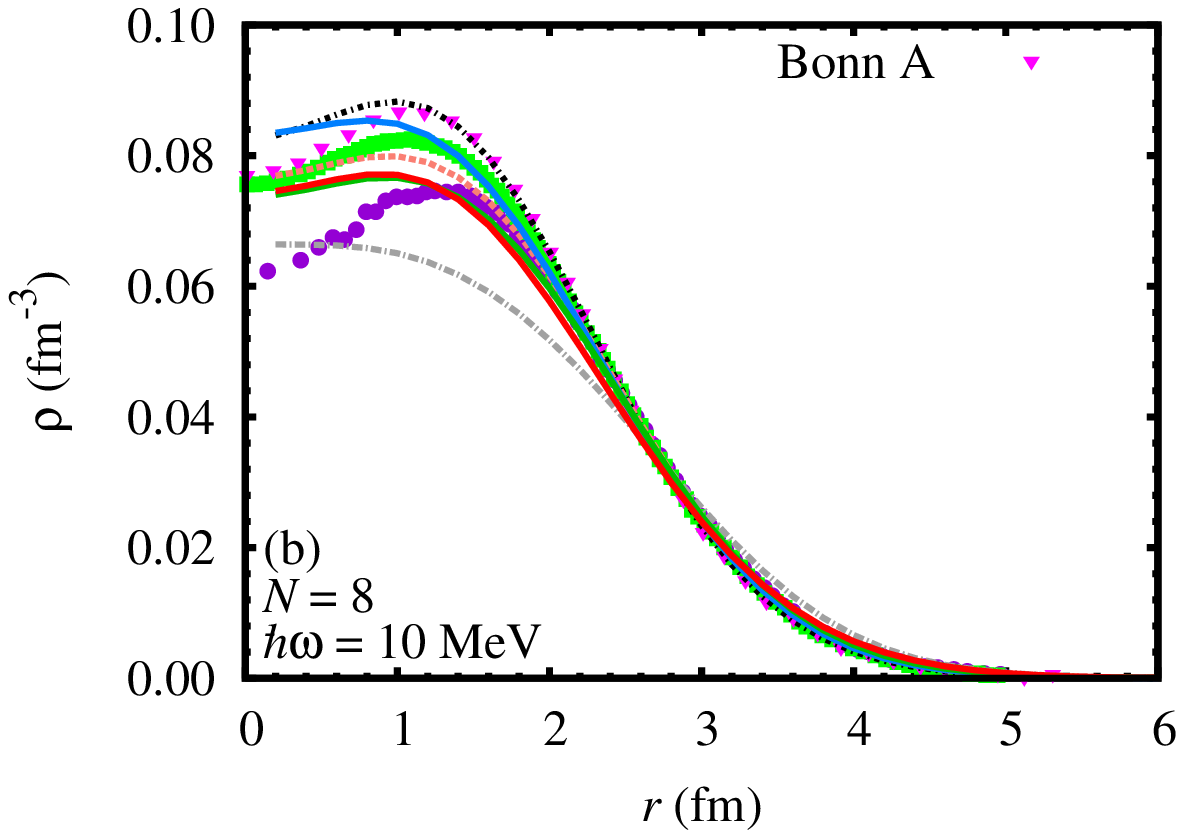}
  \caption{(Color online) (Left) Density distributions of $N$-neutron drops in a HO trap ($\hbar\omega = 10$ MeV) calculated by RBHF theory using the interaction Bonn A.
  Figure taken from Ref.~\cite{Shen2018_PRC97-054312}.
  (Right) Density distributions for $N = 8$ by RBHF with Bonn A (pink triangles), comparing to QMC with AV8'+UIX (purple circles), NCSM with JISP16 (green squares); density functionals SLy5 (black dash-dot), SkM* (orange dash), UNEDF0 (gray dash-dot), YGLO (blue line), KIDS (green line), and ELYO (red line).
  Figure taken from Ref.~\cite{Bonnard2018_PRC98-034319}.}
  \label{fig19}
\end{figure}

The left panel of Fig.~\ref{fig19} shows the density distributions of $N$-neutron drops in a HO trap ($\hbar\omega = 10$ MeV) calculated by RBHF theory using the interaction Bonn A.
With the given HO strength, the neutron density gets saturated around $0.14 \sim 0.17$ fm$^{-3}$.

A comparison between the density given by RBHF with Bonn A and other \textit{ab initio} and density functional calculations was given in Ref.~\cite{Bonnard2018_PRC98-034319}, which is shown in the right panel of Fig.~\ref{fig19}.
These calculations include: QMC with AV8' + UIX (purple circles) \cite{Maris2013_PRC87-054318}, NCSM with JISP16 (green squares) \cite{Maris2013_PRC87-054318}; density functionals SLy5 (black dash-dot) \cite{Chabanat1998_NPA635-231}, SkM* (orange dash) \cite{Bartel1982_NPA386-79}, UNEDF0 (gray dash-dot) \cite{Kortelainen2010_PRC82-024313}, YGLO (blue line) \cite{Yang2016_PRC94-031301}, KIDS (green line) \cite{Papakonstantinou2018_PRC97-014312}, and ELYO (red line) \cite{Grasso2017_PRC95-054327}.
In this figure the density distribution of $N = 8$ neutron drop is taken as an example.

It can be seen that the density distribution given by RBHF with Bonn A is very similar to that of JISP16.
Indeed (see Fig.~\ref{fig17}) the energy of these two calculations are also close to each other.
On the other hand, AV8' + UIX shows more repulsion than Bonn A from the energy point of view (Fig.~\ref{fig17}). This is consistent with the density distribution. As the neutrons feel less attraction and move more outside, AV8' + UIX has lower central density.
Among the selected functionals, the results of SLy5 and YGLO are close to RBHF with Bonn A.

\begin{figure}
  \centering
  \includegraphics[width=12cm]{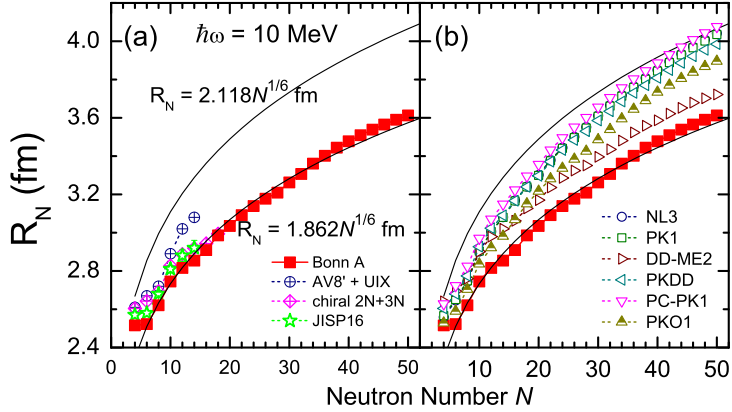}
  \caption{(Color online) Radii of $N$-neutron drops in a HO trap ($\hbar\omega = 10$ MeV) calculated by RBHF theory using the interaction Bonn A: (a) in comparison with QMC calculations using the interaction AV8' + UIX~\cite{Gandolfi2011_PRL106-012501}, with NCSM calculations~\cite{Potter2014_PLB739-445,Maris2013_PRC87-054318} using the chiral 2N + 3N and the JISP16 interactions; (b) in comparison with relativistic density functionals \cite{Zhao2016_PRC94-041302}. Further details are given in the text.
  Figure redrawn from Ref.~\cite{Shen2018_PRC97-054312}.}
  \label{fig20}
\end{figure}

Figure~\ref{fig20} shows the rms radii of $N$-neutron drops in a HO trap ($\hbar\omega = 10$ MeV) calculated by RBHF theory using the interaction Bonn A.
In panel (a) the results are compared with the QMC calculations based on the interaction AV8' + UIX~\cite{Gandolfi2011_PRL106-012501}, with the NCSM calculations~\cite{Potter2014_PLB739-445,Maris2013_PRC87-054318} based on the chiral 2N + 3N and the JISP16 forces.
In panel (b) these results are compared with calculations based on relativistic density functionals.
The black line $R_N = 2.118 N^{1/6}$ fm is obtained by solving for free Fermions in a $\hbar\omega = 10$ MeV HO trap using the Thomas-Fermi approximation as
\begin{equation}
R_N = \left(\frac{3^{4/3}}{4}\frac{\hbar}{M\omega}\right)^{1/2} N^{1/6}.
\end{equation}
For $M = 938.926$ MeV  and $\hbar\omega = 10$ MeV, one finds a factor $2.118$ fm in front of $N^{1/6}$.
The black line $R_N = 1.862 N^{1/6}$ fm is obtained by fitting to the results of Bonn A from $N$ = 6 to 50.

In general, all the calculated radii fulfill the relationship $N^{1/6}$ as a function of $N$.
In all the selected calculations, RBHF with Bonn A gives the smallest radii.
By comparing with other calculations in Fig.~\ref{fig17} (a) and Fig.~\ref{fig20} (a), it can be seen that, while it gives the smallest binding energies, AV8' + UIX also gives the largest radii.
The energies given by JISP16 are similar to those of Bonn A, but the corresponding radii are larger.
On the other hand, the radii of relativistic density functionals are much larger than those of RBHF, even though some of their binding energies are larger than RBHF before $N = 20$ as shown in Fig.~\ref{fig17} (b).
It is known that the relativistic density functionals without density-dependence in the isovector channel
(Nl3, PK1, and PD-PK1) show too large neutron radii in realistic nuclei \cite{Niksic2002_PRC66-024306}, and this can be observed for the neutron drops too.

\begin{table}
\caption{Rms radius $R_N$ of $N = 50$ neutron drop in a HO trap ($\hbar\omega = 10$ MeV) calculated by RBHF theory using the interaction Bonn A. The symmetry energy $a_{\rm sym}$ and slope parameter $L$ calculated in nuclear matter \cite{Alonso2003_PRC67-054301,VanDalen2004_NPA744-227,Katayama2013_PRC88-035805} are also listed, and they are compared with the results of relativistic functionals NL3~\cite{Lalazissis1997_PRC55-540}, PK1~\cite{Long2004_PRC69-034319}, DD-ME2~\cite{Lalazissis2005_PRC71-024312}, PKDD~\cite{Long2004_PRC69-034319}, PC-PK1~\cite{Zhao2010_PRC82-054319}, and PKO1~\cite{Long2006_PLB640-150}.}
\label{tab8}
\centering
\begin{tabular}{lccc}
\hline\hline
 & $R_{N=50}$ (fm) & $a_{\rm sym}$ (MeV) & $L$ (MeV)  \\
\hline
Bonn A & $3.61$ & $34.8$ & $\,\,\,71$ \\
NL3 & $4.04$ & $36.6$ & $119$ \\
PK1 & $4.04$ & $37.6$ & $116$ \\
DD-ME2 & $3.72$ & $32.3$ & $\,\,\,51$ \\
PKDD & $3.99$ & $36.8$ & $\,\,\,90$ \\
PC-PK1 & $4.08$ & $35.6$ & $113$ \\
PKO1 & $3.90$ & $34.4$ & $\,\,\,98$ \\
\hline\hline
\end{tabular}
\end{table}

Since the neutron matter properties are in close relation to the symmetry energy $a_{\rm sym}$ and its slope parameter $L$, Table~\ref{tab8} lists the radius of the $N = 50$ neutron drop calculated by RBHF theory using Bonn A and the corresponding symmetry energy and the slope parameter calculated at nuclear matter saturation \cite{Alonso2003_PRC67-054301,VanDalen2004_NPA744-227,Katayama2013_PRC88-035805}.
They are compared with the results of relativistic density functionals.
It can be seen that in general the radius of a neutron drop is large if $a_{\rm sym}$ or $L$ is large, although in detail small discrepancies exist.
For example, DD-ME2 gives the smallest $a_{\rm sym}$ and $L$, and its radius is indeed the smallest among those of relativistic functionals, but it is still larger than that of Bonn A.
The radius of PC-PK1 is the largest, and its $a_{\rm sym}$ or $L$ are large, but not the largest. They are slightly smaller than those of NL3 and PK1.

\begin{figure}
  \centering
  \includegraphics[width=8cm]{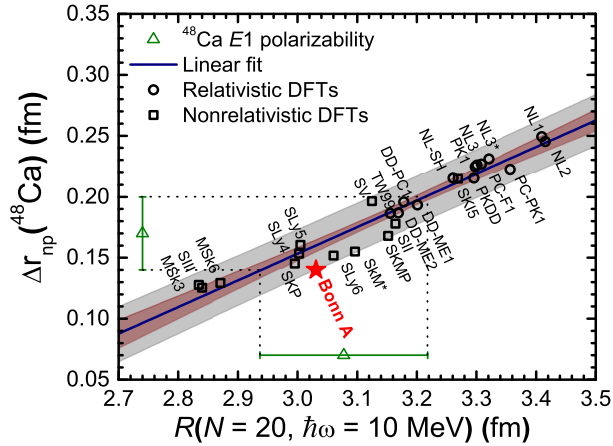}
  \caption{(Color online) Neutron-skin thickness $\Delta r_{np}$ of $^{48}$Ca and the rms radius $R$ of $N = 20$ neutron drop in a HO trap ($\hbar\omega = 10$ MeV) calculated by RBHF theory using the interaction Bonn A (red star), in comparison with the results obtained by various functionals~\cite{Zhao2016_PRC94-041302}.
  The datum of $\Delta r_{np}$ is obtained by measuring the electric dipole polarizability of $^{48}$Ca~\cite{Birkhan2017_PRL118-252501}.
  The blue line is the linear fit to the results of functionals, and the inner (outer) colored
  regions depict the 95\% confidence (prediction) intervals of the linear regression~\cite{Zhao2016_PRC94-041302}.
  Figure taken from Ref.~\cite{Shen2018_PRC97-054312}.}
  \label{fig21}
\end{figure}

The correlation between the slope parameter $L$ and the neutron-skin is well known \cite{RocaMaza2011_PRL106-252501}.
In Ref.~\cite{Zhao2016_PRC94-041302}, a strong linear correlation has been found between the neutron skin thickness $\Delta r_{np}$ and the rms radius $R_N$ of $N$-neutron drops in an external HO field.
Figure~\ref{fig21} shows the linear correlation between the neutron-skin thickness of $^{48}$Ca and the radius of $N = 20$ neutron drops in a $\hbar\omega = 10$ MeV HO external field.
The black circle and square symbols are calculated with different non-relativistic and relativistic density functionals, and the blue line is obtained by fitting to those results \cite{Zhao2016_PRC94-041302}.
The inner (outer) colored regions depict the 95\% confidence (prediction) intervals of the linear regression.

The red star in Fig.~\ref{fig21} is calculated by RBHF theory using Bonn A.
The datum of the neutron-skin thickness of $^{48}$Ca is obtained by measuring the electric dipole polarizability in Ref.~\cite{Birkhan2017_PRL118-252501}.
It can be seen that the neutron-skin thickness of $^{48}$Ca given by RBHF $\Delta r_{np} = 0.14$ fm is located within the error bar of the experimental data, which is also consistent with the $0.12 \leq \Delta r_{np} \leq 0.15$ fm given by the coupled-cluster calculations using the interaction NNLO$_{\rm sat}$ \cite{Hagen2015_NP12-186}.

\begin{figure}
  \centering
  \includegraphics[width=8cm]{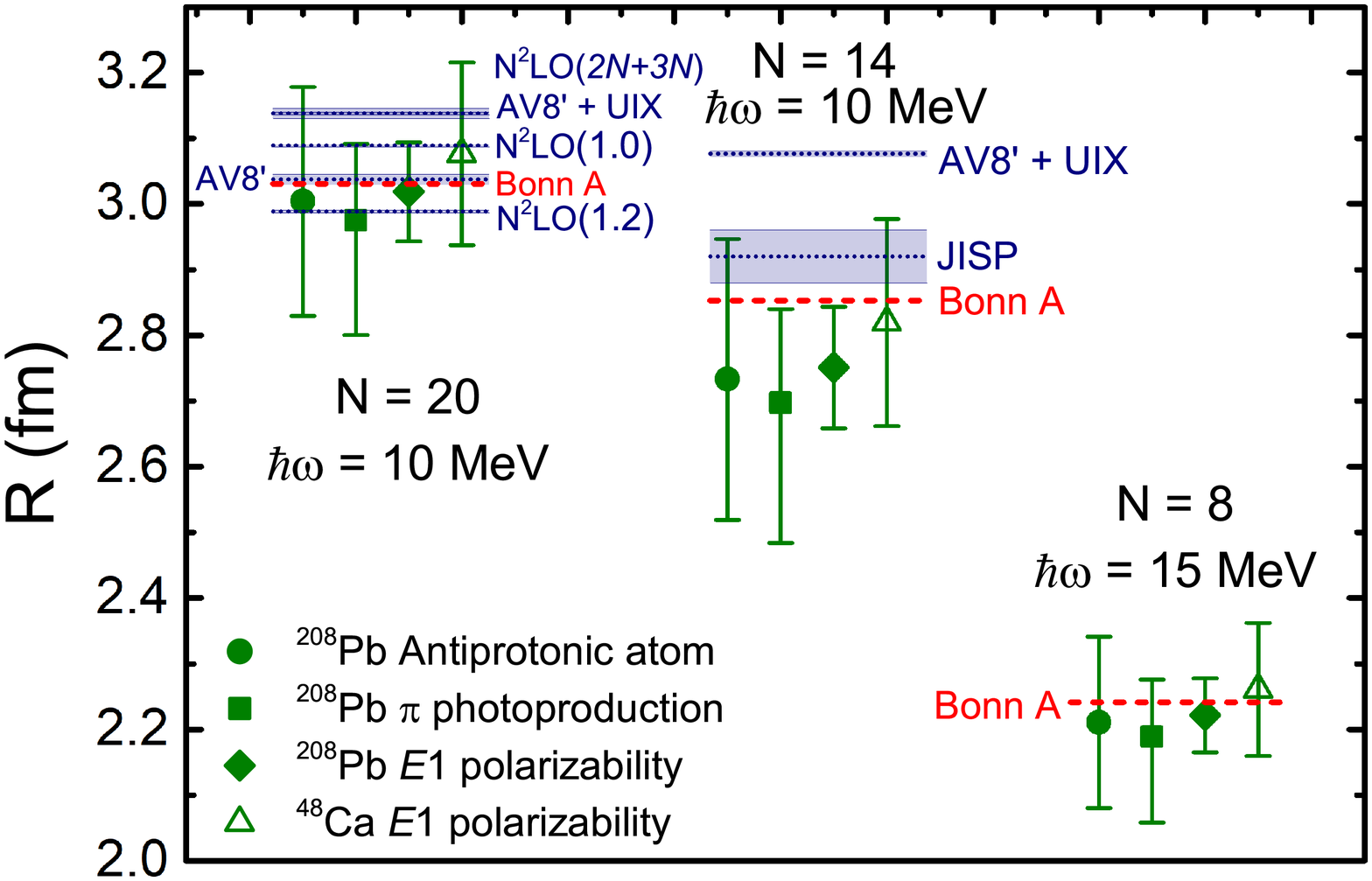}
  \caption{(Color online) Radii for $N = 20, 14$, and $8$ neutron drops calculated by RBHF theory using the interaction Bonn A (red dashed lines), in comparison with the data (green symbols) determined from the linear correlations with the neutron-skin thicknesses of $^{208}$Pb and $^{48}$Ca~\cite{Zhao2016_PRC94-041302}, and other \textit{ab initio} calculations (blue dotted lines)~\cite{Gandolfi2011_PRL106-012501,Maris2013_PRC87-054318,Tews2016_PRC93-024305}.
  Blue colored regions denote theoretical uncertainties.
  Figure taken from Ref.~\cite{Shen2018_PRC97-054312}.}
  \label{fig22}
\end{figure}

Apart from the linear correlation between $\Delta r_{np}$ of $^{48}$Ca and the radius of $N = 20$ neutron drops in Fig.~\ref{fig21}, similar correlations can be found in other cases, for example for $\Delta r_{np}$ of $^{208}$Pb or the radius of other $N$.
Using these linear correlations, the experimental data of neutron skins of $^{48}$Ca and $^{208}$Pb can be mapped to the data of radii of neutron drops with different numbers of $N$ \cite{Zhao2016_PRC94-041302}, and the results are shown with green symbols in Fig.~\ref{fig22}.
In this way, the study on the neutron skin of heavy nuclei can be linked to the study of the radius of neutron drops, while the latter is much easier to be accessed by different \textit{ab initio} calculations.

In Fig.~\ref{fig22}, the radii for $N = 20, 14$, and $8$ neutron drops calculated by RBHF theory using the interaction Bonn A (red dashed lines) are shown.
They are compared with the data (green symbols) determined from the linear correlations with the neutron-skin thicknesses of $^{208}$Pb and $^{48}$Ca~\cite{Zhao2016_PRC94-041302}, and other \textit{ab initio} calculations (blue dotted lines)~\cite{Gandolfi2011_PRL106-012501,Maris2013_PRC87-054318,Tews2016_PRC93-024305}.
For $\Delta r_{np}$ of $^{208}$Pb, the data come from different measurements with antiprotonic atoms \cite{Klos2007_PRC76-014311} (circle), pion photoproduction \cite{Tarbert2014_PRL112-242502} (square), and electric dipole polarizability \cite{RocaMaza2013_PRC88-024316} (diamond), respectively.
For $\Delta r_{np}$ of $^{48}$Ca, the datum comes from the measurement of the electric dipole polarizability~\cite{Birkhan2017_PRL118-252501} (triangle).
For the local chiral forces N$^2$LO from Refs.~\cite{Gezerlis2013_PRL111-032501,Lynn2016_PRL116-062501}, the results include a two-body force with a cutoff $R_0 = 1.0$ and $1.2$ fm, and a two-body plus three-body force ($2N+3N$) with a cutoff $R_0 = 1.2$ fm \cite{Tews2016_PRC93-024305}.
Theoretical uncertainties are denoted by the blue colored regions.
There is no particular reason to choose $N = 20, 14$, and $8$ neutron drops, as long as the central density of the neutron drop does not differ too much from the saturation density ($\approx 0.16$ fm$^{-3}$) \cite{Zhao2016_PRC94-041302}.

It can be seen that the radii obtained in the RBHF calculations with the interaction Bonn A are in a good agreement with the data determined from the linear correlations with the neutron-skin thicknesses.
In comparison with other \textit{ab initio} calculations, AV8' + UIX shows more repulsion and gives larger radii, as expected from the energies shown in Fig.~\ref{fig17}.
For the $2N$ local chiral forces N$^2$LO, the softer interaction with a cut-off radius $R_0 = 1.2$ fm gives a smaller radius and the harder one with $R_0 = 1.0$ fm gives a larger radius.
When including the $3N$ force for N$^2$LO, the radius gets larger by 0.05 fm and is similar to that of AV8' + UIX.

\subsection{Spin-orbit splitting and tensor effects}\label{Sect:4.3}

One important purpose of studying the neutron drops with RBHF theory is to pave the way to extract information for nuclear density functionals from \textit{ab initio} calculations.
Among the various properties of the neutron drops, the spin-orbit splitting is of particular interest, as, in general, it is difficult to find significant features in experimental data which are only connected to tensor forces and therefore suitable for an adjustment of their parameters.
As shown in Ref.~\cite{Shen2018_PLB778-344}, the evolution of SO splittings in neutron drops gives a clear hint of the tensor force in the nuclear density functional, which has been the topic of long debates in the past \cite{Sagawa2014_PPNP76-76}.

\begin{figure}
  \centering
  \includegraphics[width=7cm]{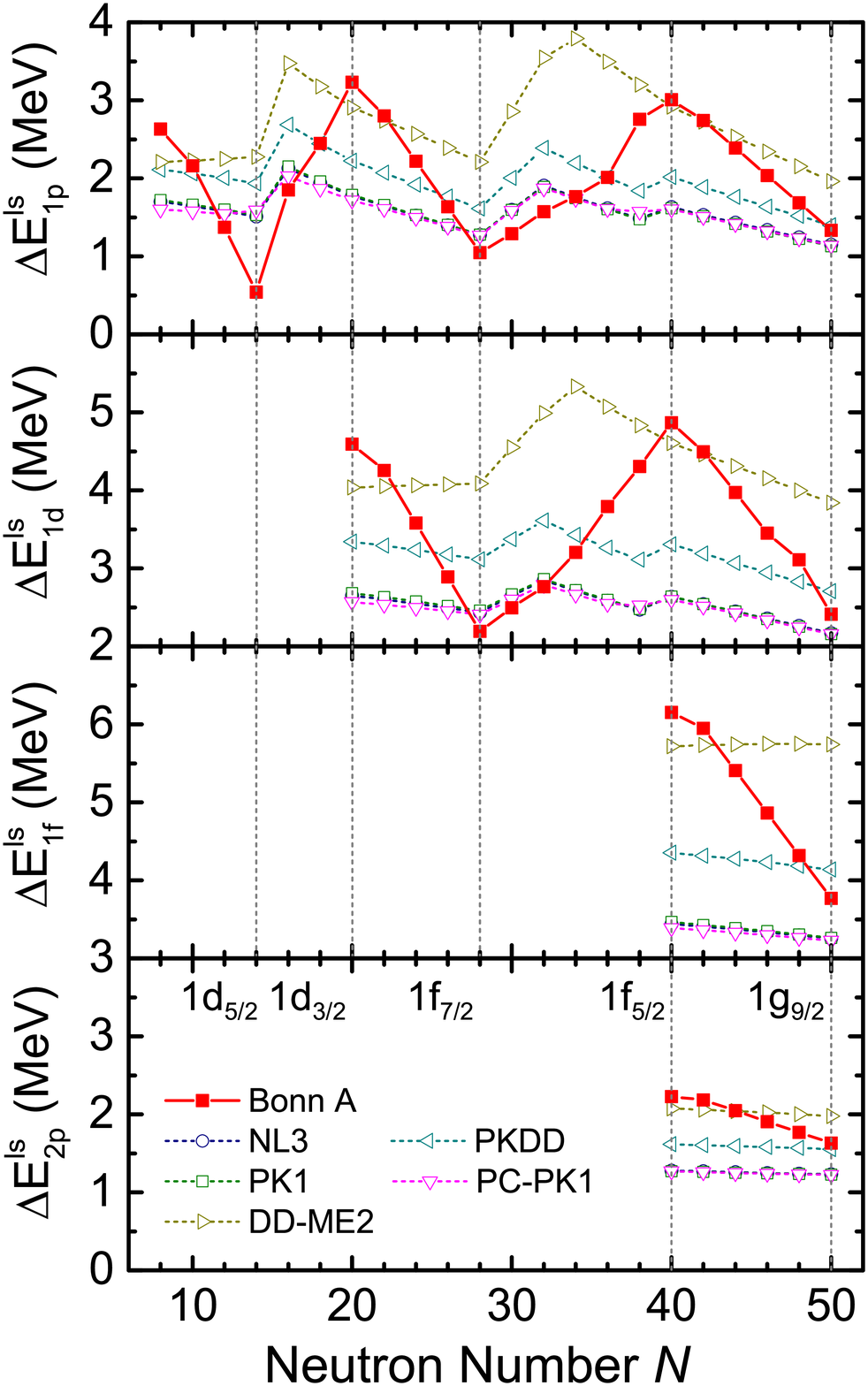}\hspace{2em}
  \includegraphics[width=7cm]{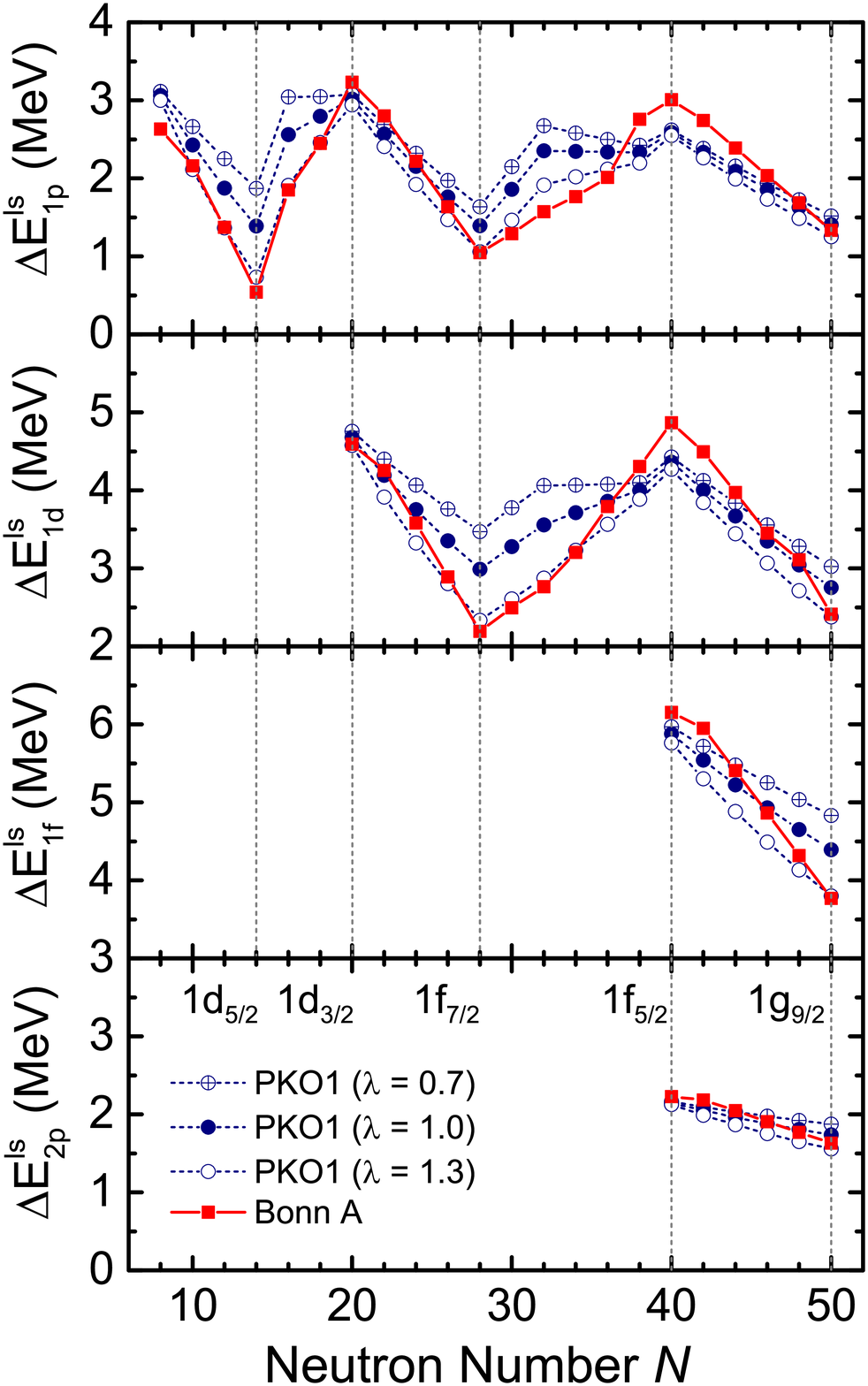}
  \caption{(Color online) From top to bottom panel, $1p,\,1d,\,1f$, and $2p$ spin-orbit splittings of $N$-neutron drops in a HO trap ($\hbar\omega = 10$ MeV) calculated by RBHF theory using the Bonn A interaction, in comparison with the results obtained by various RMF functionals (left) and RHF functional PKO1.
  Figure taken from Ref.~\cite{Shen2018_PLB778-344}.}
  \label{fig23}
\end{figure}

In Fig.~\ref{fig23}, we present the SO splittings of $N$-neutron drops in a HO trap for the $1p,\,1d,\,1f$, and $2p$ doublets calculated by RBHF theory using the Bonn A interaction.
It shows the evolution of the various SO splittings with neutron number.
In the left panel, the SO splittings are compared with the results obtained by various phenomenological relativistic mean-field (RMF) density functionals, including the nonlinear meson-exchange models NL3~\cite{Lalazissis1997_PRC55-540} and PK1~\cite{Long2004_PRC69-034319}, the density-dependent meson-exchange models DD-ME2~\cite{Lalazissis2005_PRC71-024312} and PKDD \cite{Long2004_PRC69-034319}, and the nonlinear point-coupling model PC-PK1~\cite{Zhao2010_PRC82-054319}.

For the microscopic RBHF results a clear pattern can be seen: The SO splitting of a specific orbit with orbital angular momentum $l$ decreases as the next higher $j=j_>= l + 1/2$ orbit is filled and reaches a minimum when this orbit is fully occupied. When the number of neutron increases further,the $j=j_< = l - 1/2$ orbit begins to be occupied and the SO splitting grows up again.

This pattern is similar to the monopole effect of the tensor force acting between neutrons and protons in nuclei as explained by Otsuka \textit{et al.} \cite{Otsuka2005_PRL95-232502}.
In their explanation, the tensor force produces an attractive interaction between a proton in a SO aligned orbit with $j=j_>=l + 1/2$ and a neutron in a SO anti-aligned orbit with $j'=j'_<=l' - 1/2$, and the tensor force produces a repulsive interaction between the same proton and a neutron in a SO aligned orbit with $j'=j'_>=l'+ 1/2$ (or both in a SO anti-aligned orbit with $j'=j'_<=l' - 1/2$). It was also mentioned in Ref.~\cite{Otsuka2005_PRL95-232502} that a similar mechanism, but with a smaller amplitude, exists for the tensor interaction between the same type of nucleons with $T=1$.

The behavior of the SO splittings given by RBHF in Fig.~\ref{fig23} can be understood in a qualitative way: Consider, for instance, the decrease of the $1d$ SO splitting from $N=20$ to $N=28$. Above $N=20$ additional neutrons are filled into the $1f_{7/2}$ orbit (SO aligned).
They show a repulsive interaction with the $1d_{5/2}$ neutrons (SO anti-aligned) and an attractive interaction with the $1d_{3/2}$ neutrons (SO anti-aligned).
The repulsive (attractive) interaction increases (decreases) the corresponding single-particle energy. Therefore, by filling in neutrons into the $1f_{7/2}$ shell the $1d_{5/2}$ orbit is shifted upward and the $1d_{3/2}$ is shifted downward, reducing the $1d$ SO splitting.
Above $N=28$ the neutrons are filled into $2p_{1/2}$ and $1f_{5/2}$.
They interact with the $1d$-neutrons in the opposite way and increase again the SO-splitting for the $1d$ configuration.

However, in the left panel of Fig.~\ref{fig23}, it turns out, that this specific evolution of SO splitting is not significant for any of the phenomenological RMF functionals, which do not include a tensor term.
On the other hand, as shown in the right panel of Fig.~\ref{fig23}, the same calculation but with the RHF density functional PKO1 \cite{Long2006_PLB640-150}, which includes the tensor force induced by the pion coupling through the exchange term, shows a similar pattern as the RBHF calculations, but the amplitude is smaller.
Therefore, in order to investigate the effects of the tensor force, the pion coupling has been multiplied with a factor $\lambda$ without readjusting the other parameters of this functional.
As expected, the evolution of the SO splitting is influenced by the strengths of the tensor force significantly.
For $\lambda=1$, one has the results of the original functional PKO1, which already shows the right pattern, but the size of the effect is slightly too small.
This can be understood by the fact that it is difficult to fit the strengths of the tensor force just to bulk properties such as binding energies and radii~\cite{Wang2018_PRC98-034313,Lalazissis2009_PRC80-041301}.
The general feature of these SO splittings found in the RBHF calculations can be well reproduced with PKO1 simply by multiplying a factor $\lambda = 1.3$ in front of the pion coupling.

\subsection{Connections to CDFT}\label{Sect:4.4}

Investigations of nuclear matter using RBHF theory already promotes many successful developments of CDFT.
Investigations of finite nuclear systems will provide further information to deepen our understanding of nuclear structure and to develop nuclear density functionals of higher quality.
In particular, one of the important guides is the systematic and specific pattern due to the effects of the tensor force found in the evolution of SO splittings in neutron drops.

The tensor force in non-relativistic nuclear density functionals is a topic of long debates and it is still not fully settled \cite{Sagawa2014_PPNP76-76}.
The study of the tensor force in CDFT is still at its infancy and has attracted much attraction in the past decades \cite{Long2006_PLB640-150,Long2007_PRC76-034314,Long2008_EPL82-12001,Tarpanov2008_PRC77-054316,Lalazissis2009_PRC80-041301,Moreno-Torres2010_PRC81-064327,Marcos2013_PAN76-562,Wang2013_PRC87-047301,Marcos2014_PAN77-299,Afanasjev2015_PRC92-044317,Jiang2015_PRC91-025802,Jiang2015_PRC91-034326,Li2016_PLB753-97,Li2016_PRC93-054312,Karakatsanis2017_PRC95-034318,Zong2018_CPC42-024101,Wang2018_PRC98-034313}.
In Ref.~\cite{Long2006_PLB640-150}, the first RHF density functional PKO1 has been developed, which has a comparable precision of describing finite nuclei as other successful RMF functionals.
With the inclusion of exchange terms and the contribution of the pion, the effect of the tensor force can be studied, and it improves the agreement with experimental single-particle levels \cite{Long2008_EPL82-12001}.
The tensor force from the exchange of the $\rho$-meson was further developed in the RHF functional PKA1 in Ref.~\cite{Long2007_PRC76-034314}.

\begin{figure}
  \centering
  \includegraphics[width=8cm]{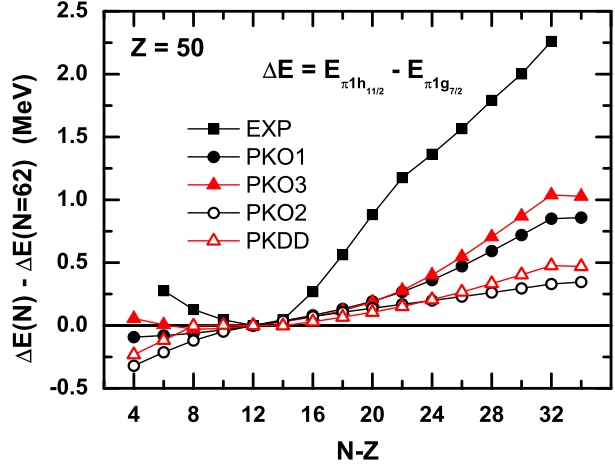}
  \caption{(Color online) Energy difference $\Delta E = E_{\pi 1h_{11/2}} - E_{\pi 1g_{7/2}}$ in $Z = 50$ isotopes as a function of neutron excess calculated by the RHF functionals PKO1, PKO2, and PKO3, and by the RMF functional PKDD.
  Different results are rescaled to zero at $N - Z = 12$.
  Figure taken from Ref.~\cite{Long2008_EPL82-12001}.}
  \label{fig24}
\end{figure}

Even though it is well known that the tensor force mainly originates from the $\pi$-exchange, it is not an easy task to identify different contributions quantitatively and study the relativistic origin of the tensor force on an equal footing as in the non-relativistic context.
The origin of the tensor force in CDFT is studied using a non-relativistic reduction and the one associated with the Fock diagrams of Lorentz scalar ($\sigma,\delta$) and vector ($\omega,\rho$) couplings are also discussed \cite{Jiang2015_PRC91-034326}.
A more detailed study has been done very recently to identify the tensor force up to order $1/M^2$ in each meson-nucleon coupling using RHF theory \cite{Wang2018_PRC98-034313}.
With this newly developed formalism, one is eventually able to make a quantitative comparisons with the corresponding tensor force in the non-relativistic density functionals.

However, in most of the previous investigations on the influence of the tensor force for CDFT (or for non-relativistic DFT), the main focus was to compare with experimental single-particle levels \cite{Lalazissis2009_PRC80-041301}.
For example, Fig.~\ref{fig24} shows the energy difference $\Delta E = E_{\pi 1h_{11/2}} - E_{\pi 1g_{7/2}}$ in $Z = 50$ isotopes as a function of neutron excess calculated by the RHF functionals PKO1, PKO2, and PKO3, and by the RMF functional PKDD \cite{Long2008_EPL82-12001}.
As the tensor force is (basically) not included in PKO2 and PKDD but included in PKO1 and PKO3, it can be seen that the tensor force can improve the agreement with the data.
Indeed, this is also consistent with the non-relativistic findings \cite{Otsuka2006_PRL97-162501,Colo2007_PLB646-227}.

\begin{figure}
  \centering
  \includegraphics[width=6cm]{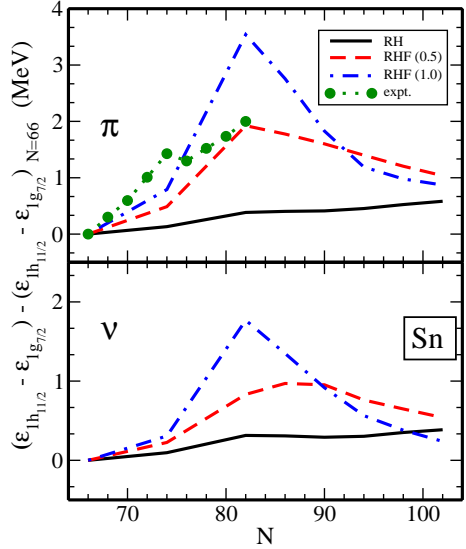}
  \vspace{-2.5cm}
  \caption{(Color online) Evolution of the $1h_{11/2}$ and $1g_{7/2}$ neutron and proton energy gaps.
  as a function of the neutron number calculated by functionals fitted to experimental ground state properties with no pion-nucleon coupling (RH), with the free pion-nucleon coupling (RHF 1.0) and with half the strength of the free pion nucleon-coupling (RHF 0.5). The single-particle levels of $N=66$ are taken as reference point.
  Figure taken from Ref.~\cite{Lalazissis2009_PRC80-041301}.}
  \label{fig25}
\end{figure}

There might be several reasons why the inclusion of tensor components in the RHF calculations shown in Fig.~\ref{fig24} is still far from reproducing the experimental results. (i) The particle-vibrational coupling, which goes beyond mean field and is not taken into account here, has a strong influence on the single-particle energies (see below). (ii) In the PKO functionals, the pion-nucleon coupling strength has an exponential density dependence.
The pion-nucleon coupling strength is fixed to the free value for vanishing density and it drops down very fast in the nuclear interior. This leads to a reduced tensor strength, because it has been shown in Ref.~\cite{Lalazissis2009_PRC80-041301} that a fit to experimental masses and radii does not like a tensor term.
In Fig.~\ref{fig25}, we show the same energy splittings as those in Fig.~\ref{fig24} in results of RHF calculations with three functionals containing a fixed pion-nucleon coupling constant $\lambda f^2_\pi$ in addition to the conventional DD-ME functional \cite{Lalazissis2009_PRC80-041301}. In these calculations, only the Fock term of pion-nucleon coupling is taken into account and the values of $\lambda$ are chosen as $0$, $0.5$, and $1.0$, respectively, and then the remaining parameters are adjusted with the same fitting protocol as DD-ME2 \cite{Lalazissis2005_PRC71-024312}.
For $\lambda = 0$, we have the usual RH theory, which does not show substantial increasing of the energy splitting.
While the splittings are much too large for $\lambda = 1.0$, one finds reasonable values for $\lambda = 0.5$, i.e., the inclusion of a tensor term allows a very reasonable fit to the single-particle levels in this region and to the remaining ground-state properties.

\begin{figure}
  \centering
  \includegraphics[angle=270,width=10cm]{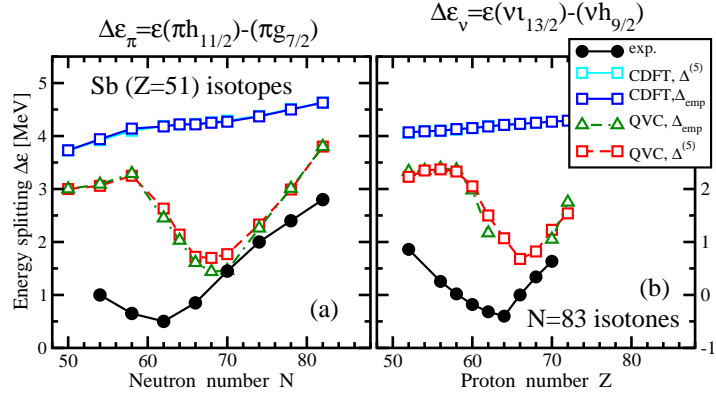}
  \caption{(Color online) (a) Energy difference $\Delta E = E_{\pi 1h_{11/2}} - E_{\pi 1g_{7/2}}$ in $Z = 51$ isotopes (which is equivalent to that of in $Z = 50$ ones), (b) energy difference $\Delta E = E_{\nu 1i_{13/2}} - E_{\nu 1h_{9/2}}$ in $N = 83$ isotones, calculated by the RMF functional NL3* (CDFT in the figure).
  In comparison with calculation of quasiparticle-vibration (QVC) model using the same functional.
  The two results, $\Delta^{(5)}$ and $\Delta_{\rm emp}$, correspond to the calculations with two pairing schemes, see Ref.~\cite{Afanasjev2015_PRC92-044317} for detail.
  Figure modified from Ref.~\cite{Afanasjev2015_PRC92-044317}.}
  \label{fig26}
\end{figure}

On the other hand, as mentioned above, the particle-vibrational coupling (PVC) plays an important role in spherical nuclei \cite{Litvinova2006_PRC73-044328,Litvinova2011_PRC84-014305}. Going beyond mean field, this effect can be included consistently without any new fitting parameter, and it produces reasonable values not only for the single-particle gap in magic configurations \cite{Litvinova2006_PRC73-044328} but also for the width of giant resonances \cite{Litvinova2007_PRC75-064308}. In Ref.~\cite{Afanasjev2015_PRC92-044317}, the energy splittings in the Sn region were calculated with the RMF functional NL3*, which does not include a tensor force, both at the mean-field level and beyond mean-field with inclusion of PVC.
It was found that the description of these energy differences can also be improved substantially when the particle-vibration coupling is included, as shown in Fig.~\ref{fig26}.

Recent experiments found in the $N=20$ isotones below $^{40}$Ca that there are considerable changes in the $1f$ and $1p$ spin-orbit splittings, in particular between $^{36}$S and $^{34}$Si. On the basis of various relativistic and non-relativistic density functionals, Karakatsanis \textit{et al.} \cite{Karakatsanis2017_PRC95-034318} investigated several reasons for these experimental observations, including the difference in the isospin dependence of relativistic and non-relativistic spin-orbit couplings, tensor forces, pairing correlations, and particle-vibrational coupling. They found that all these effects had influences on the $1f$ and $1p$ spin-orbit splittings, of course depending in size on the various parameters entering in these theories. Since these parameters have been phenomenologically adjusted in one way or another, they found it very difficult to isolate specific reasons, such as tensor forces, in the nuclear density functional by just looking into experimental single-particle levels.

In such a situation, parameter free, microscopic \textit{ab initio} calculations are definitely needed to get a clearer understanding of such phenomena and to learn more about the various parameters in the density functionals. For instance, as discussed in Section~\ref{Sect:4.3}, the effect of the tensor force can be studied clearly in RBHF calculations of the evolution of spin-orbit splittings in neutron drops. As a relativistic description, this theory includes the proper parameter-free spin-orbit coupling. Such calculations produce meta-data. On the present level, they neither include pairing correlations, nor beyond-mean-field effects such as particle-vibrational couplings. This allows to adjust the tensor force to these meta-data without the ambiguity of additional effects, since they are neither included in the present concept of DFT nor in the present RBHF calculations.


\section{Summary and Perspectives}\label{Sect:5}


%

Relativistic Brueckner-Hartree-Fock (RBHF) theory was investigated already in the 1980s for nuclear matter, and it has shown much improvement as compared with the non-relativistic calculations.
Besides the development of RBHF theory and the important progress in understanding nuclear structure on the basis of \textit{ab initio} calculations using the bare two-body nucleon-nucleon interaction, there are two other benefits of RBHF: First, it shows the relativistic origin of the three-body force in the non-relativistic framework.
It is well known that the non-relativistic \textit{ab initio} methods cannot describe the saturation property for nuclear matter using only two-body NN interactions and a three-body force has to be introduced with new adjustable parameters to reproduce the empirical values.
Of course, the microscopic origin of the full three-body force in nuclei is much more complicated, and there are also discrepancies between experimental data and the RBHF results, which should be studied in future; still, the relativistic origin of three-body forces is an important component.
Second, the studies of RBHF theory for nuclear matter stimulated the development of \textit{ab initio} covariant density functional theories in the past. Examples include the successful density-dependent relativistic mean-field and density-dependent relativistic Hartree-Fock models, or the study of the importance of scalar-isoscalar channel by DD-ME$\delta$, and so on.

Due to the complexity of the theory and the limitation of computer resources, a fully self-consistent RBHF study for finite nuclei has not been realized until recently \cite{Shen2016_CPL33-102103, Shen2017_PRC96-014316}.
Before that, RBHF studies for finite nuclei stayed at the level of the local density approximation (LDA).
These LDA studies have deepened our understanding of the nuclear system and of the structure of covariant density functional theories. However, as an indirect way to study finite nuclei, by a mapping from results of nuclear matter, it suffers from considerable ambiguities. In order to solve this problem and also from the point of view of completeness, a self-consistent RBHF study for finite nuclei was called for.
Starting from the Dirac Woods-Saxon basis and using the self-consistent RHF basis during the iteration, the coupled system RBHF equations (the relativistic Bethe-Goldstone equation and RHF equation) have been solved directly in a finite nuclear space \cite{Shen2016_CPL33-102103, Shen2017_PRC96-014316}. Convergence is studied carefully.
The results show much improvement over the non-relativistic investigations, which confirms the importance of relativistic effects in finite nuclear systems.
Moreover, full RBHF studies in finite nuclei not only solve the non-uniqueness problem of LDA, but also overcome the difficulty in decomposing the Lorentz structure of the single-particle potential which causes until today lots of troubles and uncertainties in the RBHF studies of nuclear matter.


The study of finite nuclei has further significance for the development of nuclear density functionals.
For example, there are shell effects and tensor-force effects, which do not exist in nuclear matter.

The spin-symmetry of the single-particle levels in the Dirac sea of $^{16}$O has been investigated by RBHF theory and compared with results of covariant density functional theories.
The studies on the spin- and pseudospin-symmetries beyond phenomenological functionals give us more insights of these symmetries and enables a better connection between CDFTs and RBHF calculations.

  The investigation of neutron drops in the framework of RBHF theory provides many interesting insights. As an ideal system, a neutron drop is easy to be calculated and serves as a benchmark of different methods and interactions. By looking into the evolutions of spin-orbit splitting as the number of neutrons increases, a clear signal of the tensor force is revealed. This can provide guidance for the development of nuclear density functionals. For example, a non-relativistic functional with tensor terms, SAMi-T, has already been developed along this direction. It shows promising results for the descriptions of both ground-state and excited-state properties \cite{Shen2019_PRC99-034322}. Counterparts in the relativistic framework are also in progress \cite{Wang2018_PRC98-034313, Wang2018PhD}. 



With all these quite impressive successes, the way towards an \textit{ab initio} covariant density functional theory is still at the beginning, much work is still needed.

First, the development of RBHF theory for nuclear matter is far from finished.
The Lorentz-structure decomposition of the self-energy has not been fully solved, and certain approximations, such as angle-averaging of the Pauli operator, are existing.
The study of nuclear matter by RBHF theory has important applications to nuclear astrophysics and for the construction of better nuclear density functionals through the LDA.
Therefore, a fully self-consistent calculation with high precision is needed.

Second, even though an application of RBHF theory for finite nuclear systems has now been realized, the computational demand is vast and limits the further extension of the theory.
Renormalization procedures such as $V_{{\rm low-}k}$ or the similarity renormalization group (SRG), which are commonly used in the non-relativistic \textit{ab initio} context, would be very useful to reduce this computational cost and would also allow applications for heavier systems.

Third, through RBHF applications in nuclear matter and in finite nuclei, the descriptions have been improved over the non-relativistic ones, but they are still not perfect.
As the lowest order of the hole-line expansion, the (R)BHF theory is not supposed to be the full story yet, and the next order (three hole-line) is known to have non-negligible contributions in the non-relativistic case.
Therefore, how much the three hole-line contributes in the relativistic Brueckner theory is an open and interesting question. 
 There is a possibility that, even with higher orders of the hole-line expansion added upon the current RBHF theory, discrepancies between theoretical and experimental results may still exist. Then one would have to consider the possibility to use bare three-body forces in the relativistic framework too. If the discrepancy between theoretical results with two-body interaction and experimental data can be attributed to the contribution of three-body interaction, as the relativistic effect already improves much of the description, the contribution of three-body force in RBHF should be smaller than in BHF.

Fourth, with higher orders of the hole-line expansion, diagrams beyond the ladder type are needed in principle.
Therefore, how to go beyond the ladder diagrams in a systematic and controllable way is another open and interesting question. The connection to the recent progress in the method of the so-called diagrammatic quantum Monte-Carlo \cite{Prokofev1998_PRL81-2514, Prokofev2007_PRL99-250201} may be one of the possible ways.

Last but not least, the only input for present RBHF theory, i.e., the relativistic nucleon-nucleon interactions, which can reproduce the two-body scattering data and the deuteron properties, are much less developed than their non-relativistic counterparts.
Up to now, the most commonly used relativistic NN interactions are the Groningen \cite{terHaar1987_PR149-207} and the Bonn interactions \cite{Machleidt1989_ANP19-189}, which are in some sense outdated comparing with the non-relativistic modern and high-precision interactions.
Therefore, efforts to develop relativistic chiral NN interactions are also highly demanded and work in this direction is in progress \cite{Ren2018_CPC42-014103, Li2018_CPC42-014105}.
  Of course, one needs in this context a new power counting system \cite{Ren2018_CPC42-014103}, because in a relativistic theory all the terms of the $1/M$-expansion are summed up automatically. In the leading order of the relativistic chiral interaction, a much better description of the $^1S_0$ and $^3P_0$ phase shifts was achieved comparing with the non-relativistic case, and descriptions in other partial waves are similar to non-relativistic case \cite{Ren2018_CPC42-014103}. There are indications that this improves the convergence properties of relativistic chiral perturbation theory.

\section*{Acknowledgements}

We would like to express our gratitude to all the collaborators and colleagues who contributed to the investigations presented here, in particular to L. S. Geng, J. N. Hu, B. Long, X. L. Ren, H. Tong, S. B. Wang, S. Q. Zhang, and P. W. Zhao.
This work was partly supported by the National Key R\&D Program of China (Contract No. 2017YFE0116700 and No. 2018YFA0404400),
Natural Science Foundation of China under Grants No.~11335002, and No.~11621131001,
the Overseas Distinguished Professor Project from Ministry of Education of China No.~MS2010BJDX001,
the JSPS Grant-in-Aid for Early-Career Scientists under Grant No.~18K13549,
the JSPS-NSFC Bilateral Program for Joint Research Project on Nuclear mass and life for unravelling mysteries of the r-process,
the DFG (Germany) cluster of excellence ``Origin and Structure of the Universe'' (www.universecluster.de),
the Funding from the European Union's Horizon 2020 research and innovation programme under Grant Agreement No. 654002.
H.L. would like to thank the RIKEN iTHES project and iTHEMS program.





\section*{References}
\bibliographystyle{elsarticle-num}
\bibliography{ref}

\begin{thebibliography}{100}
\expandafter\ifx\csname url\endcsname\relax
  \def\url#1{\texttt{#1}}\fi
\expandafter\ifx\csname urlprefix\endcsname\relax\def\urlprefix{URL }\fi
\expandafter\ifx\csname href\endcsname\relax
  \def\href#1#2{#2} \def\path#1{#1}\fi

\bibitem{Becquerel1896_CR122-501}
H.~Becquerel, {Sur les radiations invisibles {\'{e}}mises par les corps
  phosphorescents}, Comptes Rendus 122 (1896) 501.

\bibitem{Curie1898_CR127-175}
P.~Curie, M.~Curie, {Sur une substance nouvelle radioactive contenue dans la
  pechblende}, Comptes Rendus 127 (1898) 175.

\bibitem{Curie1898_CR127-1215}
P.~Curie, M.~Curie, G.~B{\'{e}}mont, {Sur une nouvelle substance fortement
  radio-active, contenue dans la pechblende}, Comptes Rendus 127 (1898)
  1215--1217.

\bibitem{Rutherford1911_PM21-669}
E.~Rutherford, {LXXIX. The scattering of $\alpha$ and $\beta$ particles by
  matter and the structure of the atom}, Philos. Mag. Ser. 6 21~(125) (1911)
  669--688.
\newblock \href {http://dx.doi.org/10.1080/14786440508637080}
  {\path{doi:10.1080/14786440508637080}}.

\bibitem{Chadwick1932_Nature129-312}
J.~Chadwick, Possible existence of a neutron, Nature 129 (1932) 312.
\newblock \href {http://dx.doi.org/10.1038/129312a0}
  {\path{doi:10.1038/129312a0}}.

\bibitem{Yukawa1935_PPMSJ17-48}
H.~Yukawa, {Interaction of elementary particles. Part I}, Proc. Phys. Math.
  Soc. Jpn. 17 (1935) 48--57.

\bibitem{Mayer1949_PR75-1969}
M.~G. Mayer, On closed shells in nuclei. {II}, Phys. Rev. 75 (1949) 1969--1970.
\newblock \href {http://dx.doi.org/10.1103/PhysRev.75.1969}
  {\path{doi:10.1103/PhysRev.75.1969}}.

\bibitem{Haxel1949_PR75-1766}
O.~Haxel, J.~H.~D. Jensen, H.~E. Suess, On the ``magic numbers'' in nuclear
  structure, Phys. Rev. 75 (1949) 1766--1766.
\newblock \href {http://dx.doi.org/10.1103/PhysRev.75.1766.2}
  {\path{doi:10.1103/PhysRev.75.1766.2}}.

\bibitem{Rainwater1950_PR79-432}
J.~Rainwater, {Nuclear Energy Level Argument for a Spheroidal Nuclear Model},
  Phys. Rev. 79 (1950) 432--434.
\newblock \href {http://dx.doi.org/10.1103/PhysRev.79.432}
  {\path{doi:10.1103/PhysRev.79.432}}.

\bibitem{Bohr1952_MFMDVS26-14}
A.~Bohr, {The coupling of nuclear surface oscillations to the motion of
  individual nucleons}, Mat. Fys. Medd. Dan. Vid. Selsk 26 (1952) 14.

\bibitem{Bohr1953_MFMDVS27-16}
A.~Bohr, B.~R. Mottelson, {Collective and individual-particle aspects of
  nuclear structure}, Mat. Fys. Medd. Dan. Vid. Selsk 27 (1953) 16.

\bibitem{Euler1937_ZP105-553}
H.~Euler, {\"Uber die Art der Wechselwirkung in den schweren Atomkernen},
  Zeitschrift f\"ur Phys. 105 (1937) 553--575.
\newblock \href {http://dx.doi.org/10.1007/BF01371561}
  {\path{doi:10.1007/BF01371561}}.

\bibitem{Jastrow1951_PR81-165}
R.~Jastrow, {On the Nucleon-Nucleon Interaction}, Phys. Rev. 81 (1951)
  165--170.
\newblock \href {http://dx.doi.org/10.1103/PhysRev.81.165}
  {\path{doi:10.1103/PhysRev.81.165}}.

\bibitem{Vautherin1972_PRC5-626}
D.~Vautherin, D.~M. Brink, Hartree-fock calculations with skyrme's interaction.
  i. spherical nuclei, Phys. Rev. C 5 (1972) 626--647.
\newblock \href {http://dx.doi.org/10.1103/PhysRevC.5.626}
  {\path{doi:10.1103/PhysRevC.5.626}}.

\bibitem{Walecka1974_APNY83-491}
J.~D. Walecka, A theory of highly condensed matter, Ann. Phys. (NY) 83 (1974)
  491--529.
\newblock \href {http://dx.doi.org/10.1016/0003-4916(74)90208-5}
  {\path{doi:10.1016/0003-4916(74)90208-5}}.

\bibitem{Decharge1980_PRC21-1568}
J.~Decharg\'e, D.~Gogny, {Hartree-Fock-Bogolyubov calculations with the $D1$
  effective interaction on spherical nuclei}, Phys. Rev. C 21 (1980)
  1568--1593.
\newblock \href {http://dx.doi.org/10.1103/PhysRevC.21.1568}
  {\path{doi:10.1103/PhysRevC.21.1568}}.

\bibitem{Bender2003_RMP75-121}
M.~Bender, P.-H. Heenen, P.-G. Reinhard, Self-consistent mean-field models for
  nuclear structure, Rev. Mod. Phys. 75 (2003) 121--180.
\newblock \href {http://dx.doi.org/10.1103/RevModPhys.75.121}
  {\path{doi:10.1103/RevModPhys.75.121}}.

\bibitem{Vretenar2005_PR409-101}
D.~Vretenar, A.~V. Afanasjev, G.~A. Lalazissis, P.~Ring, Relativistic
  {Hartree}-{Bogoliubov} theory: static and dynamic aspects of exotic nuclear
  structure, Phys. Rep. 409 (2005) 101--259.
\newblock \href {http://dx.doi.org/10.1016/j.physrep.2004.10.001}
  {\path{doi:10.1016/j.physrep.2004.10.001}}.

\bibitem{Meng2016}
J.~Meng (Ed.), Relativistic Density Functional for Nuclear Structure, Vol.~10
  of International Review of Nuclear Physics, World Scientific, Singapore,
  2016.
\newblock \href {http://dx.doi.org/10.1142/9872} {\path{doi:10.1142/9872}}.

\bibitem{Brueckner1954_PR95-217}
K.~Brueckner, C.~Levinson, H.~Mahmoud, {Two-Body Forces and Nuclear Saturation.
  I. Central Forces}, Phys. Rev. 95 (1954) 217--228.
\newblock \href {http://dx.doi.org/10.1103/PhysRev.95.217}
  {\path{doi:10.1103/PhysRev.95.217}}.

\bibitem{Jastrow1955_PR98-1479}
R.~Jastrow, {Many-Body Problem with Strong Forces}, Phys. Rev. 98 (1955)
  1479--1484.
\newblock \href {http://dx.doi.org/10.1103/PhysRev.98.1479}
  {\path{doi:10.1103/PhysRev.98.1479}}.

\bibitem{Coester1958_NP7-421}
F.~Coester, {Bound states of a many-particle system}, Nucl. Phys. 7 (1958)
  421--424.
\newblock \href {http://dx.doi.org/10.1016/0029-5582(58)90280-3}
  {\path{doi:10.1016/0029-5582(58)90280-3}}.

\bibitem{Kuo1966_NP85-40}
T.~Kuo, G.~Brown, {Structure of finite nuclei and the free nucleon-nucleon
  interaction}, Nucl. Phys. 85 (1966) 40--86.
\newblock \href {http://dx.doi.org/10.1016/0029-5582(66)90131-3}
  {\path{doi:10.1016/0029-5582(66)90131-3}}.

\bibitem{Carlson1988_PRC38-1879}
J.~Carlson, {Alpha particle structure}, Phys. Rev. C 38 (1988) 1879--1885.
\newblock \href {http://dx.doi.org/10.1103/PhysRevC.38.1879}
  {\path{doi:10.1103/PhysRevC.38.1879}}.

\bibitem{Ramos1989_NPA503-1}
A.~Ramos, A.~Polls, W.~Dickhoff, {Single-particle properties and short-range
  correlations in nuclear matter}, Nucl. Phys. A 503 (1989) 1--52.
\newblock \href {http://dx.doi.org/10.1016/0375-9474(89)90252-2}
  {\path{doi:10.1016/0375-9474(89)90252-2}}.

\bibitem{Zheng1993_PRC48-1083}
D.~C. Zheng, B.~R. Barrett, L.~Jaqua, J.~P. Vary, R.~J. McCarthy, {Microscopic
  calculations of the spectra of light nuclei}, Phys. Rev. C 48 (1993)
  1083--1091.
\newblock \href {http://dx.doi.org/10.1103/PhysRevC.48.1083}
  {\path{doi:10.1103/PhysRevC.48.1083}}.

\bibitem{Honma1996_PRL77-3315}
M.~Honma, T.~Mizusaki, T.~Otsuka, Nuclear shell model by the quantum monte
  carlo diagonalization method, Phys. Rev. Lett. 77 (1996) 3315--3318.
\newblock \href {http://dx.doi.org/10.1103/PhysRevLett.77.3315}
  {\path{doi:10.1103/PhysRevLett.77.3315}}.

\bibitem{Mueller2000_PRC61-044320}
H.~M. M{\"{u}}ller, S.~E. Koonin, R.~Seki, U.~van Kolck, {Nuclear matter on a
  lattice}, Phys. Rev. C 61 (2000) 044320.
\newblock \href {http://dx.doi.org/10.1103/PhysRevC.61.044320}
  {\path{doi:10.1103/PhysRevC.61.044320}}.

\bibitem{Tsukiyama2011_PRL106-222502}
K.~Tsukiyama, S.~K. Bogner, A.~Schwenk, In-medium similarity renormalization
  group for nuclei, Phys. Rev. Lett. 106 (2011) 222502.
\newblock \href {http://dx.doi.org/10.1103/PhysRevLett.106.222502}
  {\path{doi:10.1103/PhysRevLett.106.222502}}.

\bibitem{Carlson2015_RMP87-1067}
J.~Carlson, S.~Gandolfi, F.~Pederiva, S.~C. Pieper, R.~Schiavilla, K.~E.
  Schmidt, R.~B. Wiringa, {Quantum Monte Carlo methods for nuclear physics},
  Rev. Mod. Phys. 87~(3) (2015) 1067--1118.
\newblock \href {http://dx.doi.org/10.1103/RevModPhys.87.1067}
  {\path{doi:10.1103/RevModPhys.87.1067}}.

\bibitem{Hagen2014_RPP77-096302}
G.~Hagen, T.~Papenbrock, M.~Hjorth-Jensen, D.~J. Dean, {Coupled-cluster
  computations of atomic nuclei}, Reports Prog. Phys. 77~(9) (2014) 096302.
\newblock \href {http://dx.doi.org/10.1088/0034-4885/77/9/096302}
  {\path{doi:10.1088/0034-4885/77/9/096302}}.

\bibitem{Barrett2013_PPNP69-131}
B.~R. Barrett, P.~Navr{\'{a}}til, J.~P. Vary, {Ab initio no core shell model},
  Prog. Part. Nucl. Phys. 69~(1) (2013) 131--181.
\newblock \href {http://dx.doi.org/10.1016/j.ppnp.2012.10.003}
  {\path{doi:10.1016/j.ppnp.2012.10.003}}.

\bibitem{Dickhoff2004_PPNP52-377}
W.~Dickhoff, C.~Barbieri, {Self-consistent Green's function method for nuclei
  and nuclear matter}, Prog. Part. Nucl. Phys. 52~(2) (2004) 377--496.
\newblock \href {http://dx.doi.org/10.1016/j.ppnp.2004.02.038}
  {\path{doi:10.1016/j.ppnp.2004.02.038}}.

\bibitem{Lee2009_PPNP63-117}
D.~Lee, {Lattice simulations for few- and many-body systems}, Prog. Part. Nucl.
  Phys. 63~(1) (2009) 117--154.
\newblock \href {http://dx.doi.org/10.1016/j.ppnp.2008.12.001}
  {\path{doi:10.1016/j.ppnp.2008.12.001}}.

\bibitem{Hergert2016_PR621-165}
H.~Hergert, S.~Bogner, T.~Morris, A.~Schwenk, K.~Tsukiyama, {The In-Medium
  Similarity Renormalization Group: A novel ab initio method for nuclei}, Phys.
  Rep. 621 (2016) 165--222.
\newblock \href {http://dx.doi.org/10.1016/j.physrep.2015.12.007}
  {\path{doi:10.1016/j.physrep.2015.12.007}}.

\bibitem{Shen2016_CPL33-102103}
S.~H. Shen, J.~N. Hu, H.~Z. Liang, J.~Meng, P.~Ring, S.~Q. Zhang, {Relativistic
  Brueckner--Hartree--Fock Theory for Finite Nuclei}, Chin. Phys. Lett. 33~(10)
  (2016) 102103.
\newblock \href {http://dx.doi.org/10.1088/0256-307X/33/10/102103}
  {\path{doi:10.1088/0256-307X/33/10/102103}}.

\bibitem{Shen2017_PRC96-014316}
S.~Shen, H.~Liang, J.~Meng, P.~Ring, S.~Zhang, {Fully self-consistent
  relativistic Brueckner-Hartree-Fock theory for finite nuclei}, Phys. Rev. C
  96~(1) (2017) 014316.
\newblock \href {http://dx.doi.org/10.1103/PhysRevC.96.014316}
  {\path{doi:10.1103/PhysRevC.96.014316}}.

\bibitem{Negele1970_PRC1-1260}
J.~Negele, {Structure of Finite Nuclei in the Local-Density Approximation},
  Phys. Rev. C 1 (1970) 1260--1321.
\newblock \href {http://dx.doi.org/10.1103/PhysRevC.1.1260}
  {\path{doi:10.1103/PhysRevC.1.1260}}.

\bibitem{Taketani1952_PTP7-45}
M.~Taketani, S.~Machida, S.~O-numa, {The Meson Theory of Nuclear Forces, I: The
  Deuteron Ground State and Low Energy Neutron-Proton Scattering}, Prog. Theor.
  Phys. 7 (1952) 45--56.
\newblock \href {http://dx.doi.org/10.1143/ptp/7.1.45}
  {\path{doi:10.1143/ptp/7.1.45}}.

\bibitem{Brueckner1953_PR90-699}
K.~A. Brueckner, K.~M. Watson, {The Construction of Potentials in Quantum Field
  Theory}, Phys. Rev. 90 (1953) 699--708.
\newblock \href {http://dx.doi.org/10.1103/PhysRev.90.699}
  {\path{doi:10.1103/PhysRev.90.699}}.

\bibitem{Phillips1959_RPP22-314}
R.~J.~N. Phillips, {The two-nucleon interaction}, Rep. Prog. Phys. 22 (1959)
  314.
\newblock \href {http://dx.doi.org/10.1088/0034-4885/22/1/314}
  {\path{doi:10.1088/0034-4885/22/1/314}}.

\bibitem{Eisenbud1941_PNAS27-281}
L.~Eisenbud, E.~P. Wigner, {Invariant Forms of Interaction between Nuclear
  Particles}, Proc. Natl. Acad. Sci. 27 (1941) 281--289.
\newblock \href {http://dx.doi.org/10.1073/pnas.27.6.281}
  {\path{doi:10.1073/pnas.27.6.281}}.

\bibitem{Okubo1958_APNY4-166}
S.~Okubo, R.~Marshak, {Velocity dependence of the two-nucleon interaction},
  Ann. Phys. (NY) 4 (1958) 166--179.
\newblock \href {http://dx.doi.org/10.1016/0003-4916(58)90031-9}
  {\path{doi:10.1016/0003-4916(58)90031-9}}.

\bibitem{Gammel1957_PR107-291}
J.~L. Gammel, R.~M. Thaler, {Spin-Orbit Coupling in the Proton-Proton
  Interaction}, Phys. Rev. 107 (1957) 291--298.
\newblock \href {http://dx.doi.org/10.1103/PhysRev.107.291}
  {\path{doi:10.1103/PhysRev.107.291}}.

\bibitem{Lassila1962_PR126-881}
K.~E. Lassila, M.~H. Hull, H.~M. Ruppel, F.~A. McDonald, G.~Breit, {Note on a
  nucleon-nucleon potential}, Phys. Rev. 126 (1962) 881--882.
\newblock \href {http://dx.doi.org/10.1103/PhysRev.126.881}
  {\path{doi:10.1103/PhysRev.126.881}}.

\bibitem{Hamada1962_NP34-382}
T.~Hamada, I.~Johnston, {A potential model representation of two-nucleon data
  below 315 MeV}, Nucl. Phys. 34 (1962) 382--403.
\newblock \href {http://dx.doi.org/10.1016/0029-5582(62)90228-6}
  {\path{doi:10.1016/0029-5582(62)90228-6}}.

\bibitem{Reid1968_APNY50-411}
R.~V. Reid, {Local phenomenological nucleon-nucleon potentials}, Ann. Phys.
  (NY) 50 (1968) 411--448.
\newblock \href {http://dx.doi.org/10.1016/0003-4916(68)90126-7}
  {\path{doi:10.1016/0003-4916(68)90126-7}}.

\bibitem{Erwin1961_PRL6-628}
A.~R. Erwin, R.~March, W.~D. Walker, E.~West, {Evidence for a $\pi-\pi$
  Resonance in the $I = 1$, $J=1$ State}, Phys. Rev. Lett. 6 (1961) 628--630.
\newblock \href {http://dx.doi.org/10.1103/PhysRevLett.6.628}
  {\path{doi:10.1103/PhysRevLett.6.628}}.

\bibitem{Maglicc1961_PRL7-178}
B.~Maglic{\'{c}}, L.~Alvarez, A.~Rosenfeld, M.~Stevenson, {Evidence for a T=0
  Three-Pion Resonance}, Phys. Rev. Lett. 7 (1961) 178--182.
\newblock \href {http://dx.doi.org/10.1103/PhysRevLett.7.178}
  {\path{doi:10.1103/PhysRevLett.7.178}}.

\bibitem{Green1967_RMP39-594}
A.~E.~S. Green, T.~Sawada, {Meson Theoretic N-N Interactions for Nuclear
  Physics}, Rev. Mod. Phys. 39 (1967) 594--610.
\newblock \href {http://dx.doi.org/10.1103/RevModPhys.39.594}
  {\path{doi:10.1103/RevModPhys.39.594}}.

\bibitem{Erkelenz1974_PR13-191}
K.~Erkelenz, {Current status of the relativistic two-nucleon one boson exchange
  potential}, Phys. Rep. 13 (1974) 191--258.
\newblock \href {http://dx.doi.org/10.1016/0370-1573(74)90008-8}
  {\path{doi:10.1016/0370-1573(74)90008-8}}.

\bibitem{Tanabashi2018_PRD98-030001}
M.~Tanabashi, K.~Hagiwara, K.~Hikasa, K.~Nakamura, Y.~Sumino, F.~Takahashi,
  J.~Tanaka, K.~Agashe, G.~Aielli, C.~Amsler, M.~Antonelli, D.~M. Asner,
  H.~Baer, S.~Banerjee, R.~M. Barnett, T.~Basaglia, C.~W. Bauer, J.~J. Beatty,
  V.~I. Belousov, J.~Beringer, S.~Bethke, A.~Bettini, H.~Bichsel, O.~Biebel,
  K.~M. Black, E.~Blucher, O.~Buchmuller, V.~Burkert, M.~A. Bychkov, R.~N.
  Cahn, M.~Carena, A.~Ceccucci, A.~Cerri, D.~Chakraborty, M.-C. Chen, R.~S.
  Chivukula, G.~Cowan, O.~Dahl, G.~D'Ambrosio, T.~Damour, D.~de~Florian,
  A.~de~Gouv\^ea, T.~DeGrand, P.~de~Jong, G.~Dissertori, B.~A. Dobrescu,
  M.~D'Onofrio, M.~Doser, M.~Drees, H.~K. Dreiner, D.~A. Dwyer, P.~Eerola,
  S.~Eidelman, J.~Ellis, J.~Erler, V.~V. Ezhela, W.~Fetscher, B.~D. Fields,
  R.~Firestone, B.~Foster, A.~Freitas, H.~Gallagher, L.~Garren, H.-J. Gerber,
  G.~Gerbier, T.~Gershon, Y.~Gershtein, T.~Gherghetta, A.~A. Godizov,
  M.~Goodman, C.~Grab, A.~V. Gritsan, C.~Grojean, D.~E. Groom, M.~Gr\"unewald,
  A.~Gurtu, T.~Gutsche, H.~E. Haber, C.~Hanhart, S.~Hashimoto, Y.~Hayato, K.~G.
  Hayes, A.~Hebecker, S.~Heinemeyer, B.~Heltsley, J.~J. Hern\'andez-Rey,
  J.~Hisano, A.~H\"ocker, J.~Holder, A.~Holtkamp, T.~Hyodo, K.~D. Irwin, K.~F.
  Johnson, M.~Kado, M.~Karliner, U.~F. Katz, S.~R. Klein, E.~Klempt, R.~V.
  Kowalewski, F.~Krauss, M.~Kreps, B.~Krusche, Y.~V. Kuyanov, Y.~Kwon,
  O.~Lahav, J.~Laiho, J.~Lesgourgues, A.~Liddle, Z.~Ligeti, C.-J. Lin,
  C.~Lippmann, T.~M. Liss, L.~Littenberg, K.~S. Lugovsky, S.~B. Lugovsky,
  A.~Lusiani, Y.~Makida, F.~Maltoni, T.~Mannel, A.~V. Manohar, W.~J. Marciano,
  A.~D. Martin, A.~Masoni, J.~Matthews, U.-G. Mei\ss{}ner, D.~Milstead, R.~E.
  Mitchell, K.~M\"onig, P.~Molaro, F.~Moortgat, M.~Moskovic, H.~Murayama,
  M.~Narain, P.~Nason, S.~Navas, M.~Neubert, P.~Nevski, Y.~Nir, K.~A. Olive,
  S.~Pagan~Griso, J.~Parsons, C.~Patrignani, J.~A. Peacock, M.~Pennington,
  S.~T. Petcov, V.~A. Petrov, E.~Pianori, A.~Piepke, A.~Pomarol, A.~Quadt,
  J.~Rademacker, G.~Raffelt, B.~N. Ratcliff, P.~Richardson, A.~Ringwald,
  S.~Roesler, S.~Rolli, A.~Romaniouk, L.~J. Rosenberg, J.~L. Rosner, G.~Rybka,
  R.~A. Ryutin, C.~T. Sachrajda, Y.~Sakai, G.~P. Salam, S.~Sarkar, F.~Sauli,
  O.~Schneider, K.~Scholberg, A.~J. Schwartz, D.~Scott, V.~Sharma, S.~R.
  Sharpe, T.~Shutt, M.~Silari, T.~Sj\"ostrand, P.~Skands, T.~Skwarnicki, J.~G.
  Smith, G.~F. Smoot, S.~Spanier, H.~Spieler, C.~Spiering, A.~Stahl, S.~L.
  Stone, T.~Sumiyoshi, M.~J. Syphers, K.~Terashi, J.~Terning, U.~Thoma, R.~S.
  Thorne, L.~Tiator, M.~Titov, N.~P. Tkachenko, N.~A. T\"ornqvist, D.~R. Tovey,
  G.~Valencia, R.~Van~de Water, N.~Varelas, G.~Venanzoni, L.~Verde, M.~G.
  Vincter, P.~Vogel, A.~Vogt, S.~P. Wakely, W.~Walkowiak, C.~W. Walter,
  D.~Wands, D.~R. Ward, M.~O. Wascko, G.~Weiglein, D.~H. Weinberg, E.~J.
  Weinberg, M.~White, L.~R. Wiencke, S.~Willocq, C.~G. Wohl, J.~Womersley,
  C.~L. Woody, R.~L. Workman, W.-M. Yao, G.~P. Zeller, O.~V. Zenin, R.-Y. Zhu,
  S.-L. Zhu, F.~Zimmermann, P.~A. Zyla, J.~Anderson, L.~Fuller, V.~S. Lugovsky,
  P.~Schaffner, Review of particle physics, Phys. Rev. D 98 (2018) 030001.
\newblock \href {http://dx.doi.org/10.1103/PhysRevD.98.030001}
  {\path{doi:10.1103/PhysRevD.98.030001}}.

\bibitem{Jackson1975_NPA249-397}
A.~Jackson, D.~Riska, B.~Verwest, {Meson exchange model for the nucleon-nucleon
  interaction}, Nucl. Phys. A 249 (1975) 397--444.
\newblock \href {http://dx.doi.org/10.1016/0375-9474(75)90666-1}
  {\path{doi:10.1016/0375-9474(75)90666-1}}.

\bibitem{Lacombe1980_PRC21-861}
M.~Lacombe, B.~Loiseau, J.~Richard, R.~Mau, J.~C{\^{o}}t{\'{e}},
  P.~Pir{\`{e}}s, R.~de~Tourreil, {Parametrization of the Paris N-N potential},
  Phys. Rev. C 21 (1980) 861--873.
\newblock \href {http://dx.doi.org/10.1103/PhysRevC.21.861}
  {\path{doi:10.1103/PhysRevC.21.861}}.

\bibitem{Partovi1970_PRD2-1999}
M.~H. Partovi, E.~L. Lomon, {Field-Theoretical Nucleon-Nucleon Potential},
  Phys. Rev. D 2 (1970) 1999--2032.
\newblock \href {http://dx.doi.org/10.1103/PhysRevD.2.1999}
  {\path{doi:10.1103/PhysRevD.2.1999}}.

\bibitem{Machleidt1987_PR149-1}
R.~Machleidt, K.~Holinde, C.~Elster, {The bonn meson-exchange model for the
  nucleon-nucleon interaction}, Phys. Rep. 149 (1987) 1--89.
\newblock \href {http://dx.doi.org/10.1016/S0370-1573(87)80002-9}
  {\path{doi:10.1016/S0370-1573(87)80002-9}}.

\bibitem{Detar1980_NPA335-203}
C.~Detar, {Quarks and gluons in the future of nuclear physics}, Nucl. Phys. A
  335 (1980) 203--209.
\newblock \href {http://dx.doi.org/10.1016/0375-9474(80)90177-3}
  {\path{doi:10.1016/0375-9474(80)90177-3}}.

\bibitem{Myhrer1988_RMP60-629}
F.~Myhrer, J.~Wroldsen, {The nucleon-nucleon force and the quark degrees of
  freedom}, Rev. Mod. Phys. 60 (1988) 629--661.
\newblock \href {http://dx.doi.org/10.1103/RevModPhys.60.629}
  {\path{doi:10.1103/RevModPhys.60.629}}.

\bibitem{Weinberg1979_PA96-327}
S.~Weinberg, {Phenomenological Lagrangians}, Physica 96A (1979) 327--340.
\newblock \href {http://dx.doi.org/10.1016/0378-4371(79)90223-1}
  {\path{doi:10.1016/0378-4371(79)90223-1}}.

\bibitem{Weinberg1990_PLB251-288}
S.~Weinberg, {Nuclear forces from chiral lagrangians}, Phys. Lett. B 251 (1990)
  288--292.
\newblock \href {http://dx.doi.org/10.1016/0370-2693(90)90938-3}
  {\path{doi:10.1016/0370-2693(90)90938-3}}.

\bibitem{Weinberg1991_NPB363-3}
S.~Weinberg, {Effective chiral lagrangians for nucleon-pion interactions and
  nuclear forces}, Nucl. Phys. B 363 (1991) 3--18.
\newblock \href {http://dx.doi.org/10.1016/0550-3213(91)90231-L}
  {\path{doi:10.1016/0550-3213(91)90231-L}}.

\bibitem{Ordonez1992_PLB291-459}
C.~Ord{\'{o}}{\~{n}}ez, U.~van Kolck, {Chiral lagrangians and nuclear forces},
  Phys. Lett. B 291 (1992) 459--464.
\newblock \href {http://dx.doi.org/10.1016/0370-2693(92)91404-W}
  {\path{doi:10.1016/0370-2693(92)91404-W}}.

\bibitem{Epelbaum1998_NPA637-107}
E.~Epelbaoum, W.~Gl{\"{o}}ckle, U.~G. Mei{\ss}ner, {Nuclear forces from chiral
  Lagrangians using the method of unitary transformation (I): Formalism}, Nucl.
  Phys. A 637 (1998) 107--134.
\newblock \href {http://dx.doi.org/10.1016/S0375-9474(98)00220-6}
  {\path{doi:10.1016/S0375-9474(98)00220-6}}.

\bibitem{Epelbaum2000_NPA671-295}
E.~Epelbaum, W.~Gl{\"{o}}ckle, U.~G. Mei{\ss}ner, {Nuclear forces from chiral
  Lagrangians using the method of unitary transformation II: The two- nucleon
  system}, Nucl. Phys. A 671 (2000) 295--331.
\newblock \href {http://dx.doi.org/10.1016/S0375-9474(99)00821-0}
  {\path{doi:10.1016/S0375-9474(99)00821-0}}.

\bibitem{Epelbaum2002_PRC66-064001}
E.~Epelbaum, A.~Nogga, W.~Gl{\"{o}}ckle, H.~Kamada, U.~G. Mei{\ss}ner,
  H.~Wita{\l}a, {Three-nucleon forces from chiral effective field theory},
  Phys. Rev. C 66 (2002) 064001.
\newblock \href {http://dx.doi.org/10.1103/PhysRevC.66.064001}
  {\path{doi:10.1103/PhysRevC.66.064001}}.

\bibitem{Entem2003_PRC68-041001}
D.~R. Entem, R.~Machleidt, {Accurate charge-dependent nucleon-nucleon potential
  at fourth order of chiral perturbation theory}, Phys. Rev. C 68~(4) (2003)
  041001.
\newblock \href {http://dx.doi.org/10.1103/PhysRevC.68.041001}
  {\path{doi:10.1103/PhysRevC.68.041001}}.

\bibitem{Epelbaum2005_NPA747-362}
E.~Epelbaum, W.~Gl{\"{o}}ckle, U.~G. Mei{\ss}ner, {The two-nucleon system at
  next-to-next-to-next-to-leading order}, Nucl. Phys. A 747 (2005) 362--424.
\newblock \href {http://dx.doi.org/10.1016/j.nuclphysa.2004.09.107}
  {\path{doi:10.1016/j.nuclphysa.2004.09.107}}.

\bibitem{Epelbaum2015_PRL115-122301}
E.~Epelbaum, H.~Krebs, U.~G. Mei{\ss}ner, {Precision Nucleon-Nucleon Potential
  at Fifth Order in the Chiral Expansion}, Phys. Rev. Lett. 115 (2015) 122301.
\newblock \href {http://dx.doi.org/10.1103/PhysRevLett.115.122301}
  {\path{doi:10.1103/PhysRevLett.115.122301}}.

\bibitem{Entem2015_PRC91-014002}
D.~R. Entem, N.~Kaiser, R.~Machleidt, Y.~Nosyk, {Peripheral nucleon-nucleon
  scattering at fifth order of chiral perturbation theory}, Phys. Rev. C 91
  (2015) 014002.
\newblock \href {http://dx.doi.org/10.1103/PhysRevC.91.014002}
  {\path{doi:10.1103/PhysRevC.91.014002}}.

\bibitem{Entem2015_PRC92-064001}
D.~R. Entem, N.~Kaiser, R.~Machleidt, Y.~Nosyk, {Dominant contributions to the
  nucleon-nucleon interaction at sixth order of chiral perturbation theory},
  Phys. Rev. C 92 (2015) 064001.
\newblock \href {http://dx.doi.org/10.1103/PhysRevC.92.064001}
  {\path{doi:10.1103/PhysRevC.92.064001}}.

\bibitem{Ren2018_CPC42-014103}
X.~L. Ren, K.~W. Li, L.~S. Geng, B.~Long, P.~Ring, J.~Meng, {Leading order
  relativistic chiral nucleon-nucleon interaction}, Chin. Phys. C 42 (2018)
  014103.
\newblock \href {http://dx.doi.org/10.1088/1674-1137/42/1/014103}
  {\path{doi:10.1088/1674-1137/42/1/014103}}.

\bibitem{Li2018_CPC42-014105}
K.~W. Li, X.~L. Ren, L.~S. Geng, B.~W. Long, {Leading order relativistic
  hyperon-nucleon interactions in chiral effective field theory}, Chin. Phys. C
  42 (2018) 014105.
\newblock \href {http://dx.doi.org/10.1088/1674-1137/42/1/014105}
  {\path{doi:10.1088/1674-1137/42/1/014105}}.

\bibitem{Epelbaum2009_RMP81-1773}
E.~Epelbaum, H.~W. Hammer, U.~G. Mei{\ss}ner, {Modern theory of nuclear
  forces}, Rev. Mod. Phys. 81 (2009) 1773--1825.
\newblock \href {http://dx.doi.org/10.1103/RevModPhys.81.1773}
  {\path{doi:10.1103/RevModPhys.81.1773}}.

\bibitem{Machleidt2011_PR503-1}
R.~Machleidt, D.~Entem, {Chiral effective field theory and nuclear forces},
  Phys. Rep. 503 (2011) 1--75.
\newblock \href {http://dx.doi.org/10.1016/j.physrep.2011.02.001}
  {\path{doi:10.1016/j.physrep.2011.02.001}}.

\bibitem{Stoks1993_PRC48-792}
V.~G.~J. Stoks, R.~A.~M. Klomp, M.~C.~M. Rentmeester, J.~J. {de Swart},
  {Partial-wave analysis of all nucleon-nucleon scattering data below 350 MeV},
  Phys. Rev. C 48 (1993) 792--815.
\newblock \href {http://dx.doi.org/10.1103/PhysRevC.48.792}
  {\path{doi:10.1103/PhysRevC.48.792}}.

\bibitem{Stoks1994_PRC49-2950}
V.~G.~J. Stoks, R.~A.~M. Klomp, C.~P.~F. Terheggen, J.~J. de~Swart,
  {Construction of high-quality NN potential models}, Phys. Rev. C 49 (1994)
  2950--2962.
\newblock \href {http://dx.doi.org/10.1103/PhysRevC.49.2950}
  {\path{doi:10.1103/PhysRevC.49.2950}}.

\bibitem{Wiringa1995_PRC51-38}
R.~Wiringa, V.~Stoks, R.~Schiavilla, {Accurate nucleon-nucleon potential with
  charge-independence breaking}, Phys. Rev. C 51 (1995) 38--51.
\newblock \href {http://dx.doi.org/10.1103/PhysRevC.51.38}
  {\path{doi:10.1103/PhysRevC.51.38}}.

\bibitem{Machleidt2001_PRC63-024001}
R.~Machleidt, {High-precision, charge-dependent Bonn nucleon-nucleon
  potential}, Phys. Rev. C 63 (2001) 024001.
\newblock \href {http://dx.doi.org/10.1103/PhysRevC.63.024001}
  {\path{doi:10.1103/PhysRevC.63.024001}}.

\bibitem{Entem2017_PRC96-024004}
D.~R. Entem, R.~Machleidt, Y.~Nosyk, {High-quality two-nucleon potentials up to
  fifth order of the chiral expansion}, Phys. Rev. C 96 (2017) 024004.
\newblock \href {http://dx.doi.org/10.1103/PhysRevC.96.024004}
  {\path{doi:10.1103/PhysRevC.96.024004}}.

\bibitem{Luscher1991_NPB354-531}
M.~L{\"{u}}scher, {Two-particle states on a torus and their relation to the
  scattering matrix}, Nucl. Phys. B 354 (1991) 531--578.
\newblock \href {http://dx.doi.org/10.1016/0550-3213(91)90366-6}
  {\path{doi:10.1016/0550-3213(91)90366-6}}.

\bibitem{Beane2011_PPNP66-1}
S.~R. Beane, W.~Detmold, K.~Orginos, M.~J. Savage, {Nuclear physics from
  lattice QCD}, Prog. Part. Nucl. Phys. 66 (2011) 1--40.
\newblock \href {http://dx.doi.org/10.1016/j.ppnp.2010.08.002}
  {\path{doi:10.1016/j.ppnp.2010.08.002}}.

\bibitem{Ishii2007_PRL99-022001}
N.~Ishii, S.~Aoki, T.~Hatsuda, {Nuclear Force from Lattice QCD}, Phys. Rev.
  Lett. 99 (2007) 022001.
\newblock \href {http://dx.doi.org/10.1103/PhysRevLett.99.022001}
  {\path{doi:10.1103/PhysRevLett.99.022001}}.

\bibitem{Aoki2012_PTEP2012-01A105}
S.~Aoki, T.~Doi, T.~Hatsuda, Y.~Ikeda, T.~Inoue, N.~Ishii, K.~Murano,
  H.~Nemura, K.~Sasaki, {Lattice quantum chromodynamical approach to nuclear
  physics}, Prog. Theor. Exp. Phys. 2012 (2012) 01A105.
\newblock \href {http://dx.doi.org/10.1093/ptep/pts010}
  {\path{doi:10.1093/ptep/pts010}}.

\bibitem{Doi2015_LATTICE20-086}
T.~Doi, S.~Aoki, S.~Gongyo, T.~Hatsuda, Y.~Ikeda, T.~Inoue, T.~Iritani,
  N.~Ishii, T.~Miyamoto, {First results of baryon interactions from lattice QCD
  with physical masses (1) General overview and two-nucleon forces}, Proc. Sci.
  LATTICE 20 (2015) 086.

\bibitem{Johnson1955_PR98-783}
M.~H. Johnson, E.~Teller, {Classical Field Theory of Nuclear Forces}, Phys.
  Rev. 98 (1955) 783--787.
\newblock \href {http://dx.doi.org/10.1103/PhysRev.98.783}
  {\path{doi:10.1103/PhysRev.98.783}}.

\bibitem{Duerr1956_PR103-469}
H.~P. Duerr, {Relativistic Effects in Nuclear Forces}, Phys. Rev. 103 (1956)
  469--480.
\newblock \href {http://dx.doi.org/10.1103/PhysRev.103.469}
  {\path{doi:10.1103/PhysRev.103.469}}.

\bibitem{Rozsnyai1961_PR124-860}
B.~Rozsnyai, {Self-Consistent Nuclear Model}, Phys. Rev. 124 (1961) 860--867.
\newblock \href {http://dx.doi.org/10.1103/PhysRev.124.860}
  {\path{doi:10.1103/PhysRev.124.860}}.

\bibitem{Miller1972_PRC5-241}
L.~Miller, A.~Green, {Relativistic Self-Consistent Meson Field Theory of
  Spherical Nuclei}, Phys. Rev. C 5 (1972) 241--252.
\newblock \href {http://dx.doi.org/10.1103/PhysRevC.5.241}
  {\path{doi:10.1103/PhysRevC.5.241}}.

\bibitem{Serot1986_ANP16-1}
B.~D. Serot, J.~D. Walecka, The relativistic nuclear many-body problem, Adv.
  Nucl. Phys. 16 (1986) 1--327.

\bibitem{Chin1977_APNY108-301}
S.~A. Chin, A relativistic many-body theory of high density matter, Ann. Phys.
  (NY) 108~(2) (1977) 301 -- 367.
\newblock \href {http://dx.doi.org/10.1016/0003-4916(77)90016-1}
  {\path{doi:10.1016/0003-4916(77)90016-1}}.

\bibitem{Boguta1977_NPA292-413}
J.~Boguta, A.~R. Bodmer, Relativistic calculation of nuclear matter and the
  nuclear surface, Nucl. Phys. A 292 (1977) 413--428.
\newblock \href {http://dx.doi.org/10.1016/0375-9474(77)90626-1}
  {\path{doi:10.1016/0375-9474(77)90626-1}}.

\bibitem{Serot1979_PLB86-146}
B.~D. Serot, {A relativistic nuclear field theory with $\pi$ and $\rho$
  mesons}, Phys. Lett. B 86 (1979) 146--150.
\newblock \href {http://dx.doi.org/10.1016/0370-2693(79)90804-9}
  {\path{doi:10.1016/0370-2693(79)90804-9}}.

\bibitem{Gambhir1990_APNY198-132}
Y.~K. Gambhir, P.~Ring, A.~Thimet, Relativistic mean field theory for finite
  nuclei, Ann. Phys. (NY) 198 (1990) 132--179.
\newblock \href {http://dx.doi.org/10.1016/0003-4916(90)90330-Q}
  {\path{doi:10.1016/0003-4916(90)90330-Q}}.

\bibitem{Koepf1988_PLB212-397}
W.~Koepf, P.~Ring, Has the nucleus 24mg a triaxial shape? a relativistic
  investigation, Phys. Lett. B 212~(4) (1988) 397 -- 401.
\newblock \href {http://dx.doi.org/10.1016/0370-2693(88)91786-8}
  {\path{doi:10.1016/0370-2693(88)91786-8}}.

\bibitem{Kucharek1991_ZPA339-23}
H.~Kucharek, P.~Ring, Relativistic field theory of superfluidity in nuclei, Z.
  Phys. A 339~(1) (1991) 23--35.
\newblock \href {http://dx.doi.org/10.1007/BF01282930}
  {\path{doi:10.1007/BF01282930}}.

\bibitem{Koepf1989_NPA493-61}
W.~Koepf, P.~Ring, A relativistic description of rotating nuclei: The yrast
  line of 20ne, Nucl. Phys. A 493~(1) (1989) 61 -- 82.
\newblock \href {http://dx.doi.org/10.1016/0375-9474(89)90532-0}
  {\path{doi:10.1016/0375-9474(89)90532-0}}.

\bibitem{Afanasjev1996_NPA608-107}
A.~Afanasjev, J.~K\"onig, P.~Ring, Superdeformed rotational bands in the a =
  140-150 mass region: A cranked relativistic mean field description, Nucl.
  Phys. A 608~(2) (1996) 107 -- 175.
\newblock \href {http://dx.doi.org/10.1016/0375-9474(96)00272-2}
  {\path{doi:10.1016/0375-9474(96)00272-2}}.

\bibitem{Malfliet1988_PPNP21-207}
R.~Malfliet, {Relativistic theory of nuclear matter and finite nuclei}, Prog.
  Part. Nucl. Phys. 21 (1988) 207--291.
\newblock \href {http://dx.doi.org/10.1016/0146-6410(88)90034-8}
  {\path{doi:10.1016/0146-6410(88)90034-8}}.

\bibitem{Reinhard1989_RPP52-439}
P.~G. Reinhard, {The relativistic mean-field description of nuclei and nuclear
  dynamics}, Rep. Prog. Phys. 52 (1989) 439--514.
\newblock \href {http://dx.doi.org/10.1088/0034-4885/52/4/002}
  {\path{doi:10.1088/0034-4885/52/4/002}}.

\bibitem{Savushkin2004}
L.~N. Savushkin, H.~Toki, The Atomic Nucleus as a Relativistic System, Springer
  Verlag, Berlin, 2004.

\bibitem{Ring1996_PPNP37-193}
P.~Ring, Relativistic mean field theory in finite nuclei, Prog. Part. Nucl.
  Phys. 37 (1996) 193--263.
\newblock \href {http://dx.doi.org/10.1016/0146-6410(96)00054-3}
  {\path{doi:10.1016/0146-6410(96)00054-3}}.

\bibitem{Reinhard1986_ZPA323-13}
P.-G. Reinhard, M.~Rufa, J.~Maruhn, W.~Greiner, J.~Friedrich, Nuclear ground
  state properties in a relativistic meson field model, Z. Phys. A 323 (1986)
  13.
\newblock \href {http://dx.doi.org/10.1007/BF01294551}
  {\path{doi:10.1007/BF01294551}}.

\bibitem{Sharma1993_PLB312-377}
M.~M. Sharma, M.~A. Nagarajan, P.~Ring, Rho meson coupling in the relativistic
  mean field theory and description of exotic nuclei, Phys. Lett. B 312 (1993)
  377--381.
\newblock \href {http://dx.doi.org/10.1016/0370-2693(93)90970-S}
  {\path{doi:10.1016/0370-2693(93)90970-S}}.

\bibitem{Lalazissis1997_PRC55-540}
G.~A. Lalazissis, J.~Konig, P.~Ring, New parametrization for the {Lagrangian}
  density of relativistic mean field theory, Phys. Rev. C 55 (1997) 540--543.
\newblock \href {http://dx.doi.org/10.1103/PhysRevC.55.540}
  {\path{doi:10.1103/PhysRevC.55.540}}.

\bibitem{Lalazissis2009_PLB671-36}
G.~A. Lalazissis, S.~Karatzikos, R.~Fossion, D.~Pe{\~n}a~Arteaga, A.~V.
  Afanasjev, P.~Ring, The effective force nl3 revisited, Phys. Lett. B 671
  (2009) 36.
\newblock \href {http://dx.doi.org/10.1016/j.physletb.2008.11.070}
  {\path{doi:10.1016/j.physletb.2008.11.070}}.

\bibitem{Sugahara1994_NPA579-557}
Y.~Sugahara, H.~Toki, Relativistic mean-field theory for unstable nuclei with
  non-linear $\sigma$ and $\omega$ terms, Nucl. Phys. A 579 (1994) 557--572.
\newblock \href {http://dx.doi.org/10.1016/0375-9474(94)90923-7}
  {\path{doi:10.1016/0375-9474(94)90923-7}}.

\bibitem{Long2004_PRC69-034319}
W.~H. Long, J.~Meng, N.~Van~Giai, S.~G. Zhou, New effective interactions in
  relativistic mean field theory with nonlinear terms and density-dependent
  meson-nucleon coupling, Phys. Rev. C 69 (2004) 034319.
\newblock \href {http://dx.doi.org/10.1103/PhysRevC.69.034319}
  {\path{doi:10.1103/PhysRevC.69.034319}}.

\bibitem{Serot1992_RPP55-1855}
B.~D. Serot, Quantum hadrodynamics, Rep. Prog. Phys. 55~(11) (1992) 1855.
\newblock \href {http://dx.doi.org/10.1088/0034-4885/55/11/001}
  {\path{doi:10.1088/0034-4885/55/11/001}}.

\bibitem{Serot1997_IJMPE6-515}
B.~D. Serot, J.~D. Walecka, Recent progress in quantum hadro dynamics, Int. J.
  Mod. Phys. E 6 (1997) 515--631.
\newblock \href {http://dx.doi.org/10.1142/S0218301397000299}
  {\path{doi:10.1142/S0218301397000299}}.

\bibitem{Furnstahl1989_PRC40-321}
R.~J. Furnstahl, R.~J. Perry, B.~D. Serot, Two-loop corrections for nuclear
  matter in the walecka model, Phys. Rev. C 40 (1989) 321--353.
\newblock \href {http://dx.doi.org/10.1103/PhysRevC.40.321}
  {\path{doi:10.1103/PhysRevC.40.321}}.

\bibitem{Brockmann1992_PRL68-3408}
R.~Brockmann, H.~Toki, {Relativistic density-dependent Hartree approach for
  finite nuclei}, Phys. Rev. Lett. 68 (1992) 3408--3411.
\newblock \href {http://dx.doi.org/10.1103/PhysRevLett.68.3408}
  {\path{doi:10.1103/PhysRevLett.68.3408}}.

\bibitem{Typel1999_NPA656-331}
S.~Typel, H.~H. Wolter, Relativistic mean field calculations with
  density-dependent meson-nucleon coupling, Nucl. Phys. A 656 (1999) 331--364.
\newblock \href {http://dx.doi.org/10.1016/S0375-9474(99)00310-3}
  {\path{doi:10.1016/S0375-9474(99)00310-3}}.

\bibitem{Niksic2002_PRC66-024306}
T.~Nik\v{s}i\'{c}, D.~Vretenar, P.~Finelli, P.~Ring, Relativistic
  {Hartree-Bogoliubov} model with density-dependent meson-nucleon couplings,
  Phys. Rev. C 66 (2002) 024306.
\newblock \href {http://dx.doi.org/10.1103/PhysRevC.66.024306}
  {\path{doi:10.1103/PhysRevC.66.024306}}.

\bibitem{Lalazissis2005_PRC71-024312}
G.~A. Lalazissis, T.~Nik\v{s}i\'{c}, D.~Vretenar, P.~Ring, New relativistic
  mean-field interaction with density-dependent meson-nucleon couplings, Phys.
  Rev. C 71 (2005) 024312.
\newblock \href {http://dx.doi.org/10.1103/PhysRevC.71.024312}
  {\path{doi:10.1103/PhysRevC.71.024312}}.

\bibitem{Nikolaus1992_PRC46-1757}
B.~A. Nikolaus, T.~Hoch, D.~G. Madland, Nuclear ground state properties in a
  relativistic point coupling model, Phys. Rev. C 46 (1992) 1757--1781.
\newblock \href {http://dx.doi.org/10.1103/PhysRevC.46.1757}
  {\path{doi:10.1103/PhysRevC.46.1757}}.

\bibitem{Burvenich2002_PRC65-044308}
T.~B\"urvenich, D.~G. Madland, J.~A. Maruhn, P.~G. Reinhard, Nuclear ground
  state observables and {QCD} scaling in a refined relativistic point coupling
  model, Phys. Rev. C 65 (2002) 044308.
\newblock \href {http://dx.doi.org/10.1103/PhysRevC.65.044308}
  {\path{doi:10.1103/PhysRevC.65.044308}}.

\bibitem{Zhao2010_PRC82-054319}
P.~W. Zhao, Z.~P. Li, J.~M. Yao, J.~Meng, New parametrization for the nuclear
  covariant energy density functional with a point-coupling interaction, Phys.
  Rev. C 82 (2010) 054319.
\newblock \href {http://dx.doi.org/10.1103/PhysRevC.82.054319}
  {\path{doi:10.1103/PhysRevC.82.054319}}.

\bibitem{Niksic2008_PRC78-034318}
T.~Nik\v{s}i\'{c}, D.~Vretenar, P.~Ring, Relativistic nuclear energy density
  functionals: Adjusting parameters to binding energies, Phys. Rev. C 78 (2008)
  034318.
\newblock \href {http://dx.doi.org/10.1103/PhysRevC.78.034318}
  {\path{doi:10.1103/PhysRevC.78.034318}}.

\bibitem{Meng2006_PPNP57-470}
J.~Meng, H.~Toki, S.~G. Zhou, S.~Q. Zhang, W.~H. Long, L.~S. Geng, Relativistic
  continuum {Hartree} {Bogoliubov} theory for ground-state properties of exotic
  nuclei, Prog. Part. Nucl. Phys. 57 (2006) 470--563.
\newblock \href {http://dx.doi.org/10.1016/j.ppnp.2005.06.001}
  {\path{doi:10.1016/j.ppnp.2005.06.001}}.

\bibitem{Paar2007_RPP70-691}
N.~Paar, D.~Vretenar, E.~Khan, G.~Col\`o, Exotic modes of excitation in atomic
  nuclei far from stability, Rep. Prog. Phys. 70 (2007) 691.
\newblock \href {http://dx.doi.org/doi:10.1088/0034-4885/70/5/R02}
  {\path{doi:doi:10.1088/0034-4885/70/5/R02}}.

\bibitem{Ring2009_PAN72-1285}
P.~Ring, E.~Litvinova, Particle vibrational coupling in covariant density
  functional theory., Phys. At. Nucl. 72 (2009) 1285--1304, translated from
  Yadernaya Fisika {\bf 72} (2009) No. 8, 1.
\newblock \href {http://dx.doi.org/10.1134/S1063778809080055}
  {\path{doi:10.1134/S1063778809080055}}.

\bibitem{Niksic2011_PPNP66-519}
T.~Nik\v{s}i\'{c}, D.~Vretenar, P.~Ring, Relativistic nuclear energy density
  functionals: Mean-field and beyond, Prog. Part. Nucl. Phys. 66 (2011)
  519--548.
\newblock \href {http://dx.doi.org/10.1016/j.ppnp.2011.01.055}
  {\path{doi:10.1016/j.ppnp.2011.01.055}}.

\bibitem{Meng2013_FPC8-55}
J.~Meng, J.~Peng, S.~Q. Zhang, P.~W. Zhao, Progress on tilted axis cranking
  covariant density functional theory for nuclear magnetic and antimagnetic
  rotation, Front. Phys. 8 (2013) 55--79.
\newblock \href {http://dx.doi.org/10.1007/s11467-013-0287-y}
  {\path{doi:10.1007/s11467-013-0287-y}}.

\bibitem{Liang2015_PR570-1}
H.~Liang, J.~Meng, S.~G. Zhou, Hidden pseudospin and spin symmetries and their
  origins in atomic nuclei, Phys. Rep. 570 (2015) 1--84.
\newblock \href {http://dx.doi.org/10.1016/j.physrep.2014.12.005}
  {\path{doi:10.1016/j.physrep.2014.12.005}}.

\bibitem{Meng2015_JPG42-093101}
J.~Meng, S.~G. Zhou, Halos in medium-heavy and heavy nuclei with covariant
  density functional theory in continuum, J. Phys. G: Nucl. Part. Phys. 42
  (2015) 093101.
\newblock \href {http://dx.doi.org/10.1088/0954-3899/42/9/093101}
  {\path{doi:10.1088/0954-3899/42/9/093101}}.

\bibitem{Otsuka2005_PRL95-232502}
T.~Otsuka, T.~Suzuki, R.~Fujimoto, H.~Grawe, Y.~Akaishi, Evolution of nuclear
  shells due to the tensor force, Phys. Rev. Lett. 95 (2005) 232502.
\newblock \href {http://dx.doi.org/10.1103/PhysRevLett.95.232502}
  {\path{doi:10.1103/PhysRevLett.95.232502}}.

\bibitem{Agbemava2014_PRC89-054320}
S.~E. Agbemava, A.~V. Afanasjev, D.~Ray, P.~Ring, Global performance of
  covariant energy density functionals: ground state observables of even-even
  nuclei and the estimate of theoretical uncertainties, Phys. Rev. C 89 (2014)
  054320.
\newblock \href {http://dx.doi.org/10.1103/PhysRevC.89.054320}
  {\path{doi:10.1103/PhysRevC.89.054320}}.

\bibitem{Agbemava2019_PRC99-014318}
S.~E. Agbemava, A.~V. Afanasjev, A.~Taninah, The propagation of statistical
  errors in covariant density functional theory: ground state observables and
  single-particle properties, Phys. Rev. C 99 (2019) 014318.
\newblock \href {http://dx.doi.org/10.1103/PhysRevC.99.014318}
  {\path{doi:10.1103/PhysRevC.99.014318}}.

\bibitem{Afanasjev2015_PRC91-014324}
A.~V. Afanasjev, S.~E. Agbemava, D.~Ray, P.~Ring, Neutron dripline: single
  particle degrees of freedom and pairing as the sources of theoretical
  uncertainties, Phys. Rev. C 91 (2015) 014324.
\newblock \href {http://dx.doi.org/10.1103/PhysRevC.91.014324}
  {\path{doi:10.1103/PhysRevC.91.014324}}.

\bibitem{Sagawa2014_PPNP76-76}
H.~Sagawa, G.~Col\`o, Tensor interaction in mean-field and density functional
  theory approaches to nuclear structure, Prog. Part. Nucl. Phys. 76 (2014)
  76--115.
\newblock \href {http://dx.doi.org/10.1016/j.ppnp.2014.01.006}
  {\path{doi:10.1016/j.ppnp.2014.01.006}}.

\bibitem{Bogner2013_CPC184-2235}
S.~Bogner, A.~Bulgac, J.~Carlson, J.~Engel, G.~Fann, R.~Furnstahl, S.~Gandolfi,
  G.~Hagen, M.~Horoi, C.~Johnson, M.~Kortelainen, E.~Lusk, P.~Maris, H.~Nam,
  P.~Navratil, W.~Nazarewicz, E.~Ng, G.~Nobre, E.~Ormand, T.~Papenbrock,
  J.~Pei, S.~Pieper, S.~Quaglioni, K.~Roche, J.~Sarich, N.~Schunck,
  M.~Sosonkina, J.~Terasaki, I.~Thompson, J.~Vary, S.~Wild, {Computational
  nuclear quantum many-body problem: The UNEDF project}, Comput. Phys. Commun.
  184 (2013) 2235--2250.
\newblock \href {http://dx.doi.org/10.1016/j.cpc.2013.05.020}
  {\path{doi:10.1016/j.cpc.2013.05.020}}.

\bibitem{Day1967_RMP39-719}
B.~D. Day, {Elements of the Brueckner-Goldstone Theory of Nuclear Matter}, Rev.
  Mod. Phys. 39 (1967) 719--744.
\newblock \href {http://dx.doi.org/10.1103/RevModPhys.39.719}
  {\path{doi:10.1103/RevModPhys.39.719}}.

\bibitem{Lippmann1950_PR79-469}
B.~A. Lippmann, J.~Schwinger, {Variational Principles for Scattering Processes.
  I}, Phys. Rev. 79 (1950) 469--480.
\newblock \href {http://dx.doi.org/10.1103/PhysRev.79.469}
  {\path{doi:10.1103/PhysRev.79.469}}.

\bibitem{Watson1953_PR89-575}
K.~Watson, {Multiple Scattering and the Many-Body Problem---Applications to
  Photomeson Production in Complex Nuclei}, Phys. Rev. 89 (1953) 575--587.
\newblock \href {http://dx.doi.org/10.1103/PhysRev.89.575}
  {\path{doi:10.1103/PhysRev.89.575}}.

\bibitem{Bethe1956_PR103-1353}
H.~Bethe, {Nuclear Many-Body Problem}, Phys. Rev. 103 (1956) 1353--1390.
\newblock \href {http://dx.doi.org/10.1103/PhysRev.103.1353}
  {\path{doi:10.1103/PhysRev.103.1353}}.

\bibitem{Goldstone1957_PRSA239-267}
J.~Goldstone, {Derivation of the Brueckner Many-Body Theory}, Proc. R. Soc. A
  239 (1957) 267--279.
\newblock \href {http://dx.doi.org/10.1098/rspa.1957.0037}
  {\path{doi:10.1098/rspa.1957.0037}}.

\bibitem{Rajaraman1967_RMP39-745}
R.~Rajaraman, H.~Bethe, {Three-Body Problem in Nuclear Matter}, Rev. Mod. Phys.
  39 (1967) 745--770.
\newblock \href {http://dx.doi.org/10.1103/RevModPhys.39.745}
  {\path{doi:10.1103/RevModPhys.39.745}}.

\bibitem{Baranger1969_Varenna40-511}
M.~Baranger, {Recent Progress in the Understanding of Finite Nuclei from the
  Two-Nucleon Interaction}, in: M.~Jean (Ed.), Nuclear Structure and Nuclear
  Reactions, Proceedings of the International School of Physics "Enrico Fermi",
  Course XL, Academic Press, New York, 1969, pp. 511--614.

\bibitem{Brueckner1956_PR103-1008}
K.~A. Brueckner, W.~Wada, {Nuclear Saturation and Two-Body Forces:
  Self-Consistent Solutions and the Effects of the Exclusion Principle}, Phys.
  Rev. 103 (1956) 1008--1016.
\newblock \href {http://dx.doi.org/10.1103/PhysRev.103.1008}
  {\path{doi:10.1103/PhysRev.103.1008}}.

\bibitem{Bethe1957_PRSA238-551}
H.~A. Bethe, J.~Goldstone, {Effect of a Repulsive Core in the Theory of Complex
  Nuclei}, Proc. R. Soc. A 238 (1957) 551--567.
\newblock \href {http://dx.doi.org/10.1098/rspa.1957.0017}
  {\path{doi:10.1098/rspa.1957.0017}}.

\bibitem{Bethe1963_PR129-225}
H.~Bethe, B.~Brandow, A.~Petschek, {Reference Spectrum Method for Nuclear
  Matter}, Phys. Rev. 129 (1963) 225--264.
\newblock \href {http://dx.doi.org/10.1103/PhysRev.129.225}
  {\path{doi:10.1103/PhysRev.129.225}}.

\bibitem{Moszkowski1960_APNY11-65}
S.~Moszkowski, B.~Scott, {Nuclear forces and the properties of nuclear matter},
  Ann. Phys. (NY) 11 (1960) 65--115.
\newblock \href {http://dx.doi.org/10.1016/0003-4916(60)90128-7}
  {\path{doi:10.1016/0003-4916(60)90128-7}}.

\bibitem{Brown1969_NPA133-481}
G.~Brown, A.~Jackson, T.~Kuo, {Nucleon-nucleon potential and minimal
  relativity}, Nucl. Phys. A 133 (1969) 481--492.
\newblock \href {http://dx.doi.org/10.1016/0375-9474(69)90549-1}
  {\path{doi:10.1016/0375-9474(69)90549-1}}.

\bibitem{Haftel1970_NPA158-1}
M.~I. Haftel, F.~Tabakin, {Nuclear saturation and the smoothness of
  nucleon-nucleon potentials}, Nucl. Phys. A 158 (1970) 1--42.
\newblock \href {http://dx.doi.org/10.1016/0375-9474(70)90047-3}
  {\path{doi:10.1016/0375-9474(70)90047-3}}.

\bibitem{Ring1980}
P.~Ring, P.~Schuck, {The nuclear many-body problem}, Springer, 1980.

\bibitem{Coester1970_PRC1-769}
F.~Coester, S.~Cohen, B.~Day, C.~Vincent, {Variation in Nuclear-Matter Binding
  Energies with Phase-Shift-Equivalent Two-Body Potentials}, Phys. Rev. C 1
  (1970) 769--776.
\newblock \href {http://dx.doi.org/10.1103/PhysRevC.1.769}
  {\path{doi:10.1103/PhysRevC.1.769}}.

\bibitem{Day1978_RMP50-495}
B.~D. Day, {Current state of nuclear matter calculations}, Rev. Mod. Phys. 50
  (1978) 495--521.
\newblock \href {http://dx.doi.org/10.1103/RevModPhys.50.495}
  {\path{doi:10.1103/RevModPhys.50.495}}.

\bibitem{Primakoff1939_PR55-1218}
H.~Primakoff, T.~Holstein, Many-body interactions in atomic and nuclear
  systems, Phys. Rev. 55 (1939) 1218--1234.
\newblock \href {http://dx.doi.org/10.1103/PhysRev.55.1218}
  {\path{doi:10.1103/PhysRev.55.1218}}.

\bibitem{Drell1953_PR91-1527}
S.~D. Drell, K.~Huang, Many-body forces and nuclear saturation, Phys. Rev. 91
  (1953) 1527--1542.
\newblock \href {http://dx.doi.org/10.1103/PhysRev.91.1527}
  {\path{doi:10.1103/PhysRev.91.1527}}.

\bibitem{Fujita1957_PTP17-360}
J.~I. Fujita, H.~Miyazawa, {Pion Theory of Three-Body Forces}, Prog. Theor.
  Phys. 17 (1957) 360--365.
\newblock \href {http://dx.doi.org/10.1143/PTP.17.360}
  {\path{doi:10.1143/PTP.17.360}}.

\bibitem{Brown1969_NPA137-1}
G.~Brown, A.~Green, {Three-body forces in nuclear matter}, Nucl. Phys. A 137
  (1969) 1--19.
\newblock \href {http://dx.doi.org/10.1016/0375-9474(69)90068-2}
  {\path{doi:10.1016/0375-9474(69)90068-2}}.

\bibitem{Pieper2001_ARNPS51-53}
S.~C. Pieper, R.~B. Wiringa, {Quantum Monte Carlo Calculations of Light
  Nuclei}, Annu. Rev. Nucl. Part. Sci. 51 (2001) 53--90.
\newblock \href {http://dx.doi.org/10.1146/annurev.nucl.51.101701.132506}
  {\path{doi:10.1146/annurev.nucl.51.101701.132506}}.

\bibitem{Lejeune1986_NPA453-189}
A.~Lejeune, P.~Grange, M.~Martzolff, J.~Cugnon, {Hot nuclear matter in an
  extended Brueckner approach}, Nucl. Phys. A 453 (1986) 189--219.
\newblock \href {http://dx.doi.org/10.1016/0375-9474(86)90010-2}
  {\path{doi:10.1016/0375-9474(86)90010-2}}.

\bibitem{Baldo1997_AA328-274}
M.~Baldo, I.~Bombaci, G.~F. Burgio, {Microscopic Nuclear Equation of State with
  Three-Body Forces and Neutron Star Structure}, Astron. Astrophys. 328 (1997)
  274--282.

\bibitem{Zuo2002_NPA706-418}
W.~Zuo, A.~Lejeune, U.~Lombardo, J.~Mathiot, {Interplay of three-body
  interactions in the EOS of nuclear matter}, Nucl. Phys. A 706 (2002)
  418--430.
\newblock \href {http://dx.doi.org/10.1016/S0375-9474(02)00750-9}
  {\path{doi:10.1016/S0375-9474(02)00750-9}}.

\bibitem{Anastasio1980_PRL45-2096}
M.~Anastasio, L.~Celenza, C.~Shakin, {Nuclear Saturation as a Relativistic
  Effect}, Phys. Rev. Lett. 45 (1980) 2096--2099.
\newblock \href {http://dx.doi.org/10.1103/PhysRevLett.45.2096}
  {\path{doi:10.1103/PhysRevLett.45.2096}}.

\bibitem{Horowitz1984_PLB137-287}
C.~Horowitz, B.~D. Serot, {Two-nucleon correlations in a relativistic theory of
  nuclear matter}, Phys. Lett. B 137~(5-6) (1984) 287--293.
\newblock \href {http://dx.doi.org/10.1016/0370-2693(84)91717-9}
  {\path{doi:10.1016/0370-2693(84)91717-9}}.

\bibitem{Horowitz1987_NPA464-613}
C.~Horowitz, B.~D. Serot, {The relativistic two-nucleon problem in nuclear
  matter}, Nucl. Phys. A 464 (1987) 613--699.
\newblock \href {http://dx.doi.org/10.1016/0375-9474(87)90370-8}
  {\path{doi:10.1016/0375-9474(87)90370-8}}.

\bibitem{Brockmann1984_PLB149-283}
R.~Brockmann, R.~Machleidt, {Nuclear saturation in a relativistic
  Brueckner-Hartree-Fock approach}, Phys. Lett. B 149 (1984) 283--287.
\newblock \href {http://dx.doi.org/10.1016/0370-2693(84)90407-6}
  {\path{doi:10.1016/0370-2693(84)90407-6}}.

\bibitem{Brockmann1990_PRC42-1965}
R.~Brockmann, R.~Machleidt, {Relativistic nuclear structure. I. Nuclear
  matter}, Phys. Rev. C 42 (1990) 1965--1980.
\newblock \href {http://dx.doi.org/10.1103/PhysRevC.42.1965}
  {\path{doi:10.1103/PhysRevC.42.1965}}.

\bibitem{terHaar1986_PRL56-1237}
B.~ter Haar, R.~Malfliet, {Equation of State of Nuclear Matter in the
  Relativistic Dirac-Brueckner Approach}, Phys. Rev. Lett. 56 (1986)
  1237--1240.
\newblock \href {http://dx.doi.org/10.1103/PhysRevLett.56.1237}
  {\path{doi:10.1103/PhysRevLett.56.1237}}.

\bibitem{terHaar1987_PR149-207}
B.~ter Haar, R.~Malfliet, {Nucleons, mesons and deltas in nuclear matter a
  relativistic Dirac-Brueckner approach}, Phys. Rep. 149 (1987) 207--286.
\newblock \href {http://dx.doi.org/10.1016/0370-1573(87)90085-8}
  {\path{doi:10.1016/0370-1573(87)90085-8}}.

\bibitem{Anastasio1983_PR100-327}
M.~Anastasio, L.~Celenza, W.~Pong, C.~Shakin, {Relativistic nuclear structure
  physics}, Phys. Rep. 100 (1983) 327--392.
\newblock \href {http://dx.doi.org/10.1016/0370-1573(83)90060-1}
  {\path{doi:10.1016/0370-1573(83)90060-1}}.

\bibitem{Celenza1986_WSLN2}
L.~S. Celenza, C.~M. Shakin, Relativistic Nuclear Physics: Theories of
  Structure and Scattering, Vol.~2 of World Scientific Lecture Notes, World
  Scientific Singapore, 1986.

\bibitem{Machleidt1989_ANP19-189}
R.~Machleidt, {The meson theory of nuclear forces and nuclear structure}, Adv.
  Nucl. Phys. 19 (1989) 189.
\newblock \href {http://dx.doi.org/10.1007/978-1-4613-9907-0}
  {\path{doi:10.1007/978-1-4613-9907-0}}.

\bibitem{terHaar1987_PRL59-1652}
B.~ter Haar, R.~Malfliet, {Equation of state of dense asymmetric nuclear
  matter}, Phys. Rev. Lett. 59 (1987) 1652--1655.
\newblock \href {http://dx.doi.org/10.1103/PhysRevLett.59.1652}
  {\path{doi:10.1103/PhysRevLett.59.1652}}.

\bibitem{Nuppenau1990_NPA511-525}
C.~Nuppenau, A.~Mackellar, Y.~Lee, {Study of nucleon-nucleus scattering based
  on the relativistic Brueckner-Bethe-Goldstone equation}, Nucl. Phys. A 511
  (1990) 525--540.
\newblock \href {http://dx.doi.org/10.1016/0375-9474(90)90108-X}
  {\path{doi:10.1016/0375-9474(90)90108-X}}.

\bibitem{Huber1996_NPA596-684}
H.~Huber, F.~Weber, M.~Weigel, {Relativisitic calculations of neutron star
  matter}, Nucl. Phys. A 596 (1996) 684--699.
\newblock \href {http://dx.doi.org/10.1016/0375-9474(95)00393-2}
  {\path{doi:10.1016/0375-9474(95)00393-2}}.

\bibitem{Katayama2015_PLB747-43}
T.~Katayama, K.~Saito, {Hyperons in neutron stars}, Phys. Lett. B 747 (2015)
  43--47.
\newblock \href {http://dx.doi.org/10.1016/j.physletb.2015.03.039}
  {\path{doi:10.1016/j.physletb.2015.03.039}}.

\bibitem{VanGiai2010_JPG37-064043}
N.~{Van Giai}, B.~V. Carlson, Z.~Ma, H.~Wolter, {The
  Dirac-Brueckner-Hartree-Fock approach: from infinite matter to effective
  Lagrangians for finite systems}, J. Phys. G 37 (2010) 064043.
\newblock \href {http://dx.doi.org/10.1088/0954-3899/37/6/064043}
  {\path{doi:10.1088/0954-3899/37/6/064043}}.

\bibitem{Sammarruca2010_IJMPE19-1259}
F.~Sammarruca, The microscopic approach to nuclear matter and neutron star
  matter, Int. J. Mod. Phys. E 19~(07) (2010) 1259--1313.
\newblock \href {http://dx.doi.org/10.1142/S0218301310015874}
  {\path{doi:10.1142/S0218301310015874}}.

\bibitem{VanDalen2010_IJMPE19-2077}
E.~N.~E. {van Dalen}, H.~M{\"{u}}ther, {Relativistic Effects in Nuclear Matter
  and Nuclei}, Int. J. Mod. Phys. E 19 (2010) 2077--2122.
\newblock \href {http://dx.doi.org/10.1142/S0218301310016533}
  {\path{doi:10.1142/S0218301310016533}}.

\bibitem{Muther2017_IJMPE26-1730001}
H.~M{\"{u}}ther, F.~Sammarruca, Z.~Ma, {Relativistic effects and three-nucleon
  forces in nuclear matter and nuclei}, Int. J. Mod. Phys. E 26 (2017) 1730001.
\newblock \href {http://dx.doi.org/10.1142/S0218301317300016}
  {\path{doi:10.1142/S0218301317300016}}.

\bibitem{Lenske2018_PPNP98-119}
H.~Lenske, M.~Dhar, T.~Gaitanos, X.~Cao, {Baryons and baryon resonances in
  nuclear matter}, Prog. Part. Nucl. Phys. 98 (2018) 119--206.
\newblock \href {http://dx.doi.org/10.1016/j.ppnp.2017.09.001}
  {\path{doi:10.1016/j.ppnp.2017.09.001}}.

\bibitem{Brown1986_AIPC142-2}
G.~E. Brown, {Relativistic effects in nuclear physics}, in: AIP Conf. Proc.,
  Vol. 142, AIP, 1986, pp. 2--26.
\newblock \href {http://dx.doi.org/10.1063/1.35633}
  {\path{doi:10.1063/1.35633}}.

\bibitem{Brown1987_CNPP17-39}
G.~E. Brown, W.~Weise, G.~Baym, J.~Speth, {Relativistic Effects in Nuclear
  Physics}, Comments Nucl. Part. Phys. 17 (1987) 39.

\bibitem{Brown1987_PS36-209}
G.~E. Brown, {Relativistic effects in nuclear physics}, Phys. Scr. 36 (1987)
  209--213.
\newblock \href {http://dx.doi.org/10.1088/0031-8949/36/2/003}
  {\path{doi:10.1088/0031-8949/36/2/003}}.

\bibitem{Sammarruca2012_PRC86-054317}
F.~Sammarruca, B.~Chen, L.~Coraggio, N.~Itaco, R.~Machleidt,
  {Dirac-Brueckner-Hartree-Fock versus chiral effective field theory}, Phys.
  Rev. C 86~(5) (2012) 054317.
\newblock \href {http://dx.doi.org/10.1103/PhysRevC.86.054317}
  {\path{doi:10.1103/PhysRevC.86.054317}}.

\bibitem{Typel2010_PRC81-015803}
S.~Typel, G.~R{\"{o}}pke, T.~Kl{\"{a}}hn, D.~Blaschke, H.~H. Wolter,
  {Composition and thermodynamics of nuclear matter with light clusters}, Phys.
  Rev. C 81 (2010) 015803.
\newblock \href {http://dx.doi.org/10.1103/PhysRevC.81.015803}
  {\path{doi:10.1103/PhysRevC.81.015803}}.

\bibitem{Fuchs2004_LNP641-111}
C.~Fuchs, The relativistic dirac-brueckner approach to nuclear matter, in:
  G.~A. Lalazissis, P.~Ring, D.~Vretenar (Eds.), Extended Density Functionals
  in Nuclear Structure Physics, Vol. 641, Springer-Verlag, Heidelberg, 2004, p.
  111.

\bibitem{Ulrych1997_PRC56-1788}
S.~Ulrych, H.~M{\"{u}}ther, {Relativistic structure of the nucleon self-energy
  in asymmetric nuclei}, Phys. Rev. C 56 (1997) 1788--1794.
\newblock \href {http://dx.doi.org/10.1103/PhysRevC.56.1788}
  {\path{doi:10.1103/PhysRevC.56.1788}}.

\bibitem{Shen1997_PRC55-1211}
H.~Shen, Y.~Sugahara, H.~Toki, {Relativistic mean field approach with density
  dependent couplings for finite nuclei}, Phys. Rev. C 55 (1997) 1211--1217.
\newblock \href {http://dx.doi.org/10.1103/PhysRevC.55.1211}
  {\path{doi:10.1103/PhysRevC.55.1211}}.

\bibitem{Schiller2001_EPJA11-15}
E.~Schiller, H.~M{\"{u}}ther, {Correlations and the Dirac structure of the
  nucleon self-energy}, Eur. Phys. J. A 11 (2001) 15--24.
\newblock \href {http://dx.doi.org/10.1007/s100500170092}
  {\path{doi:10.1007/s100500170092}}.

\bibitem{VanDalen2004_NPA744-227}
E.~van Dalen, C.~Fuchs, A.~Faessler, {The relativistic Dirac-Brueckner approach
  to asymmetric nuclear matter}, Nucl. Phys. A 744 (2004) 227--248.
\newblock \href {http://dx.doi.org/10.1016/j.nuclphysa.2004.08.019}
  {\path{doi:10.1016/j.nuclphysa.2004.08.019}}.

\bibitem{VanDalen2005_PRC72-065803}
E.~N.~E. van Dalen, C.~Fuchs, A.~Faessler, {Momentum, density, and isospin
  dependence of symmetric and asymmetric nuclear matter properties}, Phys. Rev.
  C 72 (2005) 065803.
\newblock \href {http://dx.doi.org/10.1103/PhysRevC.72.065803}
  {\path{doi:10.1103/PhysRevC.72.065803}}.

\bibitem{VanDalen2007_EPJA31-29}
E.~N.~E. van Dalen, C.~Fuchs, A.~Faessler, Dirac-brueckner-hartree-fock
  calculations for isospin asymmetric nuclear matter based on improved
  approximation schemes, Eur. Phys. J. A 31~(1) (2007) 29--42.
\newblock \href {http://dx.doi.org/10.1140/epja/i2006-10165-x}
  {\path{doi:10.1140/epja/i2006-10165-x}}.

\bibitem{Boersma1994_PRC49-233}
H.~F. Boersma, R.~Malfliet, {From nuclear matter to finite nuclei. I.
  Parametrization of the Dirac-Brueckner $G$ matrix}, Phys. Rev. C 49 (1994)
  233--244.
\newblock \href {http://dx.doi.org/10.1103/PhysRevC.49.233}
  {\path{doi:10.1103/PhysRevC.49.233}}.

\bibitem{deJong1998_PRC58-890}
F.~de~Jong, H.~Lenske, {Relativistic Brueckner-Hartree-Fock calculations with
  explicit intermediate negative energy states}, Phys. Rev. C 58~(2) (1998)
  890--899.
\newblock \href {http://dx.doi.org/10.1103/PhysRevC.58.890}
  {\path{doi:10.1103/PhysRevC.58.890}}.

\bibitem{Huber1995_PRC51-1790}
H.~Huber, F.~Weber, M.~K. Weigel, {Symmetric and asymmetric nuclear matter in
  the relativistic approach}, Phys. Rev. C 51 (1995) 1790--1799.
\newblock \href {http://dx.doi.org/10.1103/PhysRevC.51.1790}
  {\path{doi:10.1103/PhysRevC.51.1790}}.

\bibitem{Muther1988_PLB202-483}
H.~M{\"{u}}ther, R.~Machleidt, R.~Brockmann, {Dirac-Brueckner-Hartree-Fock
  approach in finite nuclei}, Phys. Lett. B 202 (1988) 483--488.
\newblock \href {http://dx.doi.org/10.1016/0370-2693(88)91848-5}
  {\path{doi:10.1016/0370-2693(88)91848-5}}.

\bibitem{Muther1990_PRC42-1981}
H.~M{\"{u}}ther, R.~Machleidt, R.~Brockmann, {Relativistic nuclear structure.
  II. Finite nuclei}, Phys. Rev. C 42 (1990) 1981--1988.
\newblock \href {http://dx.doi.org/10.1103/PhysRevC.42.1981}
  {\path{doi:10.1103/PhysRevC.42.1981}}.

\bibitem{Muther1992-CompNuclPhys2-43}
H.~M\"uther, P.~U. Sauer, The G-Matrix in Finite Nuclei, Vol.~2, Springer
  Verlag, Heidelberg, 1992, pp. 30--54.

\bibitem{Fritz1993_PRL71-46}
R.~Fritz, H.~M{\"{u}}ther, R.~Machleidt, {Dirac effects in the Hartree-Fock
  description of finite nuclei employing realistic forces}, Phys. Rev. Lett. 71
  (1993) 46--49.
\newblock \href {http://dx.doi.org/10.1103/PhysRevLett.71.46}
  {\path{doi:10.1103/PhysRevLett.71.46}}.

\bibitem{Fritz1994_PRC49-633}
R.~Fritz, H.~M{\"{u}}ther, {NN correlations and relativistic Hartree-Fock in
  finite nuclei}, Phys. Rev. C 49 (1994) 633--644.
\newblock \href {http://dx.doi.org/10.1103/PhysRevC.49.633}
  {\path{doi:10.1103/PhysRevC.49.633}}.

\bibitem{Shi1995_PRC52-144}
H.~L. Shi, B.~Q. Chen, Z.~Y. Ma, {Relativistic density-dependent Hartree-Fock
  approach for finite nuclei}, Phys. Rev. C 52 (1995) 144--156.
\newblock \href {http://dx.doi.org/10.1103/PhysRevC.52.144}
  {\path{doi:10.1103/PhysRevC.52.144}}.

\bibitem{Fuchs1995_PRC52-3043}
C.~Fuchs, H.~Lenske, H.~H. Wolter, {Density dependent hadron field theory},
  Phys. Rev. C 52 (1995) 3043--3060.
\newblock \href {http://dx.doi.org/10.1103/PhysRevC.52.3043}
  {\path{doi:10.1103/PhysRevC.52.3043}}.

\bibitem{Hofmann2001_PRC64-034314}
F.~Hofmann, C.~M. Keil, H.~Lenske, {Density dependent hadron field theory for
  asymmetric nuclear matter and exotic nuclei}, Phys. Rev. C 64 (2001) 034314.
\newblock \href {http://dx.doi.org/10.1103/PhysRevC.64.034314}
  {\path{doi:10.1103/PhysRevC.64.034314}}.

\bibitem{Ma2002_PRC66-024321}
Z.~Y. Ma, L.~Liu, {Effective Dirac Brueckner-Hartree-Fock method for asymmetric
  nuclear matter and finite nuclei}, Phys. Rev. C 66 (2002) 024321.
\newblock \href {http://dx.doi.org/10.1103/PhysRevC.66.024321}
  {\path{doi:10.1103/PhysRevC.66.024321}}.

\bibitem{VanDalen2011_PRC84-024320}
E.~N.~E. van Dalen, H.~M{\"{u}}ther, {Relativistic description of finite nuclei
  based on realistic NN interactions}, Phys. Rev. C 84 (2011) 024320.
\newblock \href {http://dx.doi.org/10.1103/PhysRevC.84.024320}
  {\path{doi:10.1103/PhysRevC.84.024320}}.

\bibitem{Hohenberg1964_PR136-B864}
P.~Hohenberg, W.~Kohn, {Inhomogeneous Electron Gas}, Phys. Rev. 136~(3B) (1964)
  B864--B871.
\newblock \href {http://dx.doi.org/10.1103/PhysRev.136.B864}
  {\path{doi:10.1103/PhysRev.136.B864}}.

\bibitem{Kohn1965_PR140-A1133}
W.~Kohn, L.~J. Sham, {Self-Consistent Equations Including Exchange and
  Correlation Effects}, Phys. Rev. 140~(4A) (1965) A1133--A1138.
\newblock \href {http://dx.doi.org/10.1103/PhysRev.140.A1133}
  {\path{doi:10.1103/PhysRev.140.A1133}}.

\bibitem{Bouyssy1985_PRL55-1731}
A.~Bouyssy, S.~Marcos, J.~F. Mathiot, N.~Van~Giai, Isovector-meson
  contributions in the {Dirac-Hartree-Fock} approach to nuclear matter, Phys.
  Rev. Lett. 55 (1985) 1731--1733.
\newblock \href {http://dx.doi.org/10.1103/PhysRevLett.55.1731}
  {\path{doi:10.1103/PhysRevLett.55.1731}}.

\bibitem{Bouyssy1987_PRC36-380}
A.~Bouyssy, J.~F. Mathiot, N.~Van~Giai, S.~Marcos, Relativistic description of
  nuclear systems in the {Hartree-Fock} approximation, Phys. Rev. C 36 (1987)
  380--401.
\newblock \href {http://dx.doi.org/10.1103/PhysRevC.36.380}
  {\path{doi:10.1103/PhysRevC.36.380}}.

\bibitem{Bernardos1993_PRC48-2665}
P.~Bernardos, V.~N. Fomenko, N.~V. Giai, M.~L. Quelle, S.~Marcos, R.~Niembro,
  L.~N. Savushkin, Relativistic {H}artree-{F}ock approximation in a nonlinear
  model for nuclear matter and finite nuclei, Phys. Rev. C 48 (1993)
  2665--2672.
\newblock \href {http://dx.doi.org/10.1103/PhysRevC.48.2665}
  {\path{doi:10.1103/PhysRevC.48.2665}}.

\bibitem{Marcos2004_JPG30-703}
S.~Marcos, L.~N. Savushkin, V.~N. Fomenko, M.~López-Quelle, R.~Niembro,
  Description of nuclear systems within the relativistic {H}artree–{F}ock
  method with zero-range self-interactions of the scalar field, J. Phys. G 30
  (2004) 703.
\newblock \href {http://dx.doi.org/10.1088/0954-3899/30/6/002}
  {\path{doi:10.1088/0954-3899/30/6/002}}.

\bibitem{Long2006_PLB640-150}
W.~H. Long, N.~Van~Giai, J.~Meng, Density-dependent relativistic {Hartree-Fock}
  approach, Phys. Lett. B 640 (2006) 150--154.
\newblock \href {http://dx.doi.org/10.1016/j.physletb.2006.07.064}
  {\path{doi:10.1016/j.physletb.2006.07.064}}.

\bibitem{Long2007_PRC76-034314}
W.~H. Long, H.~Sagawa, N.~Van~Giai, J.~Meng, Shell structure and $\rho$-tensor
  correlations in density dependent relativistic {Hartree-Fock} theory, Phys.
  Rev. C 76 (2007) 034314.
\newblock \href {http://dx.doi.org/10.1103/PhysRevC.76.034314}
  {\path{doi:10.1103/PhysRevC.76.034314}}.

\bibitem{Long2010_PRC81-024308}
W.~H. Long, P.~Ring, N.~Van~Giai, J.~Meng, Relativistic
  {Hartree-Fock-Bogoliubov} theory with density dependent meson-nucleon
  couplings, Phys. Rev. C 81 (2010) 024308.
\newblock \href {http://dx.doi.org/10.1103/PhysRevC.81.024308}
  {\path{doi:10.1103/PhysRevC.81.024308}}.

\bibitem{Long2008_EPL82-12001}
W.~H. Long, H.~Sagawa, J.~Meng, N.~Van~Giai, Evolution of nuclear shell
  structure due to the pion exchange potential, Europhys. Lett. 82 (2008)
  12001.
\newblock \href {http://dx.doi.org/10.1209/0295-5075/82/12001}
  {\path{doi:10.1209/0295-5075/82/12001}}.

\bibitem{Long2009_PLB680-428}
W.~H. Long, T.~Nakatsukasa, H.~Sagawa, J.~Meng, H.~Nakada, Y.~Zhang, Non-local
  mean field effect on nuclei near {$Z=64$} sub-shell, Phys. Lett. B 680 (2009)
  428--431.
\newblock \href {http://dx.doi.org/10.1016/j.physletb.2009.09.034}
  {\path{doi:10.1016/j.physletb.2009.09.034}}.

\bibitem{Wang2013_PRC87-047301}
L.~J. Wang, J.~M. Dong, W.~H. Long, {Tensor effects on the evolution of the
  $N=40$ shell gap from nonrelativistic and relativistic mean-field theory},
  Phys. Rev. C 87~(4) (2013) 047301.
\newblock \href {http://dx.doi.org/10.1103/PhysRevC.87.047301}
  {\path{doi:10.1103/PhysRevC.87.047301}}.

\bibitem{Long2006_PLB639-242}
W.~H. Long, H.~Sagawa, J.~Meng, N.~Van~Giai, Pseudo-spin symmetry in
  density-dependent relativistic {Hartree-Fock} theory, Phys. Lett. B 639
  (2006) 242--247.
\newblock \href {http://dx.doi.org/10.1016/j.physletb.2006.05.065}
  {\path{doi:10.1016/j.physletb.2006.05.065}}.

\bibitem{Long2010_PRC81-031302}
W.~H. Long, P.~Ring, J.~Meng, N.~Van~Giai, C.~A. Bertulani, Nuclear halo
  structure and pseudospin symmetry, Phys. Rev. C 81 (2010) 031302(R).
\newblock \href {http://dx.doi.org/10.1103/PhysRevC.81.031302}
  {\path{doi:10.1103/PhysRevC.81.031302}}.

\bibitem{Geng2006_CPL23-1139}
L.~S. Geng, J.~Meng, H.~Toki, W.~H. Long, G.~Shen, Spurious shell closures in
  the relativistic mean field model, Chin. Phys. Lett. 23 (2006) 1139--1141.
\newblock \href {http://dx.doi.org/10.1088/0256-307X/23/5/021}
  {\path{doi:10.1088/0256-307X/23/5/021}}.

\bibitem{Jiang2015_PRC91-025802}
L.~J. Jiang, S.~Yang, J.~M. Dong, W.~H. Long, {Self-consistent tensor effects
  on nuclear matter systems within a relativistic Hartree-Fock approach}, Phys.
  Rev. C 91~(2) (2015) 025802.
\newblock \href {http://dx.doi.org/10.1103/PhysRevC.91.025802}
  {\path{doi:10.1103/PhysRevC.91.025802}}.

\bibitem{Jiang2015_PRC91-034326}
L.~J. Jiang, S.~Yang, B.~Y. Sun, W.~H. Long, H.~Q. Gu, {Nuclear tensor
  interaction in a covariant energy density functional}, Phys. Rev. C 91~(3)
  (2015) 034326.
\newblock \href {http://dx.doi.org/10.1103/PhysRevC.91.034326}
  {\path{doi:10.1103/PhysRevC.91.034326}}.

\bibitem{Wang2018_PRC98-034313}
Z.~Wang, Q.~Zhao, H.~Liang, W.~H. Long, {Quantitative analysis of tensor
  effects in the relativistic Hartree-Fock theory}, Phys. Rev. C 98~(3) (2018)
  034313.
\newblock \href {http://dx.doi.org/10.1103/PhysRevC.98.034313}
  {\path{doi:10.1103/PhysRevC.98.034313}}.

\bibitem{Zong2018_CPC42-024101}
Y.~Y. Zong, B.~Y. Sun, {Relativistic interpretation of the nature of the
  nuclear tensor force}, Chin. Phys. C 42~(2) (2018) 024101.
\newblock \href {http://dx.doi.org/10.1088/1674-1137/42/2/024101}
  {\path{doi:10.1088/1674-1137/42/2/024101}}.

\bibitem{Lu2013_PRC87-034311}
X.~L. Lu, B.~Y. Sun, W.~H. Long, Description of carbon isotopes within
  relativistic {H}artree-{F}ock-bogoliubov theory, Phys. Rev. C 87 (2013)
  034311.

\bibitem{Li2014_PLB732-169}
J.~J. Li, W.~H. Long, J.~Margueron, N.~Van~Giai, Superheavy magic structures in
  the relativistic {Hartree-Fock-Bogoliubov} approach, Phys. Lett. B 732 (2014)
  169--173.
\newblock \href
  {http://dx.doi.org/http://dx.doi.org/10.1016/j.physletb.2014.03.031}
  {\path{doi:http://dx.doi.org/10.1016/j.physletb.2014.03.031}}.

\bibitem{Sun2008_PRC78-065805}
B.~Y. Sun, W.~H. Long, J.~Meng, U.~Lombardo, Neutron star properties in
  density-dependent relativistic {Hartree-Fock} theory, Phys. Rev. C 78 (2008)
  065805.
\newblock \href {http://dx.doi.org/10.1103/PhysRevC.78.065805}
  {\path{doi:10.1103/PhysRevC.78.065805}}.

\bibitem{Long2012_PRC85-025806}
W.~H. Long, B.~Y. Sun, K.~Hagino, H.~Sagawa, Hyperon effects in covariant
  density functional theory and recent astrophysical observations, Phys. Rev. C
  85 (2012) 025806.

\bibitem{Li2018_EPJA54-133}
J.~J. Li, W.~H. Long, A.~Sedrakian, Hypernuclear stars from relativistic
  hartree-fock density functional theory, Eur. Phys. J. A 54 (2018) 133.

\bibitem{Liang2008_PRL101-122502}
H.~Liang, N.~Van~Giai, J.~Meng, Spin-isospin resonances: A self-consistent
  covariant description, Phys. Rev. Lett. 101 (2008) 122502.
\newblock \href {http://dx.doi.org/10.1103/PhysRevLett.101.122502}
  {\path{doi:10.1103/PhysRevLett.101.122502}}.

\bibitem{Liang2012_PRC85-064302}
H.~Liang, P.~Zhao, J.~Meng, Fine structure of charge-exchange spin-dipole
  excitations in $^{16}${O}, Phys. Rev. C 85 (2012) 064302.
\newblock \href {http://dx.doi.org/10.1103/PhysRevC.85.064302}
  {\path{doi:10.1103/PhysRevC.85.064302}}.

\bibitem{Niu2013_PLB723-172}
Z.~M. Niu, Y.~F. Niu, H.~Z. Liang, W.~H. Long, T.~Nik\v{s}i\'{c}, D.~Vretenar,
  J.~Meng, $\beta$-decay half-lives of neutron-rich nuclei and matter flow in
  the r-process, Phys. Lett. B 723 (2013) 172 -- 176.

\bibitem{Niu2017_PRC95-044301}
Z.~M. Niu, Y.~F. Niu, H.~Z. Liang, W.~H. Long, J.~Meng, Self-consistent
  relativistic quasiparticle random-phase approximation and its applications to
  charge-exchange excitations, Phys. Rev. C 95 (2017) 044031.

\bibitem{Ebran2011_PRC83-064323}
J.-P. Ebran, E.~Khan, D.~Pe\~na Arteaga, D.~Vretenar, Relativistic
  {Hartree-Fock-Bogoliubov} model for deformed nuclei, Phys. Rev. C 83 (2011)
  064323.
\newblock \href {http://dx.doi.org/10.1103/PhysRevC.83.064323}
  {\path{doi:10.1103/PhysRevC.83.064323}}.

\bibitem{Zhou2003_PRC68-034323}
S.~G. Zhou, J.~Meng, P.~Ring, Spherical relativistic {Hartree} theory in a
  {Woods-Saxon} basis, Phys. Rev. C 68 (2003) 034323.
\newblock \href {http://dx.doi.org/10.1103/PhysRevC.68.034323}
  {\path{doi:10.1103/PhysRevC.68.034323}}.

\bibitem{Schiller1999_PRC59-2934}
E.~Schiller, H.~M{\"{u}}ther, P.~Czerski, {Pauli exclusion operator and binding
  energy of nuclear matter}, Phys. Rev. C 59~(5) (1999) 2934--2936.
\newblock \href {http://dx.doi.org/10.1103/PhysRevC.59.2934}
  {\path{doi:10.1103/PhysRevC.59.2934}}.

\bibitem{Suzuki2000_NPA665-92}
K.~Suzuki, R.~Okamoto, M.~Kohno, S.~Nagata, {Exact treatment of the Pauli
  exclusion operator in nuclear matter calculation}, Nucl. Phys. A 665~(1-2)
  (2000) 92--104.
\newblock \href {http://dx.doi.org/10.1016/S0375-9474(99)00399-1}
  {\path{doi:10.1016/S0375-9474(99)00399-1}}.

\bibitem{Shen2018_PLB781-227}
S.~Shen, H.~Liang, J.~Meng, P.~Ring, S.~Zhang, {Spin symmetry in the Dirac sea
  derived from the bare nucleon–nucleon interaction}, Phys. Lett. B 781
  (2018) 227--231.
\newblock \href {http://dx.doi.org/10.1016/j.physletb.2018.03.080}
  {\path{doi:10.1016/j.physletb.2018.03.080}}.

\bibitem{Shen2018_PLB778-344}
S.~Shen, H.~Liang, J.~Meng, P.~Ring, S.~Zhang, {Effects of tensor forces in
  nuclear spin–orbit splittings from ab initio calculations}, Phys. Lett. B
  778 (2018) 344--348.
\newblock \href {http://dx.doi.org/10.1016/j.physletb.2018.01.058}
  {\path{doi:10.1016/j.physletb.2018.01.058}}.

\bibitem{Shen2018_PRC97-054312}
S.~Shen, H.~Liang, J.~Meng, P.~Ring, S.~Zhang, {Relativistic
  Brueckner-Hartree-Fock theory for neutron drops}, Phys. Rev. C 97~(5) (2018)
  054312.
\newblock \href {http://dx.doi.org/10.1103/PhysRevC.97.054312}
  {\path{doi:10.1103/PhysRevC.97.054312}}.

\bibitem{Shen2019_PRC99-034322}
S.~Shen, G.~Col{\`{o}}, X.~Roca-Maza, {Skyrme functional with tensor terms from
  \textit{ab initio} calculations of neutron-proton drops}, Phys. Rev. C 99
  (2019) 034322.
\newblock \href {http://dx.doi.org/10.1103/PhysRevC.99.034322}
  {\path{doi:10.1103/PhysRevC.99.034322}}.

\bibitem{Rajagopal1973_PRB7-1912}
A.~K. Rajagopal, J.~Callaway, Inhomogeneous electron gas, Phys. Rev. B 7~(5)
  (1973) 1912--1919.
\newblock \href {http://dx.doi.org/10.1103/PhysRevB.7.1912}
  {\path{doi:10.1103/PhysRevB.7.1912}}.

\bibitem{Rajagopal1978_JPC11-L943}
A.~K. Rajagopal, Inhomogeneous relativistic electron gas, J. Phys. C 11~(24)
  (1978) L943--L948.
\newblock \href {http://dx.doi.org/10.1088/0022-3719/11/24/002}
  {\path{doi:10.1088/0022-3719/11/24/002}}.

\bibitem{MacDonald1979_JPC12-2977}
A.~H. MacDonald, S.~H. Vosko, A relativistic density functional formalism, J.
  Phys. C 12~(15) (1979) 2977--2990.
\newblock \href {http://dx.doi.org/10.1088/0022-3719/12/15/007}
  {\path{doi:10.1088/0022-3719/12/15/007}}.

\bibitem{Gross1981_PLA81-447}
E.~K.~U. Gross, R.~M. Dreizler, Relativistic gradient expansion of the kinetic
  energy density, Phys. Lett. A 81~(8) (1981) 447 -- 450.
\newblock \href
  {http://dx.doi.org/https://doi.org/10.1016/0375-9601(81)90408-4}
  {\path{doi:https://doi.org/10.1016/0375-9601(81)90408-4}}.

\bibitem{Engel1995_PRA51-1159}
E.~Engel, \href{https://link.aps.org/doi/10.1103/PhysRevA.51.1159}{Alternative
  form of the linear-response contribution to the exchange-correlation energy
  functional}, Phys. Rev. A 51 (1995) 1159--1166.
\newblock \href {http://dx.doi.org/10.1103/PhysRevA.51.1159}
  {\path{doi:10.1103/PhysRevA.51.1159}}.
\newline\urlprefix\url{https://link.aps.org/doi/10.1103/PhysRevA.51.1159}

\bibitem{Engel1996_PRA53-1367}
E.~Engel, S.~Keller, R.~M. Dreizler, Generalized gradient approximation for the
  relativistic exchange-only energy functional, Phys. Rev. A 53 (1996)
  1367--1374.
\newblock \href {http://dx.doi.org/10.1103/PhysRevA.53.1367}
  {\path{doi:10.1103/PhysRevA.53.1367}}.

\bibitem{Shadwick1989_CPC54-95}
B.~Shadwick, J.~Talman, M.~Norman, A program to compute variationally optimized
  relativistic atomic potentials, Comp. Phys. Comm. 54~(1) (1989) 95 -- 102.
\newblock \href
  {http://dx.doi.org/https://doi.org/10.1016/0010-4655(89)90035-0}
  {\path{doi:https://doi.org/10.1016/0010-4655(89)90035-0}}.

\bibitem{Engel1995_PRA52-2750}
E.~Engel, S.~Keller, A.~F. Bonetti, H.~M\"uller, R.~M. Dreizler, Local and
  nonlocal relativistic exchange-correlation energy functionals: Comparison to
  relativistic optimized-potential-model results, Phys. Rev. A 52 (1995)
  2750--2764.
\newblock \href {http://dx.doi.org/10.1103/PhysRevA.52.2750}
  {\path{doi:10.1103/PhysRevA.52.2750}}.

\bibitem{Engel2011_DFT8-351}
E.~Engel, R.~M. Dreizler, Relativistic Density Functional Theory, Springer
  Berlin Heidelberg, Berlin, Heidelberg, 2011, Ch.~8, pp. 351--399.
\newblock \href {http://dx.doi.org/10.1007/978-3-642-14090-7_8}
  {\path{doi:10.1007/978-3-642-14090-7_8}}.

\bibitem{Giraud2008_PRC77-014311}
B.~G. Giraud, {Density functionals in the laboratory frame}, Phys. Rev. C
  77~(1) (2008) 014311.
\newblock \href {http://dx.doi.org/10.1103/PhysRevC.77.014311}
  {\path{doi:10.1103/PhysRevC.77.014311}}.

\bibitem{Nakatsukasa2012_PTEP01A207}
T.~Nakatsukasa, Density functional approaches to collective phenomena in
  nuclei: Time-dependent density functional theory for perturbative and
  non-perturbative nuclear dynamics, Prog. Theor. Exp. Phys. 01A207.
\newblock \href {http://dx.doi.org/10.1093/ptep/pts016}
  {\path{doi:10.1093/ptep/pts016}}.

\bibitem{Nakatsukasa2016_RMP88-045004}
T.~Nakatsukasa, K.~Matsuyanagi, M.~Matsuo, K.~Yabana, Time-dependent
  density-functional description of nuclear dynamics, Rev. Mod. Phys. 88 (2016)
  045004.
\newblock \href {http://dx.doi.org/10.1103/RevModPhys.88.045004}
  {\path{doi:10.1103/RevModPhys.88.045004}}.

\bibitem{Skyrme1958_NP9-615}
T.~Skyrme, {The effective nuclear potential}, Nucl. Phys. 9~(4) (1958)
  615--634.
\newblock \href {http://dx.doi.org/10.1016/0029-5582(58)90345-6}
  {\path{doi:10.1016/0029-5582(58)90345-6}}.

\bibitem{Perdew2004_LNP620-269}
J.~P. Perdew, S.~Kurth, Density functionals for non-relativistic coulomb
  systrems in the new century, in: C.~Fiolhais, F.~Nogueira, M.~A.~L. Marques
  (Eds.), A Primer in Density Functional Theory, Vol. 620, Springer Berlin
  Heidelberg, Berlin, 2003, pp. 1--55.
\newblock \href {http://dx.doi.org/10.1007/3-540-37072-2_1}
  {\path{doi:10.1007/3-540-37072-2_1}}.

\bibitem{Stancu1977_PLB68-108}
F.~Stancu, D.~M. Brink, H.~Flocard, {The tensor part of Skyrme's interaction},
  Phys. Lett. B 68~(2) (1977) 108.
\newblock \href {http://dx.doi.org/10.1016/0370-2693(77)90178-2}
  {\path{doi:10.1016/0370-2693(77)90178-2}}.

\bibitem{Colo1994_PRC50-1496}
G.~Col{\`{o}}, N.~{Van Giai}, P.~F. Bortignon, R.~A. Broglia, {Escape and
  spreading properties of charge-exchange resonances in $^{208}$Bi}, Phys. Rev.
  C 50~(3) (1994) 1496--1508.
\newblock \href {http://dx.doi.org/10.1103/PhysRevC.50.1496}
  {\path{doi:10.1103/PhysRevC.50.1496}}.

\bibitem{Litvinova2006_PRC73-044328}
E.~Litvinova, P.~Ring, {Covariant theory of particle-vibrational coupling and
  its effect on the single-particle spectrum}, Phys. Rev. C 73~(4) (2006)
  044328.
\newblock \href {http://dx.doi.org/10.1103/PhysRevC.73.044328}
  {\path{doi:10.1103/PhysRevC.73.044328}}.

\bibitem{Fayans1998_JETPL68-169}
S.~A. Fayans, {Towards a universal nuclear density functional}, J. Exp. Theor.
  Phys. Lett. 68~(3) (1998) 169--174.
\newblock \href {http://dx.doi.org/10.1134/1.567841}
  {\path{doi:10.1134/1.567841}}.

\bibitem{Baldo2008_PLB663-390}
M.~Baldo, P.~Schuck, X.~Vi{\~{n}}as, {Kohn–Sham density functional inspired
  approach to nuclear binding}, Phys. Lett. B 663~(5) (2008) 390--394.
\newblock \href {http://dx.doi.org/10.1016/j.physletb.2008.04.013}
  {\path{doi:10.1016/j.physletb.2008.04.013}}.

\bibitem{Vinas2009_IJMPE18-935}
X.~VI{\~{N}}AS, L.~M. ROBLEDO, M.~BALDO, P.~SCHUCK, {DEFORMED NUCLEI USING THE
  BARCELONA-CATANIA-PARIS ENERGY DENSITY FUNCTIONAL}, Int. J. Mod. Phys. E
  18~(04) (2009) 935--943.
\newblock \href {http://dx.doi.org/10.1142/S0218301309013075}
  {\path{doi:10.1142/S0218301309013075}}.

\bibitem{Robledo2010_PRC81-034315}
L.~M. Robledo, M.~Baldo, P.~Schuck, X.~Vi{\~{n}}as, {Octupole deformation
  properties of the Barcelona-Catania-Paris energy density functionals}, Phys.
  Rev. C 81~(3) (2010) 034315.
\newblock \href {http://dx.doi.org/10.1103/PhysRevC.81.034315}
  {\path{doi:10.1103/PhysRevC.81.034315}}.

\bibitem{Baldo2010_JPG37-064015}
M.~Baldo, L.~Robledo, P.~Schuck, X.~Vi{\~{n}}as, {Energy density functional on
  a microscopic basis}, J. Phys. G Nucl. Part. Phys. 37~(6) (2010) 064015.
\newblock \href {http://dx.doi.org/10.1088/0954-3899/37/6/064015}
  {\path{doi:10.1088/0954-3899/37/6/064015}}.

\bibitem{Baldo2013_PRC87-064305}
M.~Baldo, L.~M. Robledo, P.~Schuck, X.~Vi{\~{n}}as, {New Kohn-Sham density
  functional based on microscopic nuclear and neutron matter equations of
  state}, Phys. Rev. C 87~(6) (2013) 064305.
\newblock \href {http://dx.doi.org/10.1103/PhysRevC.87.064305}
  {\path{doi:10.1103/PhysRevC.87.064305}}.

\bibitem{Baldo2017_PRC95-014318}
M.~Baldo, L.~M. Robledo, P.~Schuck, X.~Vi{\~{n}}as,
  {Barcelona-Catania-Paris-Madrid functional with a realistic effective mass},
  Phys. Rev. C 95~(1) (2017) 014318.
\newblock \href {http://dx.doi.org/10.1103/PhysRevC.95.014318}
  {\path{doi:10.1103/PhysRevC.95.014318}}.

\bibitem{RocaMaza2011_PRC84-054309}
X.~Roca-Maza, X.~Vi{\~{n}}as, M.~Centelles, P.~Ring, P.~Schuck, {Relativistic
  mean-field interaction with density-dependent meson-nucleon vertices based on
  microscopical calculations}, Phys. Rev. C - Nucl. Phys. 84~(5) (2011) 054309.
\newblock \href {http://dx.doi.org/10.1103/PhysRevC.84.054309}
  {\path{doi:10.1103/PhysRevC.84.054309}}.

\bibitem{Agbemava2016_PRC93-044304}
S.~E. Agbemava, A.~V. Afanasjev, P.~Ring,
  \href{http://link.aps.org/doi/10.1103/PhysRevC.93.044304}{Octupole
  deformation in the ground states of even-even nuclei: a global analysis
  within the covariant density functional theory}, Phys. Rev. C 93 (2016)
  044304.
\newblock \href {http://dx.doi.org/10.1103/PhysRevC.93.044304}
  {\path{doi:10.1103/PhysRevC.93.044304}}.
\newline\urlprefix\url{http://link.aps.org/doi/10.1103/PhysRevC.93.044304}

\bibitem{Agbemava2015_PRC92-054310}
S.~E. Agbemava, A.~V. Afanasjev, T.~Nakatsukasa, P.~Ring,
  \href{http://link.aps.org/doi/10.1103/PhysRevC.92.054310}{Covariant density
  functional theory: Reexamining the structure of superheavy nuclei}, Phys.
  Rev. C 92 (2015) 054310.
\newblock \href {http://dx.doi.org/10.1103/PhysRevC.92-054310}
  {\path{doi:10.1103/PhysRevC.92-054310}}.
\newline\urlprefix\url{http://link.aps.org/doi/10.1103/PhysRevC.92.054310}

\bibitem{Brandow1967_RMP39-771}
B.~Brandow, {Linked-Cluster Expansions for the Nuclear Many-Body Problem}, Rev.
  Mod. Phys. 39~(4) (1967) 771--828.
\newblock \href {http://dx.doi.org/10.1103/RevModPhys.39.771}
  {\path{doi:10.1103/RevModPhys.39.771}}.

\bibitem{Bethe1971_ARNS21-93}
H.~A. Bethe, {Theory of Nuclear Matter}, Annu. Rev. Nucl. Sci. 21~(1) (1971)
  93--244.
\newblock \href {http://dx.doi.org/10.1146/annurev.ns.21.120171.000521}
  {\path{doi:10.1146/annurev.ns.21.120171.000521}}.

\bibitem{Sprung1972_ANP5-225}
D.~W.~L. Sprung, {Nuclear Matter Calculations}, in: Adv. Nucl. Phys., Vol.~5,
  Springer US, Boston, MA, 1972, pp. 225--343.
\newblock \href {http://dx.doi.org/10.1007/978-1-4615-8231-1_2}
  {\path{doi:10.1007/978-1-4615-8231-1_2}}.

\bibitem{Kohler1975_PR18-217}
H.~K{\"{o}}hler, {Brueckner theory of nuclei}, Phys. Rep. 18~(4) (1975)
  217--261.
\newblock \href {http://dx.doi.org/10.1016/0370-1573(75)90025-3}
  {\path{doi:10.1016/0370-1573(75)90025-3}}.

\bibitem{Bogner2010_PPNP65-94}
S.~K. Bogner, R.~J. Furnstahl, A.~Schwenk, From low-momentum interactions to
  nuclear structure, Prog. Part. Nucl. Phys. 65 (2010) 94--147.
\newblock \href {http://dx.doi.org/10.1016/j.ppnp.2010.03.001}
  {\path{doi:10.1016/j.ppnp.2010.03.001}}.

\bibitem{Brandow1966_PR152-863}
B.~H. Brandow, {Compact-cluster expansion for the nuclear many-body problem},
  Phys. Rev. 152 (1966) 863--882.
\newblock \href {http://dx.doi.org/10.1103/PhysRev.152.863}
  {\path{doi:10.1103/PhysRev.152.863}}.

\bibitem{Jeukenne1976_PR25-83}
J.~P. Jeukenne, a.~Lejeune, C.~Mahaux, {Many-body theory of nuclear matter},
  Phys. Rep. 25 (1976) 83.
\newblock \href {http://dx.doi.org/10.1016/0370-1573(76)90017-X}
  {\path{doi:10.1016/0370-1573(76)90017-X}}.

\bibitem{Song1998_PRL81-1584}
H.~Q. Song, M.~Baldo, G.~Giansiracusa, U.~Lombardo, {Bethe-Brueckner-Goldstone
  Expansion in Nuclear Matter}, Phys. Rev. Lett. 81 (1998) 1584--1587.
\newblock \href {http://dx.doi.org/10.1103/PhysRevLett.81.1584}
  {\path{doi:10.1103/PhysRevLett.81.1584}}.

\bibitem{Davies1969_PR177-1519}
K.~Davies, M.~Baranger, R.~Tarbutton, T.~Kuo, {Brueckner-Hartree-Fock
  Calculations of Spherical Nuclei in an Harmonic-Oscillator Basis}, Phys. Rev.
  177~(4) (1969) 1519--1526.
\newblock \href {http://dx.doi.org/10.1103/PhysRev.177.1519}
  {\path{doi:10.1103/PhysRev.177.1519}}.

\bibitem{Salpeter1951_PR84-1232}
E.~E. Salpeter, H.~A. Bethe, {A relativistic equation for bound-state
  problems}, Phys. Rev. 84~(6) (1951) 1232--1242.
\newblock \href {http://dx.doi.org/10.1103/PhysRev.84.1232}
  {\path{doi:10.1103/PhysRev.84.1232}}.

\bibitem{Yaes1971_PRD3-3086}
R.~J. Yaes, {Infinite Set of Quasipotential Equations from the Kadyshevsky
  Equation}, Phys. Rev. D 3~(12) (1971) 3086--3090.
\newblock \href {http://dx.doi.org/10.1103/PhysRevD.3.3086}
  {\path{doi:10.1103/PhysRevD.3.3086}}.

\bibitem{Thompson1970_PRD1-110}
R.~H. Thompson, {Three-dimensional Bethe-Salpeter equation applied to the
  nucleon-nucleon interaction}, Phys. Rev. D 1~(1) (1970) 110--117.
\newblock \href {http://dx.doi.org/10.1103/PhysRevD.1.110}
  {\path{doi:10.1103/PhysRevD.1.110}}.

\bibitem{Brockmann1978_PRC18-1510}
R.~Brockmann, {Relativistic Hartree-Fock description of nuclei}, Phys. Rev. C
  18~(3) (1978) 1510--1524.
\newblock \href {http://dx.doi.org/10.1103/PhysRevC.18.1510}
  {\path{doi:10.1103/PhysRevC.18.1510}}.

\bibitem{Itzykson1980}
C.~Itzykson, J.~B. Zuber, Quantum Field Theory, McGraw-Hill, New York, 1980.

\bibitem{Koepf1991_ZPA339-81}
W.~Koepf, P.~Ring, The spin-orbit field in superdeformed nuclei: a relativistic
  investigation, Z. Phys. A 339 (1991) 81--90.
\newblock \href {http://dx.doi.org/10.1007/BF01282936}
  {\path{doi:10.1007/BF01282936}}.

\bibitem{Anastasio1981_PRC23-2273}
M.~R. Anastasio, L.~S. Celenza, C.~M. Shakin, {Relativistic effects in the
  Bethe-Brueckner theory of nuclear matter}, Phys. Rev. C 23~(5) (1981)
  2273--2290.
\newblock \href {http://dx.doi.org/10.1103/PhysRevC.23.2273}
  {\path{doi:10.1103/PhysRevC.23.2273}}.

\bibitem{Huber1993_PLB317-485}
H.~Huber, F.~Weber, M.~Weigel, {Relativistic investigations of symmetric and
  asymmetric nuclear matter}, Phys. Lett. B 317~(4) (1993) 485--488.
\newblock \href {http://dx.doi.org/10.1016/0370-2693(93)91359-U}
  {\path{doi:10.1016/0370-2693(93)91359-U}}.

\bibitem{Poschenrieder1988_PRC38-471}
P.~Poschenrieder, M.~K. Weigel, {Nuclear matter problem in the relativistic
  Green's function approach}, Phys. Rev. C 38~(1) (1988) 471--486.
\newblock \href {http://dx.doi.org/10.1103/PhysRevC.38.471}
  {\path{doi:10.1103/PhysRevC.38.471}}.

\bibitem{Poschenrieder1988_PLB200-231}
P.~Poschenrieder, M.~Weigel, {Nuclear matter properties in the relativistic
  $\Lambda$-approximations}, Phys. Lett. B 200~(3) (1988) 231--234.
\newblock \href {http://dx.doi.org/10.1016/0370-2693(88)90761-7}
  {\path{doi:10.1016/0370-2693(88)90761-7}}.

\bibitem{Sehn1997_PRC56-216}
L.~Sehn, C.~Fuchs, A.~Faessler, {Nucleon self-energy in the relativistic
  Brueckner approach}, Phys. Rev. C 56~(1) (1997) 216--227.
\newblock \href {http://dx.doi.org/10.1103/PhysRevC.56.216}
  {\path{doi:10.1103/PhysRevC.56.216}}.

\bibitem{Tjon1985_PRC32-1667}
J.~A. Tjon, S.~J. Wallace, {General Lorentz-invariant representation of NN
  scattering amplitudes}, Phys. Rev. C 32~(5) (1985) 1667--1680.
\newblock \href {http://dx.doi.org/10.1103/PhysRevC.32.1667}
  {\path{doi:10.1103/PhysRevC.32.1667}}.

\bibitem{Fuchs1998_PRC58-2022}
C.~Fuchs, T.~Waindzoch, A.~Faessler, D.~S. Kosov, {Scalar and vector
  decomposition of the nucleon self-energy in the relativistic Brueckner
  approach}, Phys. Rev. C 58~(4) (1998) 2022--2032.
\newblock \href {http://dx.doi.org/10.1103/PhysRevC.58.2022}
  {\path{doi:10.1103/PhysRevC.58.2022}}.

\bibitem{Gross-Boelting1999_NPA648-105}
T.~Gross-Boelting, C.~Fuchs, A.~Faessler, Covariant representations of the
  relativistic brueckner t-matrix and the nuclear matter problem, Nucl. Phys. A
  648~(1-2) (1999) 105--137.
\newblock \href {http://dx.doi.org/10.1016/S0375-9474(99)00022-6}
  {\path{doi:10.1016/S0375-9474(99)00022-6}}.

\bibitem{deJong1998_PRC57-3099}
F.~de~Jong, H.~Lenske, {Asymmetric nuclear matter in the relativistic
  Brueckner-Hartree-Fock approach}, Phys. Rev. C 57 (1998) 3099--3107.
\newblock \href {http://dx.doi.org/10.1103/PhysRevC.57.3099}
  {\path{doi:10.1103/PhysRevC.57.3099}}.

\bibitem{Lee1997_PLB412-235}
C.~H. Lee, T.~T.~S. Kuo, G.~Q. Li, G.~E. Brown, {Momentum dependence of single
  particle potential in Dirac Brueckner approach}, Phys. Lett. B 412~(3-4)
  (1997) 235--239.
\newblock \href {http://dx.doi.org/10.1016/S0370-2693(97)01058-7}
  {\path{doi:10.1016/S0370-2693(97)01058-7}}.

\bibitem{Schiller1999_PRC60-059901}
E.~Schiller, H.~M{\"{u}}ther, P.~Czerski, {Erratum: Pauli exclusion operator
  and binding energy of nuclear matter [Phys. Rev. C 59, 2934 (1999)]}, Phys.
  Rev. C 60~(5) (1999) 059901.
\newblock \href {http://dx.doi.org/10.1103/PhysRevC.60.059901}
  {\path{doi:10.1103/PhysRevC.60.059901}}.

\bibitem{Sammarruca2000_PRC62-014614}
F.~Sammarruca, X.~Meng, E.~J. Stephenson, {Exact treatment of the Pauli
  exclusion operator in nuclear matter}, Phys. Rev. C 62~(1) (2000) 014614.
\newblock \href {http://dx.doi.org/10.1103/PhysRevC.62.014614}
  {\path{doi:10.1103/PhysRevC.62.014614}}.

\bibitem{Tong2018_PRC98-054302}
H.~Tong, X.~L. Ren, P.~Ring, S.~H. Shen, S.~B. Wang, J.~Meng, {Relativistic
  Brueckner-Hartree-Fock theory in nuclear matter without the average momentum
  approximation}, Phys. Rev. C 98~(5) (2018) 054302.
\newblock \href {http://dx.doi.org/10.1103/PhysRevC.98.054302}
  {\path{doi:10.1103/PhysRevC.98.054302}}.

\bibitem{RocaMaza2018_PPNP101-96}
X.~Roca-Maza, N.~Paar, {Nuclear equation of state from ground and collective
  excited state properties of nuclei}, Prog. Part. Nucl. Phys. 101 (2018)
  96--176.
\newblock \href {http://dx.doi.org/10.1016/j.ppnp.2018.04.001}
  {\path{doi:10.1016/j.ppnp.2018.04.001}}.

\bibitem{Li2006_PRC74-047304}
Z.~H. Li, U.~Lombardo, H.~J. Schulze, W.~Zuo, L.~W. Chen, H.~R. Ma, {Nuclear
  matter saturation point and symmetry energy with modern nucleon-nucleon
  potentials}, Phys. Rev. C 74~(4) (2006) 047304.
\newblock \href {http://dx.doi.org/10.1103/PhysRevC.74.047304}
  {\path{doi:10.1103/PhysRevC.74.047304}}.

\bibitem{Vidana2009_PRC80-045806}
I.~Vida{\~{n}}a, C.~Provid{\^{e}}ncia, A.~Polls, A.~Rios, {Density dependence
  of the nuclear symmetry energy: A microscopic perspective}, Phys. Rev. C
  80~(4) (2009) 045806.
\newblock \href {http://dx.doi.org/10.1103/PhysRevC.80.045806}
  {\path{doi:10.1103/PhysRevC.80.045806}}.

\bibitem{Baldo1999_PRC59-682}
M.~Baldo, L.~S. Ferreira, {Nuclear liquid-gas phase transition}, Phys. Rev. C
  59~(2) (1999) 682--703.
\newblock \href {http://dx.doi.org/10.1103/PhysRevC.59.682}
  {\path{doi:10.1103/PhysRevC.59.682}}.

\bibitem{Margueron2018_PRC97-025805}
J.~Margueron, R.~H. Casali, F.~Gulminelli, Equation of state for dense
  nucleonic matter from metamodeling. i. foundational aspects, Phys. Rev. C 97
  (2018) 025805.
\newblock \href {http://arxiv.org/abs/1708.06894} {\path{arXiv:1708.06894}},
  \href {http://dx.doi.org/10.1103/PhysRevC.97.025805}
  {\path{doi:10.1103/PhysRevC.97.025805}}.

\bibitem{Grange1989_PRC40-1040}
P.~Grang{\'{e}}, A.~Lejeune, M.~Martzolff, J.~F. Mathiot, {Consistent
  three-nucleon forces in the nuclear many-body problem}, Phys. Rev. C 40~(2)
  (1989) 1040--1060.
\newblock \href {http://dx.doi.org/10.1103/PhysRevC.40.1040}
  {\path{doi:10.1103/PhysRevC.40.1040}}.

\bibitem{Brueckner1958_PR110-431}
K.~Brueckner, J.~Gammel, H.~Weitzner, {Theory of Finite Nuclei}, Phys. Rev.
  110~(2) (1958) 431--445.
\newblock \href {http://dx.doi.org/10.1103/PhysRev.110.431}
  {\path{doi:10.1103/PhysRev.110.431}}.

\bibitem{Nemeth1970_PLB32-561}
J.~Nemeth, D.~Vautherin, {Study of finite nuclei in the local density
  approximation}, Phys. Lett. B 32~(7) (1970) 561--564.
\newblock \href {http://dx.doi.org/10.1016/0370-2693(70)90543-5}
  {\path{doi:10.1016/0370-2693(70)90543-5}}.

\bibitem{Sprung1971_NPA168-273}
D.~D.~P. Thkrique, D.~Sprung, P.~Banerjee, {A density-dependent local effective
  interaction between nucleons in nuclear matter}, Nucl. Phys. A 168~(2) (1971)
  273--306.
\newblock \href {http://dx.doi.org/10.1016/0375-9474(71)90794-9}
  {\path{doi:10.1016/0375-9474(71)90794-9}}.

\bibitem{Campi1972_NPA194-401}
X.~Campi, D.~Sprung, {Spherical nuclei in the local density approximation},
  Nucl. Phys. A 194~(2) (1972) 401--442.
\newblock \href {http://dx.doi.org/10.1016/0375-9474(72)91046-9}
  {\path{doi:10.1016/0375-9474(72)91046-9}}.

\bibitem{Ai1987_PRC35-2299}
H.~B. Ai, L.~S. Celenza, A.~Harindranath, C.~M. Shakin, {Effective interaction
  for relativistic mean-field theories of nuclear structure}, Phys. Rev. C
  35~(6) (1987) 2299--2309.
\newblock \href {http://dx.doi.org/10.1103/PhysRevC.35.2299}
  {\path{doi:10.1103/PhysRevC.35.2299}}.

\bibitem{Ai1989_PRC39-236}
H.~B. Ai, L.~S. Celenza, S.~F. Gao, C.~M. Shakin, {Relativistic quasiparticle
  method in nuclear physics}, Phys. Rev. C 39~(1) (1989) 236--247.
\newblock \href {http://dx.doi.org/10.1103/PhysRevC.39.236}
  {\path{doi:10.1103/PhysRevC.39.236}}.

\bibitem{Holinde1972_NPA198-598}
K.~Holinde, K.~Erkelenz, R.~Alzetta, {Relativistic one-boson exchange potential
  and nuclear-matter properties}, Nucl. Physics, Sect. A 198~(2) (1972)
  598--608.
\newblock \href {http://dx.doi.org/10.1016/0375-9474(72)90711-7}
  {\path{doi:10.1016/0375-9474(72)90711-7}}.

\bibitem{Holinde1976_NPA256-479}
K.~Holinde, R.~Machleidt, {OBEP and eikonal form factor: (I). Results for
  two-nucleon data}, Nucl. Phys. A 256~(3) (1976) 479--496.
\newblock \href {http://dx.doi.org/10.1016/0375-9474(76)90385-7}
  {\path{doi:10.1016/0375-9474(76)90385-7}}.

\bibitem{Marcos1989_PRC39-1134}
S.~Marcos, R.~Niembro, M.~L{\'{o}}pez-Quelle, N.~{Van Giai}, R.~Malfliet,
  {Parametrization of the relativistic Dirac-Brueckner G matrix}, Phys. Rev. C
  39~(3) (1989) 1134--1141.
\newblock \href {http://dx.doi.org/10.1103/PhysRevC.39.1134}
  {\path{doi:10.1103/PhysRevC.39.1134}}.

\bibitem{Brockmann1988_ZPA331-367}
R.~Brockmann, P.~G. Reinhard, {The relativistic mean-field model at large
  densities}, Zeitschrift f\"ur Phys. A 331~(3) (1988) 367--368.
\newblock \href {http://dx.doi.org/10.1007/BF01355612}
  {\path{doi:10.1007/BF01355612}}.

\bibitem{Elsenhans1990_NPA515-715}
H.~Elsenhans, H.~M{\"{u}}ther, R.~Machleidt, {Parametrization of the
  relativistic effective interaction in nuclear matter}, Nucl. Phys. A 515~(4)
  (1990) 715--735.
\newblock \href {http://dx.doi.org/10.1016/0375-9474(90)90281-P}
  {\path{doi:10.1016/0375-9474(90)90281-P}}.

\bibitem{Friedman1981_NPA361-502}
B.~Friedman, V.~Pandharipande, {Hot and cold, nuclear and neutron matter},
  Nucl. Phys. A 361~(2) (1981) 502--520.
\newblock \href {http://dx.doi.org/10.1016/0375-9474(81)90649-7}
  {\path{doi:10.1016/0375-9474(81)90649-7}}.

\bibitem{Gmuca1991_JPG17-1115}
S.~Gmuca, {Relativistic mean-field fit to microscopic results in nuclear
  matter}, J. Phys. G Nucl. Part. Phys. 17~(7) (1991) 1115--1126.
\newblock \href {http://dx.doi.org/10.1088/0954-3899/17/7/010}
  {\path{doi:10.1088/0954-3899/17/7/010}}.

\bibitem{Gmuca1992_ZPA342-387}
S.~Gmuca, {Relativistic mean-field parametrization of effective interaction in
  nuclear matter}, Zeitschrift f\"ur Phys. A 342~(4) (1992) 387--392.
\newblock \href {http://dx.doi.org/10.1007/BF01294948}
  {\path{doi:10.1007/BF01294948}}.

\bibitem{Gmuca1992_NPA547-447}
S.~Gmuca, {Finite-nuclei calculations based on relativistic mean-field
  effective interactions}, Nucl. Phys. A 547~(3) (1992) 447--458.
\newblock \href {http://dx.doi.org/10.1016/0375-9474(92)90032-F}
  {\path{doi:10.1016/0375-9474(92)90032-F}}.

\bibitem{Jackson1983_NPA407-495}
A.~Jackson, M.~Rho, E.~Krotscheck, {The $\sigma$-model and the binding energy
  of nuclear matter}, Nucl. Phys. A 407~(3) (1983) 495--506.
\newblock \href {http://dx.doi.org/10.1016/0375-9474(83)90662-0}
  {\path{doi:10.1016/0375-9474(83)90662-0}}.

\bibitem{deJong1991_PRC44-998}
F.~de~Jong, R.~Malfliet, {Conserving relativistic many-body approach: Equation
  of state, spectral function, and occupation probabilities of nuclear matter},
  Phys. Rev. C 44~(3) (1991) 998--1011.
\newblock \href {http://dx.doi.org/10.1103/PhysRevC.44.998}
  {\path{doi:10.1103/PhysRevC.44.998}}.

\bibitem{Savushkin1997_PRC55-167}
L.~N. Savushkin, S.~Marcos, M.~L. Quelle, P.~Bernardos, V.~N. Fomenko,
  R.~Niembro, {Effective interaction for relativistic theory of nuclear
  structure}, Phys. Rev. C 55~(1) (1997) 167--178.
\newblock \href {http://dx.doi.org/10.1103/PhysRevC.55.167}
  {\path{doi:10.1103/PhysRevC.55.167}}.

\bibitem{Ma1994_PRC50-3170}
Z.~Ma, H.~Shi, B.~Chen, {Isovector meson contribution in the relativistic
  Hartree-Fock approach for finite nuclei}, Phys. Rev. C 50~(6) (1994)
  3170--3173.
\newblock \href {http://dx.doi.org/10.1103/PhysRevC.50.3170}
  {\path{doi:10.1103/PhysRevC.50.3170}}.

\bibitem{Boersma1994_PRC50-1253}
H.~F. Boersma, R.~Malfliet, {Erratum: from nuclear matter to finite nuclei. I.
  Parametrization of the Dirac-Brueckner $G$ matrix}, Phys. Rev. C 50~(2)
  (1994) 1253--1253.
\newblock \href {http://dx.doi.org/10.1103/PhysRevC.50.1253}
  {\path{doi:10.1103/PhysRevC.50.1253}}.

\bibitem{Boersma1994_PRC49-1495}
H.~F. Boersma, R.~Malfliet, {From nuclear matter to finite nuclei. II.
  Relativistic theories for finite nuclei}, Phys. Rev. C 49~(3) (1994)
  1495--1515.
\newblock \href {http://dx.doi.org/10.1103/PhysRevC.49.1495}
  {\path{doi:10.1103/PhysRevC.49.1495}}.

\bibitem{Brueckner1955_PR97-1353}
K.~A. Brueckner, {Two-body forces and nuclear saturation. III. Details of the
  structure of the nucleus}, Phys. Rev. 97~(5) (1955) 1353--1366.
\newblock \href {http://dx.doi.org/10.1103/PhysRev.97.1353}
  {\path{doi:10.1103/PhysRev.97.1353}}.

\bibitem{Brueckner1960_PR117-207}
K.~A. Brueckner, D.~T. Goldman, {Single Particle Energies in the Theory of
  Nuclear Matter}, Phys. Rev. 117 (1960) 207--213.
\newblock \href {http://dx.doi.org/10.1103/PhysRev.117.207}
  {\path{doi:10.1103/PhysRev.117.207}}.

\bibitem{Brueckner1960_PR118-1438}
K.~A. Brueckner, J.~L. Gammel, J.~T. Kubis, {Calculation of Single-Particle
  Energies in the Theory of Nuclear Matter}, Phys. Rev. 118~(5) (1960)
  1438--1441.
\newblock \href {http://dx.doi.org/10.1103/PhysRev.118.1438}
  {\path{doi:10.1103/PhysRev.118.1438}}.

\bibitem{Hugenholtz1958_Physica24-363}
N.~Hugenholtz, L.~van Hove, {A theorem on the single particle energy in a Fermi
  gas with interaction}, Physica 24~(1-5) (1958) 363--376.
\newblock \href {http://dx.doi.org/10.1016/S0031-8914(58)95281-9}
  {\path{doi:10.1016/S0031-8914(58)95281-9}}.

\bibitem{Kohler1965_PR137-B1145}
H.~S. K{\"{o}}hler, {Theory of Finite Nuclei}, Phys. Rev. 137~(5B) (1965)
  B1145--B1157.
\newblock \href {http://dx.doi.org/10.1103/PhysRev.137.B1145}
  {\path{doi:10.1103/PhysRev.137.B1145}}.

\bibitem{Kohler1965_PR138-B831}
H.~S. K{\"{o}}hler, {Properties of Finite Nuclei}, Phys. Rev. 138~(1958) (1965)
  B831--B846.
\newblock \href {http://dx.doi.org/10.1103/PhysRev.138.B831}
  {\path{doi:10.1103/PhysRev.138.B831}}.

\bibitem{Kohler1966_NP88-529}
H.~K{\"{o}}hler, {Nucleon separation energy by K-matrix theory}, Nucl. Phys.
  88~(3) (1966) 529--538.
\newblock \href {http://dx.doi.org/10.1016/0029-5582(66)90412-3}
  {\path{doi:10.1016/0029-5582(66)90412-3}}.

\bibitem{Faessler1969_ZP223-192}
A.~Faessler, H.~H. Wolter, {Orbital rearrangement in thesd-shell nuclei},
  Zeitschrift f{\"{u}}r Phys. A Hadron. Nucl. 223~(2) (1969) 192--198.
\newblock \href {http://dx.doi.org/10.1007/BF01392983}
  {\path{doi:10.1007/BF01392983}}.

\bibitem{Muther1973_NPA215-213}
H.~M{\"{u}}ther, A.~Faessler, K.~Goeke, {Rearrangement in the
  Brueckner-Hartree-Fock approach}, Nucl. Phys. A 215~(1) (1973) 213--220.
\newblock \href {http://dx.doi.org/10.1016/0375-9474(73)90113-9}
  {\path{doi:10.1016/0375-9474(73)90113-9}}.

\bibitem{Brandow1963_PL4-8}
B.~Brandow, {Foundations of the nuclear shell model}, Phys. Lett. 4~(1) (1965)
  8--11.
\newblock \href {http://dx.doi.org/10.1016/0031-9163(63)90562-6}
  {\path{doi:10.1016/0031-9163(63)90562-6}}.

\bibitem{Becker1970_PRL24-400}
R.~Becker, {Renormalization of Finite Nuclear Brueckner Theory with Occupation
  Probabilities}, Phys. Rev. Lett. 24~(8) (1970) 400--404.
\newblock \href {http://dx.doi.org/10.1103/PhysRevLett.24.400}
  {\path{doi:10.1103/PhysRevLett.24.400}}.

\bibitem{Becker1970_PLB32-263}
R.~Becker, {Empirical validity of phenomenological renormalized nuclear
  Brueckner-Hartree-Fock theory}, Phys. Lett. B 32~(4) (1970) 263--266.
\newblock \href {http://dx.doi.org/10.1016/0370-2693(70)90523-X}
  {\path{doi:10.1016/0370-2693(70)90523-X}}.

\bibitem{Lenske1995_PLB345-355}
H.~Lenske, C.~Fuchs, {Rearrangement in the density dependent relativistic field
  theory of nuclei}, Phys. Lett. B 345~(4) (1995) 355--360.
\newblock \href {http://dx.doi.org/10.1016/0370-2693(94)01664-X}
  {\path{doi:10.1016/0370-2693(94)01664-X}}.

\bibitem{Haddad1993_PRC48-2740}
S.~Haddad, M.~K. Weigel, {Finite nuclear systems in a relativistic extended
  Thomas-Fermi approach with density-dependent coupling parameters}, Phys. Rev.
  C 48~(6) (1993) 2740--2745.
\newblock \href {http://dx.doi.org/10.1103/PhysRevC.48.2740}
  {\path{doi:10.1103/PhysRevC.48.2740}}.

\bibitem{Ineichen1996_PRC53-2158}
F.~Ineichen, M.~K. Weigel, D.~Von-Eiff, {Nuclear structure calculations in the
  density-dependent relativistic Hartree theory}, Phys. Rev. C 53~(5) (1996)
  2158--2162.
\newblock \href {http://dx.doi.org/10.1103/PhysRevC.53.2158}
  {\path{doi:10.1103/PhysRevC.53.2158}}.

\bibitem{Engvik1994_PRL73-2650}
L.~Engvik, M.~Hjorth-Jensen, E.~Osnes, G.~Bao, E.~{\O}stgaard, {Asymmetric
  nuclear matter and neutron star properties}, Phys. Rev. Lett. 73~(20) (1994)
  2650--2653.
\newblock \href {http://dx.doi.org/10.1103/PhysRevLett.73.2650}
  {\path{doi:10.1103/PhysRevLett.73.2650}}.

\bibitem{Engvik1996_AJ469-794}
L.~Engvik, E.~Osnes, M.~Hjorth-Jensen, G.~Bao, E.~Ostgaard, {Asymmetric Nuclear
  Matter and Neutron Star Properties}, Astrophys. J. 469 (1996) 794.
\newblock \href {http://dx.doi.org/10.1086/177827} {\path{doi:10.1086/177827}}.

\bibitem{Gogelein2008_PRC77-025802}
P.~G{\"{o}}gelein, E.~N.~E. van Dalen, C.~Fuchs, H.~M{\"{u}}ther, {Nuclear
  matter in the crust of neutron stars derived from realistic $NN$
  interactions}, Phys. Rev. C 77~(2) (2008) 025802.
\newblock \href {http://dx.doi.org/10.1103/PhysRevC.77.025802}
  {\path{doi:10.1103/PhysRevC.77.025802}}.

\bibitem{VanDalen2005_PRL95-022302}
E.~N. E.~V. Dalen, C.~Fuchs, A.~Faessler, {Effective Nucleon Masses in
  Symmetric and Asymmetric Nuclear Matter}, Phys. Rev. Lett. 95~(2) (2005)
  022302.
\newblock \href {http://dx.doi.org/10.1103/PhysRevLett.95.022302}
  {\path{doi:10.1103/PhysRevLett.95.022302}}.

\bibitem{Bogner2003_PR386-1}
S.~Bogner, T.~Kuo, A.~Schwenk, {Model-independent low momentum nucleon
  interaction from phase shift equivalence}, Phys. Rep. 386~(1) (2003) 1--27.
\newblock \href {http://dx.doi.org/10.1016/j.physrep.2003.07.001}
  {\path{doi:10.1016/j.physrep.2003.07.001}}.

\bibitem{Bogner2007_PRC75-061001}
S.~K. Bogner, R.~J. Furnstahl, R.~J. Perry, Similarity renormalization group
  for nucleon-nucleon interactions, Phys. Rev. C 75 (2007) 061001(R).
\newblock \href {http://dx.doi.org/10.1103/PhysRevC.75.061001}
  {\path{doi:10.1103/PhysRevC.75.061001}}.

\bibitem{Beck1970_ZP231-26}
R.~Beck, H.~J. Mang, P.~Ring, {Symmetry-conserving
  Hartree-Fock-Bogolyubov-theory and its application to collective nuclear
  rotation}, Zeitschrift f{\"{u}}r Phys. 231~(1) (1970) 26--47.
\newblock \href {http://dx.doi.org/10.1007/BF01394547}
  {\path{doi:10.1007/BF01394547}}.

\bibitem{Zeh1965_ZP188-361}
H.~D. Zeh, {Symmetry violating trial wave functions}, Zeitschrift f{\"{u}}r
  Phys. 188~(4) (1965) 361--373.
\newblock \href {http://dx.doi.org/10.1007/BF01326951}
  {\path{doi:10.1007/BF01326951}}.

\bibitem{Chabanat1998_NPA635-231}
E.~Chabanat, P.~Bonche, P.~Haensel, J.~Meyer, R.~Schaeffer, {A Skyrme
  parametrization from subnuclear to neutron star densities Part II. Nuclei far
  from stabilities}, Nucl. Phys. A 635~(1-2) (1998) 231--256.
\newblock \href {http://dx.doi.org/10.1016/S0375-9474(98)00180-8}
  {\path{doi:10.1016/S0375-9474(98)00180-8}}.

\bibitem{Becker1974_PRC9-1221}
R.~Becker, K.~Davies, M.~Patterson, {Renormalized Brueckner-Hartree-Fock
  calculations of $^{4}$He and $^{16}$O with center-of-mass corrections}, Phys.
  Rev. C 9~(4) (1974) 1221--1242.
\newblock \href {http://dx.doi.org/10.1103/PhysRevC.9.1221}
  {\path{doi:10.1103/PhysRevC.9.1221}}.

\bibitem{Wang2017_CPC41-030003}
M.~Wang, G.~Audi, F.~G. Kondev, W.~Huang, S.~Naimi, X.~Xu, {The AME2016 atomic
  mass evaluation (II). Tables, graphs and references}, Chin. Phys. C 41~(3)
  (2017) 030003.
\newblock \href {http://dx.doi.org/10.1088/1674-1137/41/3/030003}
  {\path{doi:10.1088/1674-1137/41/3/030003}}.

\bibitem{Angeli2013_ADNDT99-69}
I.~Angeli, K.~Marinova, {Table of experimental nuclear ground state charge
  radii: An update}, At. Data Nucl. Data Tables 99~(1) (2013) 69--95.
\newblock \href {http://dx.doi.org/10.1016/j.adt.2011.12.006}
  {\path{doi:10.1016/j.adt.2011.12.006}}.

\bibitem{Coraggio2003_PRC68-034320}
L.~Coraggio, N.~Itaco, A.~Covello, A.~Gargano, T.~T.~S. Kuo, {Ground-state
  properties of closed-shell nuclei with low-momentum realistic interactions},
  Phys. Rev. C 68~(3) (2003) 034320.
\newblock \href {http://dx.doi.org/10.1103/PhysRevC.68.034320}
  {\path{doi:10.1103/PhysRevC.68.034320}}.

\bibitem{Binder2016_PRC93-044002}
S.~Binder, A.~Calci, E.~Epelbaum, R.~J. Furnstahl, J.~Golak, K.~Hebeler,
  H.~Kamada, H.~Krebs, J.~Langhammer, S.~Liebig, P.~Maris, U.~G. Mei{\ss}ner,
  D.~Minossi, A.~Nogga, H.~Potter, R.~Roth, R.~Skibi{\'{n}}ski, K.~Topolnicki,
  J.~P. Vary, H.~Wita{\l}a, {Few-nucleon systems with state-of-the-art chiral
  nucleon-nucleon forces}, Phys. Rev. C 93~(4) (2016) 044002.
\newblock \href {http://dx.doi.org/10.1103/PhysRevC.93.044002}
  {\path{doi:10.1103/PhysRevC.93.044002}}.

\bibitem{Ozawa2001_NPA691-599}
A.~Ozawa, O.~Bochkarev, L.~Chulkov, D.~Cortina, H.~Geissel, M.~Hellstr{\"{o}}m,
  M.~Ivanov, R.~Janik, K.~Kimura, T.~Kobayashi, A.~A. Korsheninnikov,
  G.~M{\"{u}}nzenberg, F.~Nickel, Y.~Ogawa, A.~Ogloblin, M.~Pf{\"{u}}tzner,
  V.~Pribora, H.~Simon, B.~Sit{\'{a}}r, P.~Strmen, K.~S{\"{u}}mmerer,
  T.~Suzuki, I.~Tanihata, M.~Winkler, K.~Yoshida, {Measurements of interaction
  cross sections for light neutron-rich nuclei at relativistic energies and
  determination of effective matter radii}, Nucl. Phys. A 691~(3-4) (2001)
  599--617.
\newblock \href {http://dx.doi.org/10.1016/S0375-9474(01)00563-2}
  {\path{doi:10.1016/S0375-9474(01)00563-2}}.

\bibitem{Hu2017_PRC95-034321}
B.~S. Hu, F.~R. Xu, Q.~Wu, Y.~Z. Ma, Z.~H. Sun, {Brueckner-Hartree-Fock
  calculations for finite nuclei with renormalized realistic forces}, Phys.
  Rev. C 95~(3) (2017) 034321.
\newblock \href {http://dx.doi.org/10.1103/PhysRevC.95.034321}
  {\path{doi:10.1103/PhysRevC.95.034321}}.

\bibitem{Hagen2009_PRC80-021306}
G.~Hagen, T.~Papenbrock, D.~J. Dean, M.~Hjorth-Jensen, B.~V. Asokan, {Ab initio
  computation of neutron-rich oxygen isotopes}, Phys. Rev. C 80~(2) (2009)
  021306.
\newblock \href {http://dx.doi.org/10.1103/PhysRevC.80.021306}
  {\path{doi:10.1103/PhysRevC.80.021306}}.

\bibitem{Hergert2013_PRC87-034307}
H.~Hergert, S.~K. Bogner, S.~Binder, A.~Calci, J.~Langhammer, R.~Roth,
  A.~Schwenk, In-medium similarity renormalization group with chiral two- plus
  three-nucleon interactions, Phys. Rev. C 87 (2013) 034307.
\newblock \href {http://dx.doi.org/10.1103/PhysRevC.87.034307}
  {\path{doi:10.1103/PhysRevC.87.034307}}.

\bibitem{Roth2011_PRL107-072501}
R.~Roth, J.~Langhammer, A.~Calci, S.~Binder, P.~Navr\'atil,
  Similarity-transformed chiral {$NN+3N$} interactions for the \textit{Ab
  initio} description of {$^{12}$C} and {$^{16}$O}, Phys. Rev. Lett. 107 (2011)
  072501.
\newblock \href {http://dx.doi.org/10.1103/PhysRevLett.107.072501}
  {\path{doi:10.1103/PhysRevLett.107.072501}}.

\bibitem{Cipollone2013_PRL111-062501}
A.~Cipollone, C.~Barbieri, P.~Navr{\'{a}}til, {Isotopic Chains Around Oxygen
  from Evolved Chiral Two- and Three-Nucleon Interactions}, Phys. Rev. Lett.
  111~(6) (2013) 062501.
\newblock \href {http://dx.doi.org/10.1103/PhysRevLett.111.062501}
  {\path{doi:10.1103/PhysRevLett.111.062501}}.

\bibitem{Lahde2014_PLB732-110}
T.~A. L{\"{a}}hde, E.~Epelbaum, H.~Krebs, D.~Lee, U.~G. Mei{\ss}ner, G.~Rupak,
  {Lattice effective field theory for medium-mass nuclei}, Phys. Lett. B 732
  (2014) 110--115.
\newblock \href {http://dx.doi.org/10.1016/j.physletb.2014.03.023}
  {\path{doi:10.1016/j.physletb.2014.03.023}}.

\bibitem{Epelbaum2009_EPJA41-125}
E.~Epelbaum, H.~Krebs, D.~Lee, U.~G. Mei{\ss}ner, {Lattice chiral effective
  field theory with three-body interactions at next-to-next-to-leading order},
  Eur. Phys. J. A 41~(1) (2009) 125--139.
\newblock \href {http://dx.doi.org/10.1140/epja/i2009-10764-y}
  {\path{doi:10.1140/epja/i2009-10764-y}}.

\bibitem{Lonardoni2018_PRC97-044318}
D.~Lonardoni, S.~Gandolfi, J.~E. Lynn, C.~Petrie, J.~Carlson, K.~E. Schmidt,
  A.~Schwenk, {Auxiliary field diffusion Monte Carlo calculations of light and
  medium-mass nuclei with local chiral interactions}, Phys. Rev. C 97~(4)
  (2018) 044318.
\newblock \href {http://dx.doi.org/10.1103/PhysRevC.97.044318}
  {\path{doi:10.1103/PhysRevC.97.044318}}.

\bibitem{Gezerlis2013_PRL111-032501}
A.~Gezerlis, I.~Tews, E.~Epelbaum, S.~Gandolfi, K.~Hebeler, A.~Nogga,
  A.~Schwenk, {Quantum Monte Carlo Calculations with Chiral Effective Field
  Theory Interactions}, Phys. Rev. Lett. 111~(3) (2013) 032501.
\newblock \href {http://dx.doi.org/10.1103/PhysRevLett.111.032501}
  {\path{doi:10.1103/PhysRevLett.111.032501}}.

\bibitem{DeVries1987_ADNDT36-495}
H.~{De Vries}, C.~{De Jager}, C.~{De Vries}, {Nuclear
  charge-density-distribution parameters from elastic electron scattering}, At.
  Data Nucl. Data Tables 36~(3) (1987) 495--536.
\newblock \href {http://dx.doi.org/10.1016/0092-640X(87)90013-1}
  {\path{doi:10.1016/0092-640X(87)90013-1}}.

\bibitem{Lapoux2016_PRL117-052501}
V.~Lapoux, V.~Som{\`{a}}, C.~Barbieri, H.~Hergert, J.~D. Holt, S.~R. Stroberg,
  {Radii and Binding Energies in Oxygen Isotopes: A Challenge for Nuclear
  Forces}, Phys. Rev. Lett. 117~(5) (2016) 052501.
\newblock \href {http://dx.doi.org/10.1103/PhysRevLett.117.052501}
  {\path{doi:10.1103/PhysRevLett.117.052501}}.

\bibitem{Ekstrom2015_PRC91-051301}
A.~Ekstr{\"{o}}m, G.~R. Jansen, K.~A. Wendt, G.~Hagen, T.~Papenbrock, B.~D.
  Carlsson, C.~Forss{\'{e}}n, M.~Hjorth-Jensen, P.~Navr{\'{a}}til,
  W.~Nazarewicz, {Accurate nuclear radii and binding energies from a chiral
  interaction}, Phys. Rev. C 91~(5) (2015) 051301.
\newblock \href {http://dx.doi.org/10.1103/PhysRevC.91.051301}
  {\path{doi:10.1103/PhysRevC.91.051301}}.

\bibitem{Lu2013_RMP85-1383}
Z.~T. Lu, P.~Mueller, G.~W.~F. Drake, W.~N{\"{o}}rtersh{\"{a}}user, S.~C.
  Pieper, Z.~C. Yan, {Colloquium : Laser probing of neutron-rich nuclei in
  light atoms}, Rev. Mod. Phys. 85~(4) (2013) 1383--1400.
\newblock \href {http://dx.doi.org/10.1103/RevModPhys.85.1383}
  {\path{doi:10.1103/RevModPhys.85.1383}}.

\bibitem{Nogga2000_PRL85-944}
A.~Nogga, H.~Kamada, W.~Gl{\"{o}}ckle, {Modern Nuclear Force Predictions for
  the $\alpha$ Particle}, Phys. Rev. Lett. 85~(5) (2000) 944--947.
\newblock \href {http://dx.doi.org/10.1103/PhysRevLett.85.944}
  {\path{doi:10.1103/PhysRevLett.85.944}}.

\bibitem{Machleidt1996_PRC53-R1483}
R.~Machleidt, F.~Sammarruca, Y.~Song, Nonlocal nature of the nuclear force and
  its impact on nuclear structure, Phys. Rev. C 53 (1996) R1483--R1487.
\newblock \href {http://dx.doi.org/10.1103/PhysRevC.53.R1483}
  {\path{doi:10.1103/PhysRevC.53.R1483}}.

\bibitem{Navratil2007_FBS41-117}
P.~Navra, P.~Navr{\'{a}}til, {Local three-nucleon interaction from chiral
  effective field theory}, Few-Body Syst. 41~(3-4) (2007) 117--140.
\newblock \href {http://dx.doi.org/10.1007/s00601-007-0193-3}
  {\path{doi:10.1007/s00601-007-0193-3}}.

\bibitem{Roth2007_PRL99-092501}
R.~Roth, P.~Navr{\'{a}}til, {\textit{Ab Initio} Study of $^{40}$Ca with an
  Importance-Truncated No-Core Shell Model}, Phys. Rev. Lett. 99~(9) (2007)
  092501.
\newblock \href {http://dx.doi.org/10.1103/PhysRevLett.99.092501}
  {\path{doi:10.1103/PhysRevLett.99.092501}}.

\bibitem{Hagen2007_PRC76-044305}
G.~Hagen, D.~J. Dean, M.~Hjorth-Jensen, T.~Papenbrock, A.~Schwenk, {Benchmark
  calculations for $^3$H, $^4$He, $^{16}$O, and $^{40}$Ca with \textit{ab
  initio} coupled-cluster theory}, Phys. Rev. C 76~(4) (2007) 044305.
\newblock \href {http://dx.doi.org/10.1103/PhysRevC.76.044305}
  {\path{doi:10.1103/PhysRevC.76.044305}}.

\bibitem{Hagen2010_PRC82-034330}
G.~Hagen, T.~Papenbrock, D.~J. Dean, M.~Hjorth-Jensen, {Ab initio
  coupled-cluster approach to nuclear structure with modern nucleon-nucleon
  interactions}, Phys. Rev. C 82~(3) (2010) 034330.
\newblock \href {http://dx.doi.org/10.1103/PhysRevC.82.034330}
  {\path{doi:10.1103/PhysRevC.82.034330}}.

\bibitem{Arima1969_PLB30-517}
A.~Arima, M.~Harvey, K.~Shimizu, Pseudo {LS} coupling and pseudo {SU(3)}
  coupling schemes, Phys. Lett. B 30 (1969) 517--522.
\newblock \href {http://dx.doi.org/10.1016/0370-2693(69)90443-2}
  {\path{doi:10.1016/0370-2693(69)90443-2}}.

\bibitem{Hecht1969_NPA137-129}
K.~T. Hecht, A.~Adler, Generalized seniority for favored {$J\ne 0$} pairs in
  mixed configurations, Nucl. Phys. A 137 (1969) 129--143.
\newblock \href {http://dx.doi.org/10.1016/0375-9474(69)90077-3}
  {\path{doi:10.1016/0375-9474(69)90077-3}}.

\bibitem{Ginocchio1997_PRL78-436}
J.~N. Ginocchio, Pseudospin as a relativistic symmetry, Phys. Rev. Lett. 78
  (1997) 436--439.
\newblock \href {http://dx.doi.org/10.1103/PhysRevLett.78.436}
  {\path{doi:10.1103/PhysRevLett.78.436}}.

\bibitem{Meng1998_PRC58-R628}
J.~Meng, K.~Sugawara-Tanabe, S.~Yamaji, P.~Ring, A.~Arima, Pseudospin symmetry
  in relativistic mean field theory, Phys. Rev. C 58 (1998) R628--R631.
\newblock \href {http://dx.doi.org/10.1103/PhysRevC.58.R628}
  {\path{doi:10.1103/PhysRevC.58.R628}}.

\bibitem{Meng1999_PRC59-154}
J.~Meng, K.~Sugawara-Tanabe, S.~Yamaji, A.~Arima, Pseudospin symmetry in {Zr}
  and {Sn} isotopes from the proton drip line to the neutron drip line, Phys.
  Rev. C 59 (1999) 154--163.
\newblock \href {http://dx.doi.org/10.1103/PhysRevC.59.154}
  {\path{doi:10.1103/PhysRevC.59.154}}.

\bibitem{Marcos2001_PLB513-30}
S.~Marcos, M.~L\'{o}pez-Quelle, R.~Niembro, L.~N. Savushkin, P.~Bernardos, On
  the sufficient conditions for the pseudospin symmetry in relativistic models,
  Phys. Lett. B 513 (2001) 30--36.
\newblock \href {http://dx.doi.org/10.1016/S0370-2693(01)00737-7}
  {\path{doi:10.1016/S0370-2693(01)00737-7}}.

\bibitem{Chen2003_CPL20-358}
T.~S. Chen, H.~F. L\"{u}, J.~Meng, S.~Q. Zhang, S.~G. Zhou, Pseudospin symmetry
  in relativistic framework with harmonic oscillator potential and
  {Woods-Saxon} potential, Chin. Phys. Lett. 20 (2003) 358--361.
\newblock \href {http://dx.doi.org/10.1088/0256-307X/20/3/312}
  {\path{doi:10.1088/0256-307X/20/3/312}}.

\bibitem{Lisboa2004_PRC69-024319}
R.~Lisboa, M.~Malheiro, A.~S. de~Castro, P.~Alberto, M.~Fiolhais, Pseudospin
  symmetry and the relativistic harmonic oscillator, Phys. Rev. C 69 (2004)
  024319.
\newblock \href {http://dx.doi.org/10.1103/PhysRevC.69.024319}
  {\path{doi:10.1103/PhysRevC.69.024319}}.

\bibitem{Liang2011_PRC83-041301}
H.~Liang, P.~Zhao, Y.~Zhang, J.~Meng, N.~Van~Giai, Perturbative interpretation
  of relativistic symmetries in nuclei, Phys. Rev. C 83 (2011) 041301(R).
\newblock \href {http://dx.doi.org/10.1103/PhysRevC.83.041301}
  {\path{doi:10.1103/PhysRevC.83.041301}}.

\bibitem{Liang2013_PRC87-014334}
H.~Liang, S.~Shen, P.~Zhao, J.~Meng, Pseudospin symmetry in supersymmetric
  quantum mechanics: {Schr\"odinger} equations, Phys. Rev. C 87 (2013) 014334.
\newblock \href {http://dx.doi.org/10.1103/PhysRevC.87.014334}
  {\path{doi:10.1103/PhysRevC.87.014334}}.

\bibitem{Shen2013_PRC88-024311}
S.~Shen, H.~Liang, P.~Zhao, S.~Zhang, J.~Meng, Pseudospin symmetry in
  supersymmetric quantum mechanics. {II.} spin-orbit effects, Phys. Rev. C 88
  (2013) 024311.
\newblock \href {http://dx.doi.org/10.1103/PhysRevC.88.024311}
  {\path{doi:10.1103/PhysRevC.88.024311}}.

\bibitem{Guo2014_PRL112-062502}
J.~Y. Guo, S.~W. Chen, Z.~M. Niu, D.~P. Li, Q.~Liu, Probing the symmetries of
  the {Dirac Hamiltonian} with axially deformed scalar and vector potentials by
  similarity renormalization group, Phys. Rev. Lett. 112 (2014) 062502.
\newblock \href {http://dx.doi.org/10.1103/PhysRevLett.112.062502}
  {\path{doi:10.1103/PhysRevLett.112.062502}}.

\bibitem{Shi2014_PRC90-034318}
M.~Shi, D.~P. Li, S.~W. Chen, J.~Y. Guo, Examination of the pseudospin symmetry
  for the relativistic harmonic oscillator with the similarity renormalization
  group, Phys. Rev. C 90 (2014) 034318.
\newblock \href {http://dx.doi.org/10.1103/PhysRevC.90.034318}
  {\path{doi:10.1103/PhysRevC.90.034318}}.

\bibitem{Zhao2014_PRC90-054326}
Q.~Zhao, J.~M. Dong, J.~L. Song, W.~H. Long, Proton radioactivity described by
  covariant density functional theory with the similarity renormalization group
  method, Phys. Rev. C 90 (2014) 054326.
\newblock \href {http://dx.doi.org/10.1103/PhysRevC.90.054326}
  {\path{doi:10.1103/PhysRevC.90.054326}}.

\bibitem{Li2015_PRC91-024311}
D.~P. Li, S.~W. Chen, Z.~M. Niu, Q.~Liu, J.~Y. Guo, Further investigation of
  relativistic symmetry in deformed nuclei by similarity renormalization group,
  Phys. Rev. C 91 (2015) 024311.
\newblock \href {http://dx.doi.org/10.1103/PhysRevC.91.024311}
  {\path{doi:10.1103/PhysRevC.91.024311}}.

\bibitem{Li2016_PRC93-054312}
J.~J. Li, W.~H. Long, J.~L. Song, Q.~Zhao, Pseudospin-orbit splitting and its
  consequences for the central depression in nuclear density, Phys. Rev. C 93
  (2016) 054312.
\newblock \href {http://dx.doi.org/10.1103/PhysRevC.93.054312}
  {\path{doi:10.1103/PhysRevC.93.054312}}.

\bibitem{Gao2017_PLB769-77}
J.~Gao, M.~C. Zhang, Pseudospin symmetric solution of the dirac–eckart
  problem with a hulthen tensor interaction in the tridiagonal representation,
  Phys. Lett. B 769~(Supplement C) (2017) 77 -- 81.
\newblock \href
  {http://dx.doi.org/https://doi.org/10.1016/j.physletb.2017.03.030}
  {\path{doi:https://doi.org/10.1016/j.physletb.2017.03.030}}.

\bibitem{Sun2017_PRC96-044312}
T.~T. Sun, W.~L. Lu, S.~S. Zhang, {Spin and pseudospin symmetries in the
  single-$\Lambda$ spectrum}, Phys. Rev. C 96~(4) (2017) 044312.
\newblock \href {http://dx.doi.org/10.1103/PhysRevC.96.044312}
  {\path{doi:10.1103/PhysRevC.96.044312}}.

\bibitem{Zhou2003_PRL91-262501}
S.~G. Zhou, J.~Meng, P.~Ring, Spin symmetry in the antinucleon spectrum, Phys.
  Rev. Lett. 91 (2003) 262501.
\newblock \href {http://dx.doi.org/10.1103/PhysRevLett.91.262501}
  {\path{doi:10.1103/PhysRevLett.91.262501}}.

\bibitem{He2006_EPJA28-265}
X.~T. He, S.~G. Zhou, J.~Meng, E.~G. Zhao, W.~Scheid, Test of spin symmetry in
  anti-nucleon spectra, Eur. Phys. J. A 28 (2006) 265--269.
\newblock \href {http://dx.doi.org/10.1140/epja/i2006-10066-0}
  {\path{doi:10.1140/epja/i2006-10066-0}}.

\bibitem{Song2009_CPL26-122102}
C.~Y. Song, J.~M. Yao, J.~Meng, Spin symmetry for anti-{Lambda} spectrum in
  atomic nucleus, Chin. Phys. Lett. 26 (2009) 122102.
\newblock \href {http://dx.doi.org/10.1088/0256-307X/26/12/122102}
  {\path{doi:10.1088/0256-307X/26/12/122102}}.

\bibitem{Liang2010_EPJA44-119}
H.~Liang, W.~H. Long, J.~Meng, N.~Van~Giai, Spin symmetry in {Dirac}
  negative-energy spectrum in density-dependent relativistic {Hartree-Fock}
  theory, Eur. Phys. J. A 44 (2010) 119--124.
\newblock \href {http://dx.doi.org/10.1140/epja/i2010-10938-6}
  {\path{doi:10.1140/epja/i2010-10938-6}}.

\bibitem{Lisboa2010_PRC81-064324}
R.~Lisboa, M.~Malheiro, P.~Alberto, M.~Fiolhais, A.~S. de~Castro, Spin and
  pseudospin symmetries in the antinucleon spectrum of nuclei, Phys. Rev. C 81
  (2010) 064324.
\newblock \href {http://dx.doi.org/10.1103/PhysRevC.81.064324}
  {\path{doi:10.1103/PhysRevC.81.064324}}.

\bibitem{Hamzavi2014_APNY341-153}
M.~Hamzavi, S.~M. Ikhdair, B.~J. Falaye, {Dirac} bound states of anharmonic
  oscillator in external fields, Ann. Phys. (NY) 341 (2014) 153--163.
\newblock \href {http://dx.doi.org/10.1016/j.aop.2013.12.003}
  {\path{doi:10.1016/j.aop.2013.12.003}}.

\bibitem{Sun2017_IJMPE26-1750025}
M.~Sun, D.~P. Li, S.~W. Chen, J.~Y. Guo, Spin and pseudospin symmetries and
  their breaking mechanisms in antinucleon spectrum, Int. J. Mod. Phys. E
  26~(05) (2017) 1750025.
\newblock \href {http://dx.doi.org/10.1142/S0218301317500252}
  {\path{doi:10.1142/S0218301317500252}}.

\bibitem{Ginocchio2005_PR414-165}
J.~N. Ginocchio, Relativistic symmetries in nuclei and hadrons, Phys. Rep. 414
  (2005) 165--261.
\newblock \href {http://dx.doi.org/10.1016/j.physrep.2005.04.003}
  {\path{doi:10.1016/j.physrep.2005.04.003}}.

\bibitem{Litvinova2011_PRC84-014305}
E.~V. Litvinova, A.~V. Afanasjev, Dynamics of nuclear single-particle structure
  in covariant theory of particle-vibration coupling: From light to superheavy
  nuclei, Phys. Rev. C 84 (2011) 014305.
\newblock \href {http://dx.doi.org/10.1103/PhysRevC.84.014305}
  {\path{doi:10.1103/PhysRevC.84.014305}}.

\bibitem{Pudliner1996_PRL76-2416}
B.~S. Pudliner, A.~Smerzi, J.~Carlson, V.~R. Pandharipande, S.~C. Pieper, D.~G.
  Ravenhall, {Neutron Drops and Skyrme Energy-Density Functionals}, Phys. Rev.
  Lett. 76~(14) (1996) 2416--2419.
\newblock \href {http://dx.doi.org/10.1103/PhysRevLett.76.2416}
  {\path{doi:10.1103/PhysRevLett.76.2416}}.

\bibitem{Bogner2011_PRC84-044306}
S.~K. Bogner, R.~J. Furnstahl, H.~Hergert, M.~Kortelainen, P.~Maris,
  M.~Stoitsov, J.~P. Vary, {Testing the density matrix expansion against ab
  initio calculations of trapped neutron drops}, Phys. Rev. C 84~(4) (2011)
  044306.
\newblock \href {http://dx.doi.org/10.1103/PhysRevC.84.044306}
  {\path{doi:10.1103/PhysRevC.84.044306}}.

\bibitem{Maris2013_PRC87-054318}
P.~Maris, J.~P. Vary, S.~Gandolfi, J.~Carlson, S.~C. Pieper, {Properties of
  trapped neutrons interacting with realistic nuclear Hamiltonians}, Phys. Rev.
  C 87~(5) (2013) 054318.
\newblock \href {http://dx.doi.org/10.1103/PhysRevC.87.054318}
  {\path{doi:10.1103/PhysRevC.87.054318}}.

\bibitem{Potter2014_PLB739-445}
H.~Potter, S.~Fischer, P.~Maris, J.~Vary, S.~Binder, A.~Calci, J.~Langhammer,
  R.~Roth, {Ab initio study of neutron drops with chiral Hamiltonians}, Phys.
  Lett. B 739 (2014) 445--450.
\newblock \href {http://dx.doi.org/10.1016/j.physletb.2014.10.020}
  {\path{doi:10.1016/j.physletb.2014.10.020}}.

\bibitem{Zhao2016_PRC94-041302}
P.~W. Zhao, S.~Gandolfi, {Radii of neutron drops probed via the neutron skin
  thickness of nuclei}, Phys. Rev. C 94~(4) (2016) 041302.
\newblock \href {http://dx.doi.org/10.1103/PhysRevC.94.041302}
  {\path{doi:10.1103/PhysRevC.94.041302}}.

\bibitem{Pudliner1995_PRL74-4396}
B.~Pudliner, V.~Pandharipande, J.~Carlson, R.~Wiringa, {Quantum Monte Carlo
  Calculations of $A \leq 6$ Nuclei}, Phys. Rev. Lett. 74~(22) (1995)
  4396--4399.
\newblock \href {http://dx.doi.org/10.1103/PhysRevLett.74.4396}
  {\path{doi:10.1103/PhysRevLett.74.4396}}.

\bibitem{Smerzi1997_PRC56-2549}
A.~Smerzi, D.~G. Ravenhall, V.~R. Pandharipande, {Neutron drops and neutron
  pairing energy}, Phys. Rev. C 56~(5) (1997) 2549--2556.
\newblock \href {http://dx.doi.org/10.1103/PhysRevC.56.2549}
  {\path{doi:10.1103/PhysRevC.56.2549}}.

\bibitem{Pederiva2004_NPA742-255}
F.~Pederiva, A.~Sarsa, K.~Schmidt, S.~Fantoni, {Auxiliary field diffusion Monte
  Carlo calculation of ground state properties of neutron drops}, Nucl. Phys. A
  742~(1-2) (2004) 255--268.
\newblock \href {http://dx.doi.org/10.1016/j.nuclphysa.2004.06.030}
  {\path{doi:10.1016/j.nuclphysa.2004.06.030}}.

\bibitem{Gandolfi2011_PRL106-012501}
S.~Gandolfi, J.~Carlson, S.~C. Pieper, {Cold Neutrons Trapped in External
  Fields}, Phys. Rev. Lett. 106~(1) (2011) 012501.
\newblock \href {http://dx.doi.org/10.1103/PhysRevLett.106.012501}
  {\path{doi:10.1103/PhysRevLett.106.012501}}.

\bibitem{Tews2016_PRC93-024305}
I.~Tews, S.~Gandolfi, A.~Gezerlis, A.~Schwenk, {Quantum Monte Carlo
  calculations of neutron matter with chiral three-body forces}, Phys. Rev. C
  93~(2) (2015) 024305.
\newblock \href {http://dx.doi.org/10.1103/PhysRevC.93.024305}
  {\path{doi:10.1103/PhysRevC.93.024305}}.

\bibitem{Bonnard2018_PRC98-034319}
J.~Bonnard, M.~Grasso, D.~Lacroix, {Energy-density functionals inspired by
  effective-field theories: Applications to neutron drops}, Phys. Rev. C 98
  (2018) 034319.
\newblock \href {http://dx.doi.org/10.1103/PhysRevC.98.034319}
  {\path{doi:10.1103/PhysRevC.98.034319}}.

\bibitem{Li1992_PRC45-2782}
G.~Q. Li, R.~Machleidt, R.~Brockmann, {Properties of dense nuclear and neutron
  matter with relativistic nucleon-nucleon interactions}, Phys. Rev. C 45~(6)
  (1992) 2782--2794.
\newblock \href {http://dx.doi.org/10.1103/PhysRevC.45.2782}
  {\path{doi:10.1103/PhysRevC.45.2782}}.

\bibitem{Shirokov2007_PLB644-33}
A.~Shirokov, J.~Vary, A.~Mazur, T.~Weber, {Realistic nuclear Hamiltonian: Ab
  exitu approach}, Phys. Lett. B 644~(1) (2007) 33--37.
\newblock \href {http://dx.doi.org/10.1016/j.physletb.2006.10.066}
  {\path{doi:10.1016/j.physletb.2006.10.066}}.

\bibitem{Maris2009_PRC79-014308}
P.~Maris, J.~P. Vary, A.~M. Shirokov, {Ab initio no-core full configuration
  calculations of light nuclei}, Phys. Rev. C 79~(1) (2009) 014308.
\newblock \href {http://dx.doi.org/10.1103/PhysRevC.79.014308}
  {\path{doi:10.1103/PhysRevC.79.014308}}.

\bibitem{Sarsa2003_PRC68-024308}
A.~Sarsa, S.~Fantoni, K.~E. Schmidt, F.~Pederiva, {Neutron matter at zero
  temperature with an auxiliary field diffusion Monte Carlo method}, Phys. Rev.
  C 68~(2) (2003) 024308.
\newblock \href {http://dx.doi.org/10.1103/PhysRevC.68.024308}
  {\path{doi:10.1103/PhysRevC.68.024308}}.

\bibitem{Bartel1982_NPA386-79}
J.~Bartel, P.~Quentin, M.~Brack, C.~Guet, H.~B. H{\aa}kansson, {Towards a
  better parametrisation of Skyrme-like effective forces: A critical study of
  the SkM force}, Nucl. Phys. A 386~(1) (1982) 79--100.
\newblock \href {http://dx.doi.org/10.1016/0375-9474(82)90403-1}
  {\path{doi:10.1016/0375-9474(82)90403-1}}.

\bibitem{Kortelainen2010_PRC82-024313}
M.~Kortelainen, T.~Lesinski, J.~Mor{\'{e}}, W.~Nazarewicz, J.~Sarich,
  N.~Schunck, M.~V. Stoitsov, S.~Wild, {Nuclear energy density optimization},
  Phys. Rev. C 82~(2) (2010) 024313.
\newblock \href {http://dx.doi.org/10.1103/PhysRevC.82.024313}
  {\path{doi:10.1103/PhysRevC.82.024313}}.

\bibitem{Yang2016_PRC94-031301}
C.~J. Yang, M.~Grasso, D.~Lacroix, {From dilute matter to the equilibrium point
  in the energy-density-functional theory}, Phys. Rev. C 94~(3) (2016) 031301.
\newblock \href {http://dx.doi.org/10.1103/PhysRevC.94.031301}
  {\path{doi:10.1103/PhysRevC.94.031301}}.

\bibitem{Papakonstantinou2018_PRC97-014312}
P.~Papakonstantinou, T.~S. Park, Y.~Lim, C.~H. Hyun, {Density dependence of the
  nuclear energy-density functional}, Phys. Rev. C 97~(1) (2018) 014312.
\newblock \href {http://dx.doi.org/10.1103/PhysRevC.97.014312}
  {\path{doi:10.1103/PhysRevC.97.014312}}.

\bibitem{Grasso2017_PRC95-054327}
M.~Grasso, D.~Lacroix, C.~J. Yang, {Lee-Yang--inspired functional with a
  density-dependent neutron-neutron scattering length}, Phys. Rev. C 95~(5)
  (2017) 054327.
\newblock \href {http://dx.doi.org/10.1103/PhysRevC.95.054327}
  {\path{doi:10.1103/PhysRevC.95.054327}}.

\bibitem{Alonso2003_PRC67-054301}
D.~Alonso, F.~Sammarruca, {Microscopic calculations in asymmetric nuclear
  matter}, Phys. Rev. C 67~(5) (2003) 054301.
\newblock \href {http://dx.doi.org/10.1103/PhysRevC.67.054301}
  {\path{doi:10.1103/PhysRevC.67.054301}}.

\bibitem{Katayama2013_PRC88-035805}
T.~Katayama, K.~Saito, {Properties of dense, asymmetric nuclear matter in
  Dirac-Brueckner-Hartree-Fock approach}, Phys. Rev. C 88~(3) (2013) 035805.
\newblock \href {http://dx.doi.org/10.1103/PhysRevC.88.035805}
  {\path{doi:10.1103/PhysRevC.88.035805}}.

\bibitem{RocaMaza2011_PRL106-252501}
X.~Roca-Maza, M.~Centelles, X.~Vi{\~{n}}as, M.~Warda, {Neutron Skin of
  $^{208}$Pb, Nuclear Symmetry Energy, and the Parity Radius Experiment}, Phys.
  Rev. Lett. 106~(25) (2011) 252501.
\newblock \href {http://dx.doi.org/10.1103/PhysRevLett.106.252501}
  {\path{doi:10.1103/PhysRevLett.106.252501}}.

\bibitem{Birkhan2017_PRL118-252501}
J.~Birkhan, M.~Miorelli, S.~Bacca, S.~Bassauer, C.~A. Bertulani, G.~Hagen,
  H.~Matsubara, P.~von Neumann-Cosel, T.~Papenbrock, N.~Pietralla, V.~Y.
  Ponomarev, A.~Richter, A.~Schwenk, A.~Tamii, {Electric Dipole Polarizability
  of $^{48}$Ca and Implications for the Neutron Skin}, Phys. Rev. Lett.
  118~(25) (2017) 252501.
\newblock \href {http://dx.doi.org/10.1103/PhysRevLett.118.252501}
  {\path{doi:10.1103/PhysRevLett.118.252501}}.

\bibitem{Hagen2015_NP12-186}
G.~Hagen, A.~Ekstr{\"{o}}m, C.~Forss{\'{e}}n, G.~R. Jansen, W.~Nazarewicz,
  T.~Papenbrock, K.~A. Wendt, S.~Bacca, N.~Barnea, B.~Carlsson, C.~Drischler,
  K.~Hebeler, M.~Hjorth-Jensen, M.~Miorelli, G.~Orlandini, A.~Schwenk,
  J.~Simonis, {Neutron and weak-charge distributions of the 48Ca nucleus}, Nat.
  Phys. 12~(2) (2015) 186--190.
\newblock \href {http://dx.doi.org/10.1038/nphys3529}
  {\path{doi:10.1038/nphys3529}}.

\bibitem{Klos2007_PRC76-014311}
B.~K{\l}os, A.~Trzci{\'{n}}ska, J.~Jastrz{\c{e}}bski, T.~Czosnyka,
  M.~Kisieli{\'{n}}ski, P.~Lubi{\'{n}}ski, P.~Napiorkowski,
  L.~Pie{\'{n}}kowski, F.~J. Hartmann, B.~Ketzer, P.~Ring, R.~Schmidt, T.~von
  Egidy, R.~Smola{\'{n}}czuk, S.~Wycech, K.~Gulda, W.~Kurcewicz, E.~Widmann,
  B.~A. Brown, {Neutron density distributions from antiprotonic $^{208}$Pb and
  $^{209}$Bi atoms}, Phys. Rev. C 76~(1) (2007) 014311.
\newblock \href {http://dx.doi.org/10.1103/PhysRevC.76.014311}
  {\path{doi:10.1103/PhysRevC.76.014311}}.

\bibitem{Tarbert2014_PRL112-242502}
C.~M. Tarbert, D.~P. Watts, D.~I. Glazier, P.~Aguar, J.~Ahrens, J.~R.~M.
  Annand, H.~J. Arends, R.~Beck, V.~Bekrenev, B.~Boillat, A.~Braghieri,
  D.~Branford, W.~J. Briscoe, J.~Brudvik, S.~Cherepnya, R.~Codling, E.~J.
  Downie, K.~Foehl, P.~Grabmayr, R.~Gregor, E.~Heid, D.~Hornidge, O.~Jahn,
  V.~L. Kashevarov, A.~Knezevic, R.~Kondratiev, M.~Korolija, M.~Kotulla,
  D.~Krambrich, B.~Krusche, M.~Lang, V.~Lisin, K.~Livingston, S.~Lugert,
  I.~J.~D. MacGregor, D.~M. Manley, M.~Martinez, J.~C. McGeorge, D.~Mekterovic,
  V.~Metag, B.~M.~K. Nefkens, A.~Nikolaev, R.~Novotny, R.~O. Owens, P.~Pedroni,
  A.~Polonski, S.~N. Prakhov, J.~W. Price, G.~Rosner, M.~Rost, T.~Rostomyan,
  S.~Schadmand, S.~Schumann, D.~Sober, A.~Starostin, I.~Supek, A.~Thomas,
  M.~Unverzagt, T.~Walcher, L.~Zana, F.~Zehr, {Neutron Skin of $^{208}$Pb from
  Coherent Pion Photoproduction}, Phys. Rev. Lett. 112~(24) (2014) 242502.
\newblock \href {http://dx.doi.org/10.1103/PhysRevLett.112.242502}
  {\path{doi:10.1103/PhysRevLett.112.242502}}.

\bibitem{RocaMaza2013_PRC88-024316}
X.~Roca-Maza, M.~Brenna, G.~Col{\`{o}}, M.~Centelles, X.~Vi{\~{n}}as, B.~K.
  Agrawal, N.~Paar, D.~Vretenar, J.~Piekarewicz, {Electric dipole
  polarizability in $^{208}$Pb: Insights from the droplet model}, Phys. Rev. C
  88~(2) (2013) 024316.
\newblock \href {http://dx.doi.org/10.1103/PhysRevC.88.024316}
  {\path{doi:10.1103/PhysRevC.88.024316}}.

\bibitem{Lynn2016_PRL116-062501}
J.~E. Lynn, I.~Tews, J.~Carlson, S.~Gandolfi, A.~Gezerlis, K.~E. Schmidt,
  A.~Schwenk, {Chiral Three-Nucleon Interactions in Light Nuclei,
  Neutron-$\alpha$ Scattering, and Neutron Matter}, Phys. Rev. Lett. 116~(6)
  (2016) 062501.
\newblock \href {http://dx.doi.org/10.1103/PhysRevLett.116.062501}
  {\path{doi:10.1103/PhysRevLett.116.062501}}.

\bibitem{Lalazissis2009_PRC80-041301}
G.~A. Lalazissis, S.~Karatzikos, M.~Serra, T.~Otsuka, P.~Ring, {Covariant
  density functional theory: The role of the pion}, Phys. Rev. C 80~(4) (2009)
  041301.
\newblock \href {http://dx.doi.org/10.1103/PhysRevC.80.041301}
  {\path{doi:10.1103/PhysRevC.80.041301}}.

\bibitem{Tarpanov2008_PRC77-054316}
D.~Tarpanov, H.~Liang, N.~Van~Giai, C.~Stoyanov, Mean-field study of
  single-particle spectra evolution in {$Z = 14$} and {$N = 28$} chains, Phys.
  Rev. C 77 (2008) 054316.
\newblock \href {http://dx.doi.org/10.1103/PhysRevC.77.054316}
  {\path{doi:10.1103/PhysRevC.77.054316}}.

\bibitem{Moreno-Torres2010_PRC81-064327}
M.~Moreno-Torres, M.~Grasso, H.~Liang, V.~De~Donno, M.~Anguiano, N.~Van~Giai,
  Tensor effects in shell evolution at {$Z,N=8, 20$}, and $28$ using
  nonrelativistic and relativistic mean-field theory, Phys. Rev. C 81 (2010)
  064327.
\newblock \href {http://dx.doi.org/10.1103/PhysRevC.81.064327}
  {\path{doi:10.1103/PhysRevC.81.064327}}.

\bibitem{Marcos2013_PAN76-562}
S.~Marcos, M.~L{\'{o}}pez-Quelle, R.~Niembro, L.~N. Savushkin, {Nuclear
  relativistic Hartree-Fock approximation with weakened pion tensor force},
  Phys. At. Nucl. 76~(5) (2013) 562--576.
\newblock \href {http://dx.doi.org/10.1134/S1063778813050074}
  {\path{doi:10.1134/S1063778813050074}}.

\bibitem{Marcos2014_PAN77-299}
S.~Marcos, M.~L{\'{o}}pez-Quelle, R.~Niembro, L.~N. Savushkin, {Pion tensor
  force and nuclear binding energy in the relativistic Hartree-Fock formalism},
  Phys. At. Nucl. 77~(3) (2014) 299--309.
\newblock \href {http://dx.doi.org/10.1134/S1063778814020136}
  {\path{doi:10.1134/S1063778814020136}}.

\bibitem{Afanasjev2015_PRC92-044317}
A.~V. Afanasjev, E.~Litvinova, {Impact of collective vibrations on
  quasiparticle states of open-shell odd-mass nuclei and possible interference
  with the tensor force}, Phys. Rev. C 92~(4) (2015) 044317.
\newblock \href {http://dx.doi.org/10.1103/PhysRevC.92.044317}
  {\path{doi:10.1103/PhysRevC.92.044317}}.

\bibitem{Li2016_PLB753-97}
J.~J. Li, J.~Margueron, W.~H. Long, N.~{Van Giai}, {Magicity of neutron-rich
  nuclei within relativistic self-consistent approaches}, Phys. Lett. B 753
  (2016) 97--102.
\newblock \href {http://dx.doi.org/10.1016/j.physletb.2015.12.004}
  {\path{doi:10.1016/j.physletb.2015.12.004}}.

\bibitem{Karakatsanis2017_PRC95-034318}
K.~Karakatsanis, G.~A. Lalazissis, P.~Ring, E.~Litvinova, {Spin-orbit
  splittings of neutron states in $N=20$ isotones from covariant density
  functionals and their extensions}, Phys. Rev. C 95~(3) (2017) 034318.
\newblock \href {http://dx.doi.org/10.1103/PhysRevC.95.034318}
  {\path{doi:10.1103/PhysRevC.95.034318}}.

\bibitem{Otsuka2006_PRL97-162501}
T.~Otsuka, T.~Matsuo, D.~Abe, Mean field with tensor force and shell structure
  of exotic nuclei, Phys. Rev. Lett. 97 (2006) 162501.
\newblock \href {http://dx.doi.org/10.1103/PhysRevLett.97.162501}
  {\path{doi:10.1103/PhysRevLett.97.162501}}.

\bibitem{Colo2007_PLB646-227}
G.~Col\`o, H.~Sagawa, S.~Fracasso, P.~F. Bortignon, Spin-orbit splitting and
  the tensor component of the {Skyrme} interaction, Phys. Lett. B 646 (2007)
  227--231.
\newblock \href {http://dx.doi.org/10.1016/j.physletb.2007.01.033}
  {\path{doi:10.1016/j.physletb.2007.01.033}}.

\bibitem{Litvinova2007_PRC75-064308}
E.~Litvinova, P.~Ring, V.~I. Tselyaev, Particle-vibration coupling within
  covariant density function theory., Phys. Rev. C 75 (2007) 064308.
\newblock \href {http://dx.doi.org/10.1103/PhysRevC.75.064308}
  {\path{doi:10.1103/PhysRevC.75.064308}}.

\bibitem{Wang2018PhD}
Z.~Wang, Analysis of nuclear tensor-force effects in covariant density
  functional theory, Ph.D. thesis, Lanzhou University (2018).

\bibitem{Prokofev1998_PRL81-2514}
N.~V. Prokof'ev, B.~V. Svistunov, {Polaron Problem by Diagrammatic Quantum
  Monte Carlo}, Phys. Rev. Lett. 81~(12) (1998) 2514--2517.
\newblock \href {http://dx.doi.org/10.1103/PhysRevLett.81.2514}
  {\path{doi:10.1103/PhysRevLett.81.2514}}.

\bibitem{Prokofev2007_PRL99-250201}
N.~Prokof'ev, B.~Svistunov, {Bold Diagrammatic Monte Carlo Technique: When the
  Sign Problem Is Welcome}, Phys. Rev. Lett. 99~(25) (2007) 250201.
\newblock \href {http://dx.doi.org/10.1103/PhysRevLett.99.250201}
  {\path{doi:10.1103/PhysRevLett.99.250201}}.

\end{thebibliography}


%
%
%
\end{document}